\documentclass[nofootinbib, aps, pra, twocolumn,  longbibliography]{revtex4-1}
\usepackage{amsmath}
\usepackage{amssymb}
\usepackage{graphicx}
\usepackage[usenames,dvipsnames]{xcolor}
\usepackage{tikz}
\usepackage{pgffor}
\usepackage{verbatim}
\usepackage{bbm}
\usepackage{float}
\usepackage{mathrsfs}
\usepackage[colorlinks, breaklinks=true,linkcolor=red, citecolor=blue, linktocpage=true]{hyperref}

\newcommand{\<}{\langle}
\renewcommand{\>}{\rangle}

\newcommand{\mac}{\mathcal}

\begin{document}

\title{Generalized string-net models: A thorough exposition}

\author{Chien-Hung Lin$^1$}
\author{Michael Levin$^2$}
\author{Fiona J. Burnell$^{1,3}$}
 \affiliation{$^1$ School of Physics and Astronomy, University of Minnesota, Minneapolis, Minnesota 55455, USA}
\affiliation{$^2$ James Franck Institute and Department of Physics, University of Chicago, Chicago, Illinois 60637, USA}
\affiliation{$^3$ Institute for Advanced Study, Princeton, New Jersey 08540, USA}

\begin{abstract}
We describe how to construct generalized string-net models, a class of exactly solvable lattice models that realize a large family of 2D topologically ordered phases of matter. The ground states of these models can be thought of as superpositions of different ``string-net configurations'', where each string-net configuration is a trivalent graph with labeled edges, drawn in the $xy$ plane. What makes this construction more general than the original string-net construction is that, unlike the original construction, tetrahedral reflection symmetry is not assumed, nor is it assumed that the ground state wave function $\Phi$ is ``isotropic'': i.e. in the generalized setup, two string-net configurations $X_1, X_2$ that can be continuously deformed into one another can have different ground state amplitudes, $\Phi(X_1) \neq \Phi(X_2)$. As a result, generalized string-net models can realize topological phases that are inaccessible to the original construction. In this paper, we provide a more detailed discussion of ground state wave functions, Hamiltonians, and minimal self-consistency conditions for generalized string-net models than what exists in the previous literature. We also show how to construct string operators that create anyon excitations in these models, and we show how to compute the braiding statistics of these excitations. Finally, we derive necessary and sufficient conditions for generalized string-net models to have isotropic ground state wave functions on the plane or the sphere -- a property that may be useful in some applications.
\end{abstract}

\maketitle
\section{Introduction}

In recent decades, profound connections between the physics of strongly interacting 2-dimensional systems and the mathematics of unitary modular tensor categories (UMTCs) has emerged, through an increasingly well-developed understanding of  topologically ordered phases of matter.  This understanding has culminated in a comprehensive picture of how the defining properties of topological order, such as topological ground state degeneracies and non-trivial braiding statistics of the low-energy point-like excitations (or anyons), are mathematically described by the theory of UMTCs\cite{KitaevHoneycomb,FrolichTQFT,WangBook,BondersonThesis,Bakalov,Kong2017}.  This connection has enabled a very complete understanding of the mathematical structure of 2D topologically ordered phases, and fostered recent developments in our understanding of the interplay between symmetry and topology both in 2D \cite{LongQ, teo2015theory, tarantino2016symmetry, LanKongWen} and 3D \cite{SenthilVishwanath,CBVF} interacting systems. 

In studying the properties of topologically ordered phases of matter, exactly soluble lattice models realizing these phases\cite{KitaevToric,KitaevHoneycomb,LevinWenstrnet,HuWanWu12, WalkerWang} have been an indispensable tool, playing a role analogous to that of tight-binding Hamiltonians in the study of Fermi liquids. A particularly useful class of models, known as string-net models, were introduced by Ref.~\onlinecite{LevinWenstrnet}. The string-net construction describes the low energy physics of a 2D topologically ordered phase through the dynamics of networks of 1-dimensional, or string-like, objects. Though the resulting Hamiltonians are quite different from the low-energy Hamiltonians typically studied in the context of real materials, they are a powerful theoretical tool as they provide a  systematic way of constructing exactly soluble lattice Hamiltonians realizing a large class of (bosonic) topological orders.

Following the original string-net construction, several works\cite{KitaevKong,LanWen13,LinLevinstrnet, LakeWu, HahnWolf}  introduced ``generalized" string-net models, capable of realizing additional topological orders beyond those of Ref.~\onlinecite{LevinWenstrnet}.  These generalizations allow string-net models to realize any topological order associated with the Drinfeld center of a fusion category \cite{KitaevKong}, which are believed to be the most general class of (bosonic) topological orders compatible with gapped boundaries \cite{KitaevKong,KongCondensation,LinLevinstrnet,freed2020gapped}.  Since such gapped boundaries are a generic feature of string-net Hamiltonians, these generalized string-net models therefore comprise the most general possible construction of this type.  Examples of topological orders that can be realized by these models include (i) discrete gauge theories \cite{KitaevToric}; (ii) Dijkgraaf-Witten theories \cite{HuWanWu12}; and ``doubled'' topological orders of the form $\mathcal{T} \times \mathcal{T}^{op}$ where $\mathcal{T}$ are $\mathcal{T}^{op}$ are two topological orders related by time-reversal.

One shortcoming of the previous literature on generalized string-net models is that it has mostly focused on higher level properties of the models, such as their excitations and boundaries, while omitting a detailed discussion of the models themselves. For example, while Ref.~\onlinecite{KitaevKong} (see also \cite{Kong2012}) and Ref.~\onlinecite{LanWen13} sketched the construction of general string-net models, both papers primarily focused on understanding the dictionary between generalized string-net models and the theory of unitary fusion categories, as well as the systematic construction of excitations and gapped boundaries. A more detailed discussion of generalized string-net models was given in Ref.~\onlinecite{LinLevinstrnet}, but this discussion was restricted to \emph{Abelian} string-net models. More recently, Ref.~\onlinecite{HahnWolf} worked out the explicit form of the Hamiltonians and ground state wave functions of generalized non-Abelian string-net models but did not obtain general expressions for the string operators that create the anyon excitations in these models.

In this paper, we fill in this gap by providing a concrete and detailed discussion of all the basic aspects of generalized string-net models in the general non-Abelian case, including ground state wave functions, Hamiltonians, and string operators for these models. We also derive necessary and sufficient conditions for string-net models to have ``isotropic'' (i.e. topologically invariant) ground state wave functions on the plane or the sphere -- a property that may be useful in some applications. An important feature of our approach is that we derive all the properties of these models using simple algebraic calculations that do not require any knowledge of tensor category formalism. We also discuss the relationship between generalized string-net models and the original string-net construction of Ref.~\onlinecite{LevinWenstrnet} and we show that the original string-net models correspond to a subset of the models discussed in this paper.  We believe that our more explicit discussion may be useful in situations where exactly solvable models are a primary tool for studying properties of the associated topological phases, such as how topological order interplays with symmetry \cite{Heinrich16,ChengSET}, or the possible phase transitions into and out of these states \cite{SlingerlandBais,TSBShort,TSBLong,Schulz16}.

To carry out our construction, we adopt a philosophy similar to that of the original string-net construction of Ref.~\onlinecite{LevinWenstrnet}. Namely, we first specify the ground state for our model, and then show how to construct an exactly solvable parent Hamiltonian for this ground state. To define our ground states, we begin by defining a ``string-net" as a trivalent labeled graph, where the labels must satisfy certain conditions (or branching rules) at each trivalent vertex.  Next, we specify a set of relations, expressed in terms of a choice of parameters that we will call $F$, $\tilde{F}$, and $Y$, that fix the relative coefficients of different string-net configurations in our model's ground state.  
By requiring that our relations fix the amplitudes of the ground-state wave function in a consistent manner, we then obtain a set of consistency conditions that the parameters $F$, $\tilde{F}$, and $Y$ must satisfy.  These consistency conditions turn out to force us to choose our parameters to be associated with a unitary pivotal fusion category $\mac{F}$. Thus, we \emph{derive} the mathematical structure of unitary fusion categories, rather than assuming it from the start.


The key difference between our generalized string-net models and the original string-net construction is that our models relax certain restrictions that  Ref.~\onlinecite{LevinWenstrnet} imposed on the string-net data.  As a consequence, our string-nets may not be \emph{isotropic} in the sense that two string-net configurations $X_1,X_2$ which can be continuously deformed into one another may not have the same ground state amplitude: $\Phi(X_1)\neq \Phi(X_2)$.   We show that some non-isotropic string-net states can be transformed to isotropic ones via local unitary (gauge) transformations of the string-net data.  However, this is not always the case, and we identify some obstructions to obtaining fully isotropic string-net ground states.
Finally, we identify extra conditions on the input data $F$, $\tilde{F}$, and $Y$ required  to ensure that our string-nets are isotropic.  We show that these conditions, together with a tetrahedral reflection symmetry condition, produce string-nets that are equivalent to those of Ref.~\onlinecite{LevinWenstrnet}.

One notable consequence of the lower symmetry that we require of our string-net states is that our generalized models can realize topological phases that break time reversal symmetry.   We illustrate this with several examples whose quasiparticle statistics are not time-reversal symmetric.

The paper is organized as follows. In Sec. \ref{sec:model}, we construct ground state wave functions for general string-net models.  In Sec. \ref{sec:h} we construct lattice Hamiltonians. We analyze the low energy quasiparticle excitations of these models in Sec. \ref{sec:string}.  In Sec. \ref{sec:iso}, we derive the additional constraints for isotropic string-net models.   We then discuss the relation between our construction and the models of Ref.~\onlinecite{LevinWenstrnet}  in Sec. \ref{oldmodels}. We illustrate our construction with concrete examples in Sec. \ref{sec:exp}.  Several technical details can be found in the appendices.

\section{String-net ground states \label{sec:model}}

Before discussing model Hamiltonians, we will first describe how to construct a class of ground-state wave functions which we will call generalized string-net ground states.  We require these wave functions to satisfy certain conditions which, as we will show in the later sections of this paper, ensure the following properties.  First, they are ground states of exactly solvable lattice Hamiltonians that can be expressed as sums of commuting projectors.  Second, low-lying excitations of this Hamiltonian above the string-net ground states are anyons, and this Hamiltonian describes a zero-correlation length fixed point of a topological phase.

Our string-net ground states are similar to those of Levin and Wen \cite{LevinWenstrnet}, but with several important differences.  Both constructions lead to liquid-like ground states expressed as superpositions over many different labeled trivalent graphs (i.e. string-nets).  Additionally, in both cases the string-net wave function is required to be invariant under certain transformations.  For example, our string-net wave function is scale invariant, in the sense that if two string-nets differ only by an overall scale, they appear in our string-net ground state with the same amplitude.  Unlike the string-nets of Ref.~\onlinecite{LevinWenstrnet}, however, we do not require our ground state to be invariant under arbitrary bendings of the string-nets, or under rotations or reflections. This is the sense in which our string-nets are ``generalized"; we will explore its implications in more detail below.

\subsection{String-net Hilbert space \label{section:hilbert}}

\begin{figure}[ptb]
\begin{center}
\includegraphics[width=0.7\columnwidth]{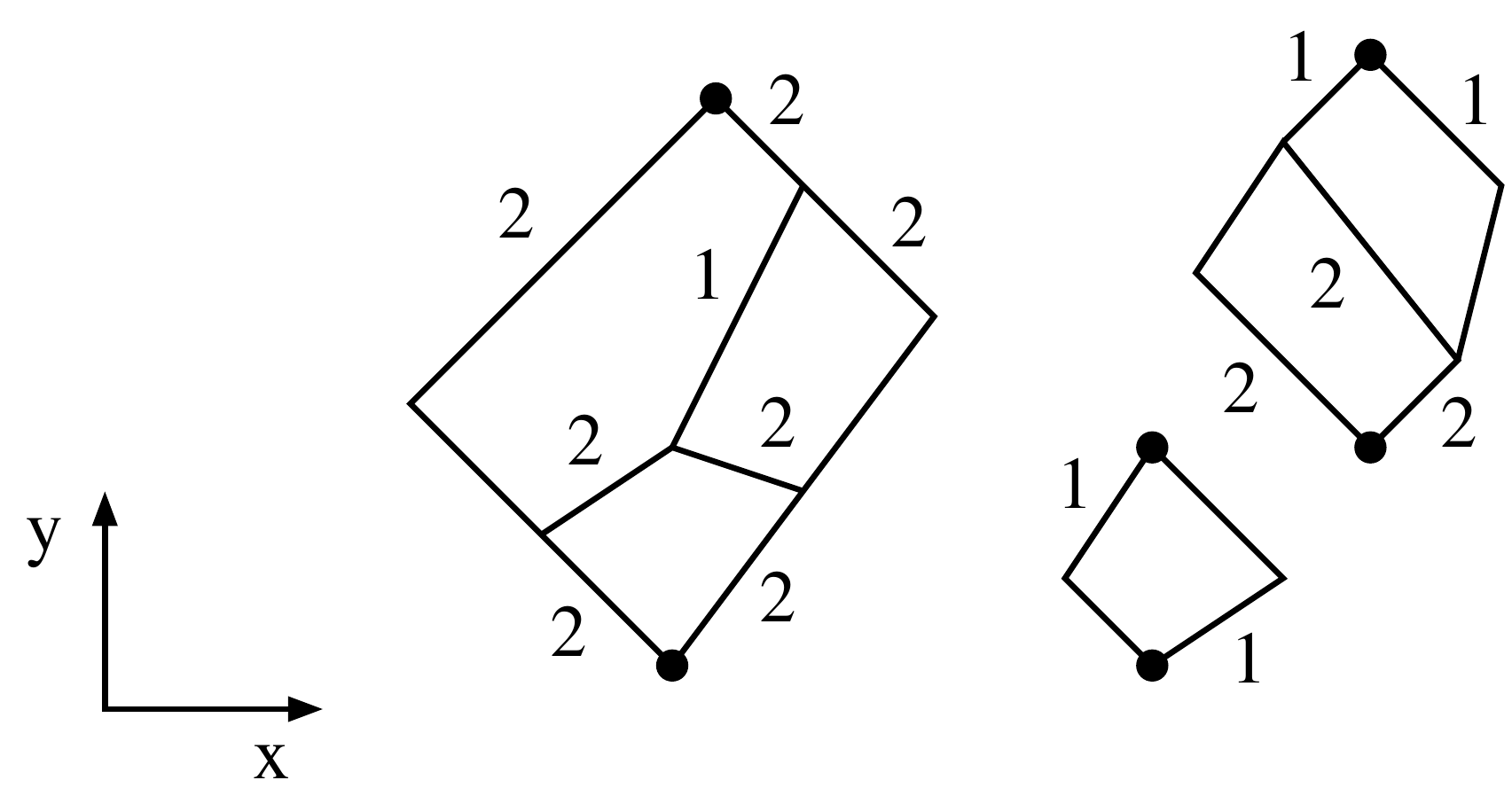}
\end{center}
\caption{A typical example of a string-net with string types $\{1,2\}$, with dual string types defined by $\bar{1} = 1$ and $\bar{2} = 2$, and branching rules $\{(1,2; 2), (2,1;2), (2,2;1), (2,2; 2)\}$. Bivalent vertices are marked with dots for clarity. Note that other (unmarked) corners are not bivalent vertices, but rather kinks in the piecewise differentiable strings.
} 
\label{fig:state0}
\end{figure}

A \emph{string-net} is a special type of planar graph with labeled edges and with vertices that are either bivalent or trivalent, i.e. of degree $2$ or degree $3$ (Fig.~\ref{fig:state0}). We will often refer to the edges that make up a string-net as ``strings'', and the fixed, finite set of edge labels $\{a,b,c,...\}$ as ``string types''. What makes a string-net different from an ordinary planar graph is that it satisfies the following additional properties. First, the strings/edges in a string-net are piecewise differentiable curves, drawn in the $xy$ plane. Second, when we traverse a string from one endpoint to the other, the tangent vector $\hat{v}$ has either a strictly positive or strictly negative $y$-component throughout the string, without any sign changes. Here, ``$y$'' denotes the vertical direction, so we will refer to this requirement as the ``no vertical bending'' property. One consequence of this property is that each string carries a natural orientation, which we always take to be in  the $+\hat{y}$ direction.

Third, every trivalent vertex is of one of the two types shown in Eq.~(\ref{vertex3}), i.e. with either one incoming and two outgoing strings or two incoming and one outgoing string with respect to the $\hat{y}$ direction. Similarly, every bivalent vertex is of one of the two types shown in Eq.~(\ref{vertex2}). 

The final property of string-nets is that only certain special combinations of strings (or edges) can meet at the vertices. In the case of the bivalent vertices, the allowed branchings are determined by an additional piece of data: an involution $a \rightarrow \bar{a}$ on the set of string types. We will refer to the string type $\bar{a}$ as the \emph{dual} of $a$. Once we fix this definition of dual string types, the allowed bivalent vertices are those of the form
\begin{equation}
	\raisebox{-0.22in}{\includegraphics[height=0.28in]{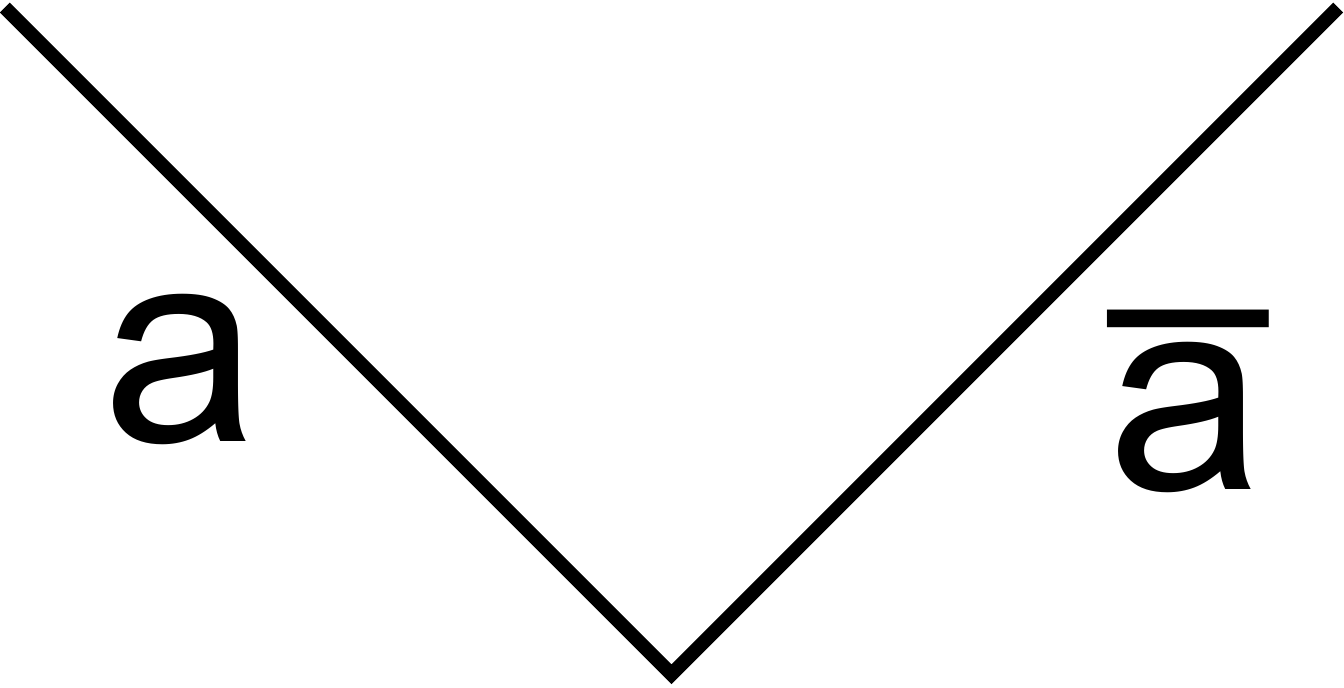}} \qquad
	\raisebox{-0.22in}{\includegraphics[height=0.28in]{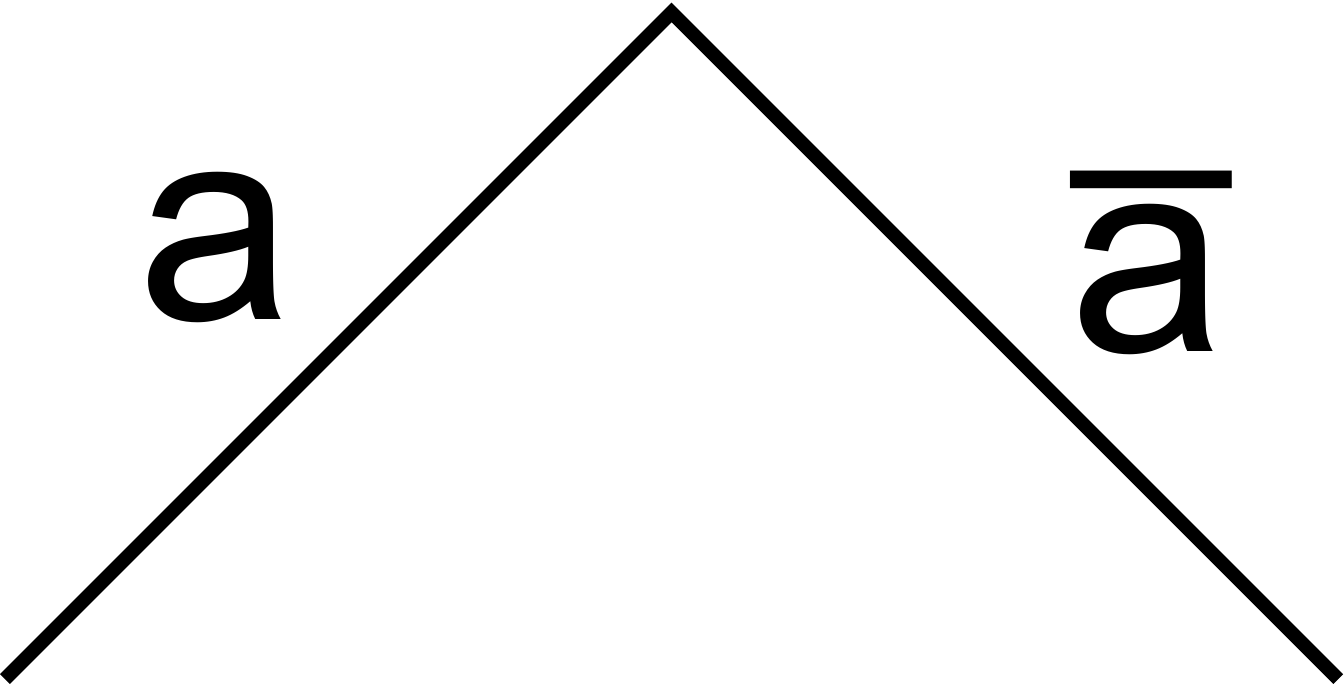}}.
	\label{vertex2}
\end{equation}

For the trivalent vertices, the allowed branchings are specified by a set of \emph{branching} (or fusion) rules --- a collection of (ordered) triplets $\{(a, b : c)\}$. The same branching rules apply to both ``upward'' and ``downward'' vertices:
\begin{equation}
	\raisebox{-0.22in}{\includegraphics[height=0.5in]{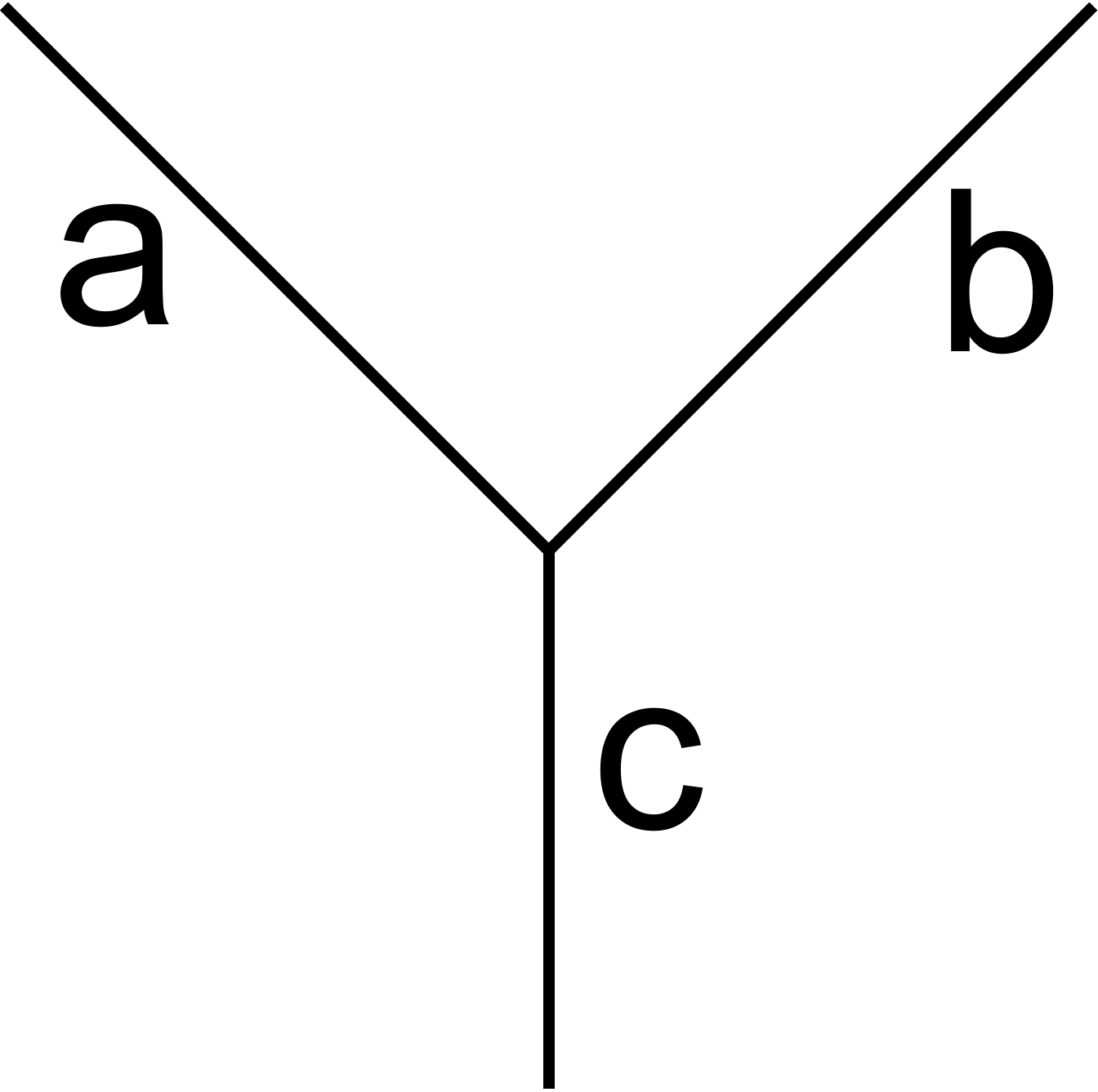}} \qquad
	\raisebox{-0.22in}{\includegraphics[height=0.5in]{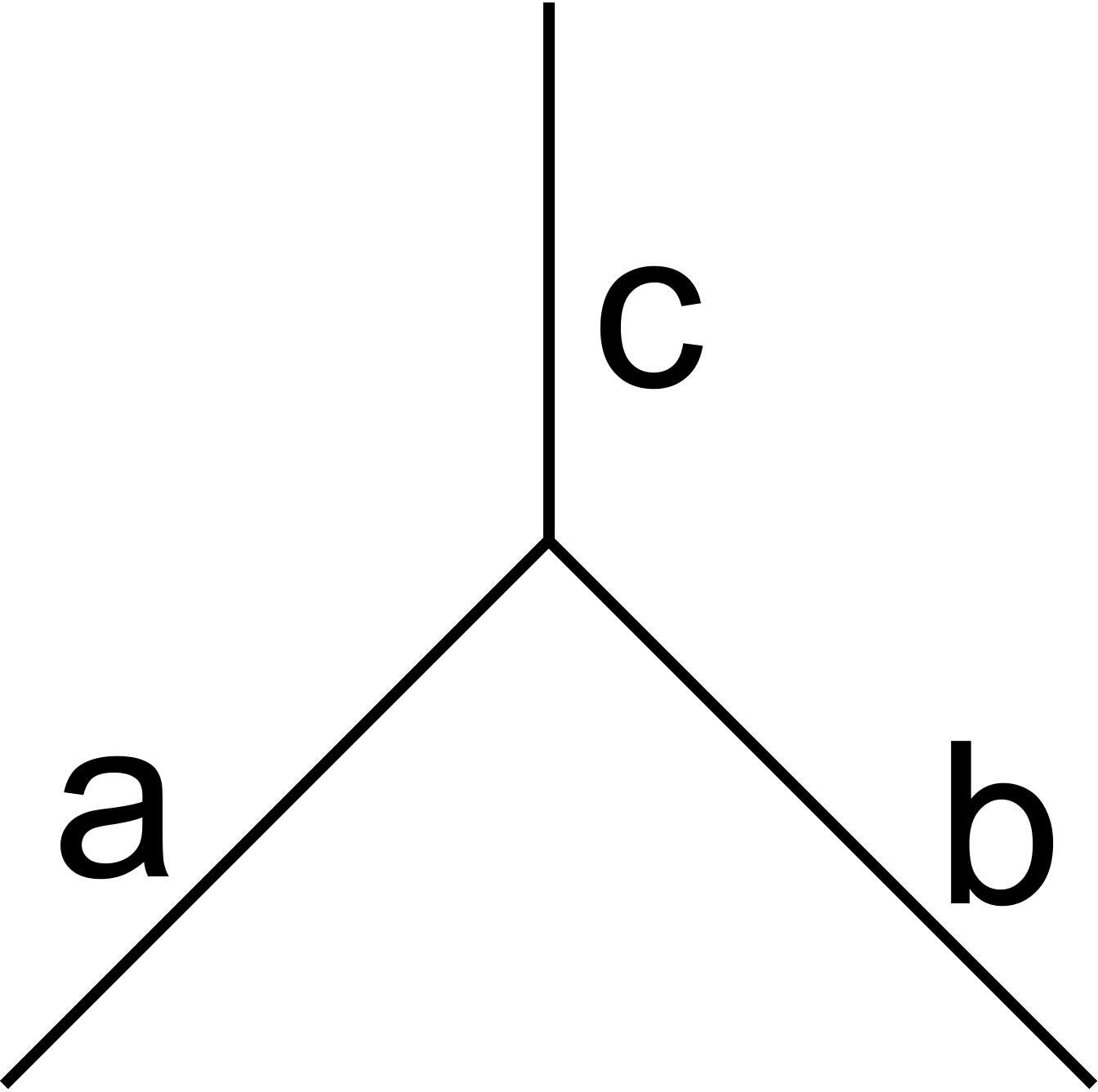}}.
	\label{vertex3}
\end{equation}

 The branching rules cannot be chosen arbitrarily: we will require that they obey the following associativity condition:
\begin{align}
\sum_e \delta^{ab}_e \delta^{ec}_d = \sum_f \delta^{bc}_f \delta^{af}_d
\label{fusionassoc}
\end{align}
where $\delta^{ab}_c$ is defined by
\begin{equation}
	\delta^{ab}_c = 
	\begin{cases}
		1, & \text{ if }(a,b:c) \text{ is allowed} \\
		0, & \text{ otherwise}.
	\end{cases}
	\label{delta}
\end{equation}
The motivation for (\ref{fusionassoc}) will become clear below when we define the $F$-symbol: we will see that Eq.~\ref{fusionassoc} guarantees that $(F^{abc}_d)_{ef}$ is a \emph{square} matrix. 

Note that the branching rules need not be symmetric with respect to $a,b$: $\delta^{ab}_c \neq \delta^{ba}_c$ in general. However, one can show\footnote{Cyclical symmetry follows from associativity (\ref{fusionassoc}) together with the branching rules for the null string (defined below).} that the branching rules are always \emph{cyclically} symmetric: $\delta^{ab}_c =\delta^{\bar{c}a}_{\bar{b}} = \delta^{b \bar{c}}_{\bar{a}}$.

Expert readers may notice that our definition of string-net does not allow for the possibility of {\it fusion multiplicity}, i.e. our string-nets have the property that there is a unique way to combine the labels $a$ and $b$ to obtain the label $c$. We focus on the unique fusion case throughout this paper for notational simplicity, but it is straightforward to generalize all of our constructions to string-nets with general fusion multiplicity. In the latter case, string-nets carry an additional label that lives at each vertex (see Appendix \ref{App:FusionMults} for details).  

To see an example of a string-net, consider a string-net model with two string types, $\{1,2\}$, with dual string types defined by $\bar{1} = 1$ and $\bar{2} = 2$, and branching rules given by $\{(1,2; 2), (2,1;2), (2,2;1), (2,2; 2)\}$. A typical example of a string-net with this data is shown in Fig.~\ref{fig:state0}.  Note that, unlike the original string-net construction of Ref.~\onlinecite{LevinWenstrnet}, we do not draw orientations on the strings: this is not necessary because we use the convention that every string is oriented in the \emph{upward} ($+\hat{y}$) direction so there is no need to explicitly show orientations in our figures.

At this point it is useful to introduce the notion of the \emph{null} string, which we will denote by $0$ or by a dashed line. Formally, the null string is a special string type with the property that (i) $\bar{0} = 0$ and (ii) the allowed branchings involving the null string are $\{(0,a :a), (a,0:a), (a,\bar{a}:0)\}$. More physically, the null string is equivalent to having no string at all: for any string-net, we can erase or add null strings wherever we want and it does not change the physical state. Thus, the null string can be thought of as an accounting trick for treating bivalent and trivalent vertices in a unified fashion.

We are now ready to define the string-net Hilbert space $\mathcal{H}$: an orthonormal basis for the string-net Hilbert space $\mathcal{H}$ is given by all possible string-net configurations which satisfy the branching rules and other conditions. Note that the spatial configuration of the string-net is important here: two string-nets that are geometrically distinct correspond to orthogonal states whether or not they are topologically equivalent.

\subsection{Ground state wave function}
The ground state $|\Phi\>=\sum_{X\in\mathcal{H}}\Phi(X)|X\>$ of our models is a superposition of different string-net configurations $|X \rangle$ in $\mathcal{H}$. The state $|\Phi\>$ is described implicitly by the following local constraint equations:
\begin{subequations}
\begin{align}
	\Phi\left(\raisebox{-0.22in}{\includegraphics[height=0.5in]{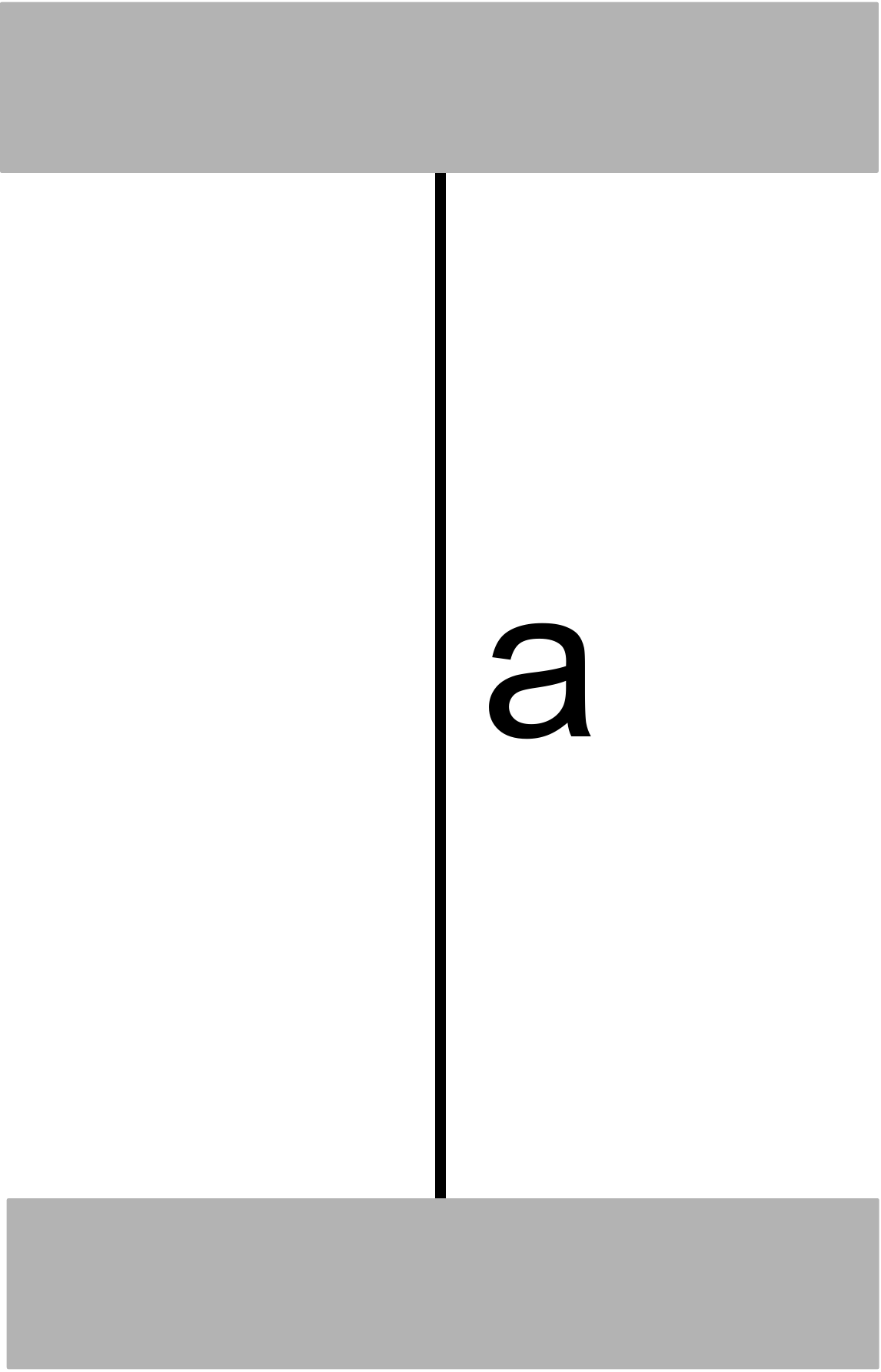}} \right)&=\Phi \left(\raisebox{-0.22in}{\includegraphics[height=0.5in]{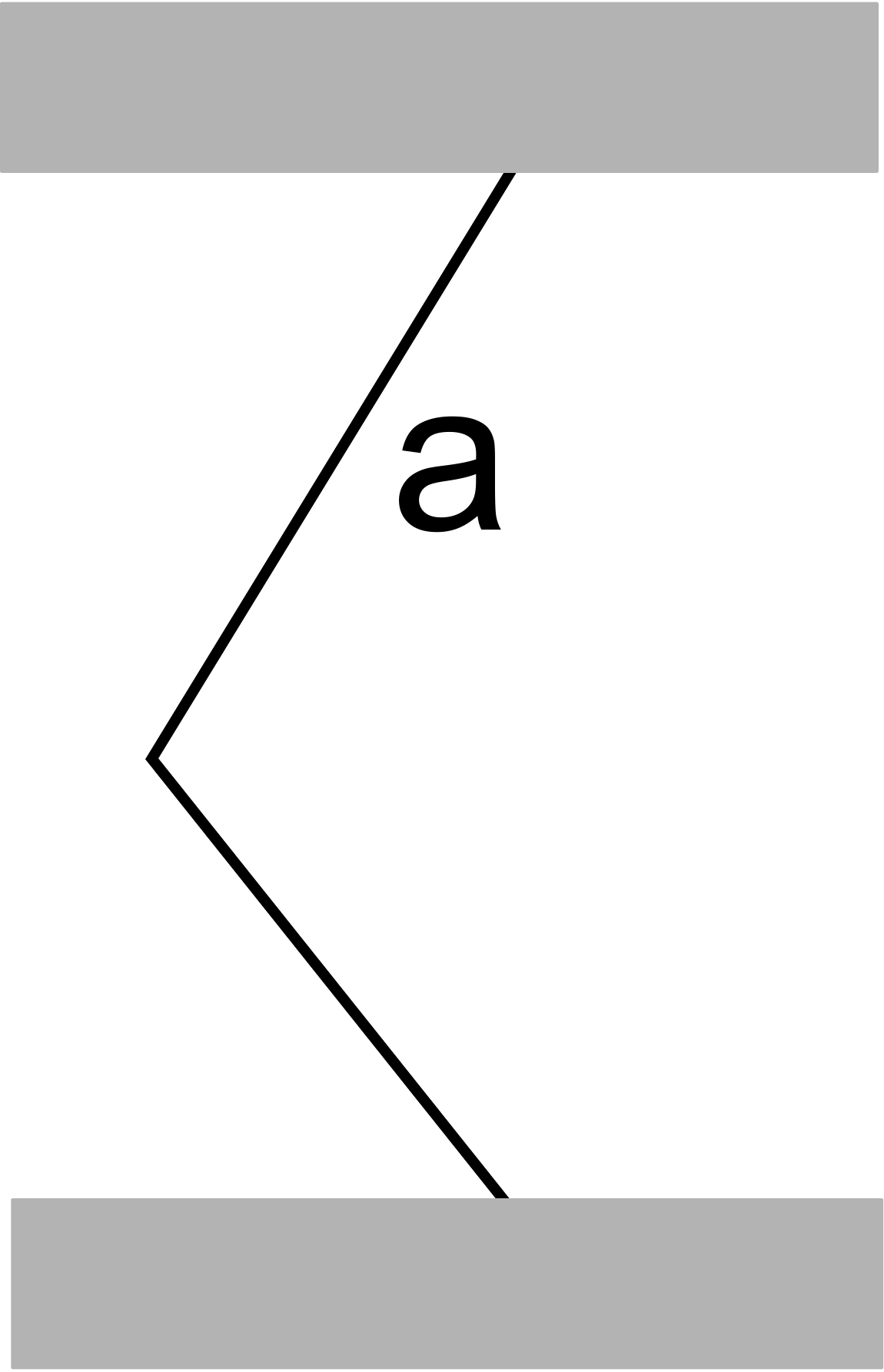}} \right) \label{1g}\\
	\Phi\left(\raisebox{-0.22in}{\includegraphics[height=0.5in]{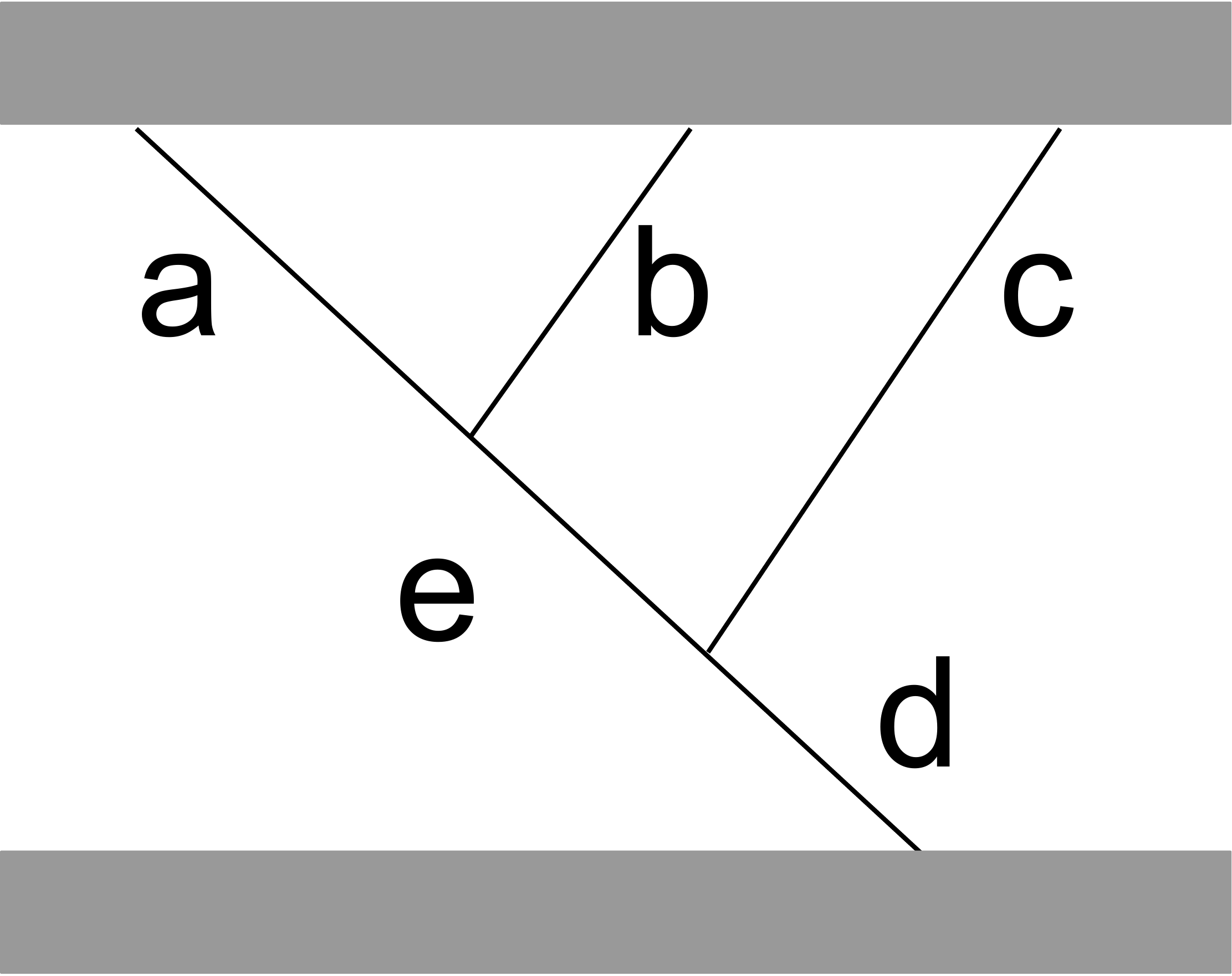}} \right)&=\sum_f F^{abc}_{def}\Phi \left(\raisebox{-0.22in}{\includegraphics[height=0.5in]{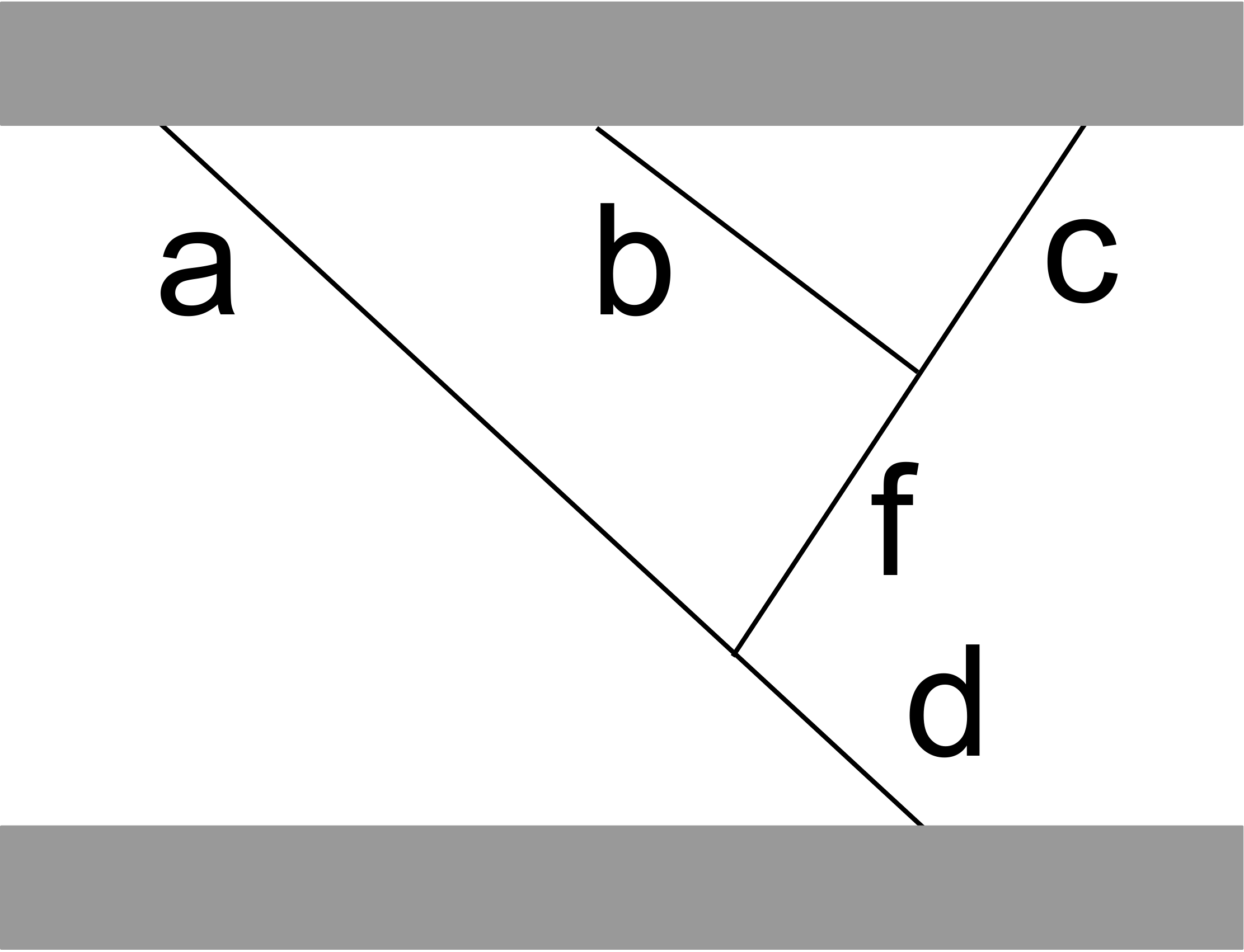}} \right) \label{1a}\\
	\Phi\left(\raisebox{-0.22in}{\includegraphics[height=0.5in]{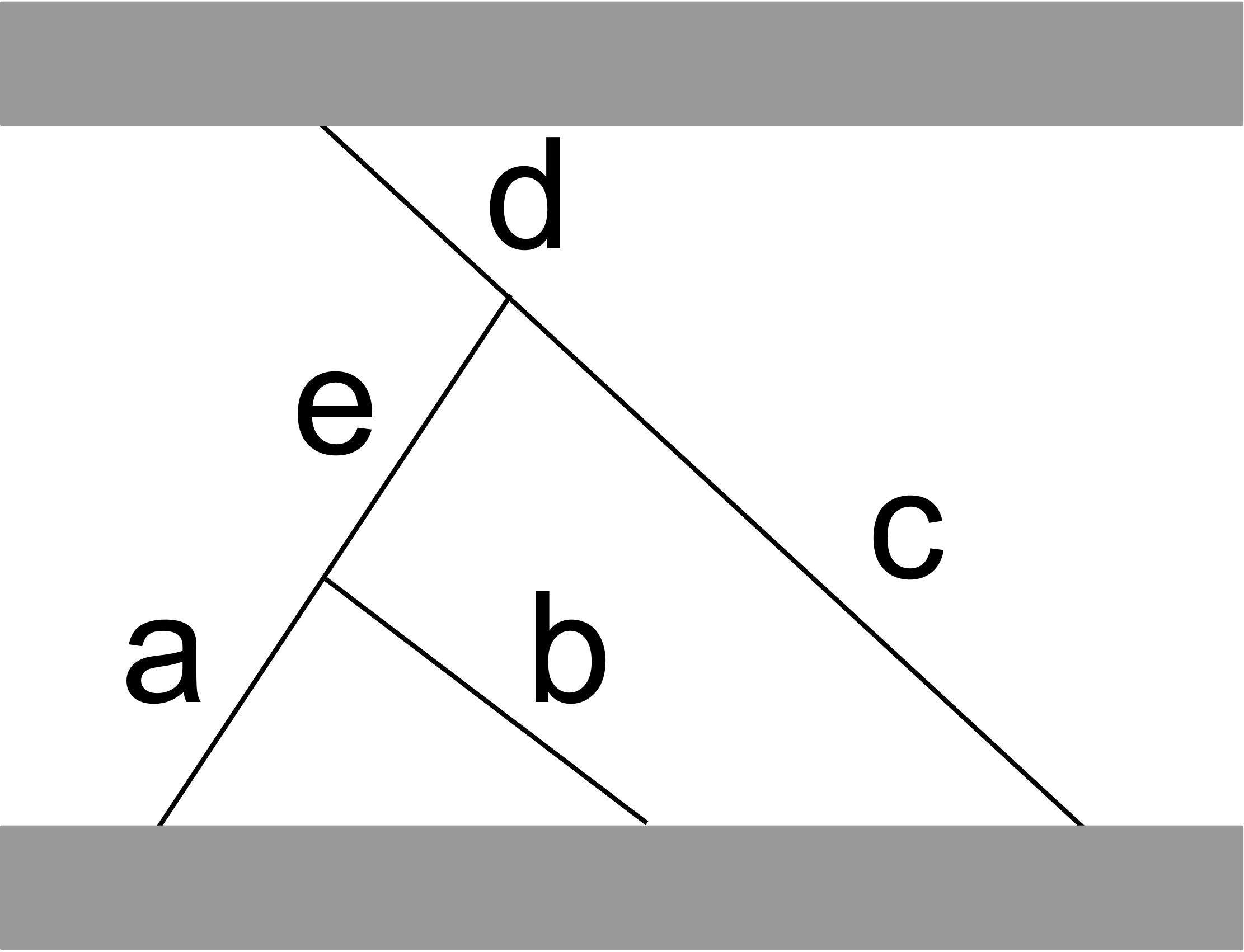}} \right)&=\sum_f \tilde{F}^{abc}_{def}\Phi \left(\raisebox{-0.22in}{\includegraphics[height=0.5in]{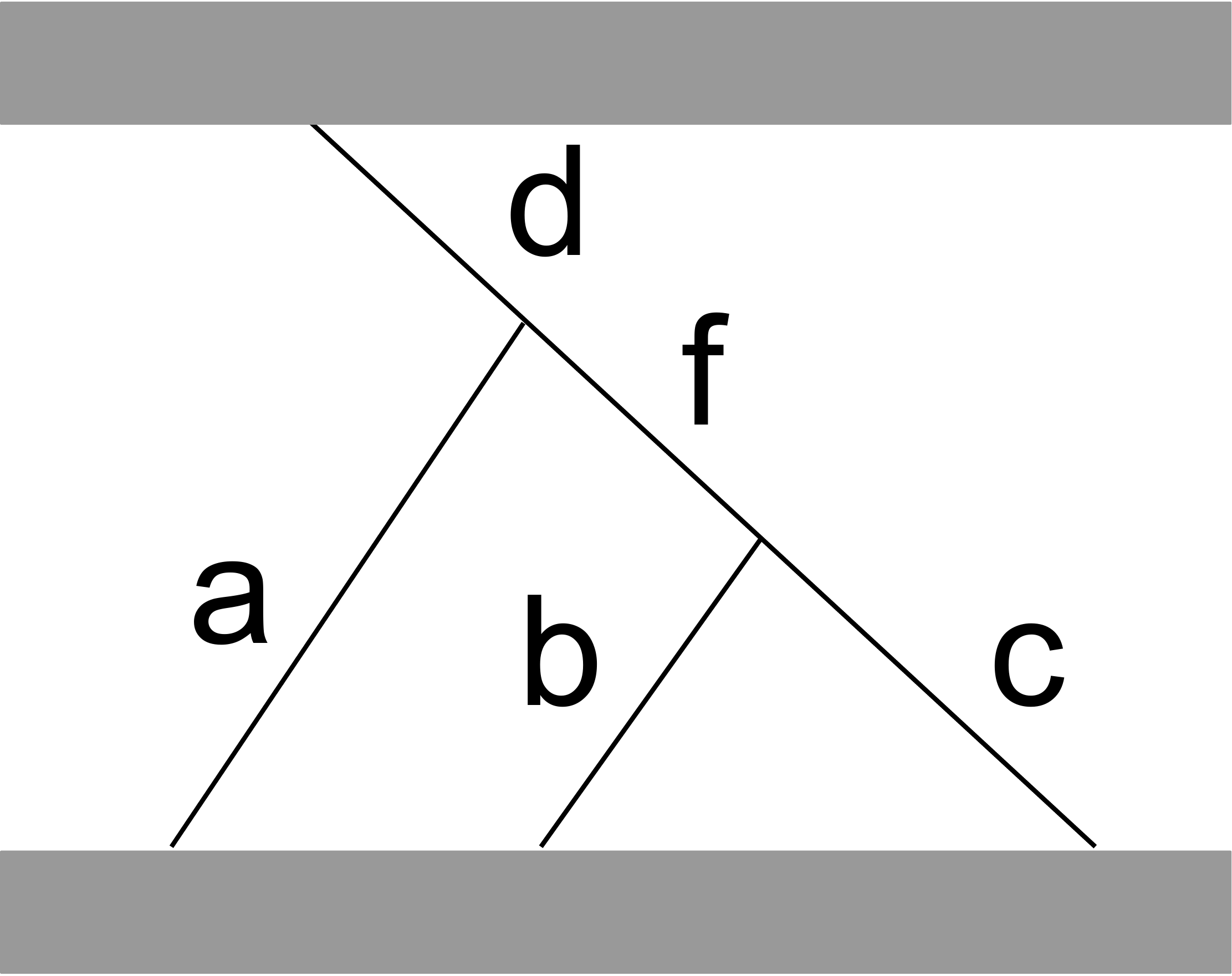}} \right) \label{1d} \\	
	\Phi\left(\raisebox{-0.22in}{\includegraphics[height=0.5in]{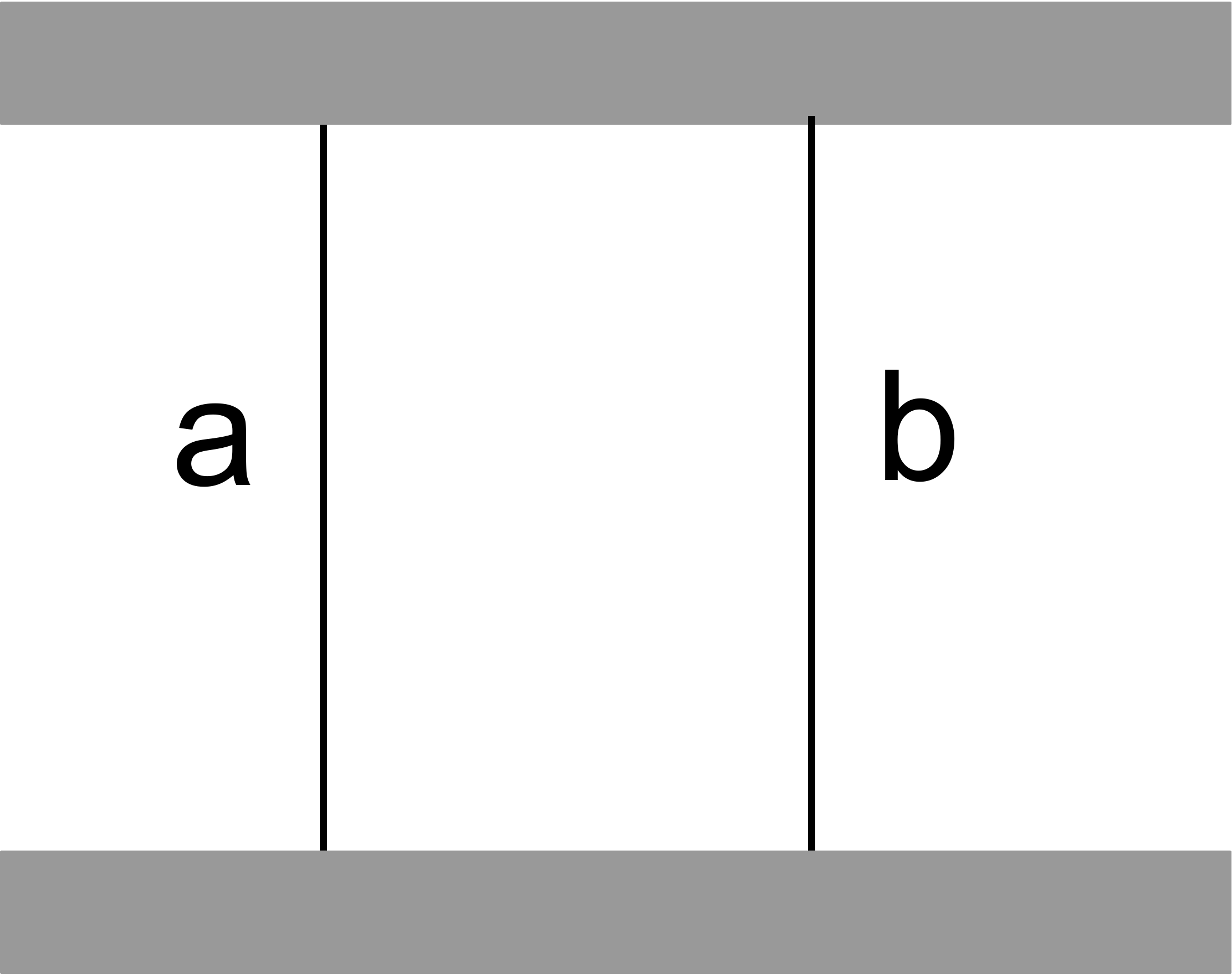}} \right)&=\sum_c  \frac{1}{Y^{ab}_c}
	\Phi \left(\raisebox{-0.22in}{\includegraphics[height=0.5in]{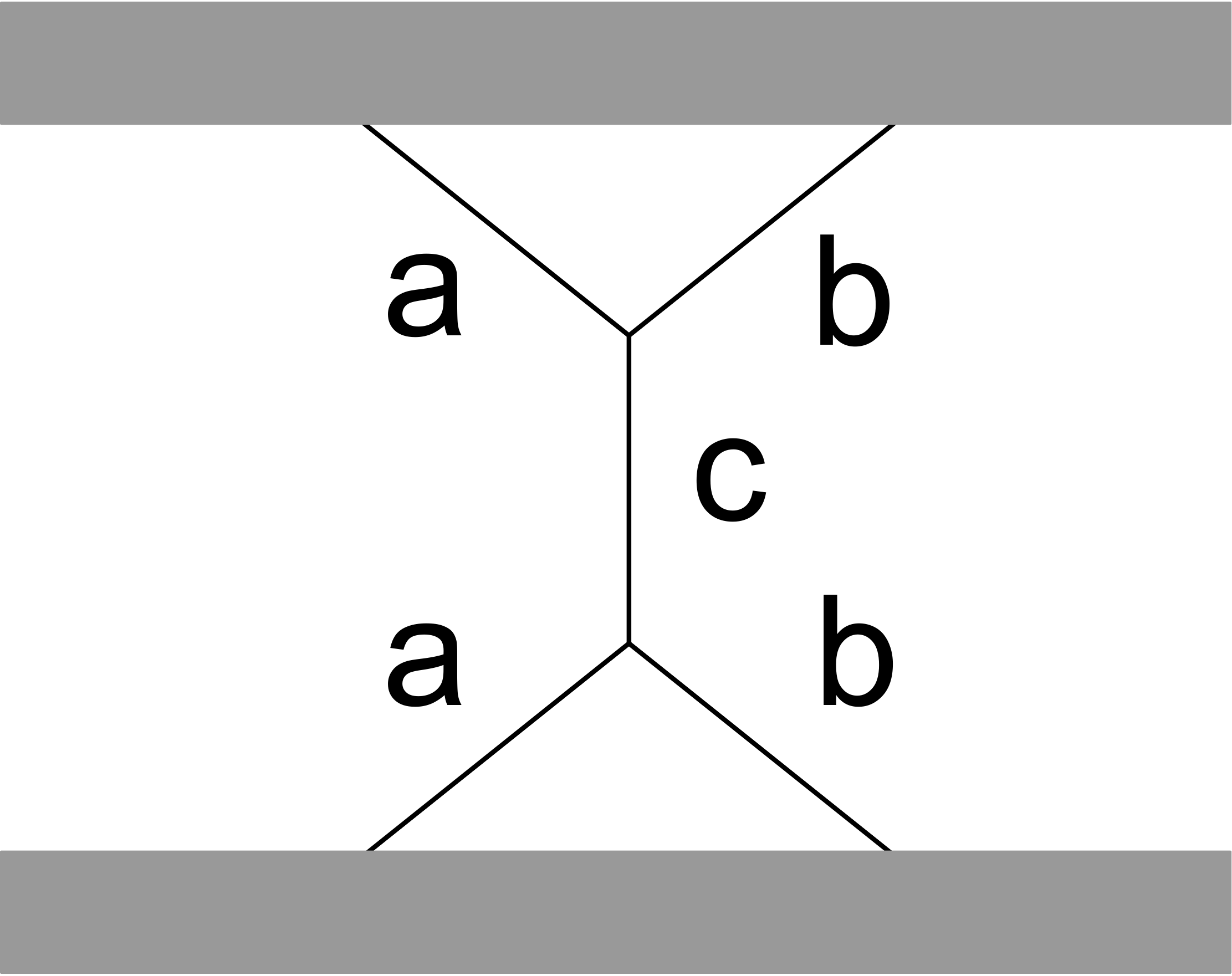}} \right) \label{1b} \\
	\Phi\left(\raisebox{-0.22in}{\includegraphics[height=0.5in]{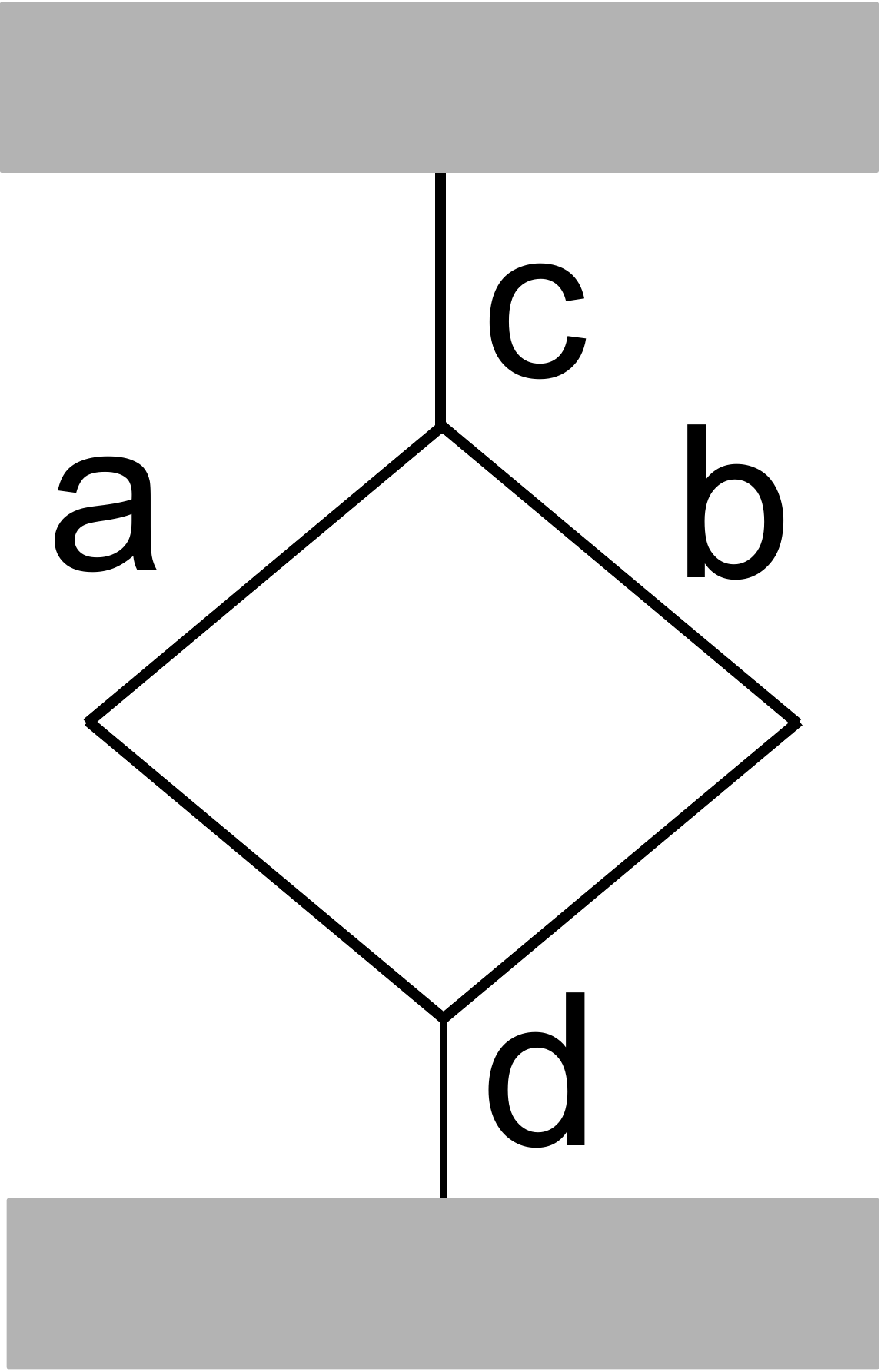}} \right)&=\delta_{c,d} Y^{ab}_c
	\Phi\left(\raisebox{-0.22in}{\includegraphics[height=0.5in]{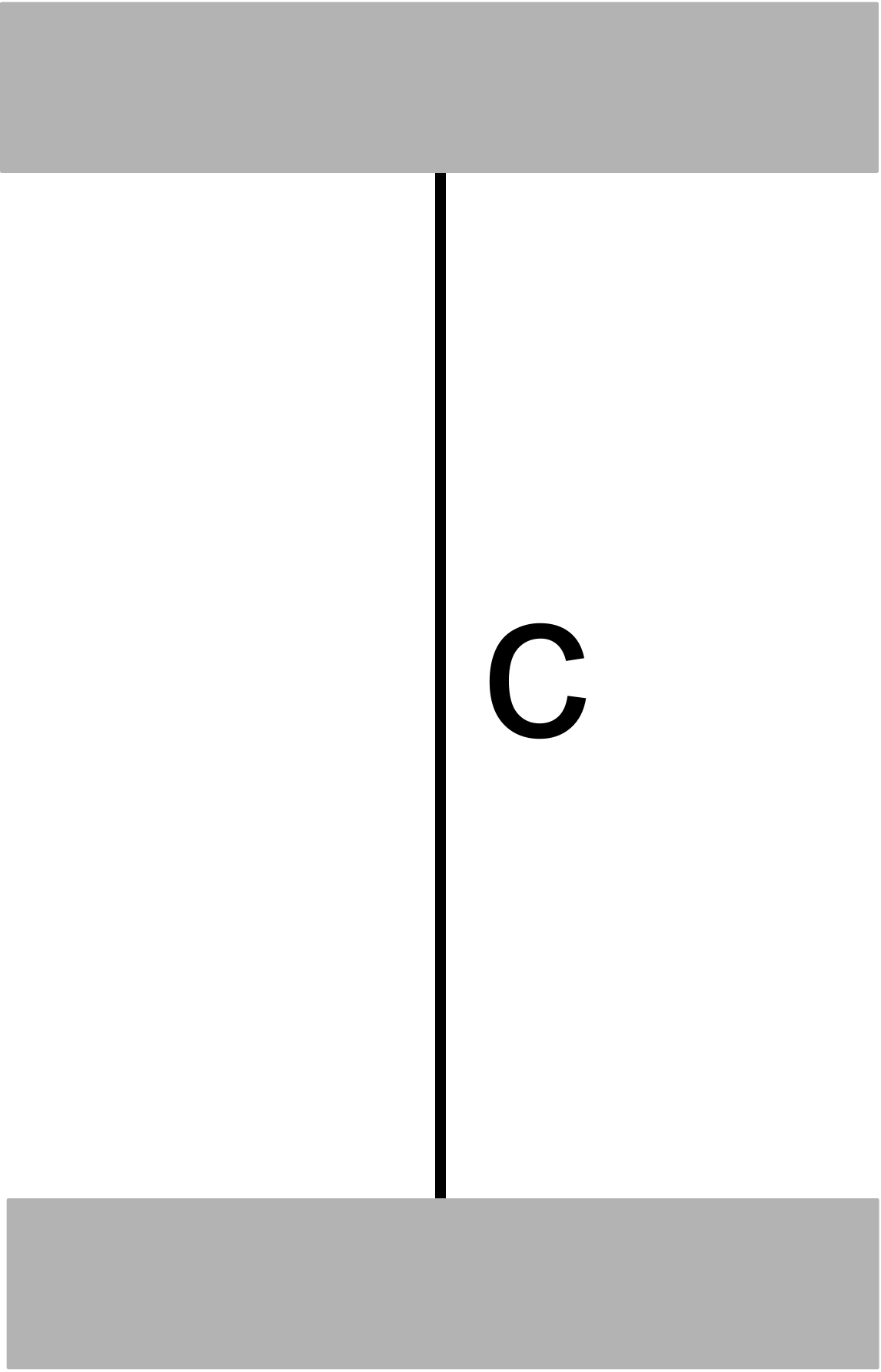}} \right). \label{1c}
\end{align}
\label{localrules}
\end{subequations}
These equations are defined in the Hilbert space $\mathcal{H}$ where the configurations on both sides of the equations satisfy branching rules at every vertex.
Here $a,b,c,\dots$ are arbitrary string types (including the null string types) and the shaded regions represent arbitrary string-net configurations which are not changed from one side of the equation to the other.
The symbol $\delta_{c,d}=1$ if $c=d$ and $\delta_{c,d}=0$ otherwise.
 The parameters $F^{abc}_{def},\tilde{F}^{abc}_{def}$ are complex numbers that depend on 6 string types $a,b,..,f$ obeying the appropriate branching rules: $\delta^{ab}_e = \delta^{ec}_d = \delta^{bc}_f = \delta^{af}_d = 1$. Likewise, $Y^{ab}_c$ is a complex number that depends on three string types $a,b,c$ obeying the branching rule $\delta^{ab}_c = 1$.
For the moment, the parameters \{$F^{abc}_{def},\tilde{F}^{abc}_{def},Y^{ab}_c$\} are arbitrary except for two minor restrictions: we require that (i) $Y^{ab}_c \neq 0$, and (ii) the matrices defined by $(F^{abc}_{d})_{ef}$ and
$(\tilde{F}^{abc}_d)_{ef}$ are \emph{invertible}.\footnote{Note that $(F^{abc}_{d})_{ef}$ and
$(\tilde{F}^{abc}_d)_{ef}$ are \emph{square} matrices due to the associativity constraint (\ref{fusionassoc}).}
However, we will soon see that these parameters have to satisfy nontrivial algebraic equations (\ref{consistency}) for the above constraints to be \emph{self-consistent}.

We now explain the meaning of these local constraints or ``rules.'' The first rule (\ref{1g}) has been drawn schematically.  This rule means that any two string-net configurations that can be deformed continuously into one another must have the same amplitude. Here, for a deformation to qualify as ``continuous'', it must be continuous in a geometric sense and also preserve the graph structure of the string-net: i.e. the deformation is not allowed to introduce or delete vertices (either bivalent or trivalent) or change the orientation along any of the strings. For example, Eq.~(\ref{1g}) implies
\begin{equation}
	\Phi\left(\raisebox{-0.22in}{\includegraphics[height=0.5in]{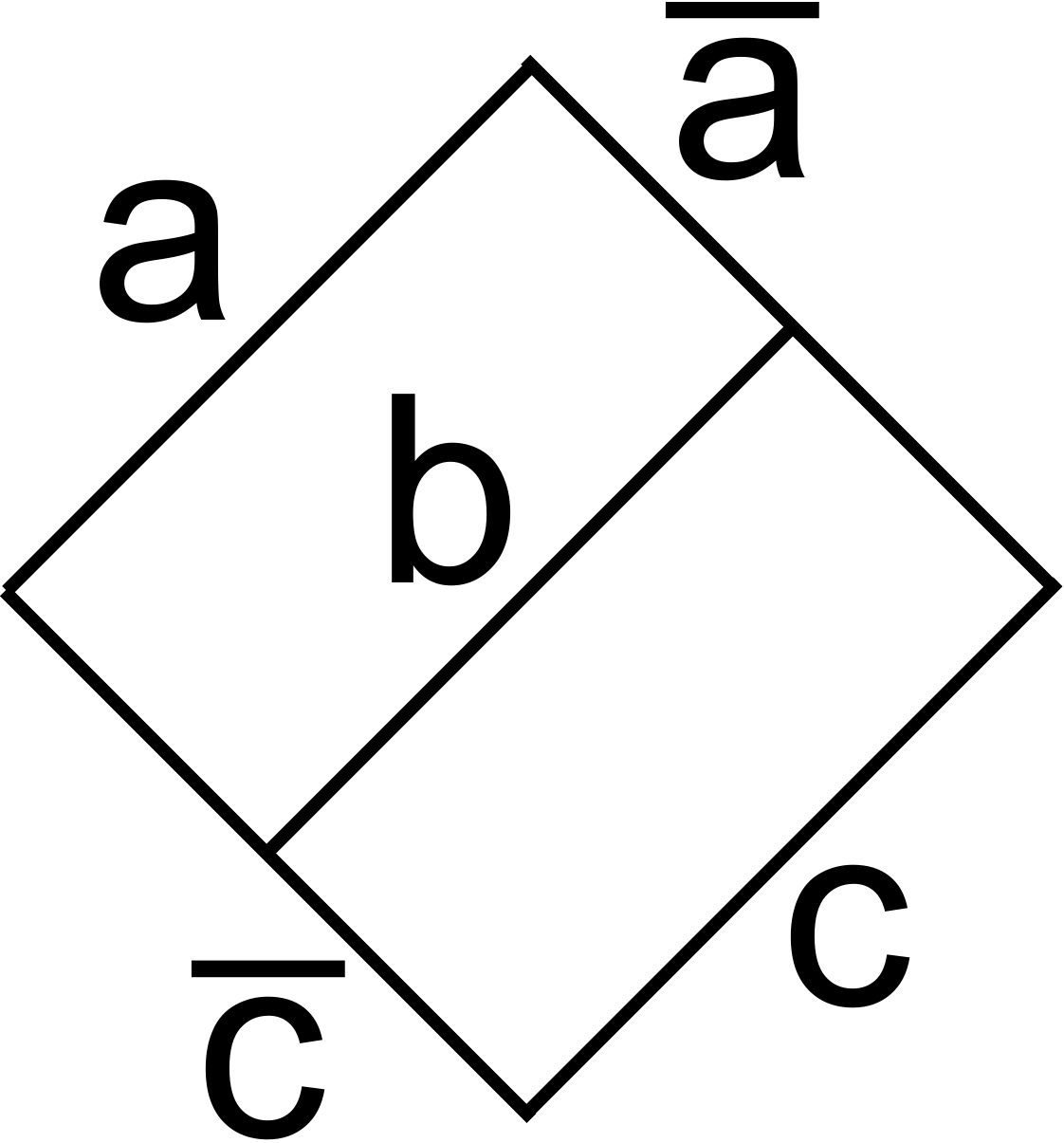}} \right)
	=\Phi \left(\raisebox{-0.22in}{\includegraphics[height=0.5in]{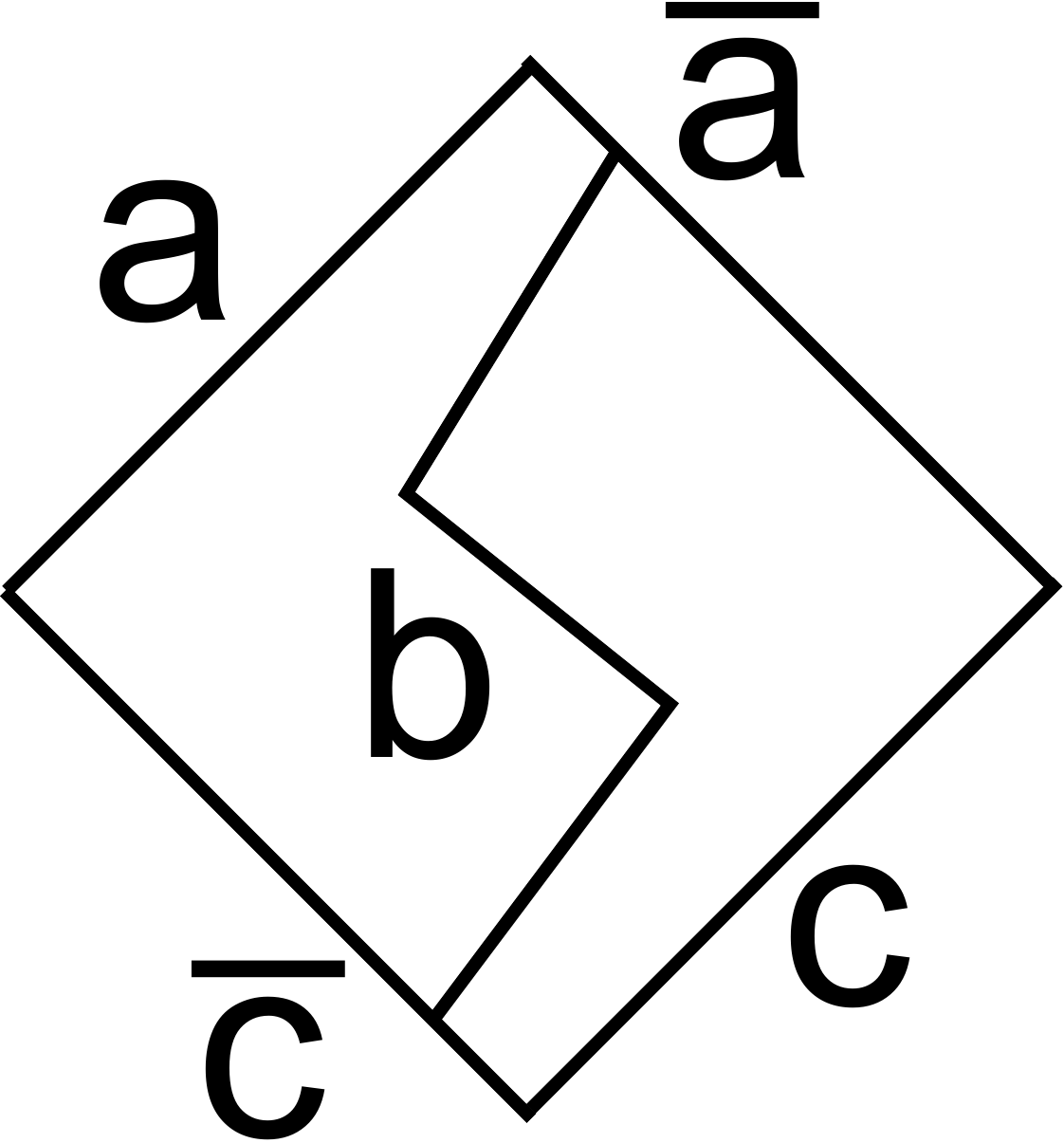}} \right)
	=\Phi \left(\raisebox{-0.22in}{\includegraphics[height=0.5in]{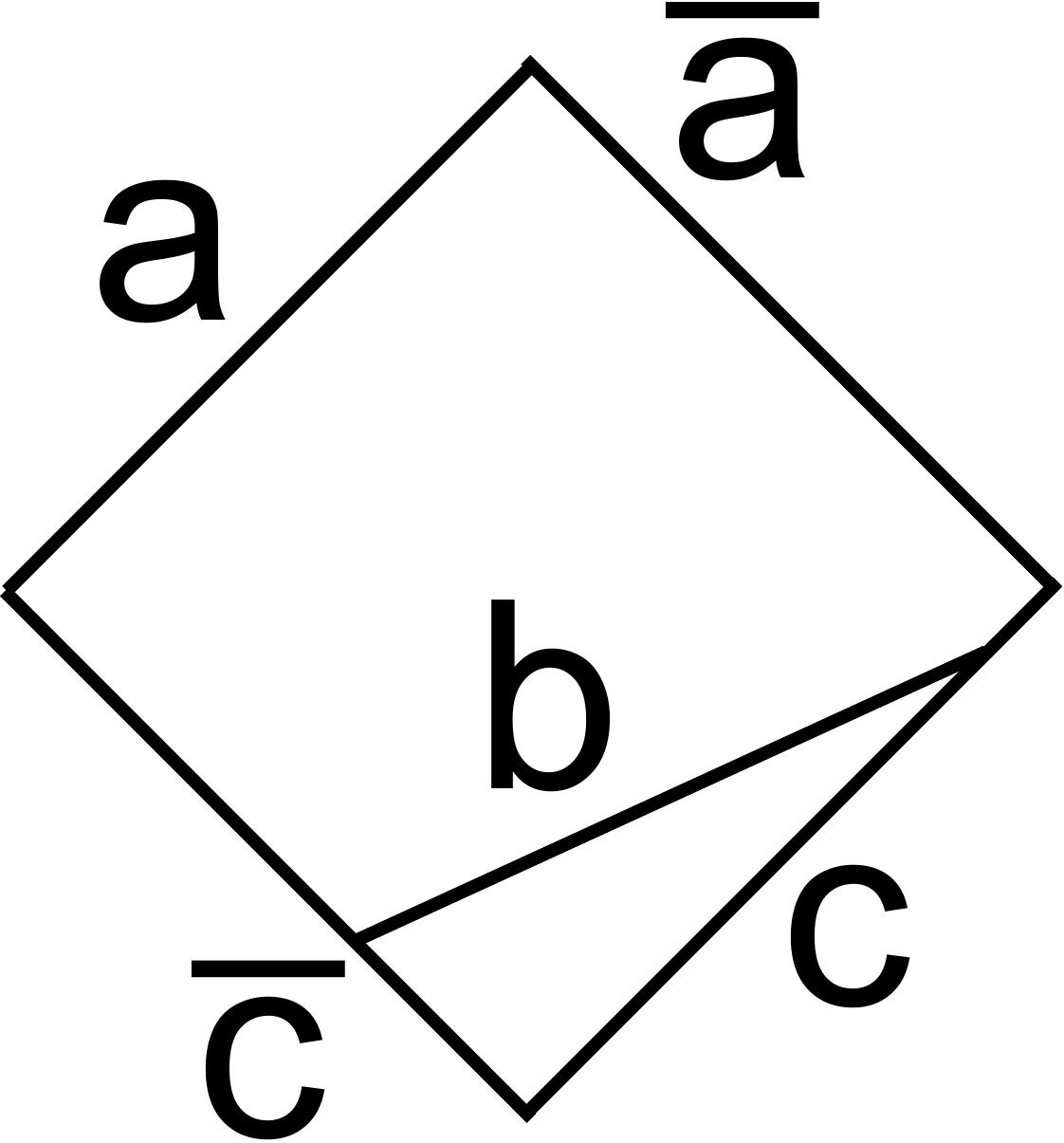}} \right).
\end{equation}
In contrast, 
\begin{equation}
	\Phi\left(\raisebox{-0.22in}{\includegraphics[height=0.5in]{allow1.pdf}} \right)
	\neq \Phi \left(\raisebox{-0.22in}{\includegraphics[height=0.5in]{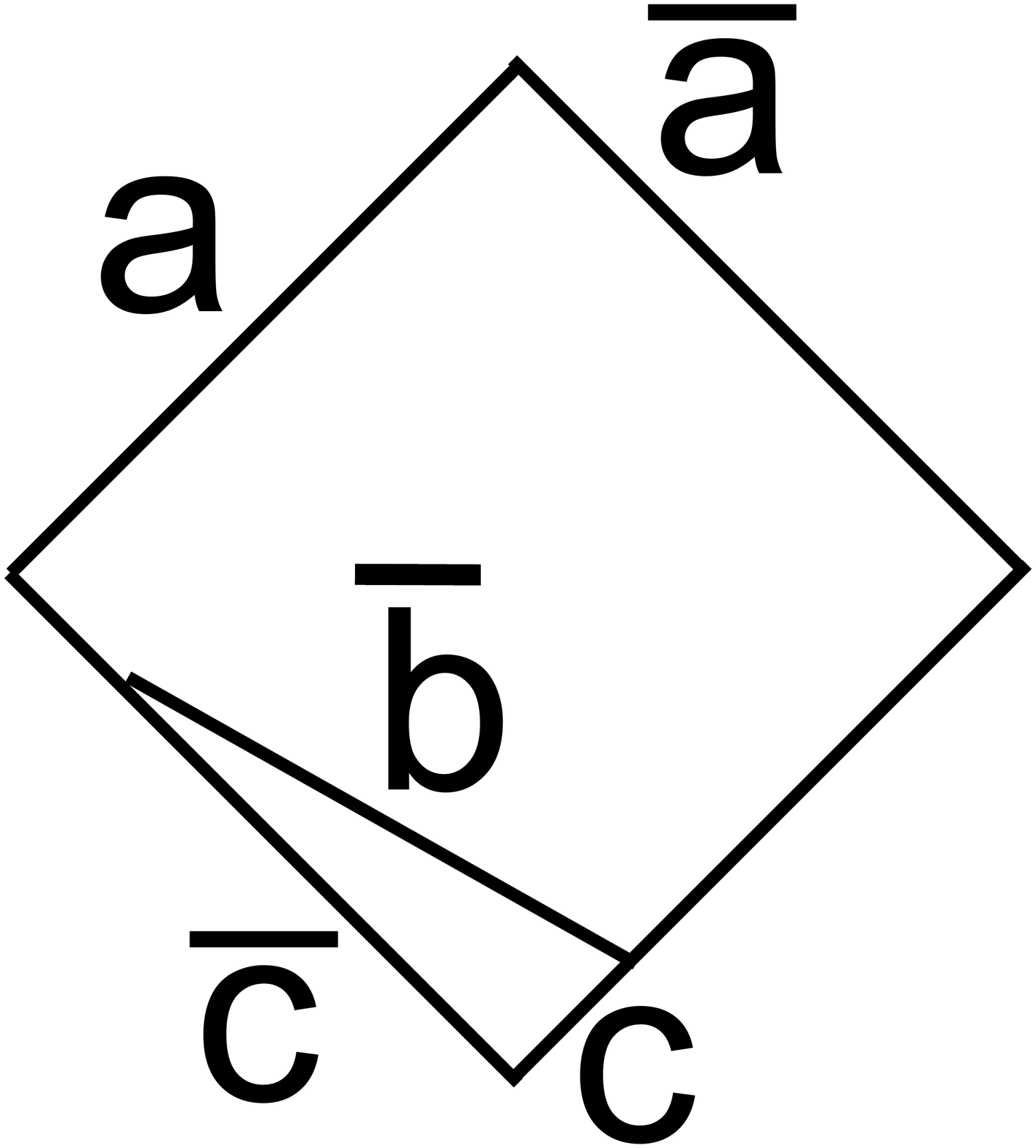}} \right)
	\neq \Phi \left(\raisebox{-0.22in}{\includegraphics[height=0.5in]{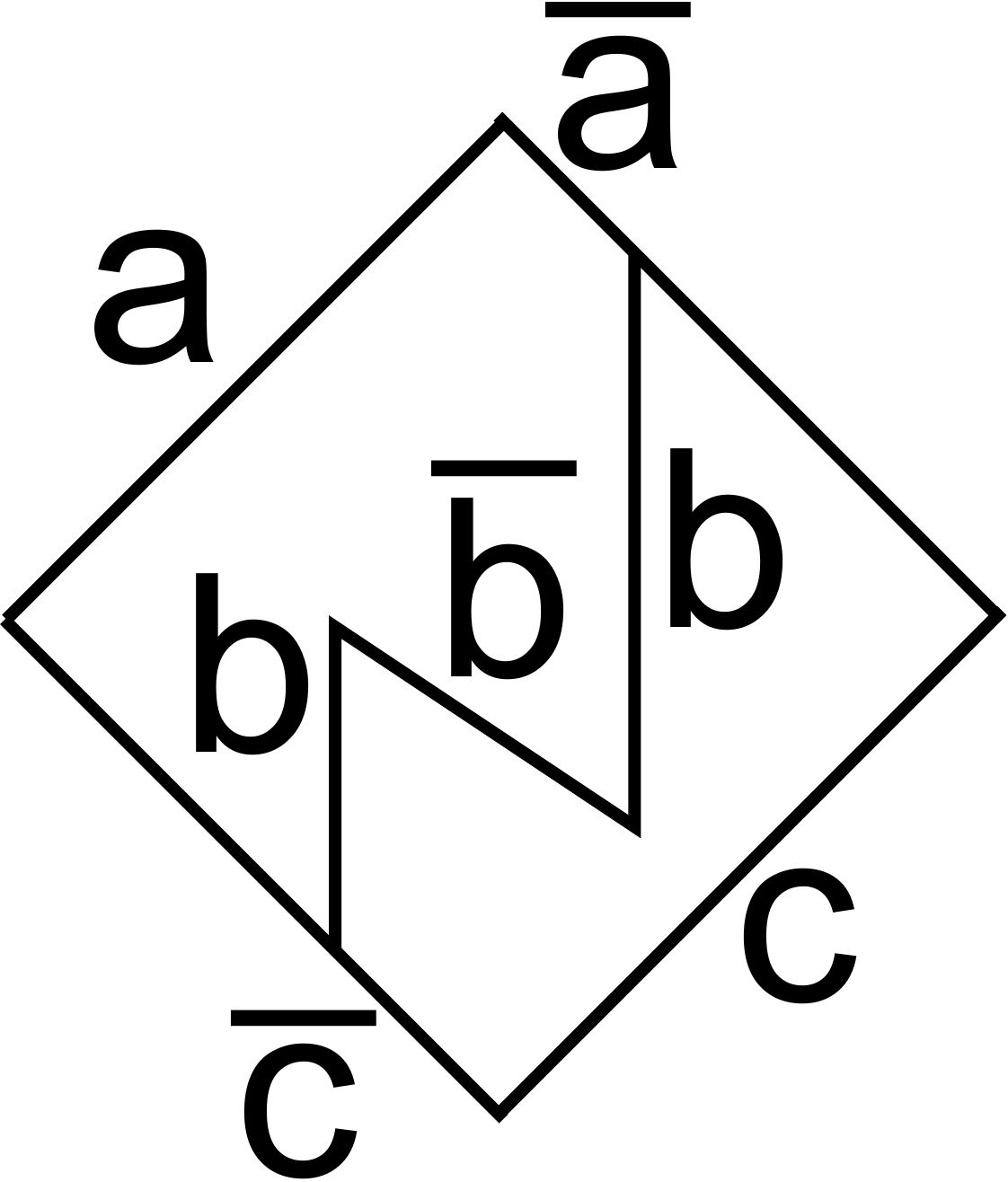}} \right).
	\label{nomove}
\end{equation}
Here the first equality is not valid because the $b$ string in the first configuration has been replaced by $\bar{b}$ in the second configuration. Likewise, the second equality is invalid because the third configuration has two extra bivalent vertices along the $b$ string.

Moving on to the next two rules (\ref{1a})-(\ref{1d}), these tell us that when evaluating an amplitude of a string-net, we can replace any tree-like configuration of the type shown on the left hand side with the corresponding configuration shown on the right hand side, up to factors of $F^{abc}_{def}$ or $\tilde{F}^{abc}_{def}$ and taking a sum over the internal index  $f$. Similarly, rules (\ref{1b}) and (\ref{1c}) imply that we can replace the configuration on the left hand side with the corresponding configuration on the right hand side, up to factors of $1/Y^{ab}_c$ and $Y^{ab}_c \delta_{c,d}$, respectively.

The basic idea of (\ref{localrules}) is that by applying these local rules multiple times, one can relate the amplitude of any string-net configurations to the amplitude of the vacuum or no-string configuration. Then, by using the convention\footnote{This is a natural normalization convention when we consider infinite-dimensional Hilbert space, e.g. the string-net Hilbert space on the whole two-dimensional plane.}
that 
\begin{equation}
	\Phi(\text{vacuum})=1,
	\label{vac0}
\end{equation}
the amplitude of any configuration is fully determined.
Thus, once the parameters 
$\{F^{abc}_{cde},\tilde{F}^{abc}_{def},Y^{ab}_c\}$
are given, the rules determine the wave function completely.

An important point is that when applying the above rules, we are allowed to freely erase null strings or draw additional ones without affecting the amplitude of a string-net state. (As we mentioned earlier, the null strings are essentially a redundancy in our notation so erasing them or adding them doesn't change the physical state at all). This freedom is crucial because erasing the null string is the main way that we can simplify string-net configurations and reduce them to the vacuum configuration. For example, by erasing the vacuum string we can remove any vertex of the type shown in (\ref{vertex3}) with $a=0$ or $b=0$:
\begin{equation}
\Phi\left(\raisebox{-0.22in}{\includegraphics[height=0.5in]{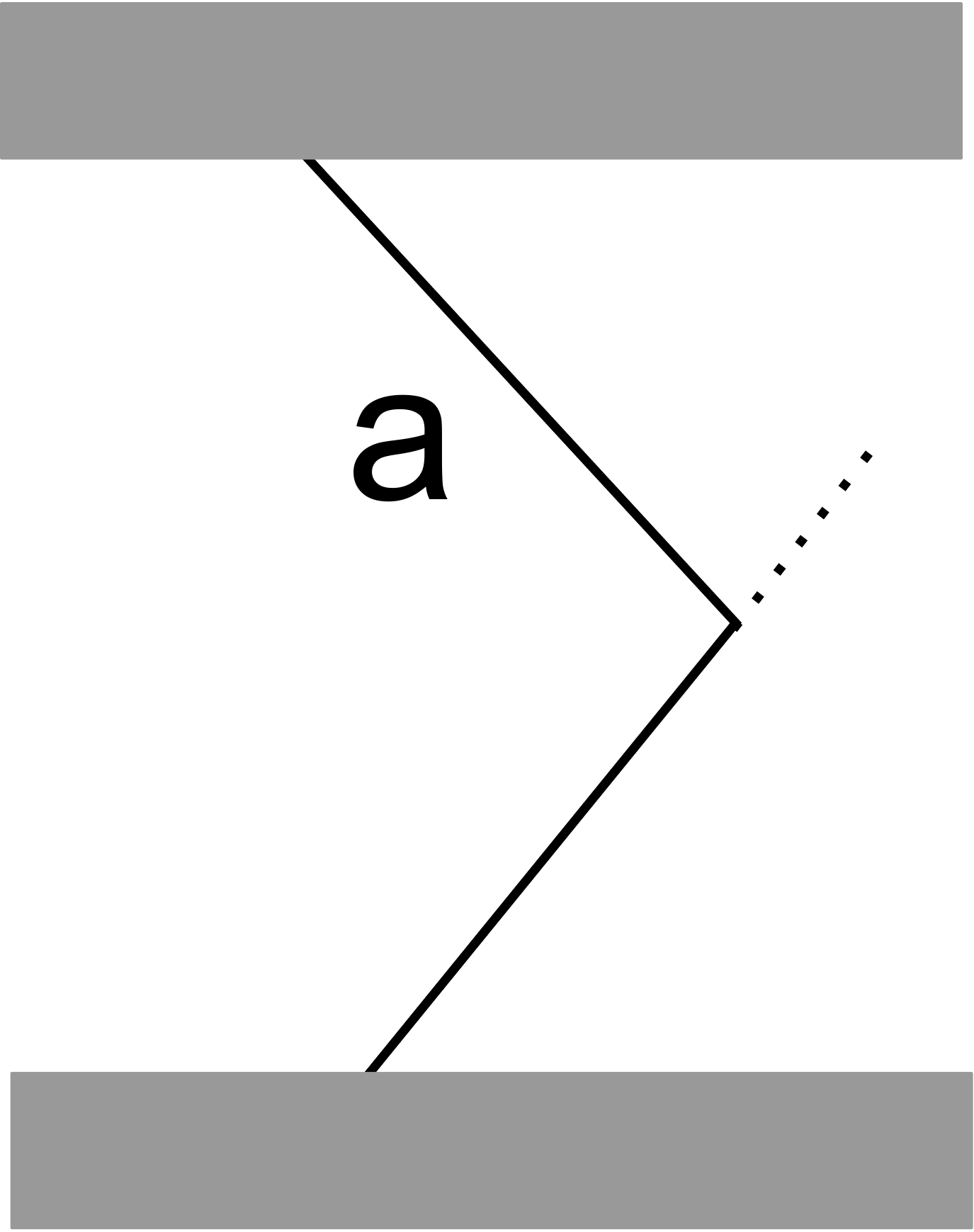}} \right)
	=\Phi\left(\raisebox{-0.22in}{\includegraphics[height=0.5in]{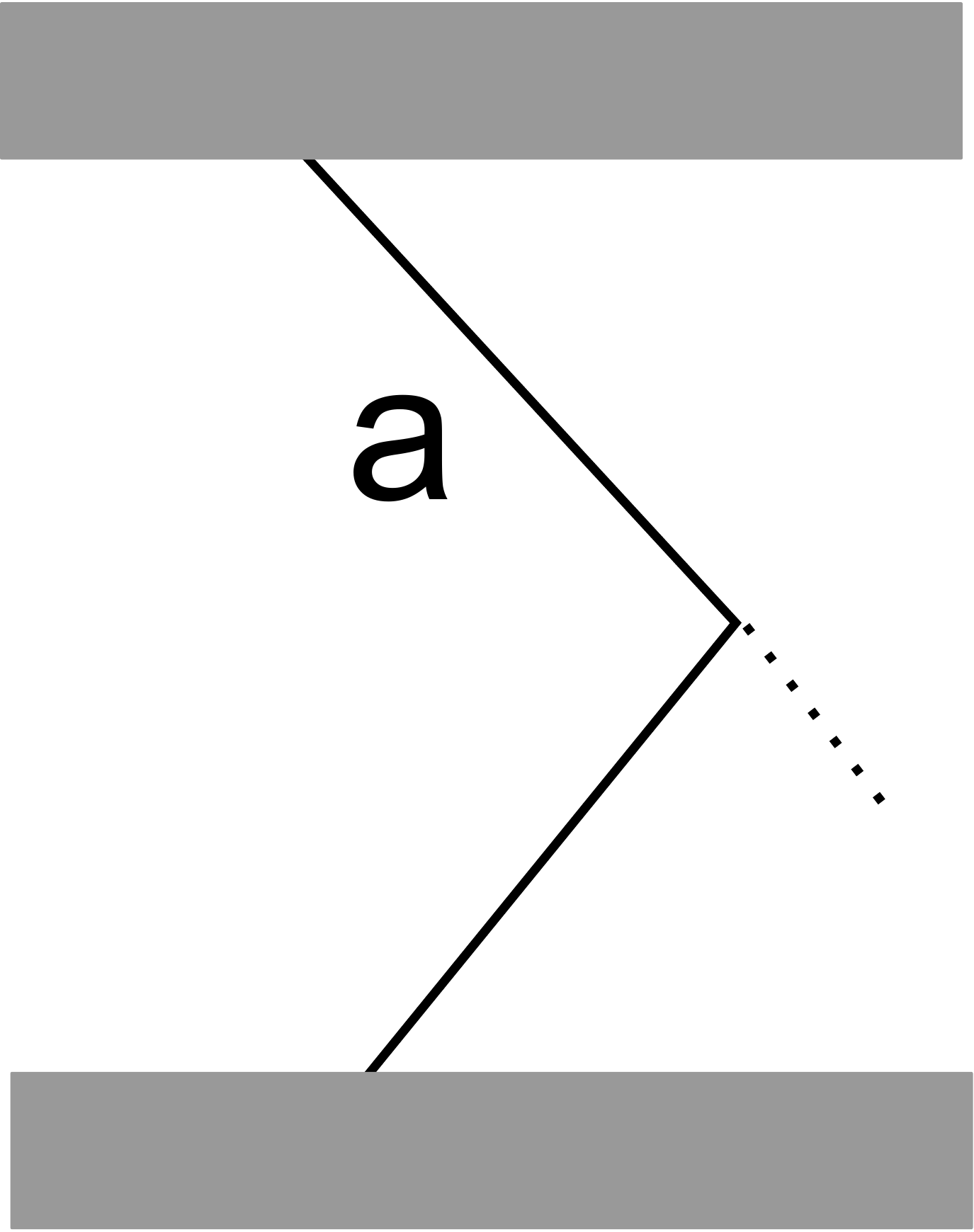}} \right)
	=\Phi\left(\raisebox{-0.22in}{\includegraphics[height=0.5in]{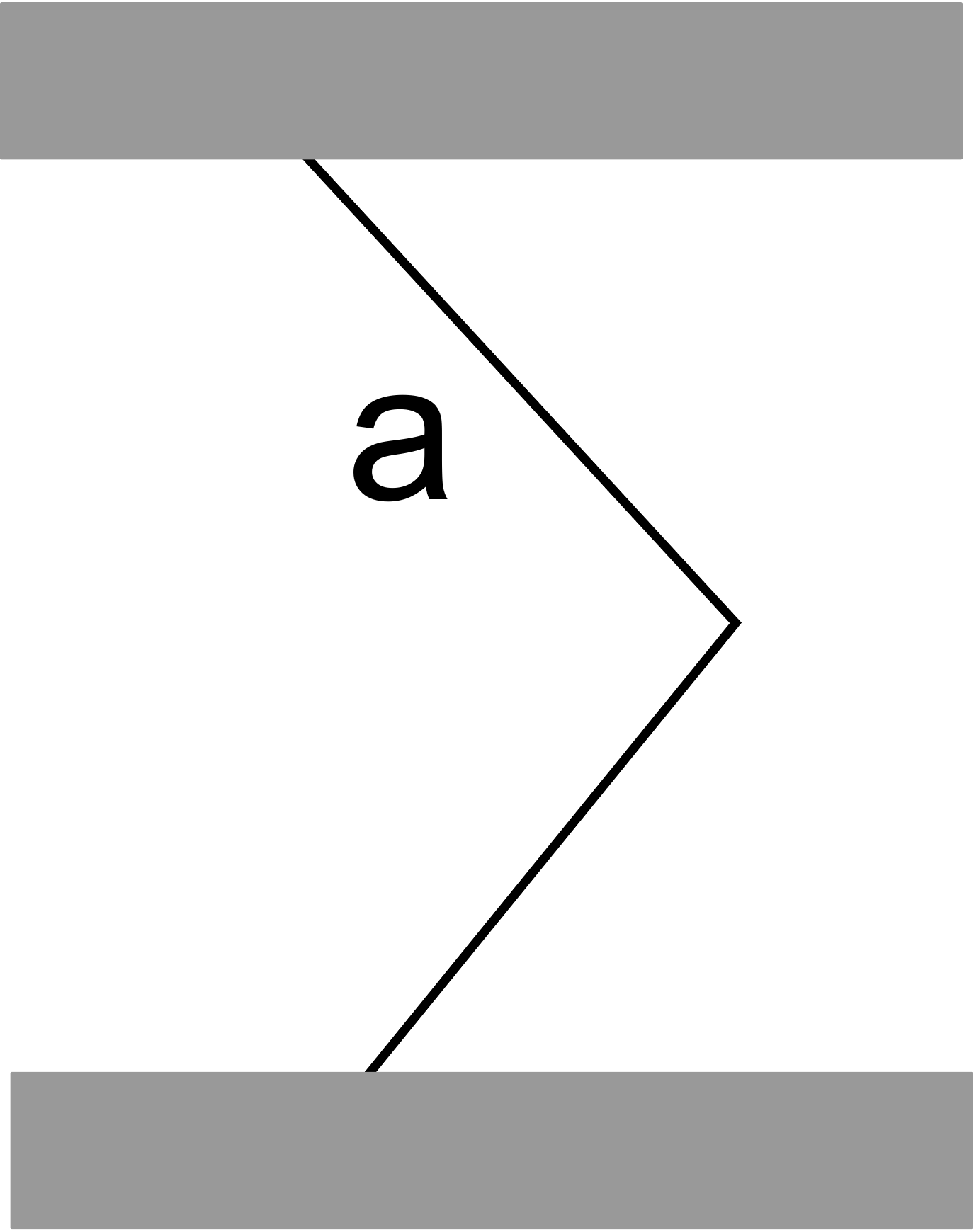}} \right)
	=\Phi\left(\raisebox{-0.22in}{\includegraphics[height=0.5in]{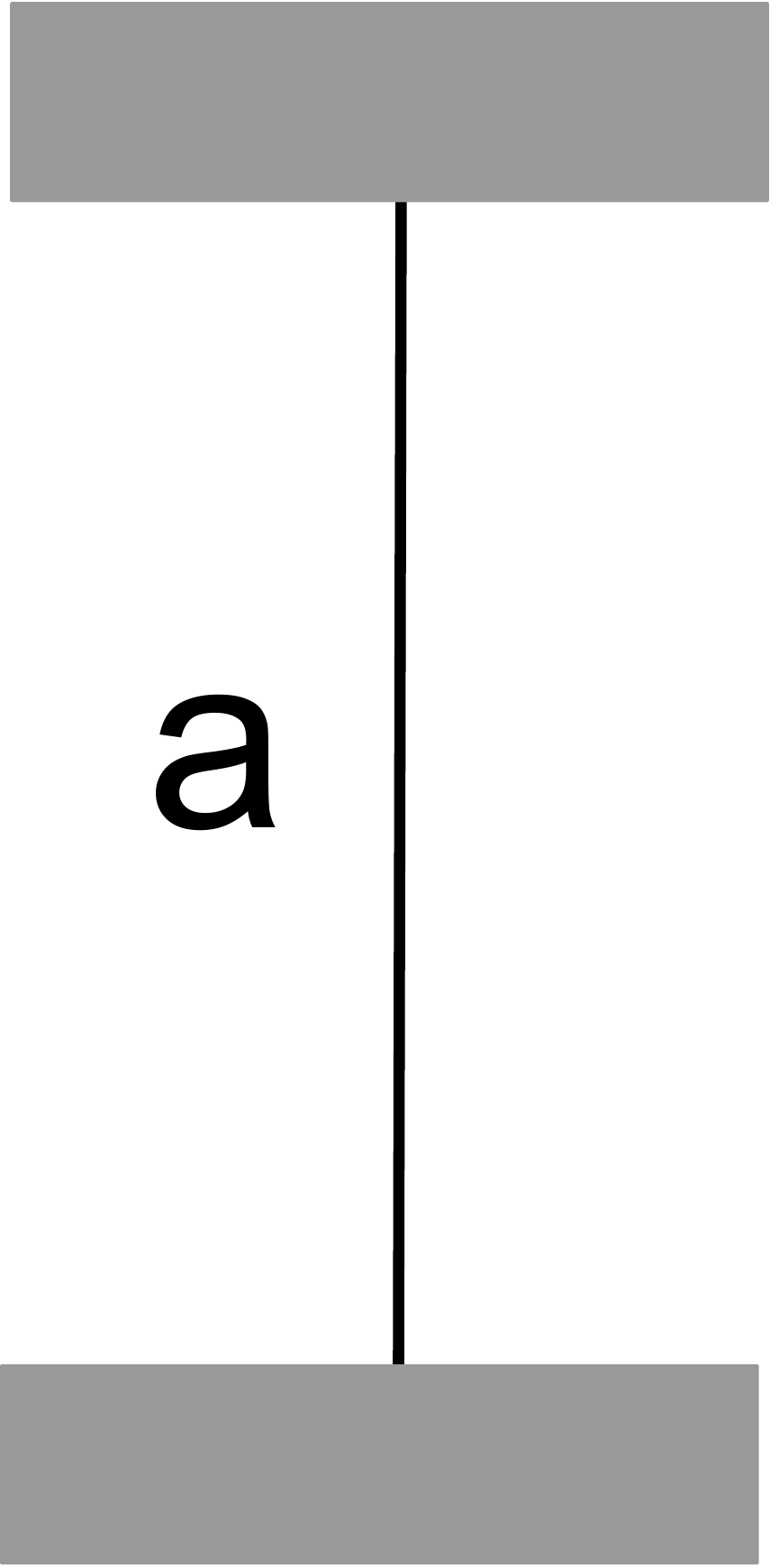}} \right).
	\label{hbend}
\end{equation}

We now present some examples illustrating how we can compute the amplitude of any string-net configuration using the local rules (\ref{localrules}). First we evaluate the following string-net amplitude:
\begin{equation}
\begin{split}
		\Phi\left(\raisebox{-0.15in}{\includegraphics[height=0.4in]{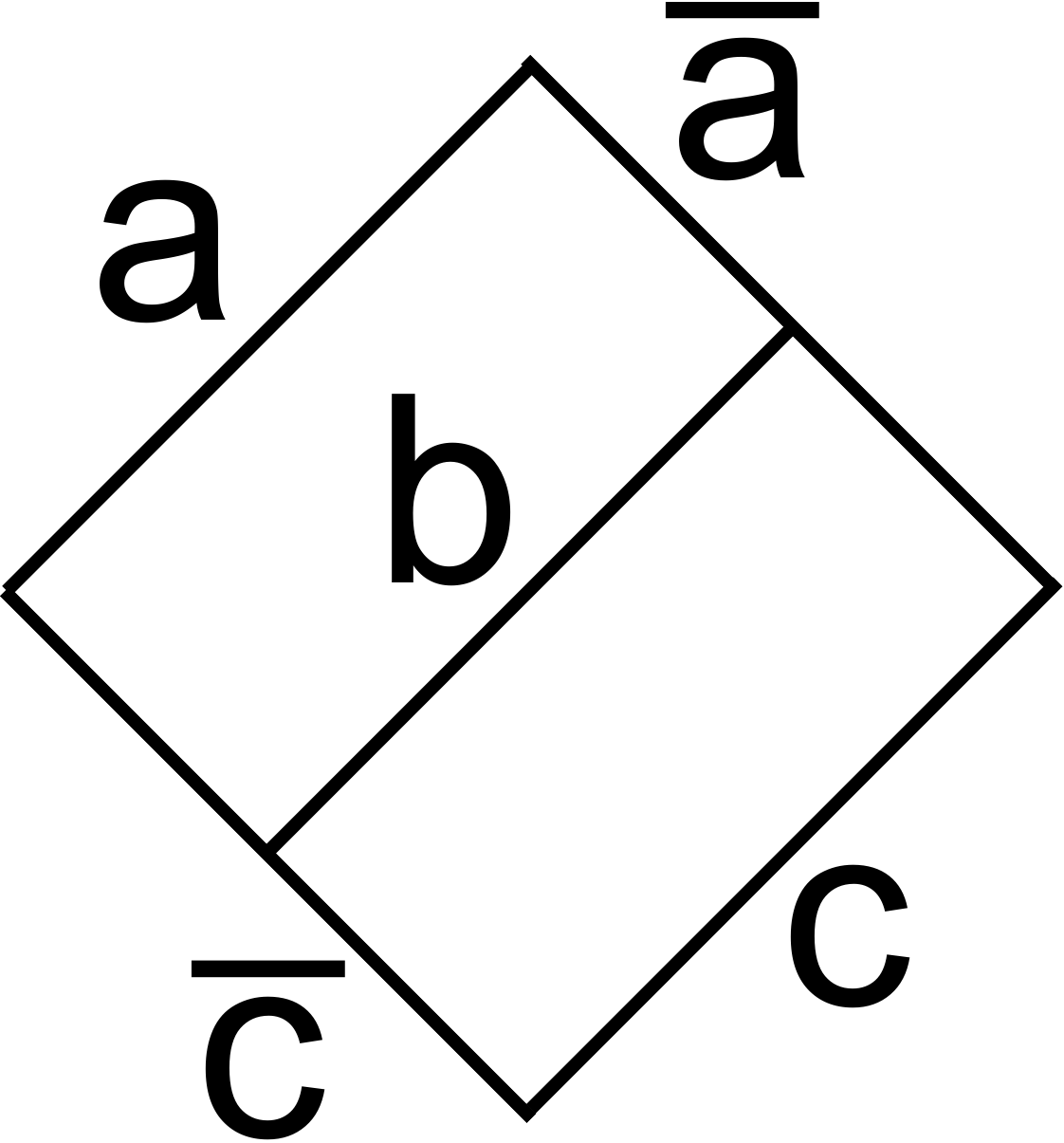}} \right)
		&=\Phi\left(\raisebox{-0.15in}{\includegraphics[height=0.4in]{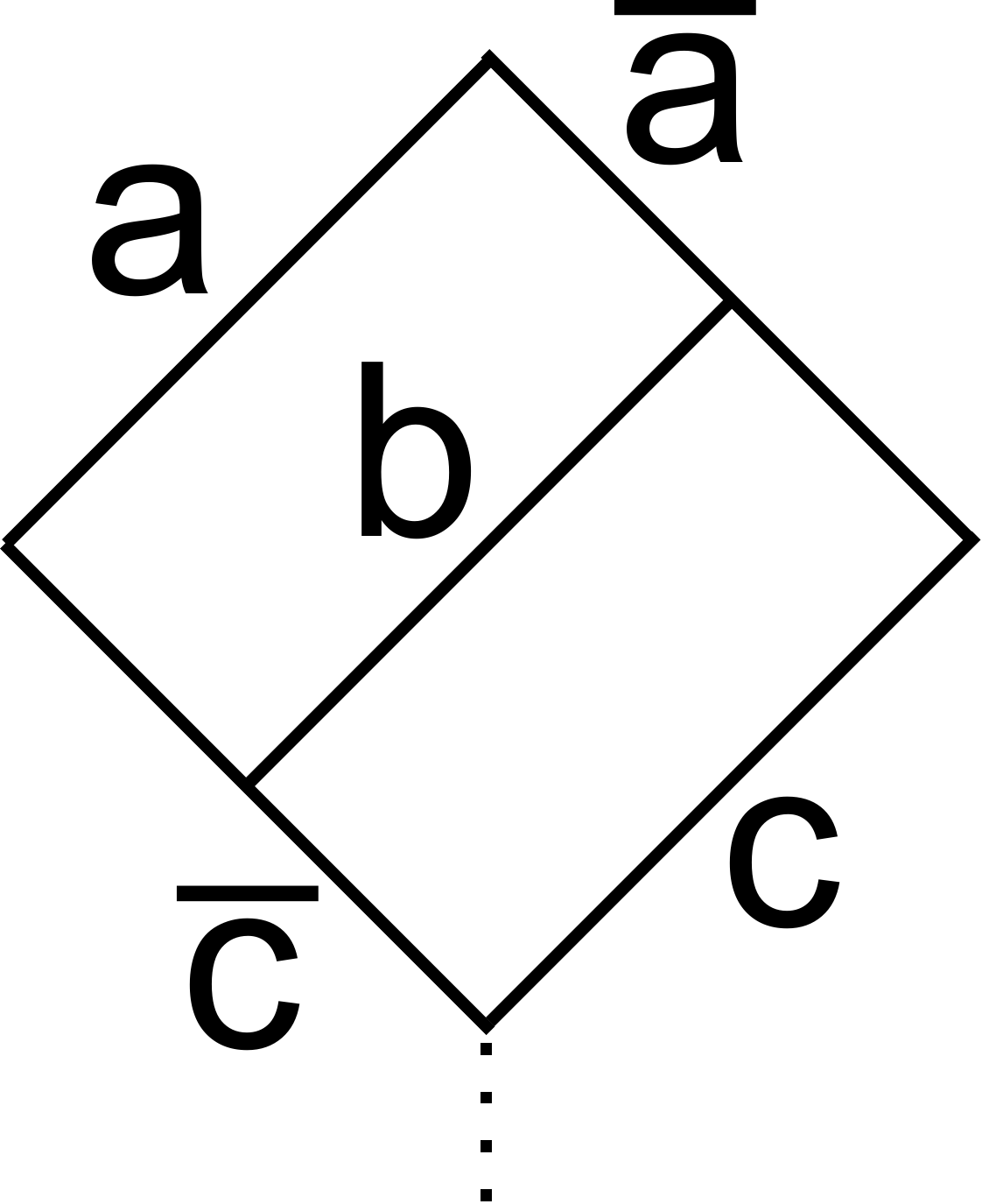}} \right)\\
		&=F^{abc}_{0\bar{c}\bar{a}}\Phi\left(\raisebox{-0.15in}{\includegraphics[height=0.4in]{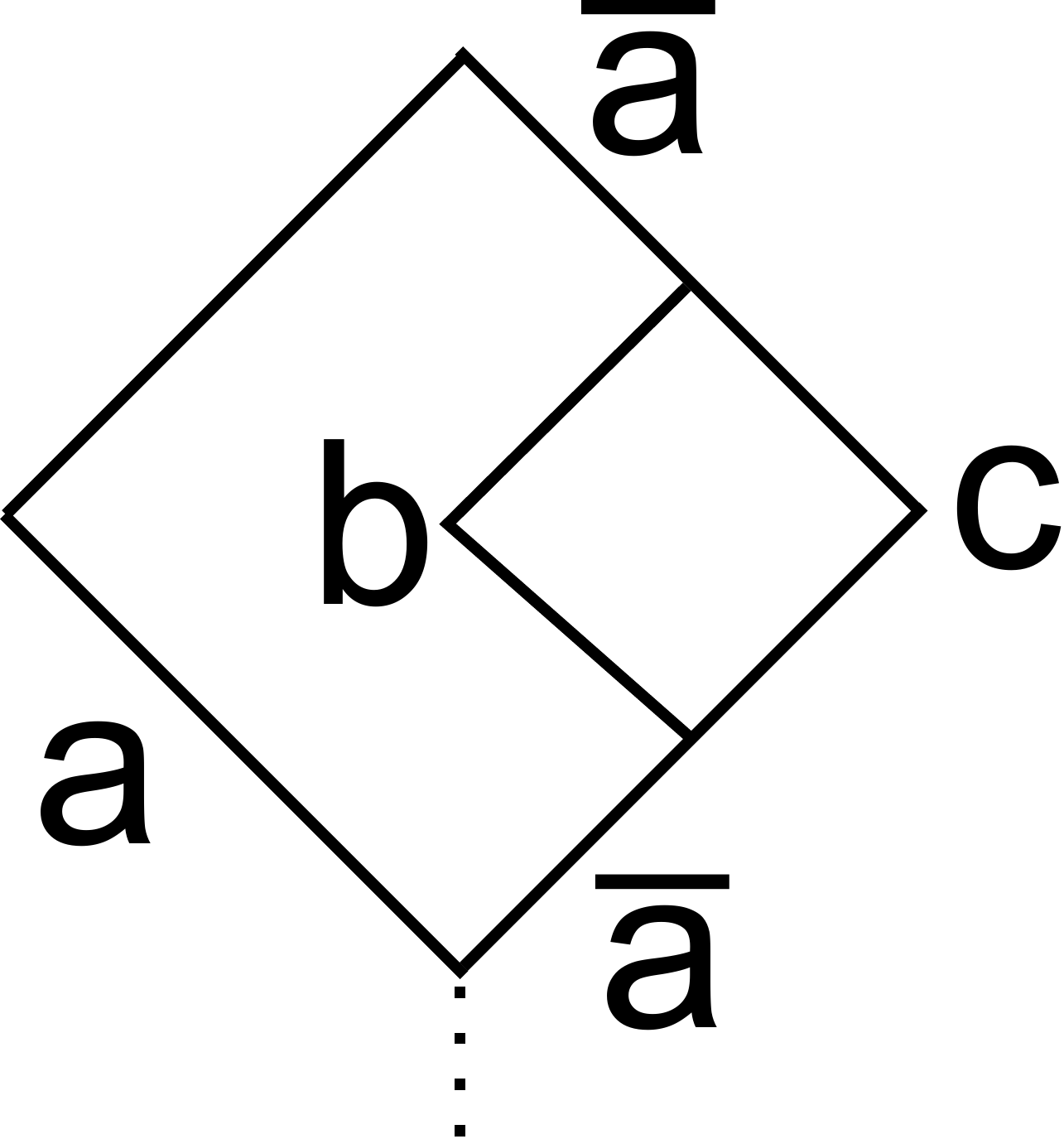}} \right)\\
		&=F^{abc}_{0\bar{c}\bar{a}} Y^{bc}_{\bar{a}}\Phi\left(\raisebox{-0.15in}{\includegraphics[height=0.4in]{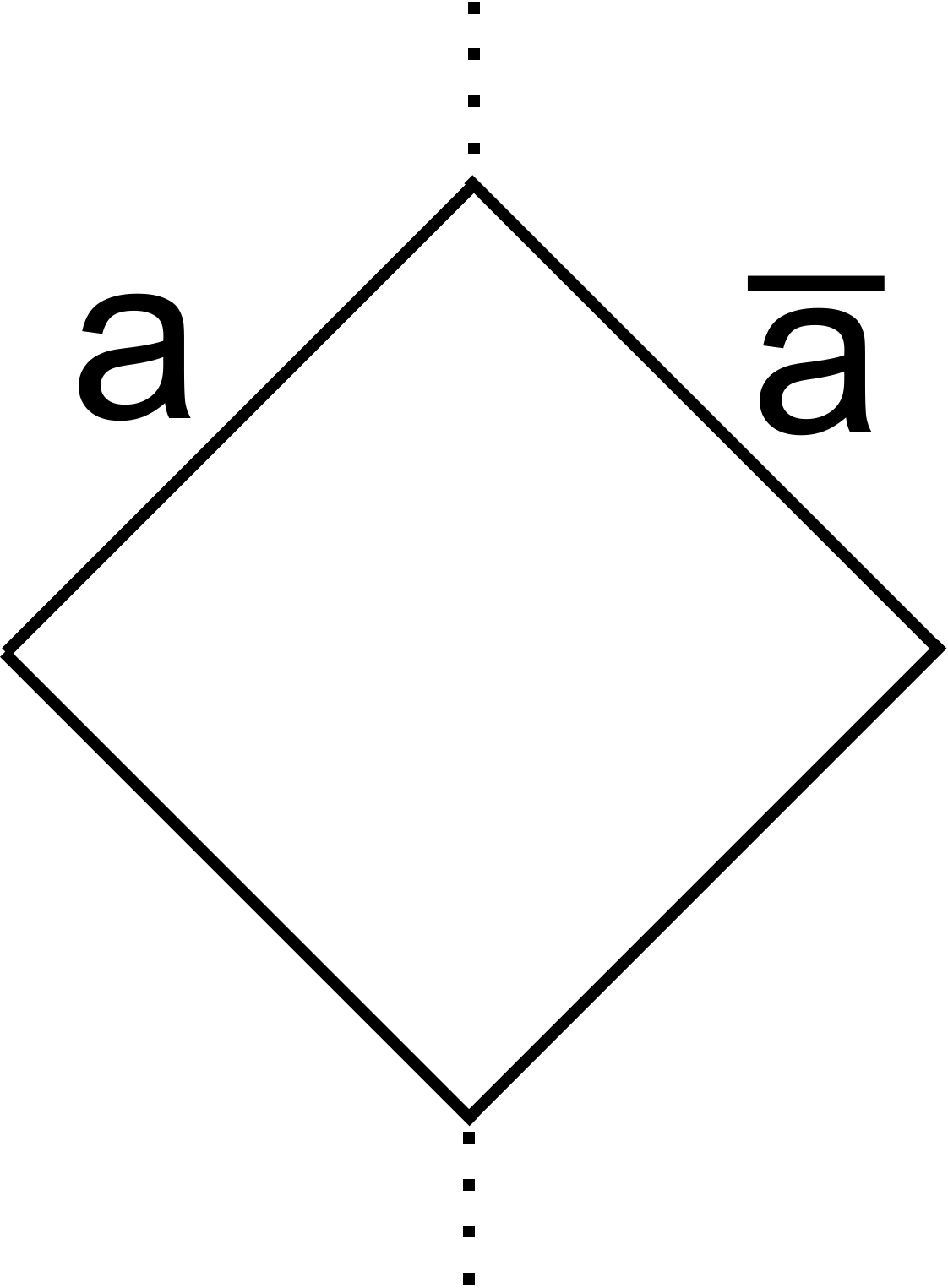}} \right)\\
		&=F^{abc}_{0\bar{c}\bar{a}} Y^{bc}_{\bar{a}} Y^{a\bar{a}}_0 \Phi(\text{vacuum}) \\
		&= F^{abc}_{0\bar{c}\bar{a}} Y^{bc}_{\bar{a}} Y^{a\bar{a}}_0
			\end{split}
	\label{}
\end{equation}
In the first step, we add one null string and then use Eq.~(\ref{1a}) in the second step. Next we use Eq.~(\ref{1c}) twice to reduce the graph to the vacuum. Finally, we use the normalization convention (\ref{vac0}).

Next we consider a slightly more complicated example:
\begin{equation}
	\begin{split}
		\Phi\left(\raisebox{-0.15in}{\includegraphics[height=0.4in]{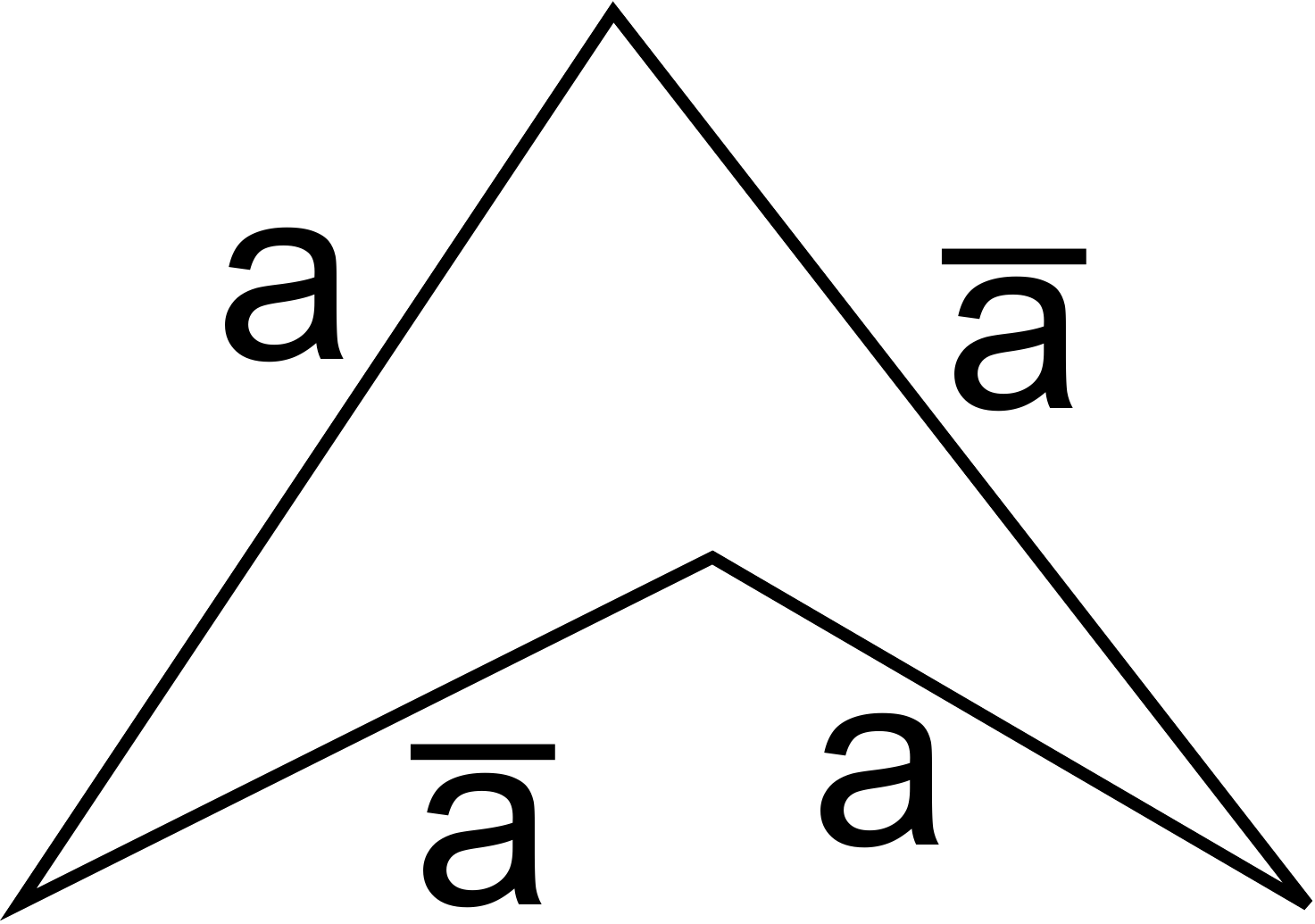}} \right)&=\Phi \left(\raisebox{-0.15in}{\includegraphics[height=0.4in]{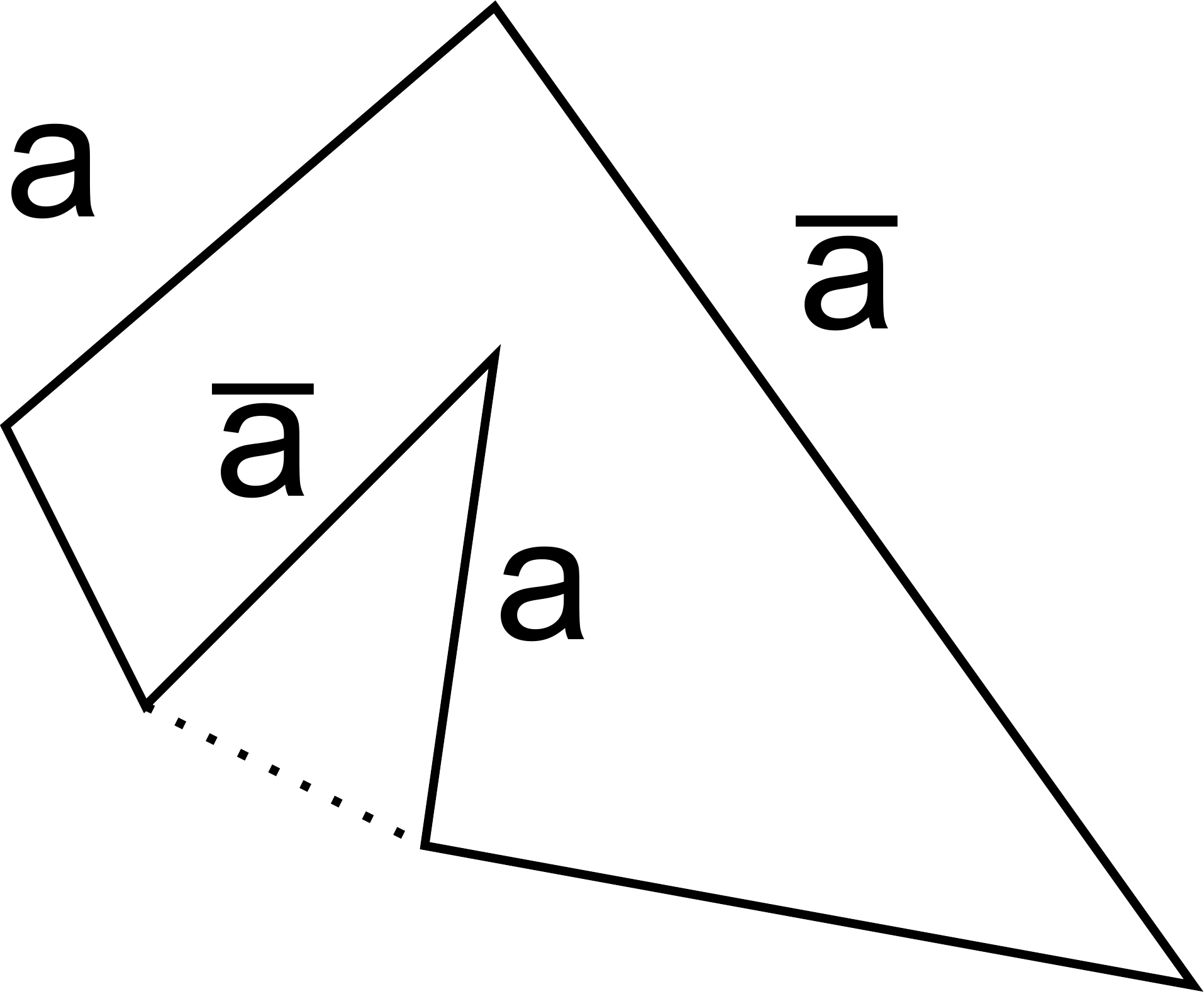}} \right)\\
		&=F^{a\bar{a}a}_{a00} \Phi \left(\raisebox{-0.15in}{\includegraphics[height=0.4in]{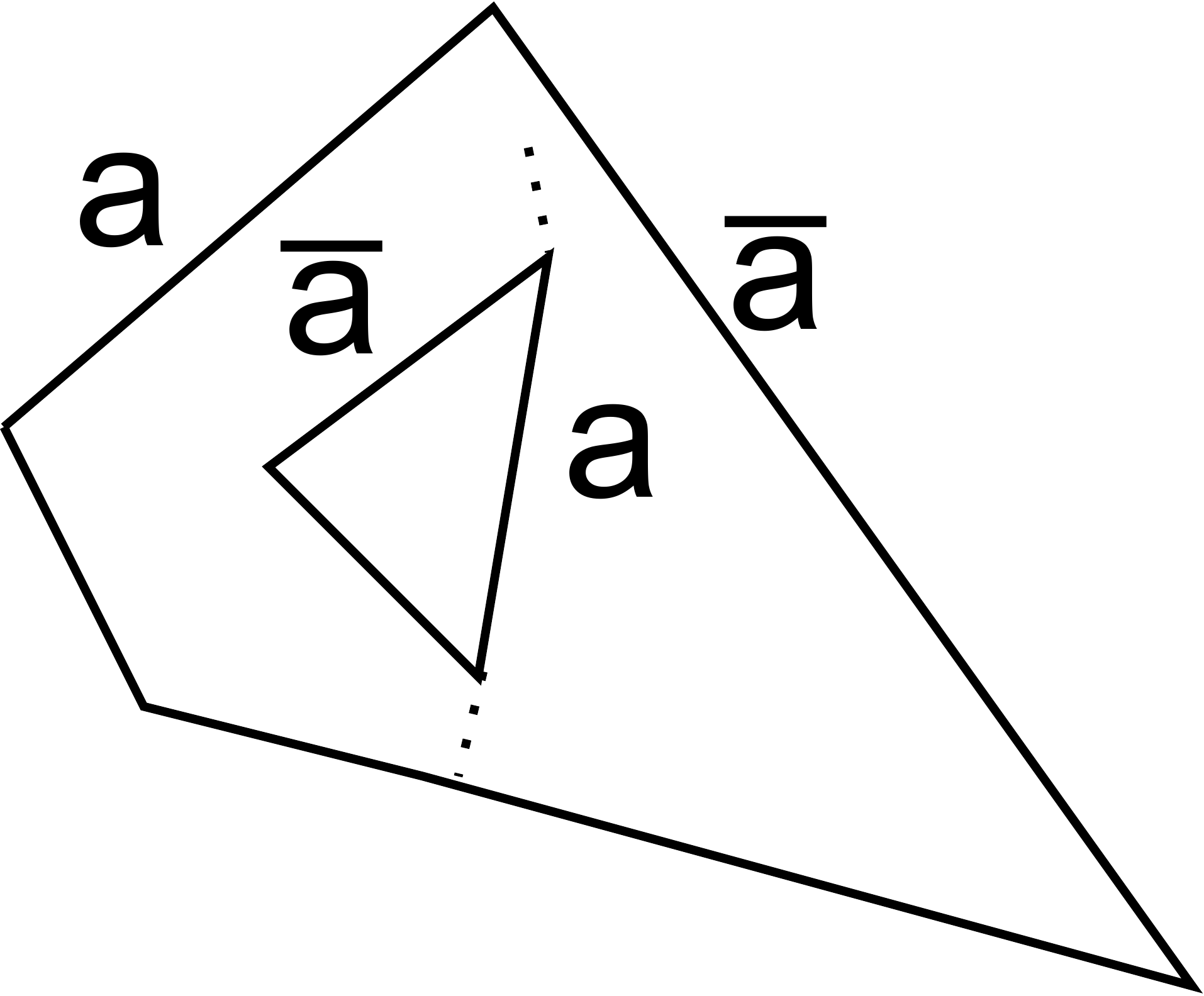}} \right)\\
		&=F^{a\bar{a}a}_{a00} Y^{\bar{a}a}_0 \Phi \left(\raisebox{-0.15in}{\includegraphics[height=0.35in]{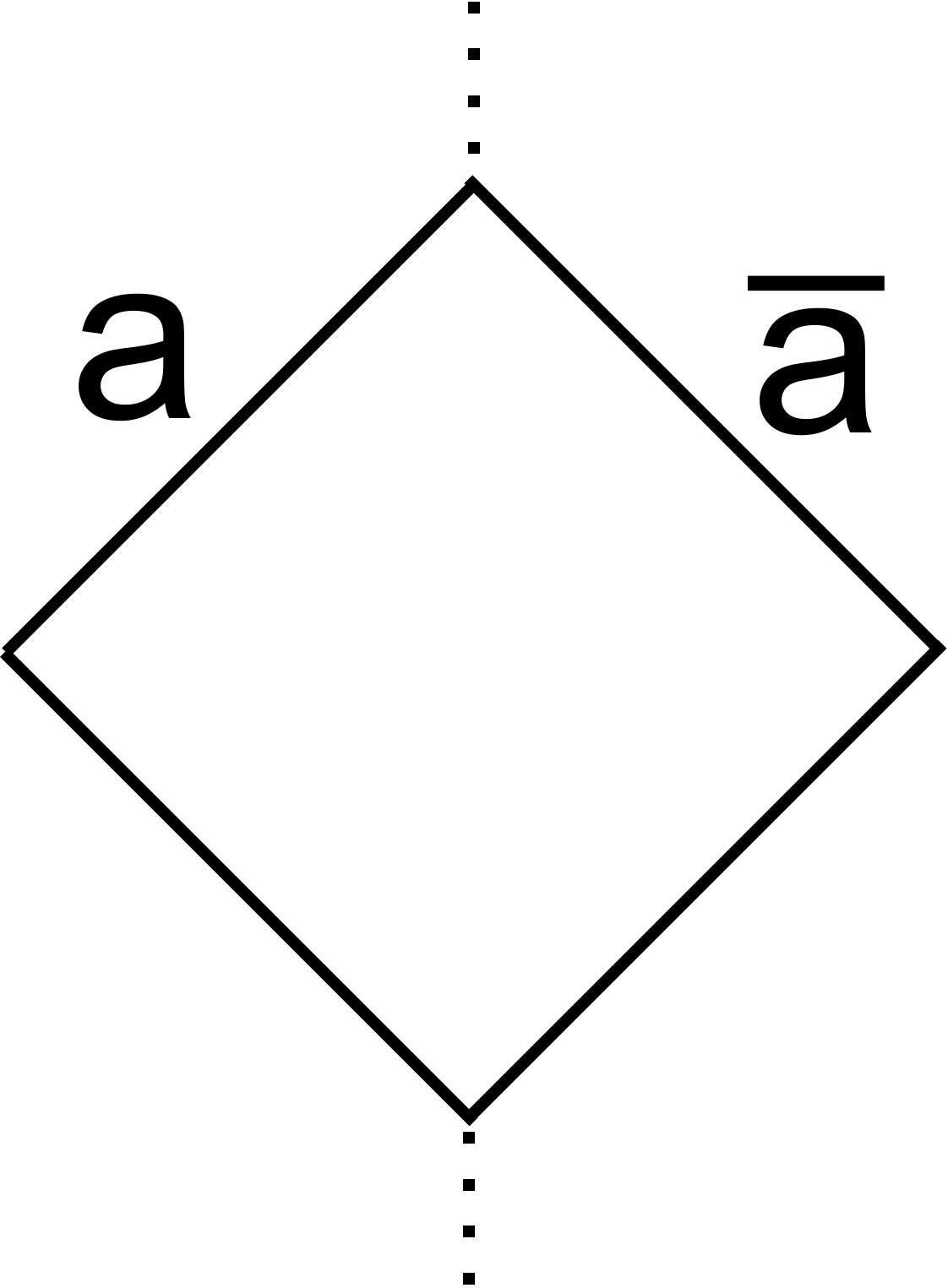}} \right)\\
		&=F^{a\bar{a}a}_{a00} Y^{\bar{a}a}_0  Y^{a\bar{a}}_0 \Phi (\text{vacuum}) \\
		&=F^{a\bar{a}a}_{a00} Y^{\bar{a}a}_0  Y^{a\bar{a}}_0 .
	\end{split}
	\label{ex2}
\end{equation}
In the first step, we continuously deform the graph and then add a null string. In the second step we use Eq.~(\ref{1a}). Next, we use Eq.~(\ref{1c}) twice to reduce the graph to the vacuum. Finally, we use the normalization convention (\ref{vac0}).

As the above examples demonstrate, the quantity $Y^{a \bar{a}}_0$ often appears in amplitudes for string-net configurations. The absolute value of this quantity, $|Y^{a \bar{a}}_0|$, will play an important role below, so we give it its own name:
\begin{equation}
d_a \equiv |Y^{a \bar{a}}_0|
\label{qdim}
\end{equation}
We will refer to $d_a$ as the \emph{quantum dimension} of the string type $a$.

\subsection{Auxiliary rules}
 Although the local rules (\ref{localrules}) are sufficient, by themselves, to evaluate any string-net amplitude, it is useful to introduce two auxiliary rules to simplify calculations:  
\begin{subequations}
	\begin{align}
\Phi\left(\raisebox{-0.22in}{\includegraphics[height=0.5in]{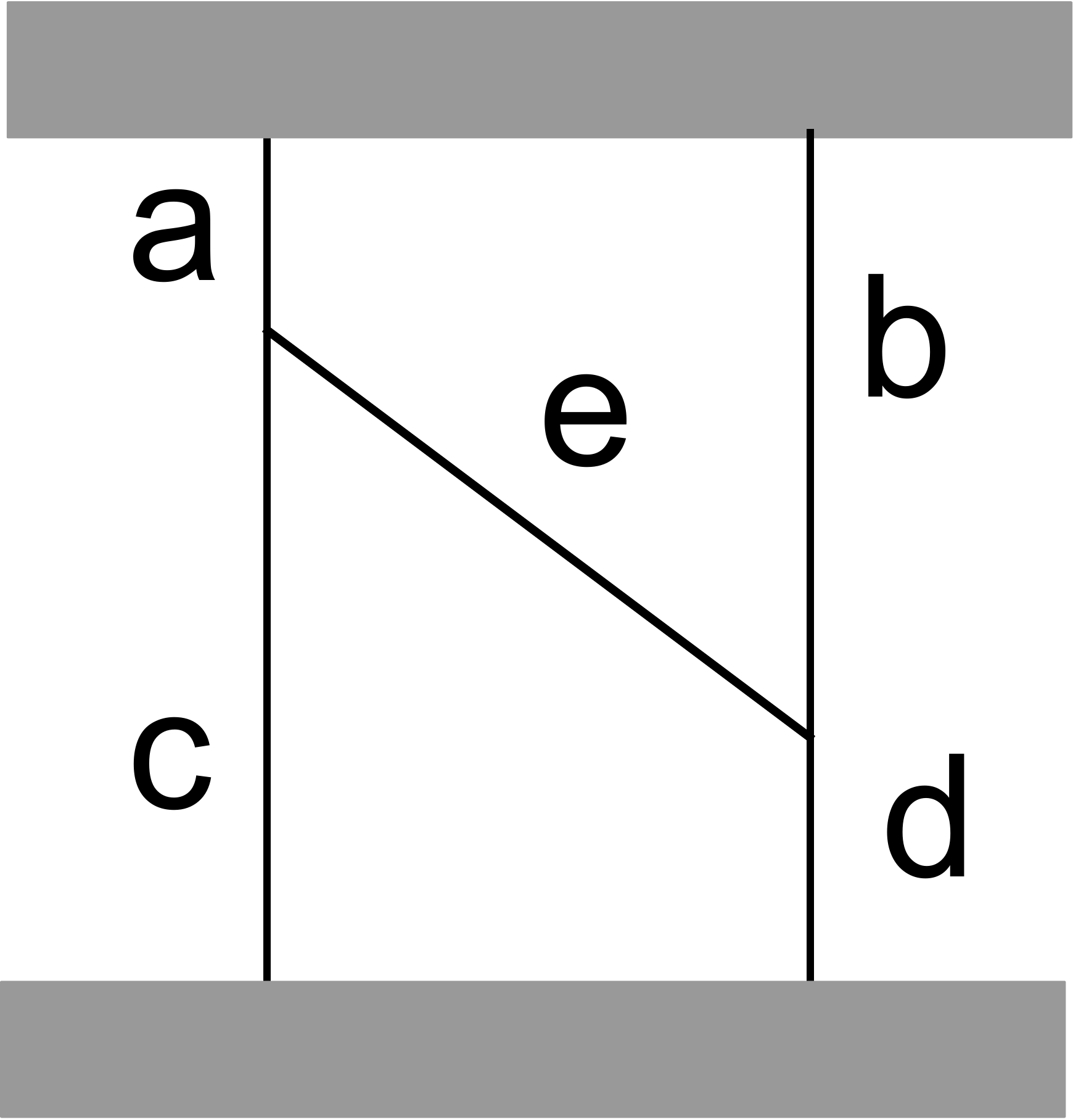}} \right)&=\sum_f [F^{ab}_{cd}]_{ef}\Phi\left(\raisebox{-0.22in}{\includegraphics[height=0.5in]{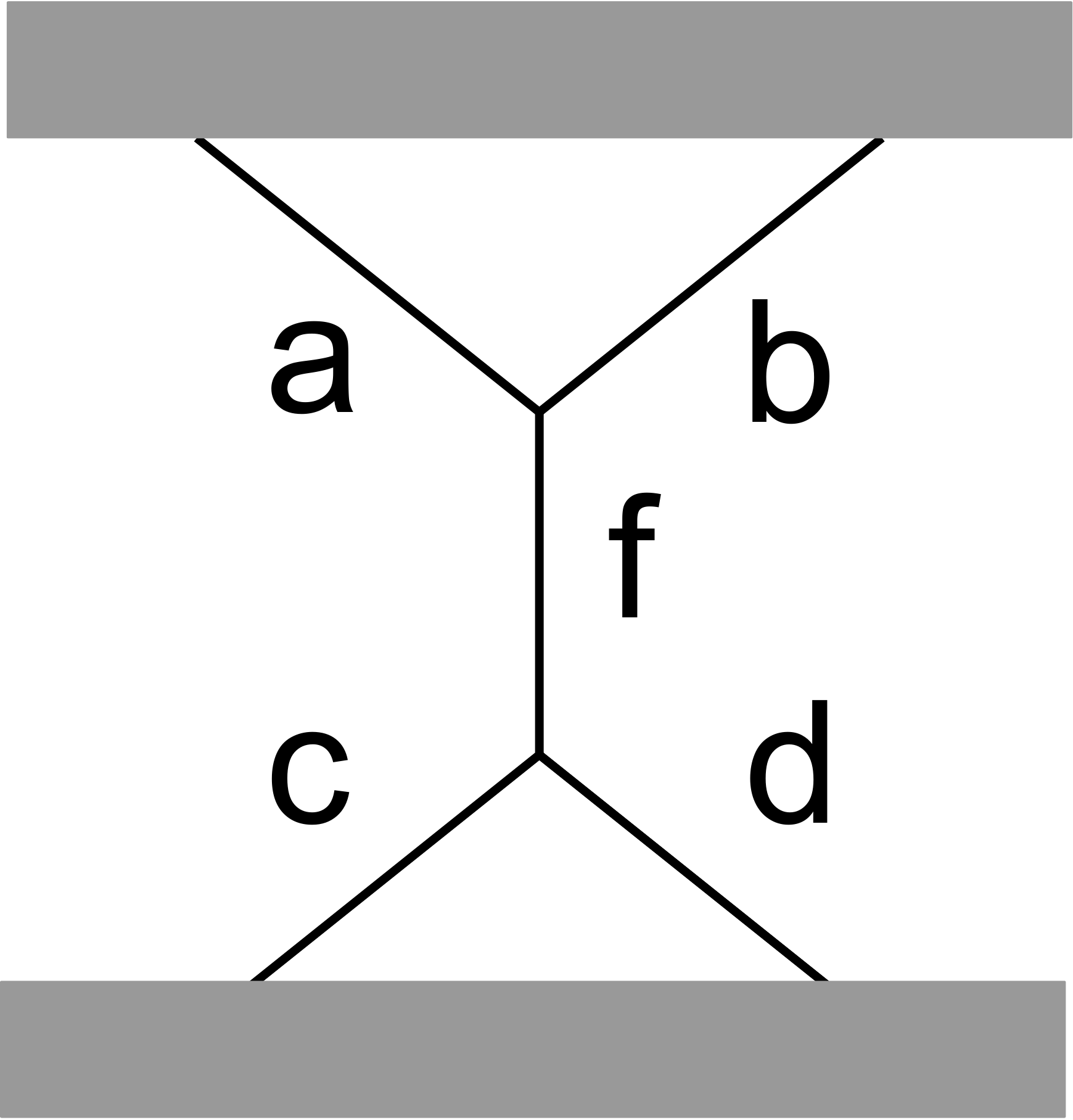}} \right) \label{1e}\\
	\Phi\left(\raisebox{-0.22in}{\includegraphics[height=0.5in]{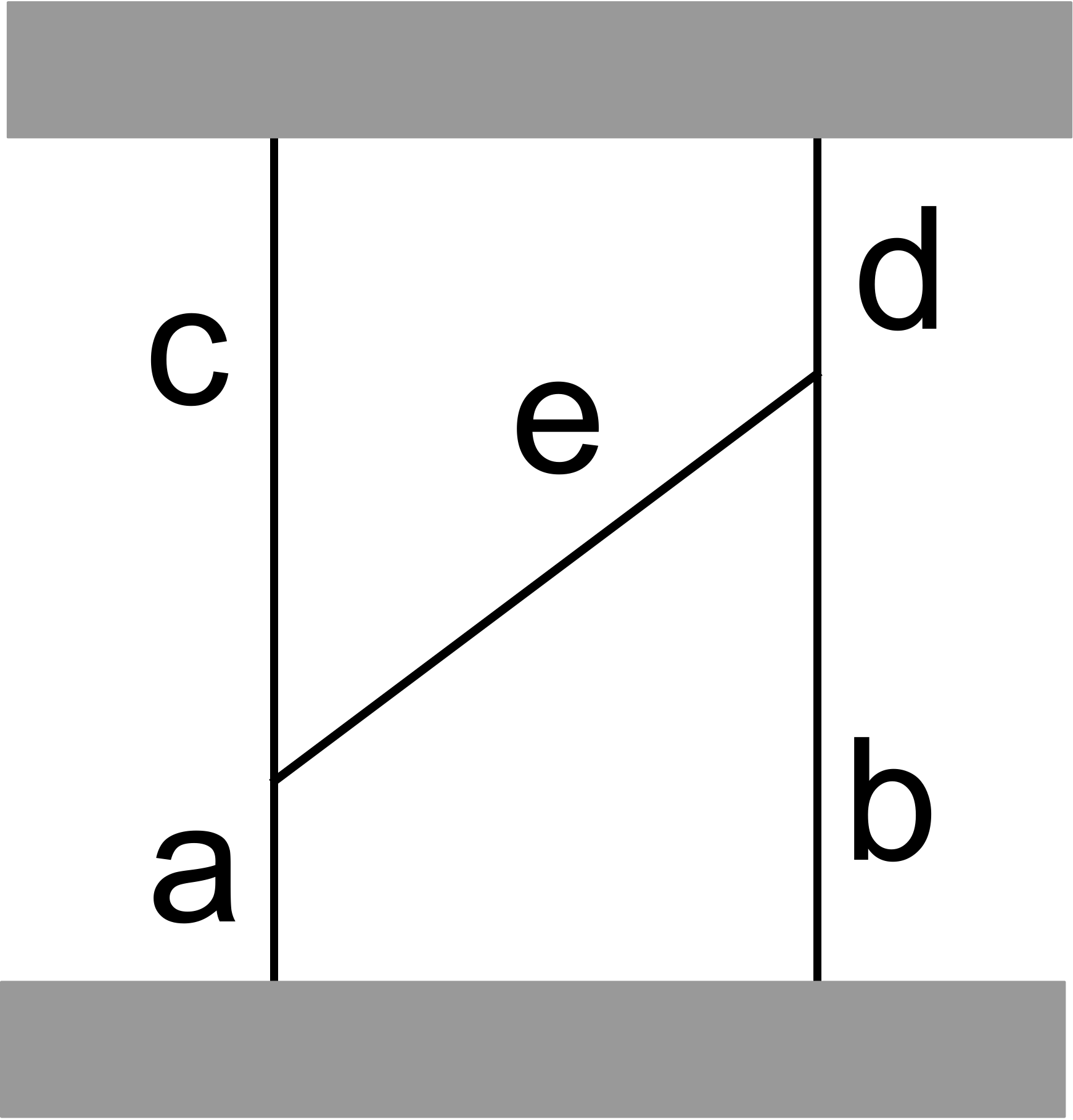}} \right)&=\sum_f [\tilde{F}^{ab}_{cd}]_{ef} \Phi\left(\raisebox{-0.22in}{\includegraphics[height=0.5in]{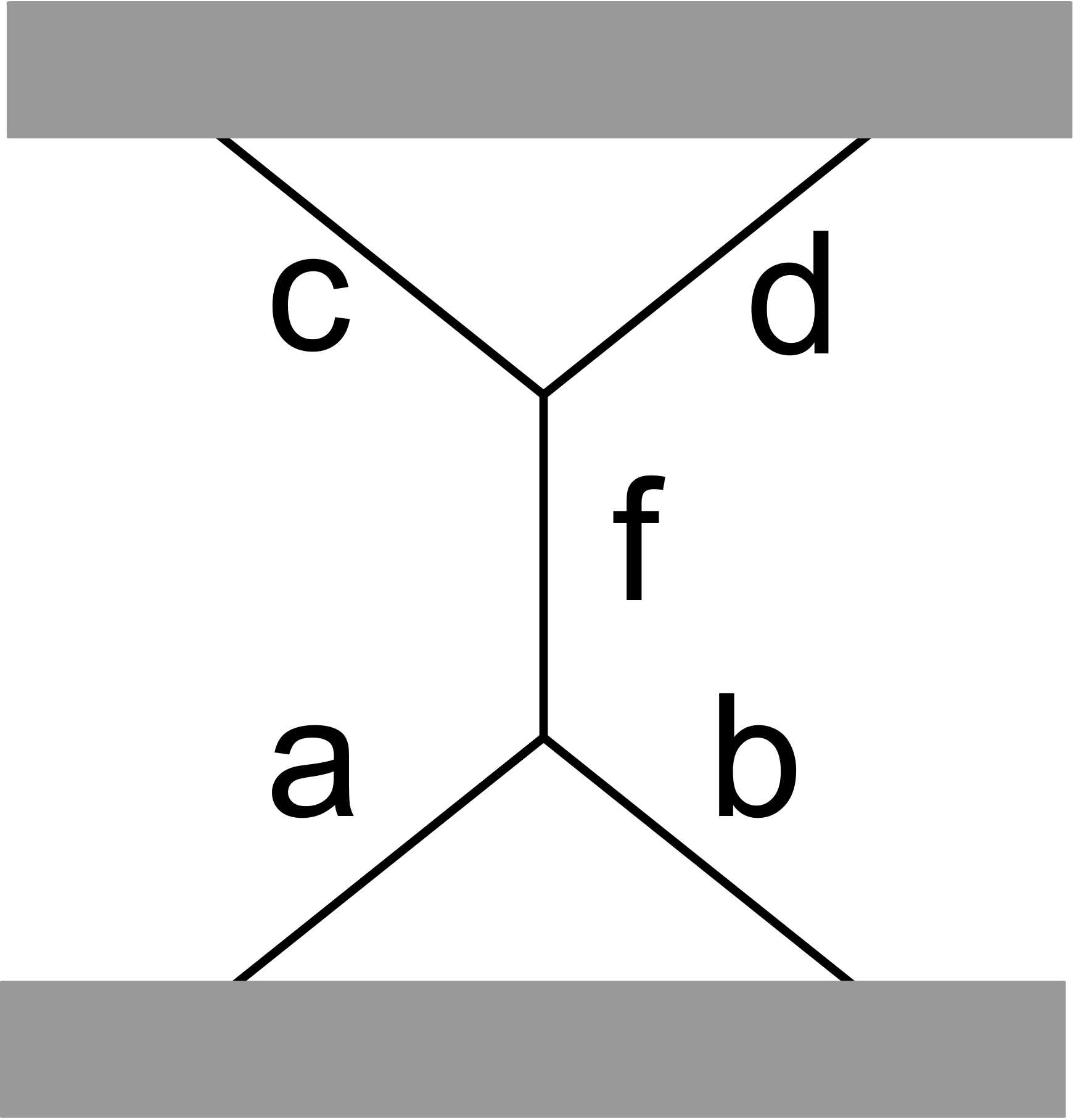}} \right) \label{1f}		
	\end{align}
	\label{localrules1}
\end{subequations}
with 
\begin{subequations}
	\begin{align}
		[F^{ab}_{cd}]_{ef} &
		=(F^{ceb}_f)^{-1}_{da} \frac{Y^{ce}_a}{Y^{cd}_f}
		\label{3c}\\
	[\tilde{F}^{ab}_{cd}]_{ef} &= F^{ceb}_{fad}  \frac{Y^{eb}_d}{Y^{ab}_f} \label{3d}
	\end{align}
	\label{consistency1}
\end{subequations}
Here, $(F^{abc}_d)^{-1}_{fe}$ is the matrix element of the \emph{inverse} of $(F^{abc}_d)$ where $F^{abc}_d$ is the matrix defined by $(F^{abc}_{d})_{ef} \equiv F^{abc}_{def}$.
 These two rules (\ref{3c}-\ref{3d}) can be derived from the basic rules (\ref{localrules}) (see Appendix \ref{app:sfc}).

To see how (\ref{localrules1}) facilitates the computation of $\Phi$, we re-evaluate the second example:
\begin{equation}
	\begin{split}
		\Phi\left(\raisebox{-0.15in}{\includegraphics[height=0.4in]{e2a.pdf}} \right)&=\Phi \left(\raisebox{-0.15in}{\includegraphics[height=0.4in]{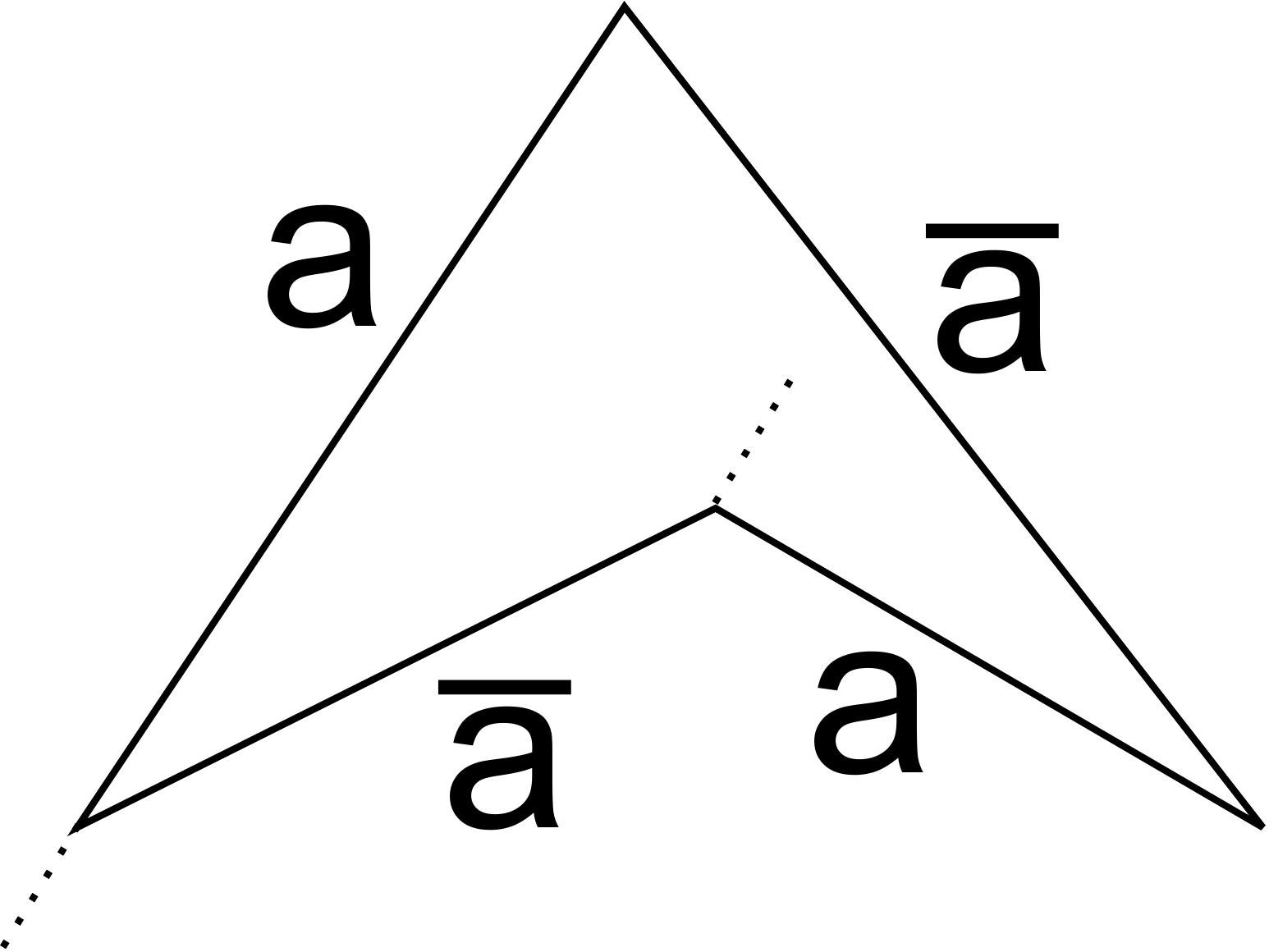}} \right)\\
		&=[\tilde{F}^{0a}_{a0}]_{\bar{a}a}\Phi \left(\raisebox{-0.15in}{\includegraphics[height=0.4in]{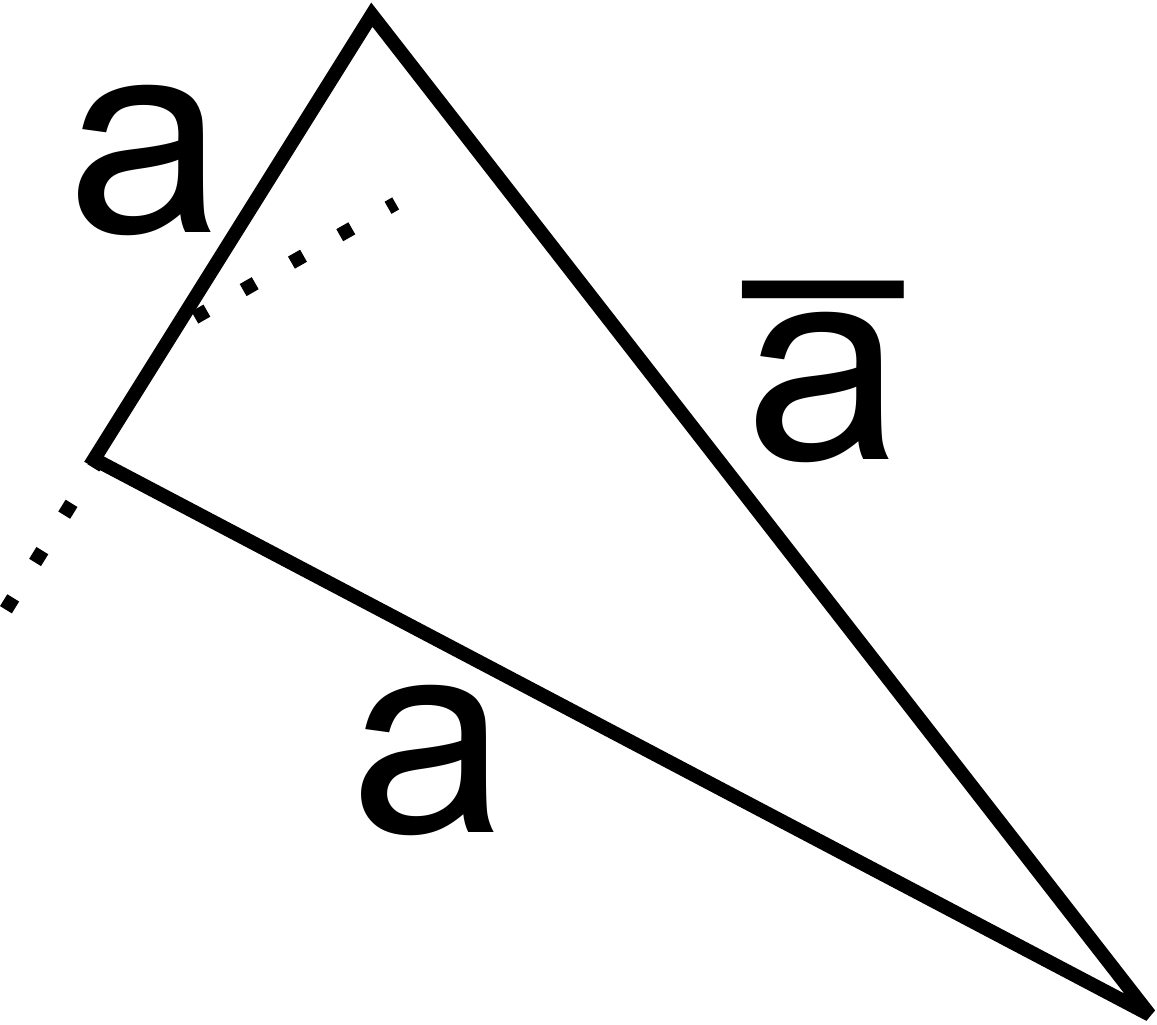}} \right)\\
		&=[\tilde{F}^{0a}_{a0}]_{\bar{a}a} Y^{a\bar{a}}_0 \Phi (\text{vacuum}) \\
		&= [\tilde{F}^{0a}_{a0}]_{\bar{a}a} Y^{a\bar{a}}_0.
	\end{split}
	\label{}
\end{equation}
In the first step, we add two null strings and then use Eq.~(\ref{1f}) in the second step. Next, we erase the null strings and use (\ref{1c}) to relate the amplitude of a loop to the amplitude of the vacuum. Finally, we use the normalization convention (\ref{vac0}). 
Notice that in terms of the auxiliary rules (\ref{localrules1}), we do not need to continuously deform the graph as in (\ref{ex2}) in order to use (\ref{localrules}).  This is useful, since in practice it may not be obvious how to properly deform the graph to use (\ref{localrules}) in more complicated  configurations.

\subsection{Self-consistency conditions}
\begin{figure}[ptb]
\begin{center}
\includegraphics[width=0.7\columnwidth]{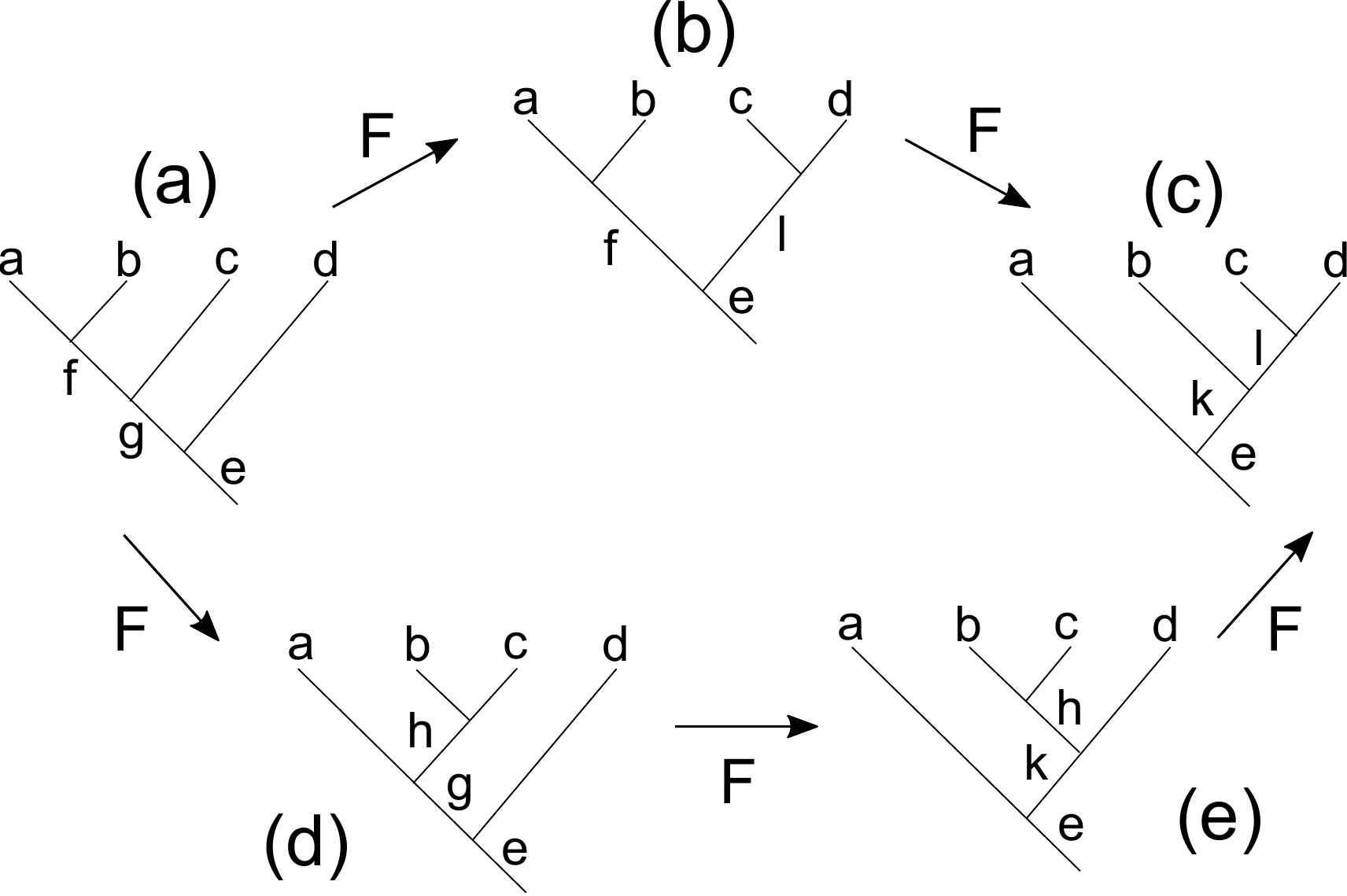}
\end{center}
\caption{Two different ways to relate the amplitude of (a) to the amplitude of (c). Consistency requires the two sequences of operations give the same result.
} 
\label{fig:pi0}
\end{figure}

In general there are multiple ways to compute the amplitude of each string-net configuration, since there are multiple ways to resolve a diagram using the local rules (\ref{localrules}). If we choose the data $\{F^{abc}_{cde},\tilde{F}^{abc}_{cde},Y^{ab}_c\}$ in an arbitrary way then these different computations will give different answers, i.e. the rules/constraints will not be \emph{self-consistent}. Thus we must impose special conditions on $\{ F^{abc}_{cde}, \tilde{F}^{abc}_{cde}, Y^{ab}_c \}$ to get self-consistent rules and a well-defined wave function $\Phi$. In particular, we claim that the following conditions are both necessary and sufficient for the rules to be self-consistent:
\begin{subequations}
	\begin{align}
		F^{fcd}_{egl}F^{abl}_{efk}&=\sum_h F^{abc}_{gfh} F^{ahd}_{egk} F^{bcd}_{khl} \label{3a}\\
		\tilde{F}^{abc}_{def} & =(F^{abc}_d)^{-1}_{fe}\frac{Y^{ab}_e Y^{ec}_d}{Y^{bc}_f Y^{af}_d} \label{3b} \\
		F^{abc}_{def} &=\tilde{F}^{abc}_{def}=1 \quad \text{if }a\text{ or }b \text{ or }c=0 \label{f1a}\\
		Y^{ab}_c &=1 \quad \text{if }a\text{ or }b=0 \label{f1b}
	\end{align}
	\label{consistency}
\end{subequations}
Equation (\ref{3a}) is known as the ``pentagon identity'' in fusion category theory. To see why it is \emph{necessary} for the rules to be self-consistent, consider the sequence of manipulations shown in Fig. \ref{fig:pi0}. We can see that the amplitude of the string-net configurations (a) and (c) can be related to one another in two different ways: (a)$\rightarrow$(b)$\rightarrow$(c) and (a)$\rightarrow$(d)$\rightarrow$(e)$\rightarrow$(c). Clearly $F$ must satisfy equation (\ref{3a}) for these two relations to be consistent with one another. The necessity of equation (\ref{3b}) follows from a similar consistency requirement (see Appendix \ref{app:sfc}). As for equations (\ref{f1a},\ref{f1b}), the necessity of these conditions follows from our convention that we can freely add or erase a null string. Proving that equations (\ref{consistency}) are \emph{sufficient} to ensure self-consistency is harder; we discuss this issue in Appendix \ref{app:localrules}.

While the conditions (\ref{consistency}) are sufficient to construct a well-defined wave function, our construction aims to do more: we wish to construct a wave function that is the ground state of an exactly solvable, Hermitian parent Hamiltonian. To do this, we impose four more conditions  on $F^{abc}_{def}$ and $Y^{ab}_c$:
\begin{subequations}
	\begin{gather}
		(F^{abc}_d)^{-1}_{ef}=(F^{abc}_{dfe})^* \label{unitary} \\
		|F^{ab\bar{b}}_{ac0}| =\sqrt{\frac{d_c}{d_a d_b}}  
		\label{y0}\\
		|Y^{ab}_c|=\sqrt{\frac{d_a d_b}{d_c}}  
		\label{ynorm}\\
		Y^{a\bar{a}}_0=(Y^{\bar{a}a}_0)^*. \label{y1}
	\end{gather}
	\label{hermicity}
\end{subequations}
Here $d_a$ is defined in (\ref{qdim}).  The significance of the above constraints (\ref{hermicity}) is that they ensure the \emph{Hermiticity} of the parent Hamiltonian (\ref{hsn0}) that has $|\Phi\>$ as its ground state (see Appendix \ref{app:property}).  Conversely, violating the constraints (\ref{hermicity}) can sometimes lead to a $|\Phi\>$ that is not the ground state of any gapped Hermitian Hamiltonian \cite{FreedmanWangGalois}.

 Eqs.~(\ref{consistency}) and (\ref{hermicity}) are the only conditions that we will impose on $\{F^{abc}_{cde},\tilde{F}^{abc}_{cde},Y^{ab}_c\}$. We will see that, for every solution $\{F^{abc}_{cde},\tilde{F}^{abc}_{cde},Y^{ab}_c\}$ to Eqs.~(\ref{consistency}) and (\ref{hermicity}), we can construct both a string-net wave function $\Phi$ and an exactly solvable Hermitian parent Hamiltonian that has $\Phi$ as its ground state.

\subsection{Local unitary transformations and gauge equivalence}
Given a solution $\{F^{abc}_{def}, \tilde{F}^{abc}_{def}, Y^{ab}_c\}$ to Eqs.~(\ref{consistency},\ref{hermicity}), we can construct an infinite class of other solutions $\{\hat{F}^{abc}_{def}, \hat{\tilde{F}}^{abc}_{def}, \hat{Y}^{ab}_c\}$ by defining
\begin{equation}
	\begin{split}
		\hat{F}^{abc}_{def}&=F^{abc}_{def}\cdot \frac{f^{ab}_ef^{ec}_d}{f^{bc}_f f^{af}_d}  \\
		\hat{\tilde{F}}^{abc}_{def}&=\tilde{F}^{abc}_{def}\cdot \frac{f^{e}_{ab}f^{d}_{ec}}{f^{f}_{bc} f^{d}_{af}}\\
		\hat{Y}^{ab}_c & = Y^{ab}_c
	\end{split}
	\label{fgauge}
\end{equation}
Here $f^{ab}_c, f^c_{ab}$ are complex numbers that depend on $a,b,c$ and that satisfy
\begin{equation}
\begin{split}
		|f^{ab}_c|&=1, \quad f^c_{ab}=\frac{1}{f^{ab}_c}, \\
		f^{ab}_c&=1 \text{ if }a \text{ or }b=0.
	\end{split}
	\label{gt}
\end{equation}
In addition, we can construct solutions by defining
\begin{equation}
	\begin{split}
		\hat{F}^{abc}_{def}&={F}^{abc}_{def} \\
		\hat{\tilde{F}}^{abc}_{def}&=\tilde{F}^{abc}_{def} \cdot \frac{g_{ab}^e g_{ec}^d}{g_{bc}^f g_{af}^d}\\
		\hat{Y}^{ab}_c & = Y^{ab}_c \cdot g_{ab}^c\\
	\end{split}
	\label{ggauge}
\end{equation}
where $g^{c}_{ab}$ satisfies
\begin{equation}
	\begin{split}
	 |g_{ab}^c|&=1, \quad g_{a\bar{a}}^0 = (g_{\bar{a}a}^0)^*,\\
	g_{ab}^c &=1 \text{ if } a \text{ or }b=0.
	\end{split}
	\label{}
\end{equation}

If two solutions $\{F^{abc}_{def}, \tilde{F}^{abc}_{def}, Y^{ab}_c\}$ and $\{\hat{F}^{abc}_{def}, \hat{\tilde{F}}^{abc}_{def}, \hat{Y}^{ab}_c\}$, are related by one of the above transformations, we will say that they are ``gauge equivalent''. The reason for this terminology is that the corresponding wave functions,  $\Phi$ and $\hat{\Phi}$, are very closely related: there exists a local unitary transformation $U$ such that $U |\hat{\Phi}\> = |\Phi\>$.  In the case of the $f$-transformation, this local unitary is defined by
\begin{equation}
\begin{split}
    U \left| \raisebox{-0.22in}{\includegraphics[height=0.5in]{branching1.pdf}}\right>  
    &= f^{ab}_c  \left| \raisebox{-0.22in}{\includegraphics[height=0.5in]{branching1.pdf}}\right> 
    \\
	U \left| \raisebox{-0.22in}{\includegraphics[height=0.5in]{branching2.pdf}}\right>
	&= f^{c}_{ab}\left| \raisebox{-0.22in}{\includegraphics[height=0.5in]{branching2.pdf}}\right> .
\end{split}
\end{equation}
Similarly, the $U$ associated with the $g$-transformation is defined by
\begin{equation}
\begin{split}
    U\left |\raisebox{-0.22in}{\includegraphics[height=0.5in]{branching1.pdf}}\right>  
    &= \left |\raisebox{-0.22in}{\includegraphics[height=0.5in]{branching1.pdf}}\right>  
    \\
	U\left |\raisebox{-0.22in}{\includegraphics[height=0.5in]{branching2.pdf}}\right> 
	&= g^{c}_{ab} \left|\raisebox{-0.22in}{\includegraphics[height=0.5in]{branching2.pdf}}\right> .
\end{split}    
\end{equation}
Here, the above notation means that $U$ multiplies each string-net basis state by a product of $f^{ab}_c$'s and $f^c_{ab}$'s --- one for each of the above (trivalent) vertices.

It is worth noting that the $f$ and $g$-gauge transformations have a different status in the fusion category literature: while the $f$-gauge transformations (\ref{fgauge}) are well-known, the $g$-gauge transformations (\ref{ggauge}) are largely absent. The reason for this is that $Y^{ab}_c$ is usually chosen to have a fixed value in the fusion category literature (e.g. see Eq.~(\ref{ygauge}) below), thus ruling out non-trivial $g$-gauge transformations.

\subsection{Simplifying the string-net consistency conditions \label{solve}}

Given that our string-net modes are in one-to-one correspondence with solutions $\{F^{abc}_{def}, \tilde{F}^{abc}_{def}, Y^{ab}_c\}$ to (\ref{consistency},\ref{hermicity}),  it is worth pausing to note some simplifications that facilitate finding a solution. First, notice that Eq.~(\ref{3b}) completely determines $\tilde{F}^{abc}_{def}$ in terms of $F^{abc}_{def}$ and $Y^{ab}_{c}$. This means that we can essentially forget about $\tilde{F}^{abc}_{def}$ and focus on finding $\{F^{abc}_{def}, Y^{ab}_{c}\}$ that obey the remaining equations: (\ref{3a}), (\ref{f1a}), (\ref{f1b}) and (\ref{hermicity}).

Second, the quantum dimensions $d_a$ are in fact completely fixed by the branching rules. To see this, take the square of both sides of (\ref{y0}) and then sum over $c$. Using (\ref{unitary}) gives
\begin{equation}
d_a d_b =\sum_c \delta^{ab}_c d_c
	\label{did}
\end{equation}
Eq.~(\ref{did}) can be thought of as an eigenvalue equation for the matrix $N(a)$ defined by $[N(a)]_{bc} \equiv \delta^{ab}_c$: from this point of view, Eq.~(\ref{did}) tells us that $N(a)$ has an eigenvector $v$ whose components are $v_c \equiv d_c$, and whose corresponding eigenvalue is $d_a$. Given that $N(a)$ is a non-negative matrix and $v_c$ is strictly positive, the Perron-Frobenius theorem implies that $d_a$ is the largest eigenvalue of $N(a)$\cite{KitaevHoneycomb}.  In particular, $d_a$ is completely determined by the branching rules, as we wished to show.

For the last simplification, notice that we can always make $Y^{ab}_c$ real and positive using an appropriate $g$-gauge transformation (\ref{ggauge}). After we make such a transformation, then (\ref{ynorm}) implies that
\begin{equation}
	Y^{ab}_c= \sqrt{\frac{d_a d_b}{d_c}} 
	\label{ygauge}
\end{equation}
Hence we can take $Y^{ab}_c =\sqrt{\frac{d_a d_b}{d_c}}$
without loss of generality. Notice that this choice for $Y^{ab}_c$ automatically satisfies Eqs.~(\ref{ynorm},\ref{y1}).  Other convenient gauge choices for $Y^{ab}_c$ are discussed in Appendix \ref{app:yabc}.

Putting this all together, we conclude that $F^{abc}_{def}$ is the only quantity that needs to be determined. Thus, the problem of solving the consistency equations reduces to finding all $F^{abc}_{def}$ that obey (\ref{3a}), (\ref{f1a}), (\ref{unitary}), and (\ref{y0}) where $d_a$ is fixed by the branching rules as discussed above. 
Finding such solutions is not trivial; see Refs. \onlinecite{BondersonThesis,Rowell2009}  for a discussion of how such solutions can be found in practice, as well as a discussion of many interesting examples.

\subsection{Examples of solutions to consistency conditions}

We now discuss three general classes of solutions to the consistency conditions (\ref{consistency},\ref{hermicity}). \\

{\bf 1}. For any finite group $G$, we can construct a solution to the consistency conditions (\ref{consistency},\ref{hermicity}) by defining the string types to be the irreducible representations of $G$, the dual string type $\bar{a}$ to be the dual representation of $a$, and the branching rules to be the set of all triplets $\{(a,b ; c)\}$ such that $c$ appears in the tensor product $a \otimes b$. (Here we assume that $c$ appears with multiplicity of at most $1$ for simplicity). Next, we define $F^{abc}_{def}$ to be the $6j$ symbol corresponding to $G$, and we define $Y^{ab}_c = \sqrt{d_a d_b/d_c}$ where $d_a$ is the dimension of the representation $a$. Like any solution to the consistency conditions, this solution can be used to construct an exactly soluble lattice Hamiltonian $H$ with anyon excitations, as we explain later. The topological order in this model is identical to that of a discrete gauge theory with gauge group $G$ -- also known as the ``quantum double'' of $G$ \cite{KitaevToric} (see Sec.~\ref{z2example} for the example $G = \mathbb{Z}_2$). \\
\\
{\bf 2}. For any finite group $G$ and any cocycle $\omega \in H^3(G, U(1))$, we can set the string types to be group elements $g \in G$, and the dual string type $\bar{a}$ to be the inverse $a^{-1}$, and the branching rules to be the set of all $\{(a,b; c)\}$ such that $c = ab$.  We define $Y^{ab}_c = 1$ and $F^{abc}_{def} = \omega(a,b,c)$  with $d, e, f$ determined by $a, b, c$ according to $d = abc$, and $e= ab$, and $f = bc$. In this case, the corresponding lattice model realizes a Dijkgraaf-Witten theory with group $G$ and cocycle $\omega$ -- also known as the ``twisted quantum double'' of $G$ \cite{HuWanWu12} (see Secs.~\ref{z2example}-\ref{z4example} for examples). \\
\\
{\bf 3}. Given any topological order $\mathcal{T}$, we define the string types to be the anyons in $\mathcal{T}$, and the dual string type $\bar{a}$ to be the antiparticle of $a$, and the branching rules to be the set of all $\{(a,b ; c)\}$ such that $c$ appears in the fusion product $a \times b$. (Here we are assuming that $\mathcal{T}$ has no fusion multiplicity for simplicity). Next, we define $F^{abc}_{def}$ to be the $F$-symbol of the anyons in $\mathcal{T}$, and we define $Y^{ab}_c = \sqrt{d_a d_b/d_c}$ where $d_a$ is the quantum dimension of anyon $a$. In this case, the string-net model realizes a ``doubled'' topological order of the form $\mathcal{T} \times \mathcal{T}^{op}$ where $\mathcal{T}^{op}$ is the time reversal of $\mathcal{T}$ (see Sec.~\ref{z2example} and \ref{FibonacciExample} for examples).

\section{Lattice Hamiltonian \label{sec:h}}

So far we have shown that each solution $\{F^{abc}_{def}, \tilde{F}^{abc}_{def}, Y^{ab}_c\}$ to equations (\ref{consistency},\ref{hermicity}) defines a string-net wave function $|\Phi\>$ via the local rules (\ref{localrules}). In this section, we show how to construct a corresponding exactly solvable lattice Hamiltonian whose ground state is a lattice version of $|\Phi \rangle$.

\subsection{Definition of Hamiltonian}

Our construction takes three pieces of input: (i) a set of string types and branching rules; (ii) a definition of dual string types; and (iii) a solution $\{F^{abc}_{def}, \tilde{F}^{abc}_{def}, Y^{ab}_c\}$ to the consistency conditions (\ref{consistency}), (\ref{hermicity}). The output of our construction is an exactly solvable lattice Hamiltonian whose ground state is the string-net wave function $\Phi$ (\ref{localrules}) restricted to the lattice.

To construct our lattice model, we first assign a spin to each link of the honeycomb lattice.  Each spin can be in one of $N$ states, where $N$ is the number of string types (including the null string $0$). We will label these states by $|a\>, |b\>, |c\>$, etc., where $\{a,b,c,...\}$ are the string types. With this notation, we can associate a string-net configuration to each spin configuration in the obvious way: if a spin is in state $|a\>$, we regard the link as being occupied by a string of type $a$. Likewise, if a spin is in the state $|0\>$, we think of the link as being empty or occupied by the null string.

\begin{figure}[ptb]
\begin{center}
\includegraphics[width=0.65\columnwidth]{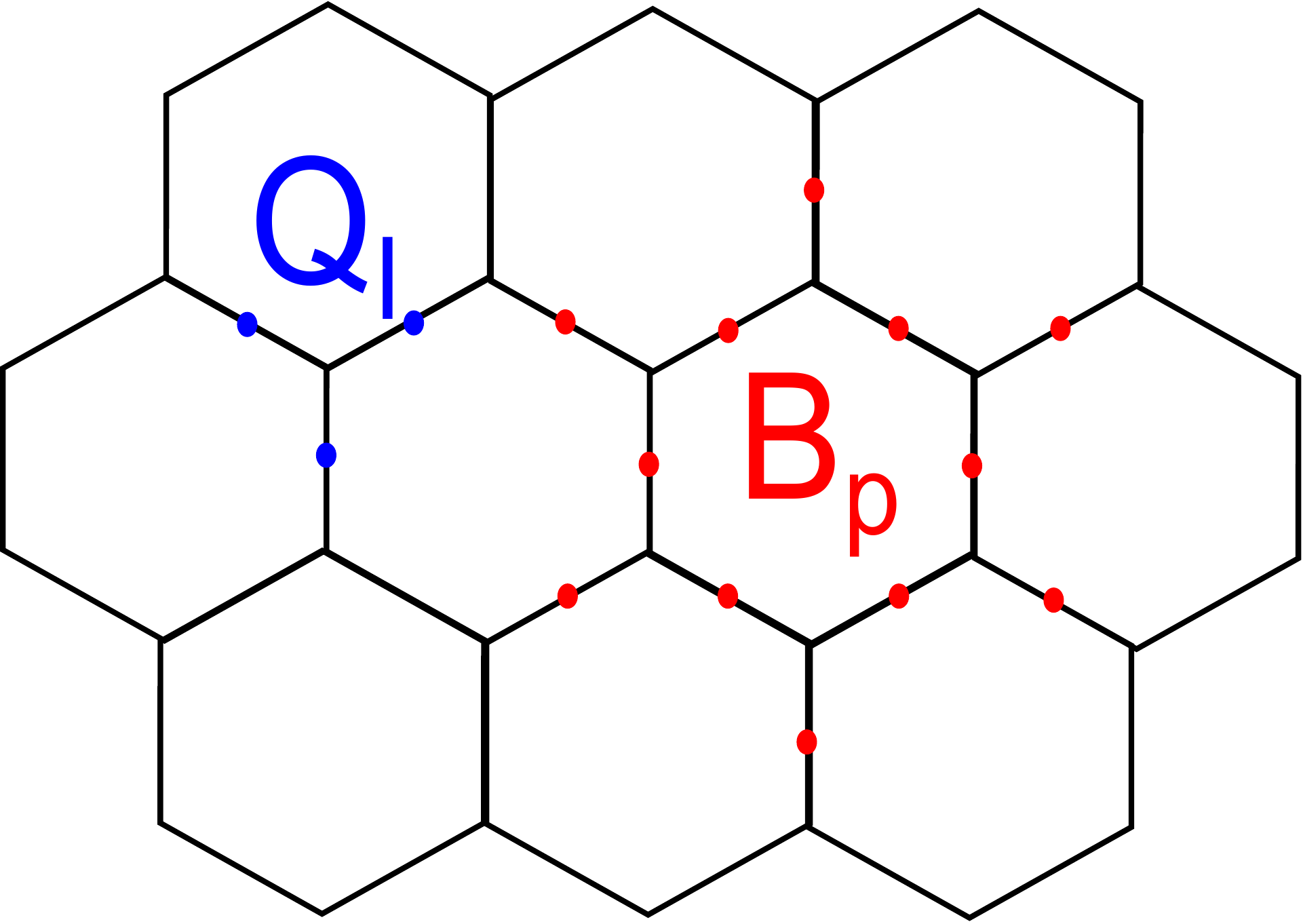}
\end{center}
\caption{The string-net Hamiltonian (\ref{hsn0}). The $Q_I$ operator acts on 3 spins around each vertex (blue dots). The $B_p$ operator acts on 12 spins adjacent to the plaquette $p$ (red dots).
} 
\label{fig:lattice0}
\end{figure}

The Hamiltonian is of the form
\begin{equation}
	H=-\sum_I Q_I-\sum_p B_p.
	\label{hsn0}
\end{equation}
Here, the two sums run over all vertices $I$ and plaquettes $p$ of the honeycomb lattice.

The $Q_I$ operator acts on the 3 spins adjacent to the vertex $I$  (Fig.~\ref{fig:lattice0}):
\begin{align}
	\begin{split}
	Q_I\left|\raisebox{-0.22in}{\includegraphics[height=0.5in]{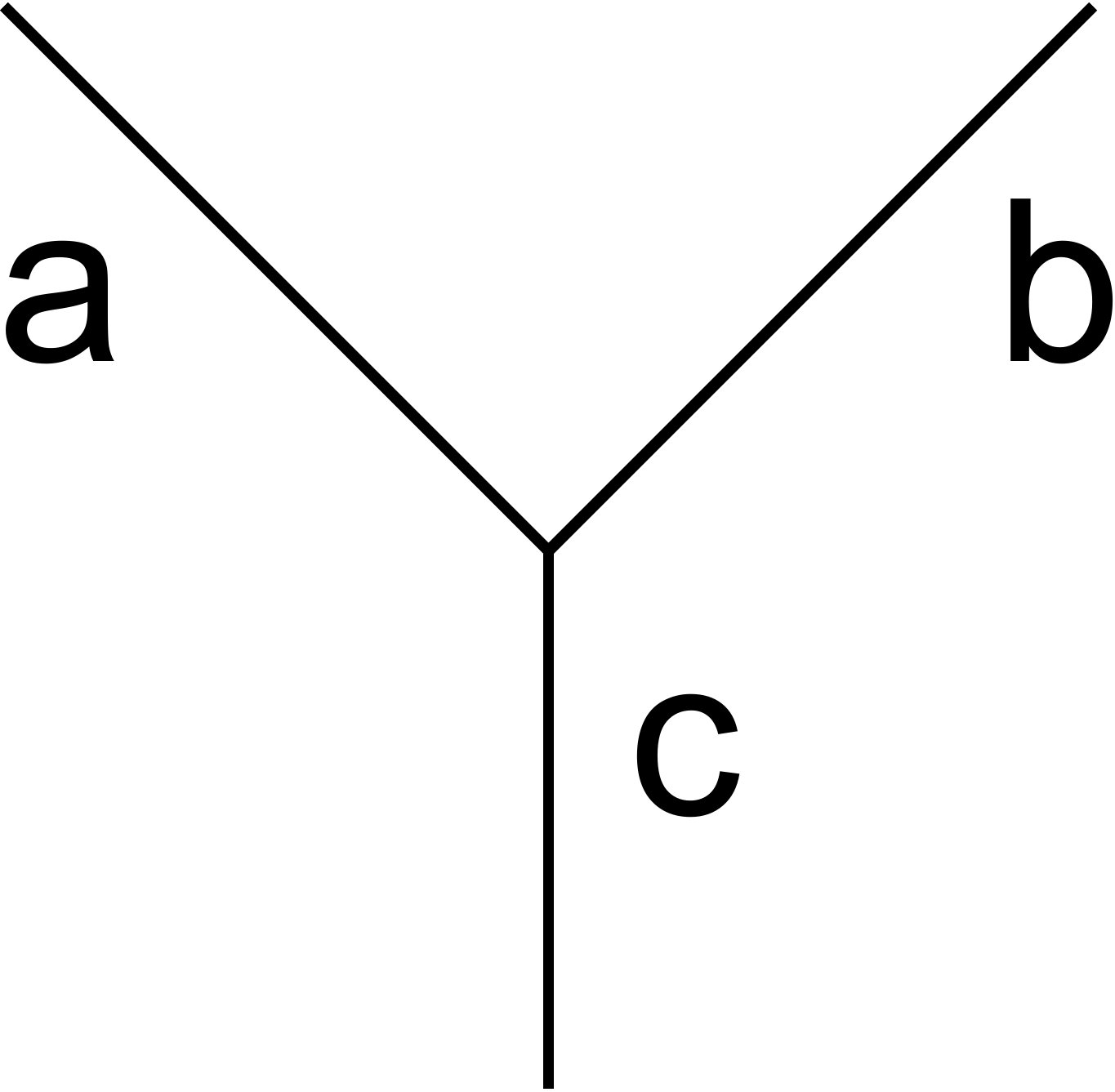}} \right\>=\delta^{ab}_c\left|\raisebox{-0.22in}{\includegraphics[height=0.5in]{Qa.pdf}} \right\> \\
	Q_I\left|\raisebox{-0.22in}{\includegraphics[height=0.5in]{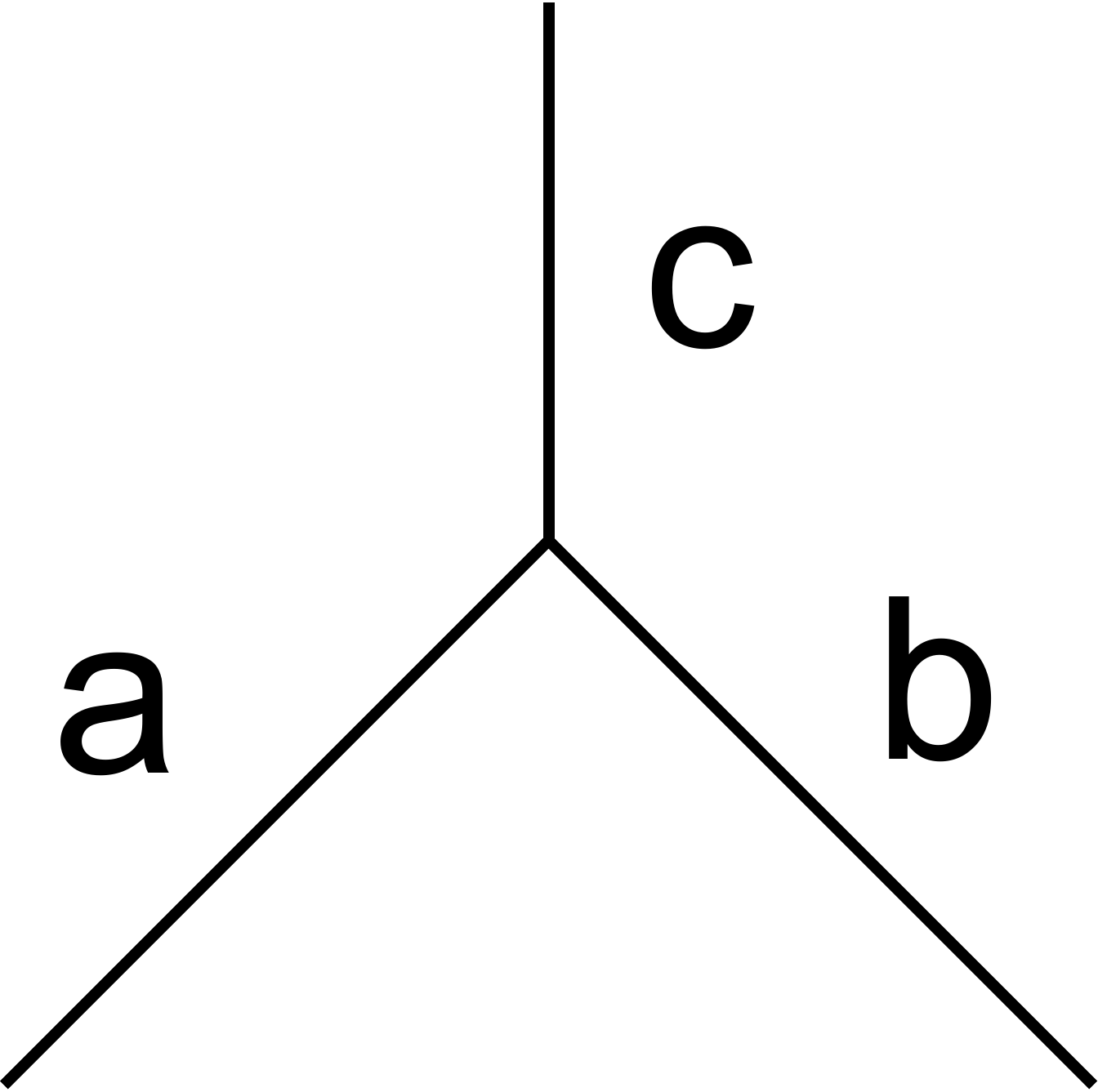}} \right\>=\delta^{ab}_c\left|\raisebox{-0.22in}{\includegraphics[height=0.5in]{Qb.pdf}} \right\>.
	\end{split}
	\label{qi}
\end{align} 
Note that the $Q_I$ term penalizes the states that do not satisfy the branching rules.

The $B_p$ operator has a more complicated structure. It is a linear combination of more basic operators, $B_p^s$:
\begin{equation}
	B_p = \sum_{s} a_s B_p^s
	\label{bp0}
\end{equation}
where the index $s$ runs over the different string types (including $s=0$) and where the coefficient $a_s$ is defined by
\begin{equation}
	a_s = \frac{Y^{\bar{s}{s}}_0}{\sum_{t} d_t^2}.
	\label{as}
\end{equation}
Each operator $B_p^s$ describes a 12 spin interaction involving the spins on the 12 links that are adjacent to the vertices of the plaquette $p$. The operator $B_p^s$ has a special structure: First, it annihilates any state that does not obey the branching rules at the 6 vertices surrounding the plaquette. Second, it acts non-trivially on the inner 6 spins but does not affect the outer 6 spins. Thus the matrix element of $B_p^s$ between two inner spin configurations $\<i_1\dots i_6|$ and $|i_1'\dots i_6\>$ depends on the state of the outer spins $(e_1\dots e_6)$. The matrix elements are defined by
\begin{equation}
	\begin{split}
	&\left <\raisebox{-0.35in}{\includegraphics[height=0.7in]{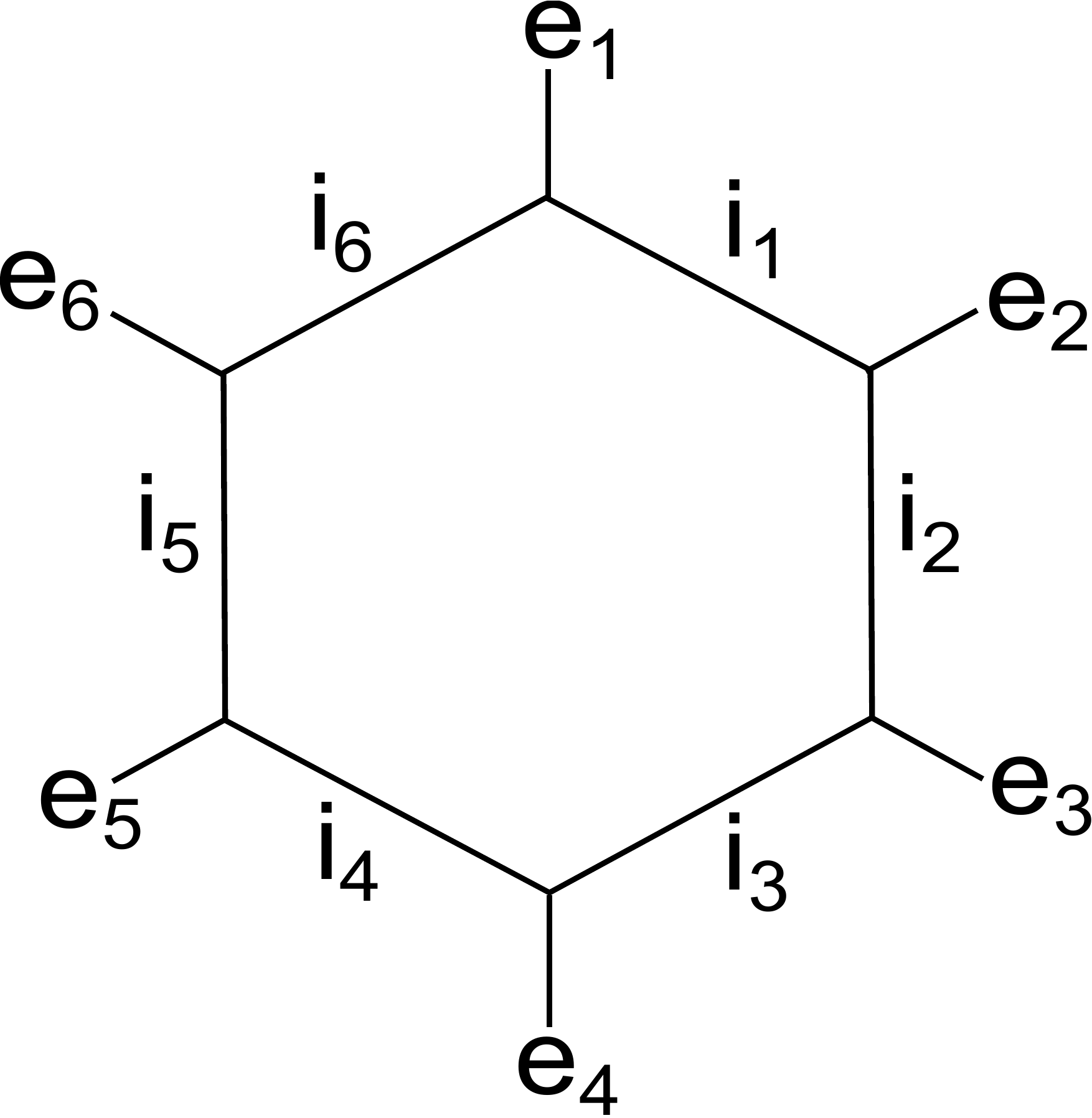}} \right|B_p^s   \left| \raisebox{-0.35in}{\includegraphics[height=0.7in]{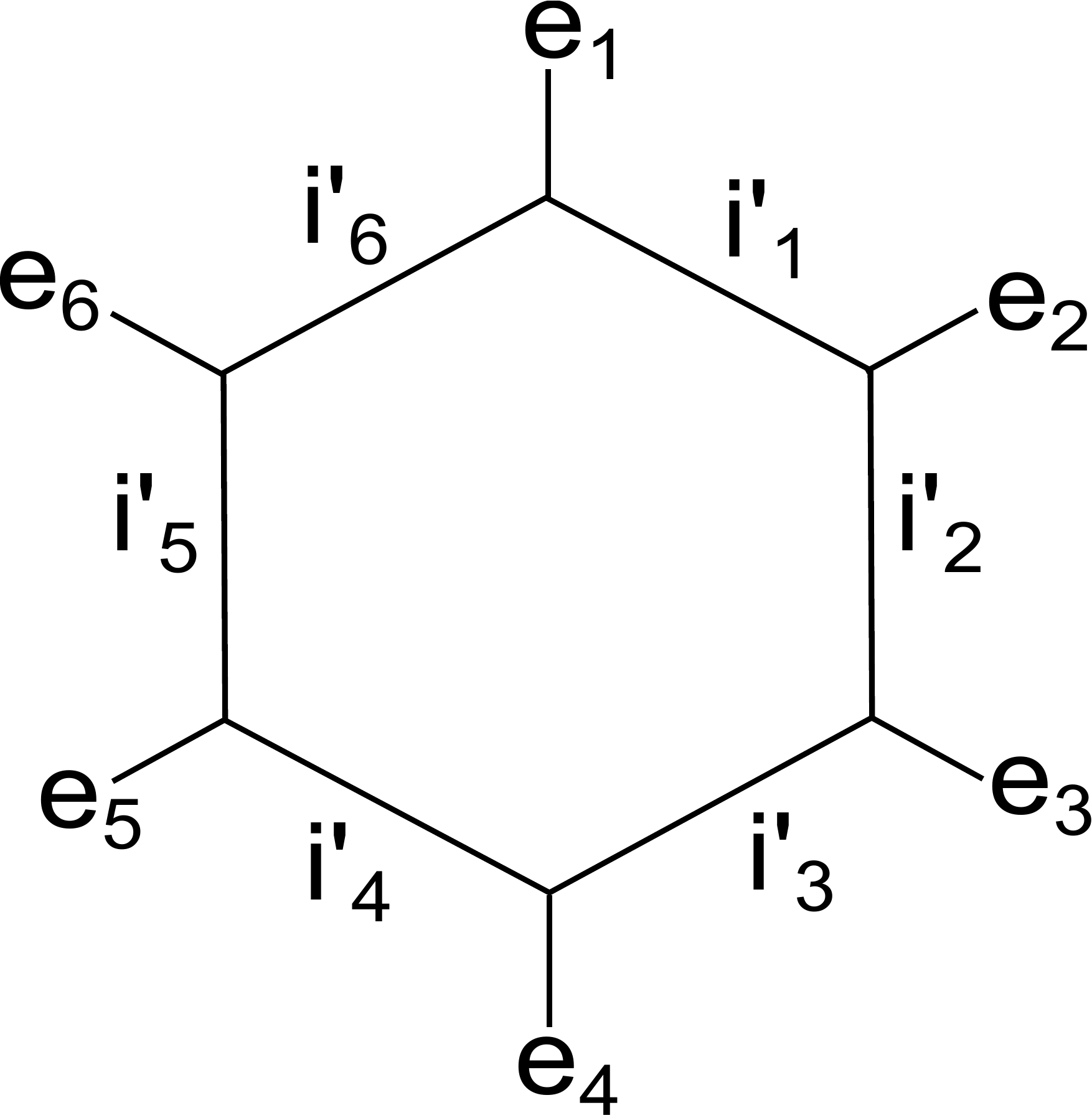}} \right>
	=
	 B_{p,i'_1i'_2\dots i'_6}^{s,i_1i_2\dots i_6}(e_1e_2\dots e_6)
	\end{split}
	\label{}
\end{equation}
where
\begin{equation}
	\begin{split}
	&B_{p,i'_1i'_2\dots i'_6}^{s,i_1i_2\dots i_6} (e_1e_2\dots e_6)=
	\frac{Y^{s\bar{s}}_0 Y^{i_6 i_1}_{e_1} Y^{i_3 e_3}_{i_2} Y^{e_5 i_4}_{i_5} }
	{Y^{i'_6 i'_1}_{e_1}  Y^{i'_3 e_3}_{i'_2} Y^{e_5 i_4'}_{i'_5}} \times \\
	&F^{\bar{s}i_3 e_3}_{i'_2 i'_3 i_2} F^{e_6 i_6 s}_{i'_5 i_5 i'_6}   F^{i'_4 \bar{s} i_3}_{e_4 i_4 i'_3}  F^{i_6 s\bar{s}}_{i_6 i'_60}
	(F^{\bar{s}i_1e_2}_{i'_2 i'_1 i_2} F^{e_5i_4s}_{i'_5i_5i'_4} F^{i_4 s \bar{s}}_{i_4 i_4'0} F^{i'_6 \bar{s}i_1}_{e_1 i_6 i'_1})^* 
	\end{split}
	\label{bps0}
\end{equation}
We emphasize that the above expression is only valid if the initial and final states obey the branching rules at each vertex; if either state violates the branching rules, the matrix element of $B_p^s$ vanishes.

We should mention that there is an alternative graphical representation for $B_p^s$ which is much simpler. It is convenient to describe this graphical representation in terms of the action of $B_p^s$ on a \emph{bra} $\<X|$ rather than a ket $|X\>$. Specifically, $B_p^s$ can be thought of as adding a loop of type-$s$ string around the boundary of $p$:
\begin{equation}
	\left\<\raisebox{-0.35in}{\includegraphics[height=0.7in]{bp0a.pdf}} \right|B_p^s=
	\left\<\raisebox{-0.35in}{\includegraphics[height=0.7in]{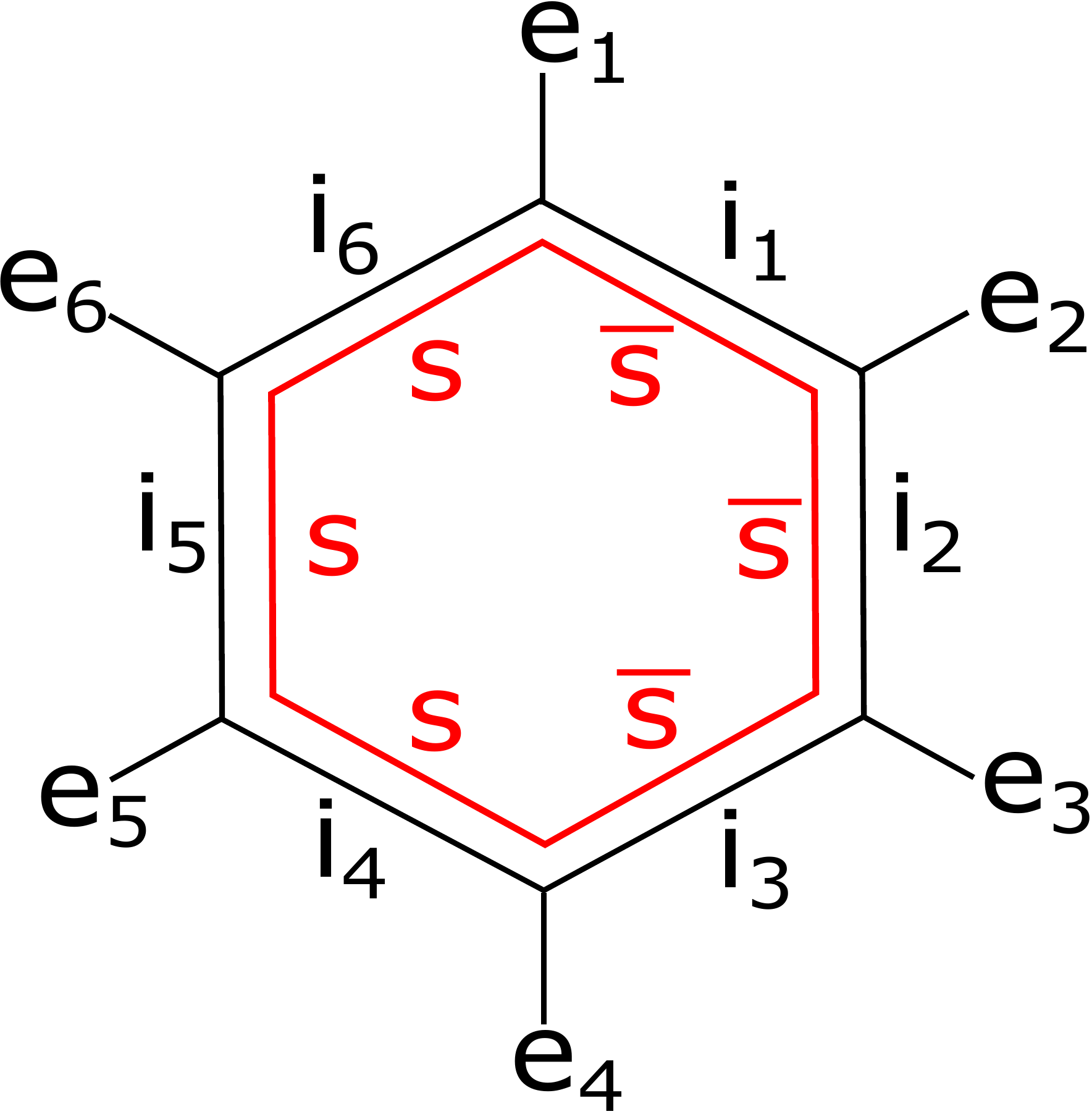}} \right|.	
	\label{graphbps}
\end{equation}
Then the matrix elements in Eq.~(\ref{bps0}) can be obtained by using the local rules (\ref{localrules}) to fuse the string $s$ onto the links along the boundary of the plaquette:
\begin{widetext}
\begin{equation} \label{Eq:Hdiagram}
\begin{split}
	&\left \<\raisebox{-0.35in}{\includegraphics[height=0.7in]{bp0a.pdf}} \right|B_p^s
	=\left \<\raisebox{-0.35in}{\includegraphics[height=0.7in]{bp0b.pdf}} \right|
	=\sum_{i'_1,\dots,i'_6}  \left \<\raisebox{-0.35in}{\includegraphics[height=0.7in]{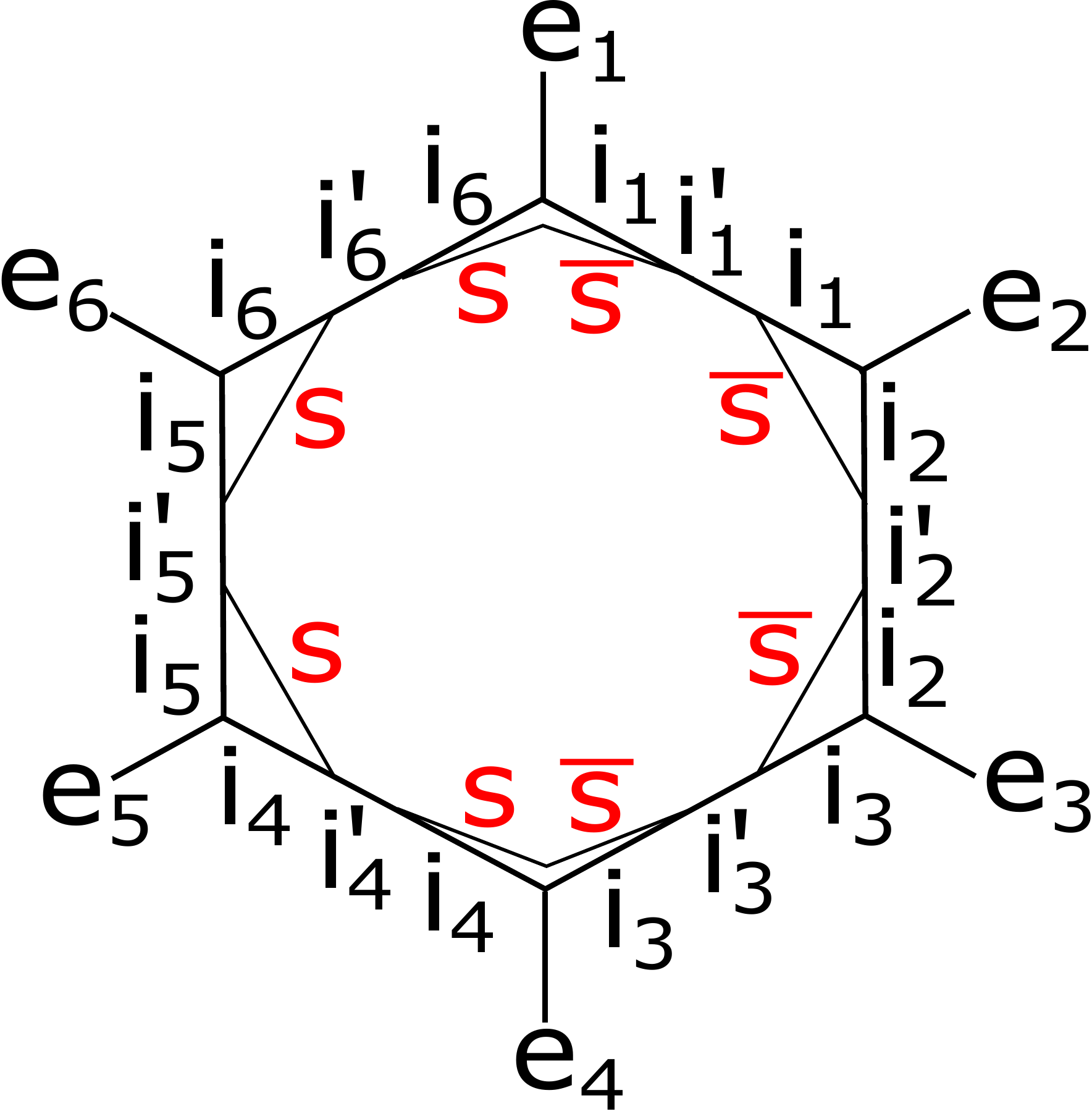}} \right| C_1 \\
	&=\sum_{i'_1,\dots,i'_6} 
	\left\< \raisebox{-0.35in}{\includegraphics[height=0.7in]{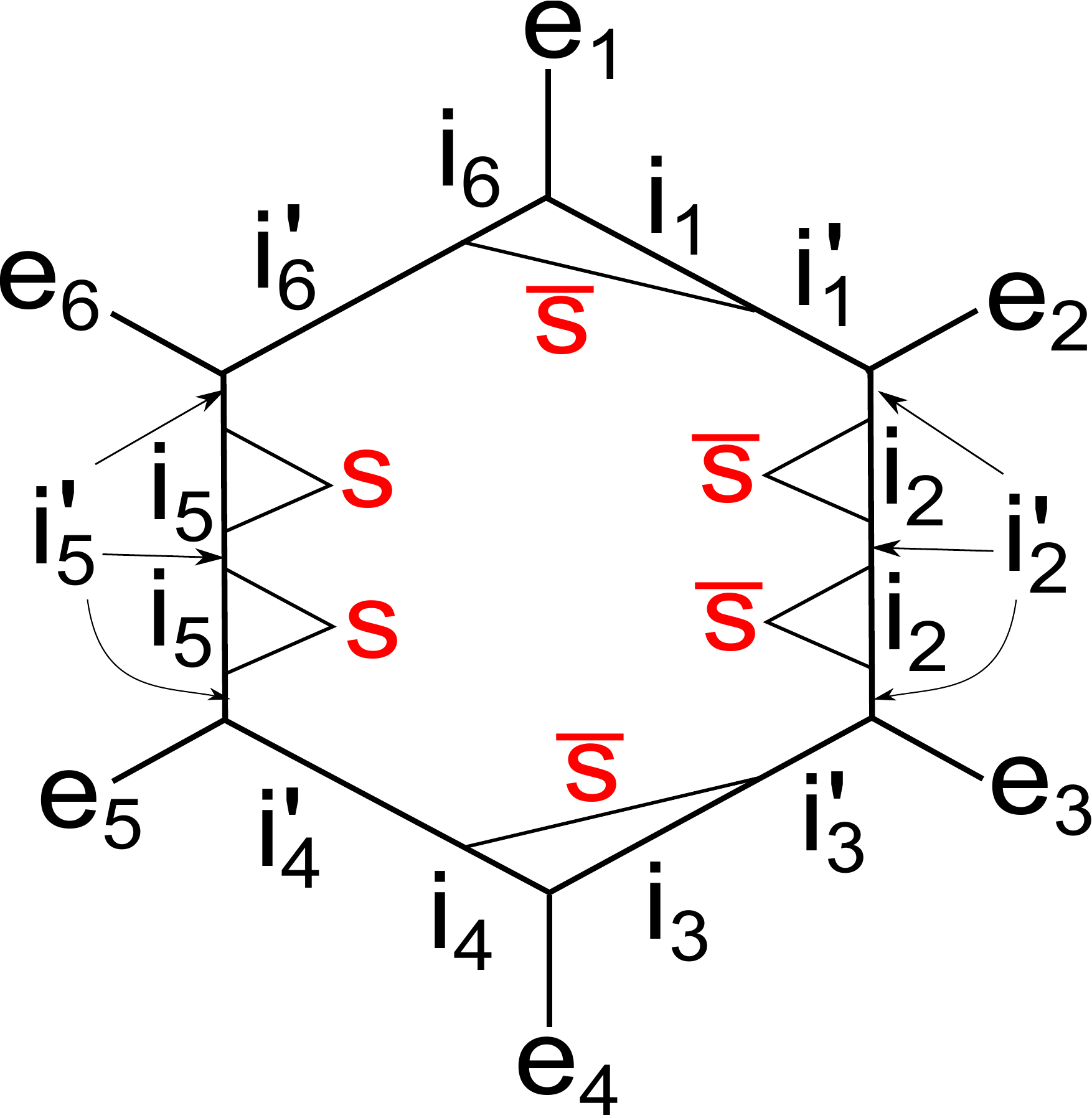}} \right| 	[F^{i'_1e_2}_{\bar{s}i_2}]_{i_1i'_2}
	[\tilde{F}_{\bar{s}i_2}^{i'_3e_3}]_{i_3i'_2}
	[F^{i_5s}_{e_5i'_4}]_{i_4i'_5}
	[\tilde{F}_{e_6i'_6}^{i_5s}]_{i_6i'_5}
	[F^{i'_4\bar{s}}_{i_40}]_{si_4}
	[\tilde{F}_{i_60}^{i'_6\bar{s}}]_{si_6} C_1 \\
	&=\sum_{i'_1,\dots,i'_6}
	\left\< \raisebox{-0.35in}{\includegraphics[height=0.7in]{bp0c.pdf}} \right| 
	[F^{i'_1e_2}_{\bar{s}i_2}]_{i_1i'_2}
	[\tilde{F}_{\bar{s}i_2}^{i'_3e_3}]_{i_3i'_2}
	[F^{i_5s}_{e_5i'_4}]_{i_4i'_5}
	[\tilde{F}_{e_6i'_6}^{i_5s}]_{i_6i'_5} 
	[F^{i'_4\bar{s}}_{i_40}]_{si_4}
	[\tilde{F}_{i_60}^{i'_6\bar{s}}]_{si_6} 
	[F^{i_6i_1}_{i'_6i'_1}]_{\bar{s}e_1}
	[\tilde{F}_{i'_4i'_3}^{i_4i_3}]_{\bar{s}e_4} C_1 C_2\\
	&\equiv \sum_{i'_1,\dots,i'_6}
	\left\< \raisebox{-0.35in}{\includegraphics[height=0.7in]{bp0c.pdf}} \right| B^{s,i_1i_2\dotsi_6}_{p,i'_1i'_2\dots i'_6}(e_1e_2\dots e_6)
\end{split}
\end{equation}
\end{widetext}
where
\begin{equation}
	\begin{split}
		C_1&=(Y^{\bar{s}i_1}_{i'_1} Y^{\bar{s}i_2}_{i'_2} Y^{\bar{s}i_3}_{i'_3}Y^{i_4 s}_{i'_4} Y^{i_5 s}_{i'_5} Y^{i_6 s}_{i'_6})^{-1}\\
		C_2 &= (Y^{\bar{s}i_2}_{i_2'}Y^{i_5s}_{i'_5})^2 Y^{i_4i_3}_{e_4} Y^{i_6 i_1}_{e_1} 
	\end{split}
	\label{}
\end{equation}

By using (\ref{consistency1},\ref{consistency}), 
we obtain
\begin{equation}
	\begin{split}
	&B^{s,i_1i_2\dotsi_6}_{p,i'_1i'_2\dots i'_6}(e_1e_2\dots e_6)
	=\\
	&
	\frac{Y^{s\bar{s}}_0 Y^{i_6 i_1}_{e_1} Y^{i_3e_3}_{i_2} Y^{e_5 i_4}_{i_5} }
	{Y^{i'_6 i'_1}_{e_1} Y^{i_3' e_3}_{i_2'} Y^{e_5 i'_4}_{i'_5} }
	F^{\bar{s}i_3 e_3}_{i'_2 i'_3 i_2} F^{i'_4 \bar{s} i_3}_{e_4i_4i'_3} F^{e_6 i_6 s}_{i'_5 i_5 i'_6}
	F^{i_6s\bar{s}}_{i_6 i'_60} \times \\
	&(F^{\bar{s}i_1 e_2}_{i'_2})^{-1}_{i_2i'_1} 
	(F^{i_4s\bar{s}}_{i_4})^{-1}_{0i'_4}
	(F^{e_5 i_4 s}_{i'_5})^{-1}_{i'_4 i_5}
	(F^{i'_6 \bar{s}i_1}_{e_1})^{-1}_{i'_1i_6}.
\end{split}
	\label{bpraw}
\end{equation}
 Using the constraint (\ref{unitary}), we can rewrite (\ref{bpraw}) as Eq.~(\ref{bps0}).

\subsection{Properties of the Hamiltonian} \label{propHsect}
The first property of the Hamiltonian (\ref{hsn0}) is that it is \emph{Hermitian}. This result follows from two identities:
\begin{align}
a_s^* = a_{\bar{s}}, \quad \quad (B_p^s)^\dagger = B_p^{\bar{s}}
\end{align}
Here the first identity follows immediately from the definition (\ref{as}); the second identity is less obvious and is derived in appendix \ref{app:property}.

In addition to being Hermitian, the Hamiltonian has several other nice properties:
\begin{enumerate}	
\item{The $Q_I$ and $B_p$ operators commute with each other:
\begin{equation}
	[Q_I,Q_J]=0,\quad [Q_I,B_p]=0,\quad [B_p,B_{p'}]=0.
\end{equation}}
\item{$Q_I$ and $B_p$ are projection operators.}
\end{enumerate}
The first two commutation relations in property 1 follow immediately from the definitions of $Q_I, B_p$. The third relation, $[B_p, B_{p'}] = 0$, is non-trivial and is derived in Appendix \ref{app:commute}.  Likewise, it is easy to see that $Q_I$ is a projector, but the fact that $B_p$ is also a projector is non-trivial and is derived in Appendix \ref{app:property}.

The above properties allow for the exact solution of $H$. To see this, note that $Q_I, B_p$ commute with one another and hence we can simultaneously diagonalize them. Denoting these simultaneous eigenstates by $|\{q_I, b_p\}\>$ where $q_I, b_p = 0,1$ are the eigenvalues, it is clear that $|\{q_I, b_p\}\>$ is an energy eigenstate with eigenvalue
\begin{align*}
E = -\sum_I q_I - \sum_p b_p.
\end{align*}
Using this expression, we can read off the complete energy spectrum of $H$ (up to determining degeneracies). In particular, we can see that the ground state(s) of $H$ have $q_I = b_p = 1$, while the excited states have $q_I = 0$ or $b_p = 0$ for at least one site $I$ or plaquette $p$. It follows that there is finite energy gap ($\Delta \geq 1$) separating the ground state(s) from the excited states.

The only remaining  task is to prove the existence of at least one state with $q_I = b_P = 1$, and determine the degeneracy of these states. We focus on the simplest case: a lattice with a disk-like geometry of the type described in Appendix G of Ref.~\onlinecite{LinLevinstrnet}. In this case, we can show that there is exactly one state with $q_I = b_p = 1$. To see that there is at \emph{least} one such state, note that $|\Phi\>=  \prod_p B_p |\text{vacuum} \>$ has $q_I = b_p=1$ everywhere, and furthermore one can check that $\<\text{vacuum} | \Phi\> \neq 0$ so $|\Phi\> \neq 0$. To see that there is at \emph{most} one such state, we use a result derived in Appendix \ref{app:localrules}: there we show that any state with $q_I = b_p = 1$ obeys a lattice version of the local rules (\ref{localrules}). Then, since the local rules can be used to relate any string-net configuration in a disk geometry to the vacuum configuration, it follows that there is at most one state with $q_I = b_p = 1$.\footnote{While this argument is suggestive, strictly speaking it is incomplete since we only know that the \emph{continuum} local rules are sufficient for relating string-net configurations to the vacuum configuration. To complete the proof, we would need to establish a similar result for the lattice local rules, which we will not undertake here.}  More generally, the ground state degeneracy depends on the global topology (or boundary conditions) of our lattice.  This topological ground state degeneracy has been discussed in a number of works\cite{BravyiKitaev,LevinWenstrnet}, and Ref.~\onlinecite{WuDegeneracy} gives a prescription for computing it on a given spatial topology.

So far we have shown that the Hamiltonian $H$ has a unique ground state and an energy gap in a disk geometry. To complete the picture, we now argue that this ground state is exactly the wave function $|\Phi\>$ defined by (\ref{localrules}), restricted to string-net configurations that live on the lattice. 

To prove that the ground state is $|\Phi\>$, it suffices to show that $Q_I |\Phi\> = B_p |\Phi\> = |\Phi\>$. The first equality, $Q_I |\phi\> = |\Phi\>$, is obvious since $|\Phi\>$ is a linear combination of string-net configurations, all of which obey the branching rules. To prove the second equality, $B_p |\Phi\> = |\Phi\>$, we use the following identity which we will derive below:
\begin{align}
B_p^s |\Phi\> = Y^{s \bar{s}}_0 |\Phi\>
\label{bpsphi}
\end{align}
Substituting this identity into the definition of $B_p$ (\ref{bp0}) and observing that $\sum_s a_s Y^{s \bar{s}}_0 = 1$, it follows that $B_p |\Phi\> = |\Phi\>$. All that remains is to prove (\ref{bpsphi}). To derive this identity, we multiply both sides of Eq.~(\ref{graphbps}) by $|\Phi\>$ and then use the local rule (\ref{1c}) to trade the type-$s$ loop on the right hand side for an extra factor of $Y^{s \bar{s}}_0$.The identity (\ref{bpsphi}) follows immediately.

\section{Quasiparticle excitations \label{sec:string}}
Having described the string-net Hamiltonian and its ground state in the previous section, we now turn to its low-lying excitations. Specifically, we describe so-called string operators which create point-like quasiparticles at their endpoints -- but no excitations anywhere else -- when acting on the ground state.    We show how to extract the braiding statistics of these quasiparticles by computing certain ground-state matrix elements associated with the corresponding string operators.

\subsection{Finding the quasiparticle string operators}

We start with finding the quasiparticles in the model (\ref{hsn0}).
The basic logic is as follows.
We will identify a set of {\it string operators} $\{ W_\alpha(P)\}$, which act along  oriented paths $P$. 
We require each string operator to act on the string-net ground state in a way that is path independent, i.e. it must give the same state for any choice of path $P$ connecting the same two endpoints.  More formally, path independence is the requirement that
\begin{equation}
	W_\alpha(P)|\Phi\> \propto W_\alpha(P')|\Phi\> \ \ .
	\label{pathind}
\end{equation}
Path independence is important because it ensures that, when acting on the ground state, open string operators only create excitations near their \emph{endpoints}. More specifically, if $P$ is an open path oriented from $i$ to $f$, then acting on the ground state with the open string operator $W_\alpha(P)$ creates a  quasiparticle $\alpha$ at the string's endpoint $f$, and the corresponding antiparticle $\bar{\alpha}$ at the string's starting point $i$.  Likewise, for a closed contractible path $P$, path independence implies that $W_\alpha(P)|\Phi\> \propto |\Phi\>$ -- that is,  closed string operators do not create any excitations when acting on the ground state. 

To construct string operators, we follow the strategy of Ref.~\onlinecite{LevinWenstrnet}: we describe a general ansatz for constructing string operators $W_\alpha(P)$ in terms of certain input data $(\Omega_\alpha, \bar{\Omega}_{\alpha}, n_\alpha)$, and we work out the conditions under which the resulting string operators obey the path independence condition (\ref{pathind}). 

First, we explain the input data in more detail. We start with the third piece of data, $n_\alpha$. This piece of data is shorthand for a collection of non-negative integers $n_{\alpha, s}$ where $s$ runs over the different string types. Each integer $n_{\alpha,s}$ describes the ``multiplicity'' of the string type $s$ within the string operator $\alpha$. The remaining data, $\Omega_\alpha$ and $\bar{\Omega}_{\alpha}$, is shorthand for two collections of complex (rectangular) matrices,  $(\Omega^{a,rsb}_{\alpha})_{\sigma_r\sigma_s}$ and $(\bar{\Omega}^{a,rsb}_{\alpha})_{\sigma_r\sigma_s}$, parameterized by four string types $a, r, s, b$. Here the two matrix indices $\sigma_r, \sigma_s$ can take $n_{\alpha, r}$ and $n_{\alpha, s}$ values, respectively. Like the $F$-symbol, the matrix elements $(\Omega^{a,rsb}_{\alpha})_{\sigma_r\sigma_s}$ are only defined when $a, r, s, b$ obey certain branching rules, specifically, $\delta^{ra}_b = \delta^{as}_b = 1$, and when $n_{\alpha, r}$ and $n_{\alpha, s}$ are both nonzero. Likewise, the matrix elements $(\bar{\Omega}^{a,rsb}_{\alpha})_{\sigma_r\sigma_s}$ are only defined if $\delta^{ar}_b = \delta^{sa}_b = 1$. We should also mention that we require that the matrix elements $(\Omega^{a,rsb}_{\alpha})_{\sigma_r\sigma_s}$ and $(\bar{\Omega}^{a,rsb}_{\alpha})_{\sigma_r\sigma_s}$ take particular values when $a=0$: in that case $(\Omega^{0,rsb}_{\alpha})_{\sigma_r\sigma_s}=(\bar{\Omega}^{0,rsb}_{\alpha})_{\sigma_r\sigma_s}=\delta_{s,r}\delta_{b,r}\delta_{\sigma_r,\sigma_s}$. This is necessary to ensure that our string operator has a trivial action when crossing a vacuum string, $a=0$, as will become clear below.

We now explain our ansatz for constructing string operators, $W_\alpha(P)$ from the above input data. We first specialize to the case that $P$ is an \emph{upward}-oriented path, which is sufficient to identify the quasiparticle types.
For simplicity, we will work in the  gauge 
\begin{equation}
	Y^{ab}_c =\sqrt{\frac{d_a d_b}{d_c}} 
\end{equation}
in the following sections.

When $W_\alpha(P)$ is applied to a string-net state $\<X|$, its action is described graphically by adding a string labeled by $\alpha$ along the path $P$ \emph{under} the preexisting string-nets:
\begin{equation}
	\left\<\raisebox{-0.22in}{\includegraphics[height=0.5in]{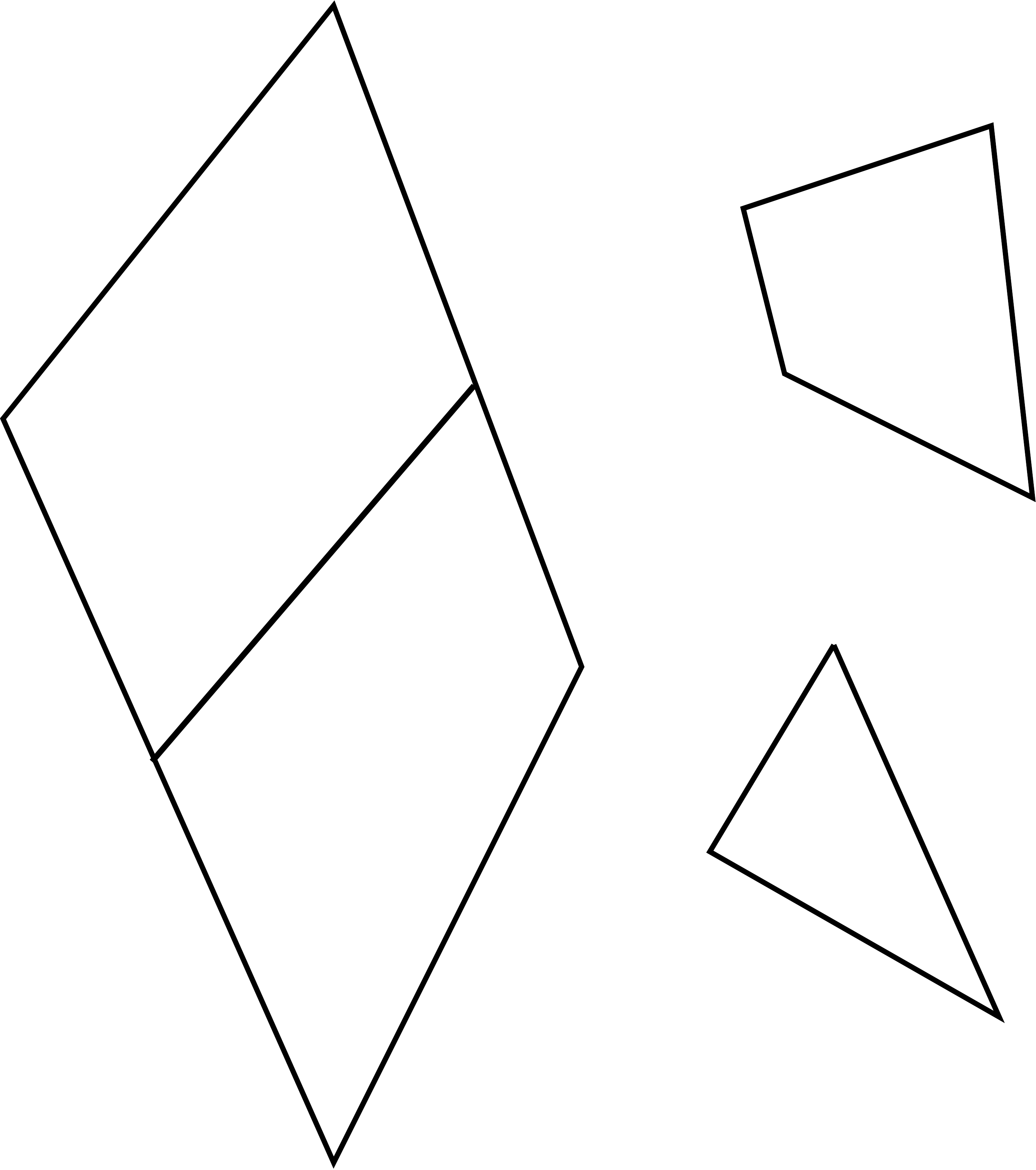}} \right|W_{\alpha}(P)=\left<\raisebox{-0.22in}{\includegraphics[height=0.5in]{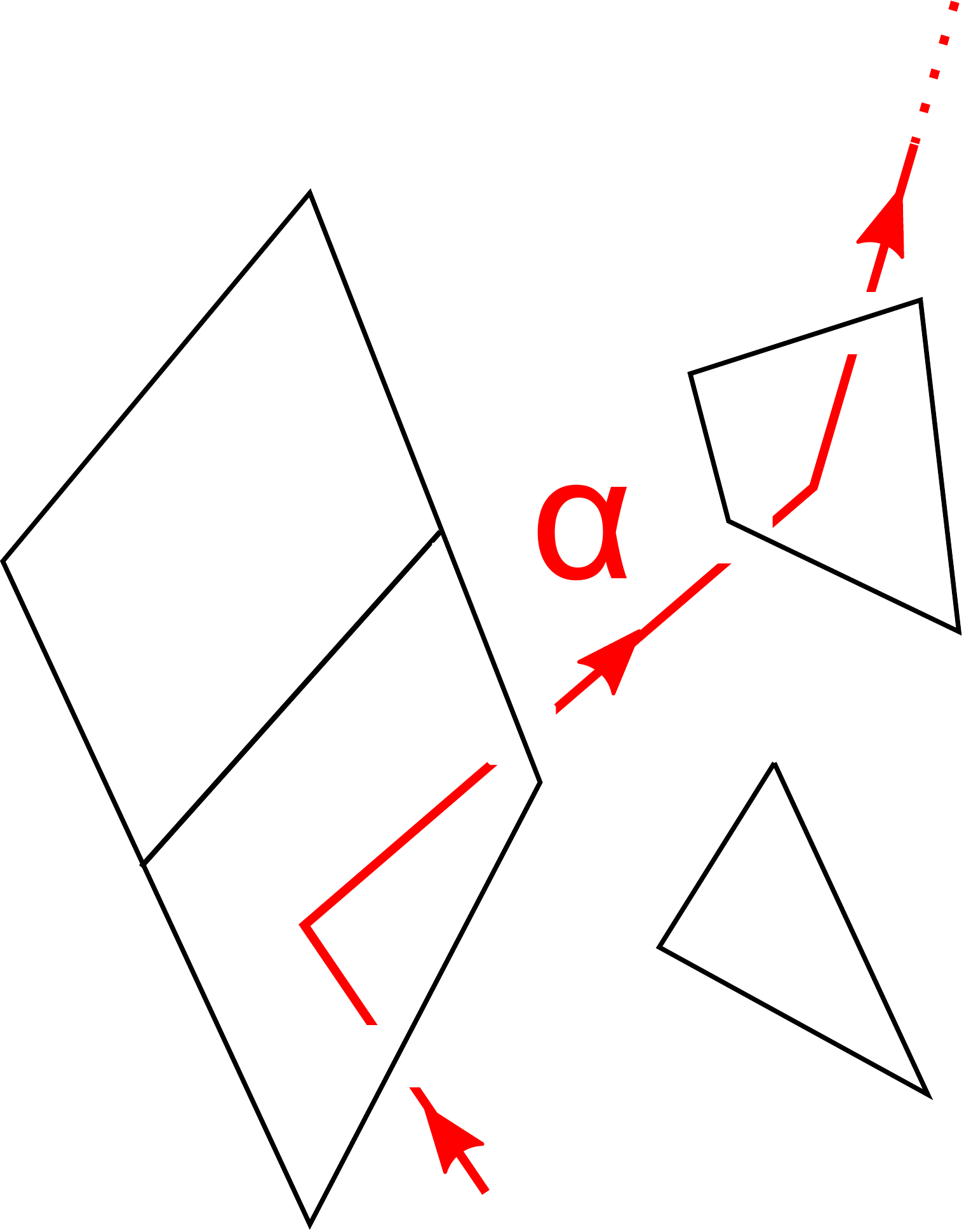}} \right|.
	\label{path1}
\end{equation}
We then replace the $\alpha$-string at every crossing with a sum over string labels $r,b$, and $s$, using the rules
\begin{align}
		\left<\raisebox{-0.22in}{\includegraphics[height=0.5in]{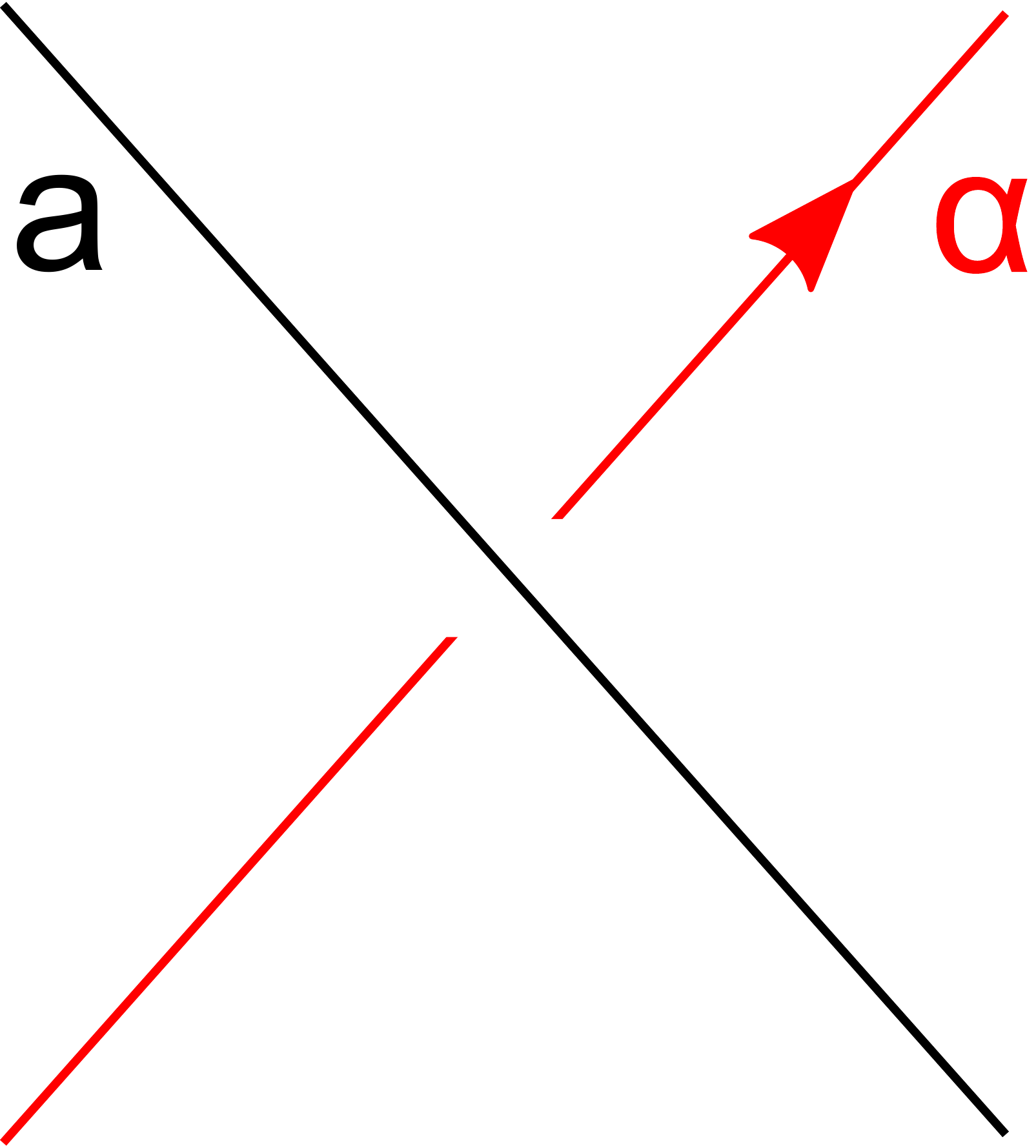}} \right|=&\sum_{b,s,r} ({\Omega}^{a,rsb}_{\alpha})_{\sigma_r\sigma_s} \sqrt{\frac{d_b}{d_a \sqrt{d_r d_s}}} \left<\raisebox{-0.22in}{\includegraphics[height=0.5in]{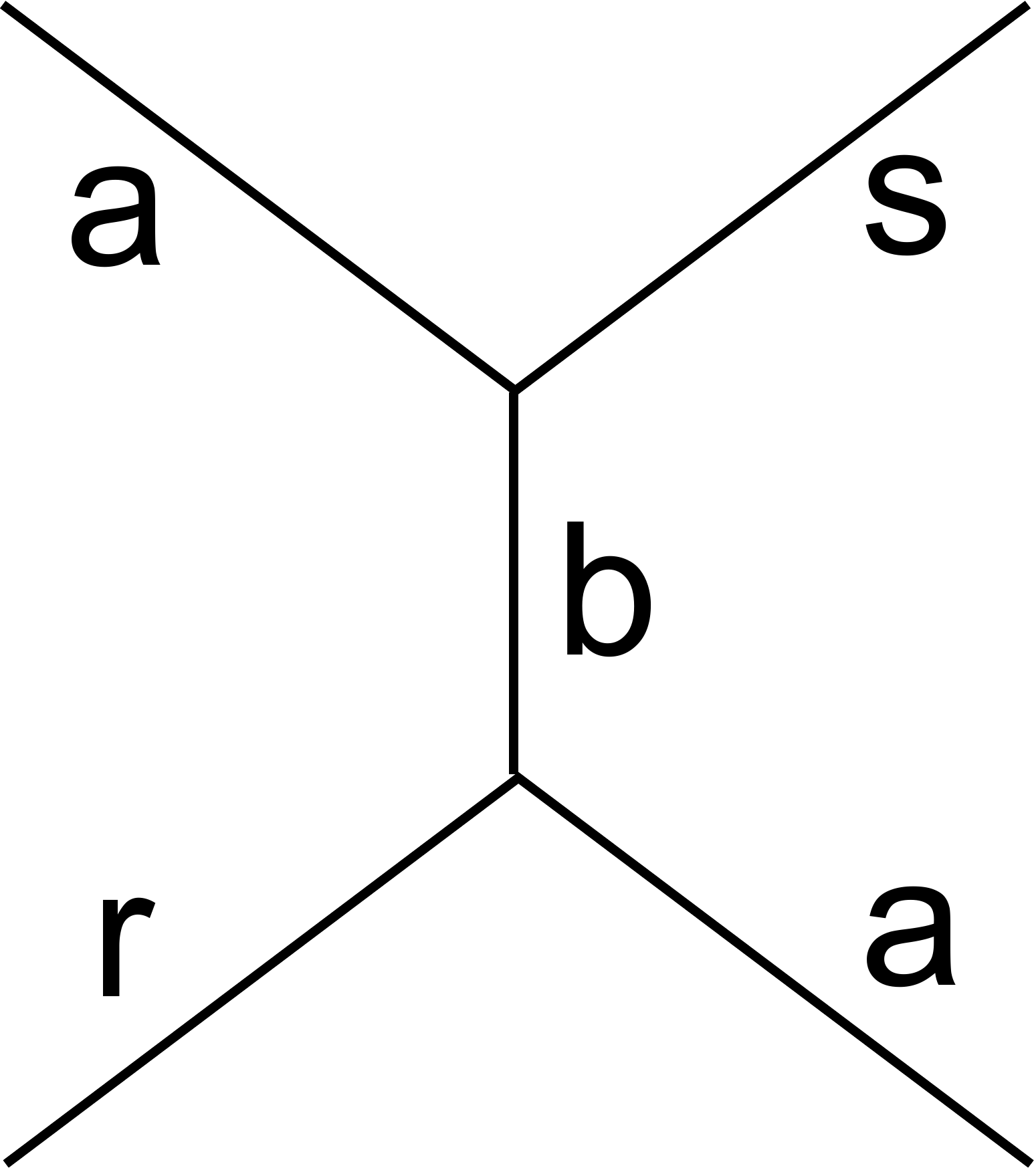}} \right|\\
		\left<\raisebox{-0.22in}{\includegraphics[height=0.5in]{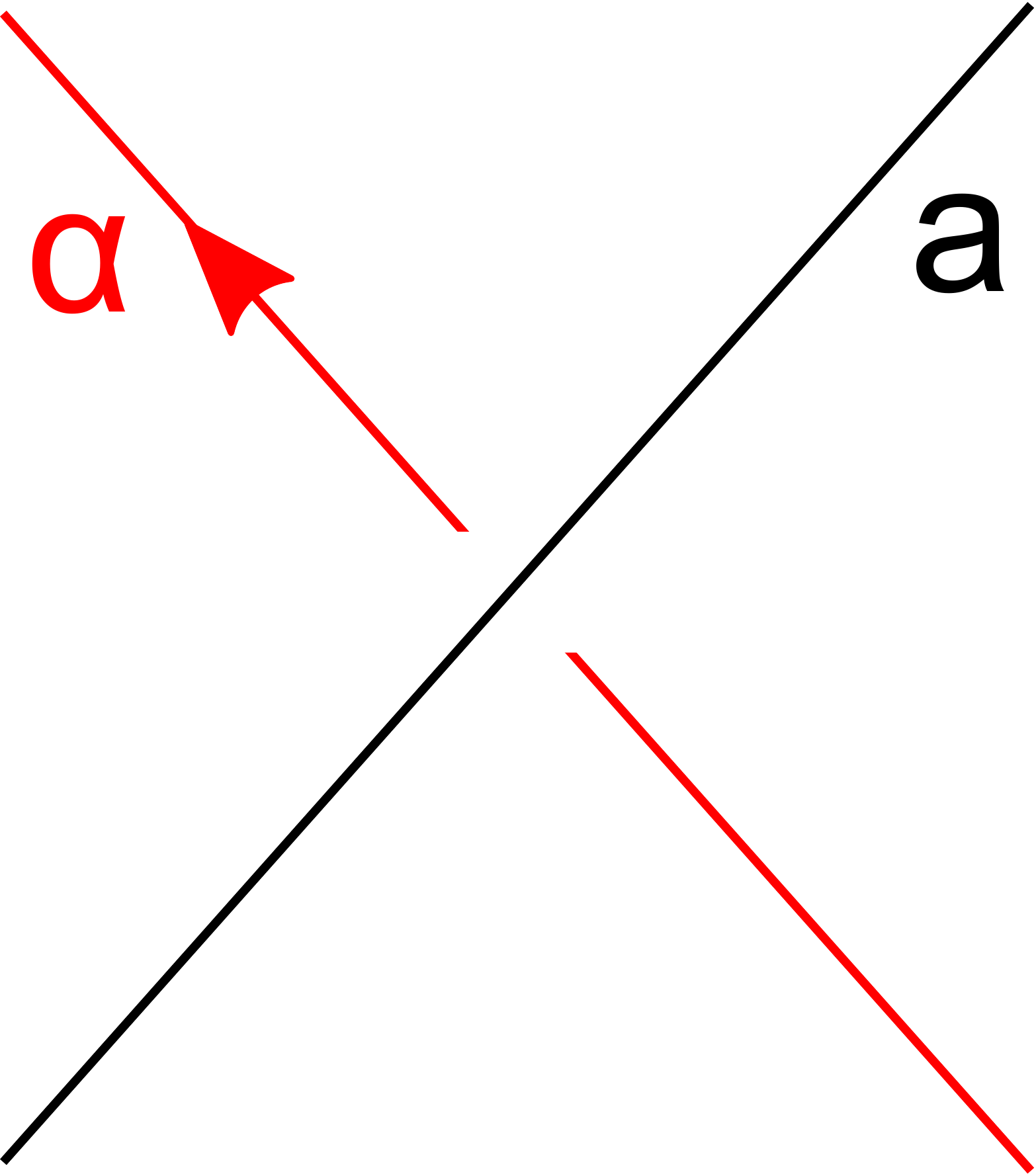}} \right|=&\sum_{b,s,r}({\bar{\Omega}}^{a,rsb}_{\alpha})_{\sigma_r\sigma_s}  \sqrt{\frac{d_b}{d_a \sqrt{d_r d_s}}} \left<\raisebox{-0.22in}{\includegraphics[height=0.5in]{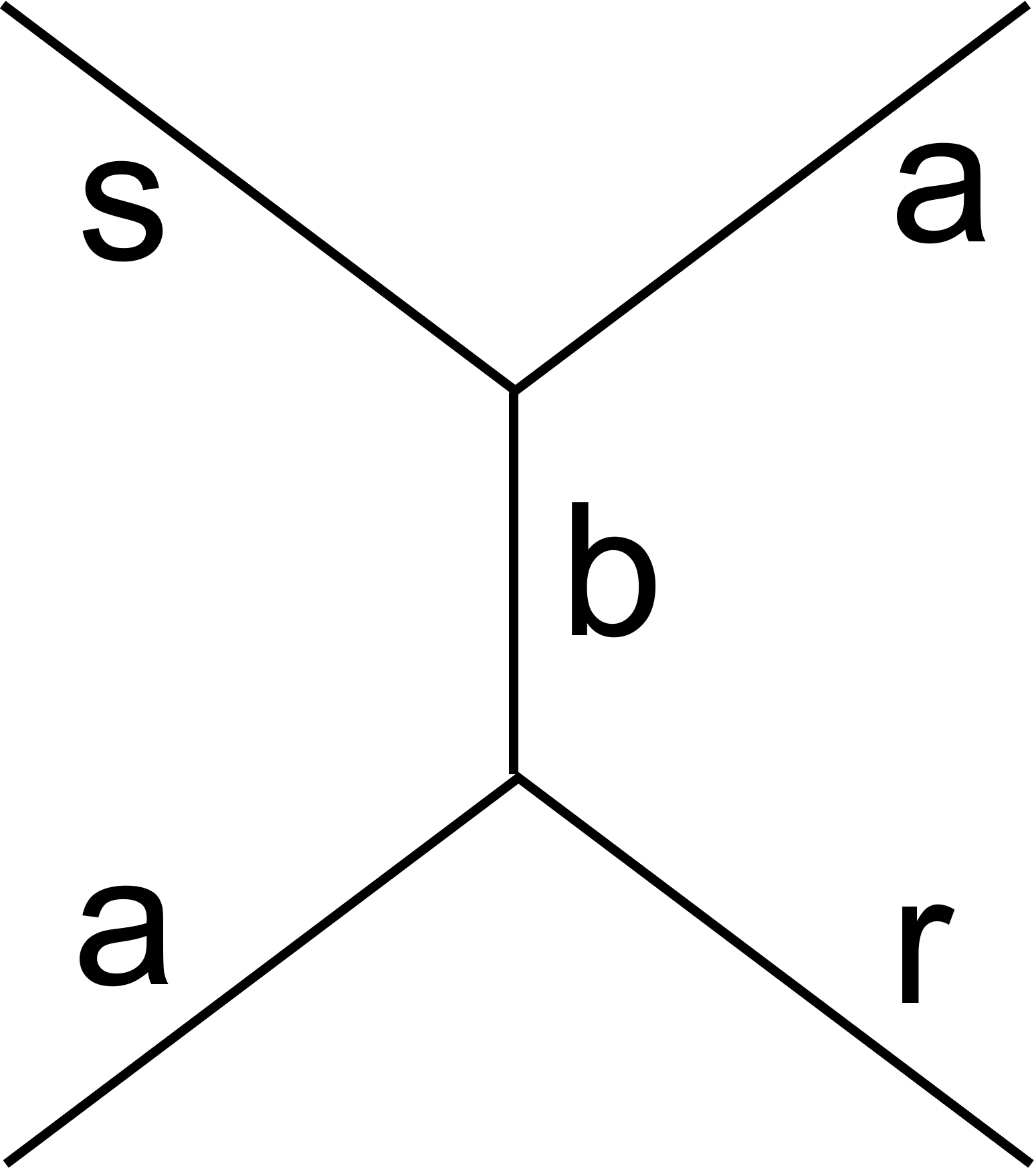}} \right|
	\label{stringrules}
\end{align}
Here, $(\Omega^{a,rsb}_{\alpha})_{\sigma_r\sigma_s}$ and $(\bar{\Omega}^{a,rsb}_{\alpha})_{\sigma_r\sigma_s}$ are the complex matrices of dimension $n_{r, \alpha} \times n_{s, \alpha}$ that define our string operators (see discussion above). The two indices $\sigma_r, \sigma_s$ should be thought of as living on the $r$ and $s$ string respectively. The factors of $d_a$ are included to simplify the constraints satisfied by $\Omega_\alpha ,\bar{\Omega}_\alpha$, which we present shortly.\footnote{Because of these factors of $d_a$, the $\Omega_\alpha ,\bar{\Omega}_\alpha$ in this paper have a different normalization than in Ref.~\onlinecite{LevinWenstrnet}.}  

 After making the replacements in (\ref{stringrules}), we obtain the action of the string operator on any string-net state as follows.  First, we require the string labels $r,s$ to be the same throughout any region where the path $P$ does not cross any edges of the initial string-net.  Second, along  each such path segment we contract the corresponding matrix indices $\sigma_r, \sigma_s, etc.$. 
For example,
\begin{equation}
	\left \langle \raisebox{-0.22in}{\includegraphics[height=0.5in]{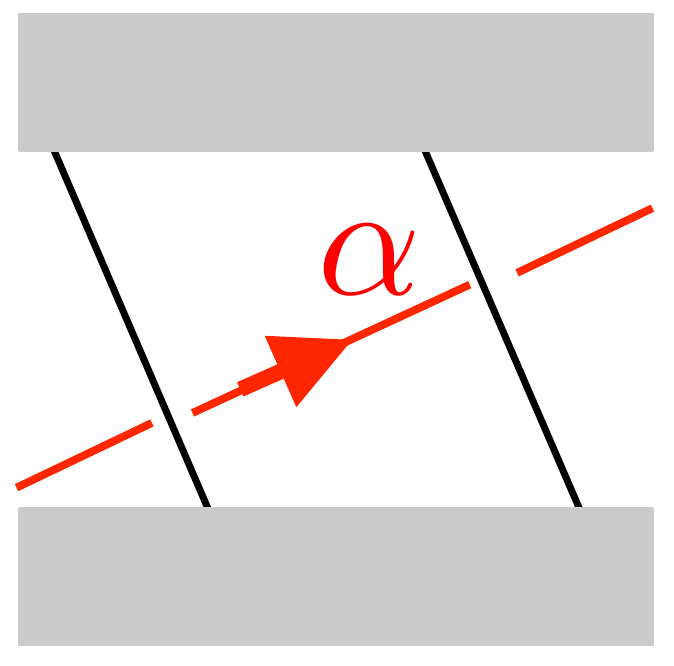}} \right |=\sum_{b,d,s,t,u,v} \Omega^{a,stb}_{\alpha}\Omega^{c,uvd}_{\alpha}\delta_{t,u} \left \<\raisebox{-0.22in}{\includegraphics[height=0.5in]{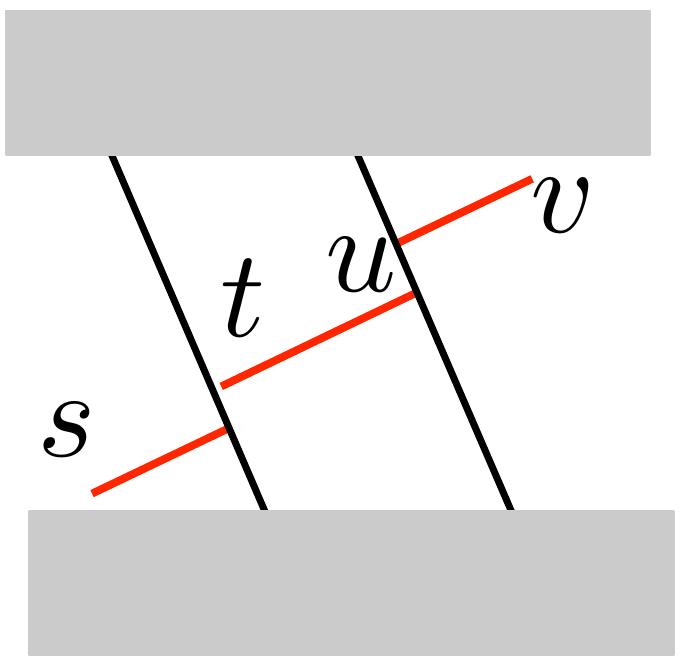}} \right |.
	\label{Eq:example}
\end{equation} 
where the matrix product is taken along the index associated with the shared edge label $t$, i.e. 
\begin{equation}  \label{Eq:MatrixMult} \Omega^{a,stb}_{\alpha}\Omega^{c,uvd}_{\alpha} \delta_{t,u} =  \sum_{\sigma_t} (\Omega^{a,stb}_{\alpha})_{\sigma_s, \sigma_t} (\Omega^{c,tvd}_{\alpha})_{\sigma_t, \sigma_v} \ .
\end{equation}
The end result is a superposition of new states of the form $\<X|W_\alpha(P)=\sum_{X'}C(X,X')\<X'|$, where $\<X'|$ is a string-net state everywhere except near the endpoints of $P$, and $C(X,X')$ is a product of matrices $\Omega_\alpha, \bar{\Omega}_{\alpha}$, with each matrix corresponding to a crossing between the path $P$ and a string in the string-net ket $\< X |$.   
 
Finally, to define the action of the string operator on the honeycomb lattice, away from the endpoints of $P$ we use the local rules to reduce these new string-nets to string-nets on the honeycomb lattice, as shown for the plaquette operator in Eq.~(\ref{Eq:Hdiagram}).  In this way, the ansatz $(\Omega_\alpha, \bar{\Omega}_\alpha, n_{\alpha})$ fully defines the lattice action of the string operator $W(P)$.\footnote{While there is some ambiguity in defining the action of the string operator $W_\alpha(P)$ near the endpoints of $P$, this ambiguity is not important for our purposes since it does not affect on the quasiparticle statistics of the excitation created by $W_\alpha(P)$.}

Before continuing, we should clarify one point about the string operator multiplicity $n_{\alpha, s}$. As discussed in Appendix \ref{App:FusionMults}, for the most general class of string-nets, every vertex carries a matrix index to account for the fact that there may be more than one state in the string-net Hilbert space that satisfies the branching rules. This phenomenon is known in the mathematical literature as {\it fusion multiplicity}. We emphasize that fusion multiplicity should not be confused with the string operator multiplicity $n_{\alpha, s}$. In particular, it is possible for $n_{\alpha, s}$ to be larger than $1$ even in string-net models that do not have any fusion multiplicity (i.e. models with $\delta^{ab}_c \leq 1$ for all $a, b, c$). An example where this occurs is given by the string-net whose labels correspond to group elements of the symmetric group $S_3$; the resulting string-net model contains an excitation $B$ for which $n_{B,0} = 2$.

To proceed, we must identify which $(\Omega_\alpha,\bar{\Omega}_\alpha, n_{\alpha})$ satisfy the path independence condition (\ref{pathind}).
Without loss of generality, we assume that the proportionality constant in (\ref{pathind}) for two upward-oriented paths $P$, $P'$ is exactly $1$, so that the path independence condition takes the form:
\begin{equation}
	\<X|W_\alpha(P)|\Phi\>=\<X|W_\alpha(P')|\Phi\>
	\label{pathind0}
\end{equation}
where away from the end-points of $P$, $\<X|$ is an arbitrary string-net state.  
 To ensure that Eq.~(\ref{pathind0}) is satisfied, it suffices to check path independence for some elementary deformations between upward-oriented paths $P$ and $ P'$, because larger deformations that fix the points $i$ and $f$ can be built out of these elementary ones.  For upward-oriented paths, the elementary deformations are:
\begin{subequations}
	\begin{align}
		\left< \raisebox{-0.22in}{\includegraphics[height=0.5in]{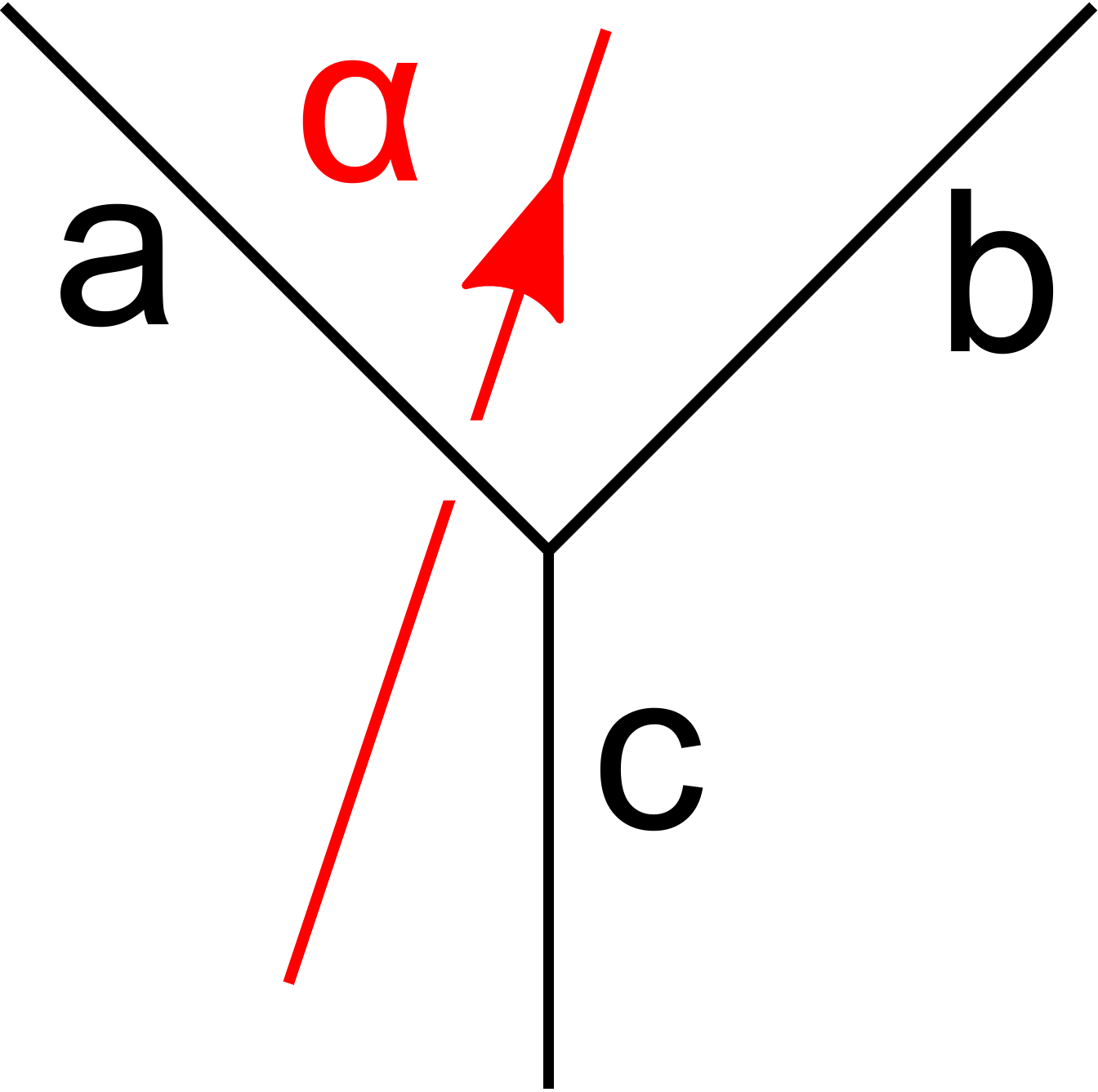}} \middle|\Phi \right \>&=\left< \raisebox{-0.22in}{\includegraphics[height=0.5in]{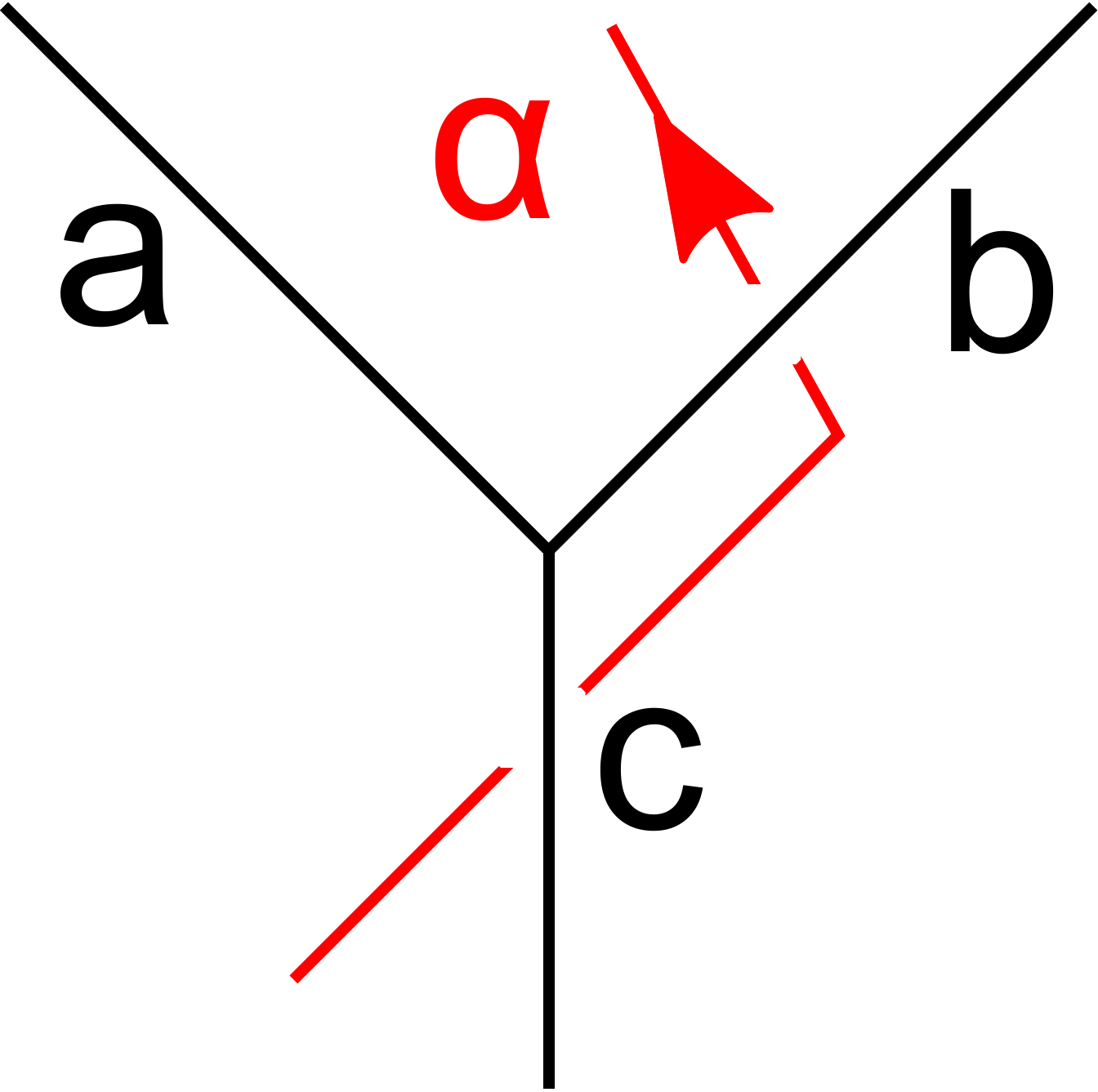}} \middle|\Phi\right\> \label{stringa}\\
		\left< \raisebox{-0.22in}{\includegraphics[height=0.5in]{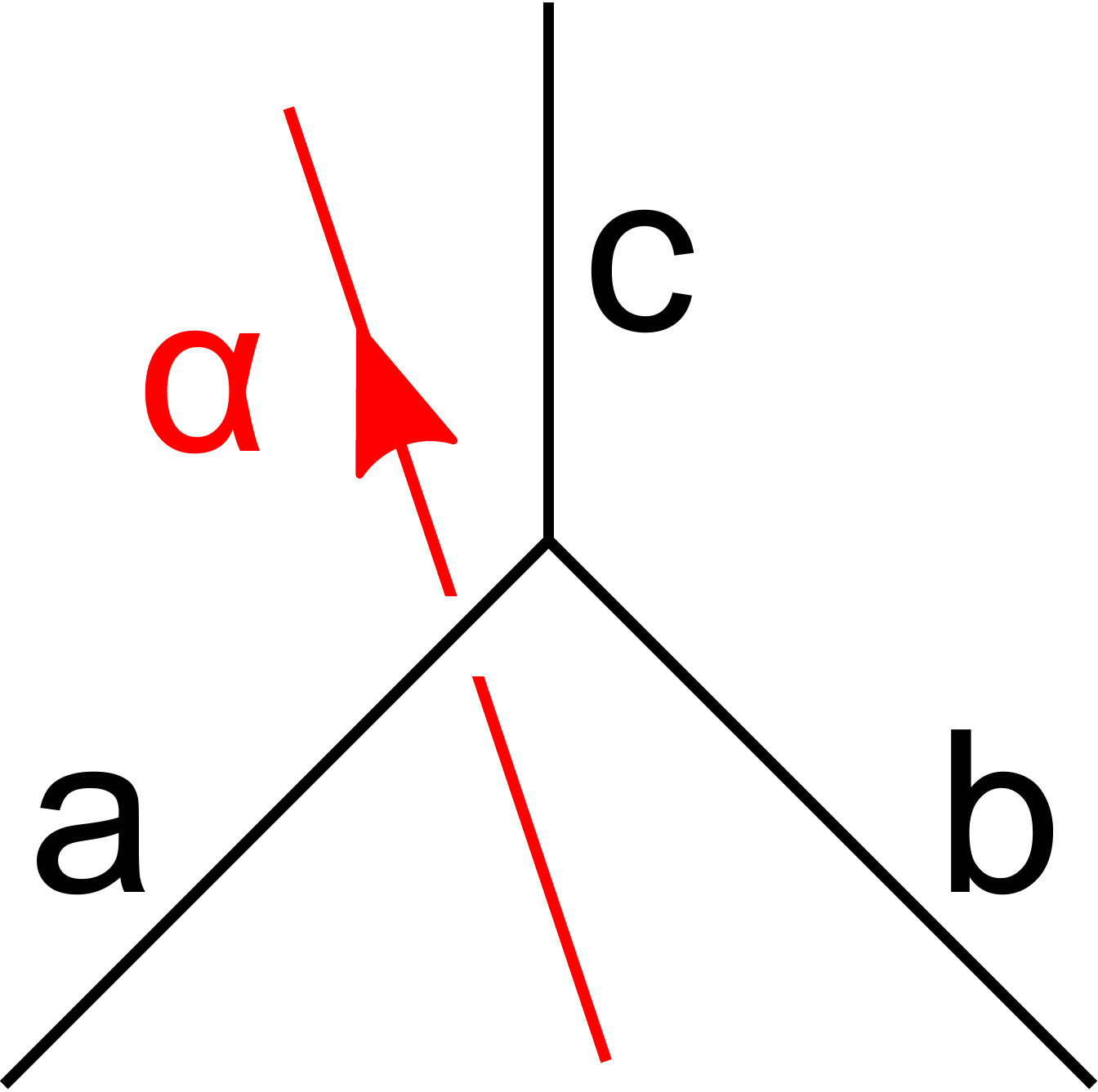}} \middle|\Phi \right \>&=\left< \raisebox{-0.22in}{\includegraphics[height=0.5in]{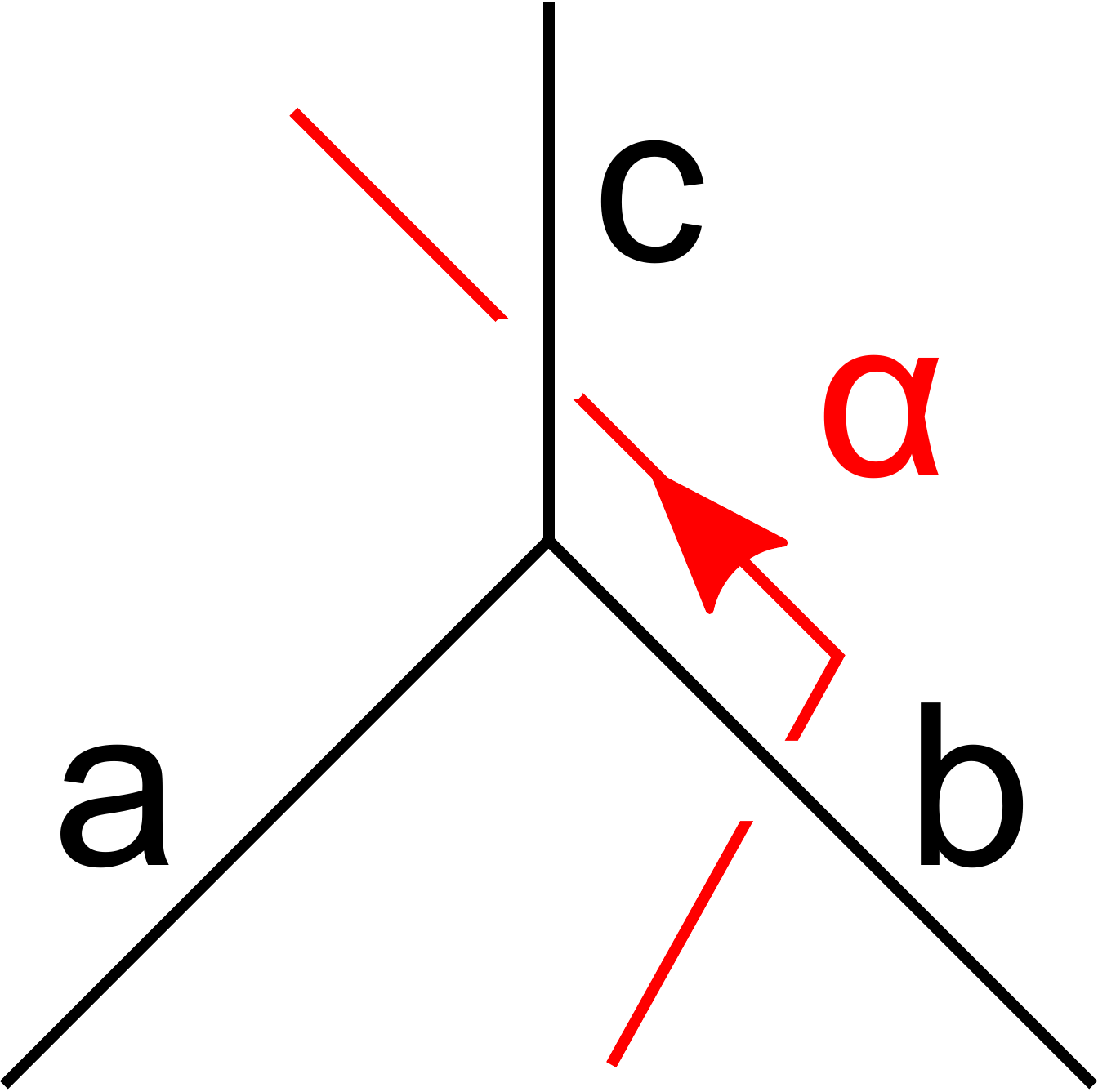}} \middle|\Phi\right\> \label{stringb}\\
		\left< \raisebox{-0.22in}{\includegraphics[height=0.5in]{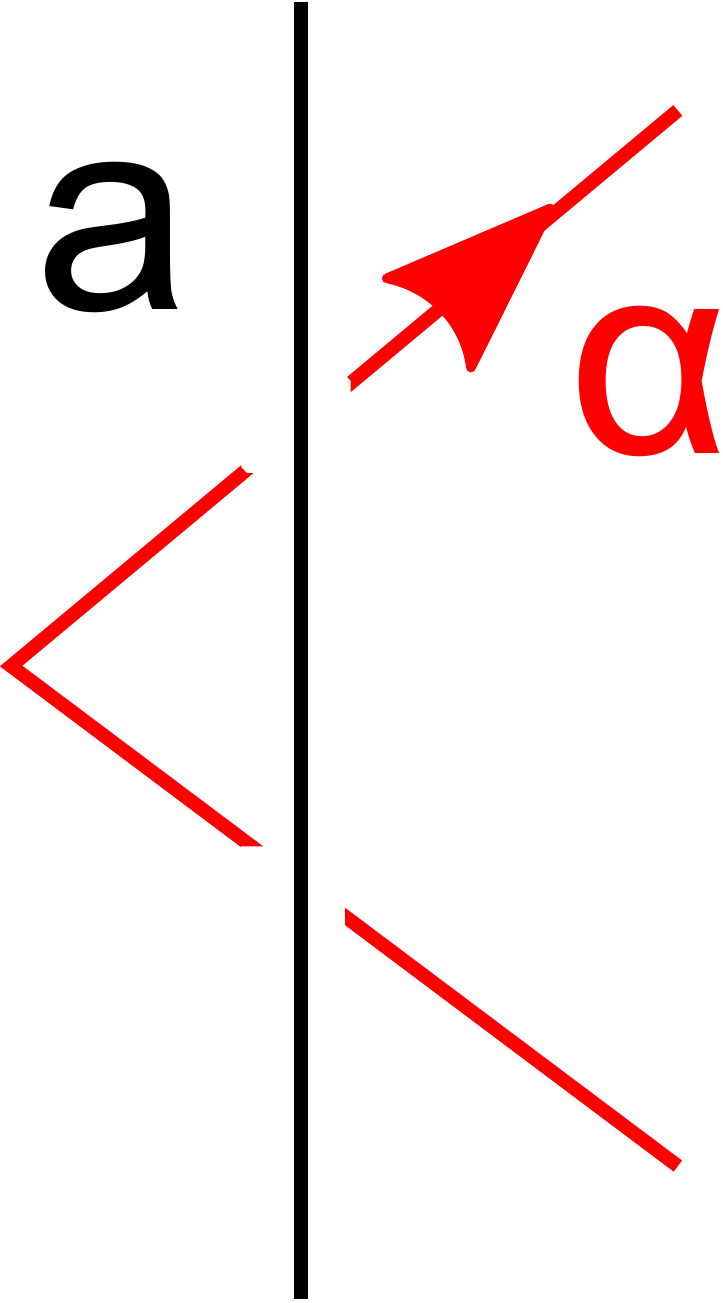}} \middle|\Phi \right \>&=\left< \raisebox{-0.22in}{\includegraphics[height=0.5in]{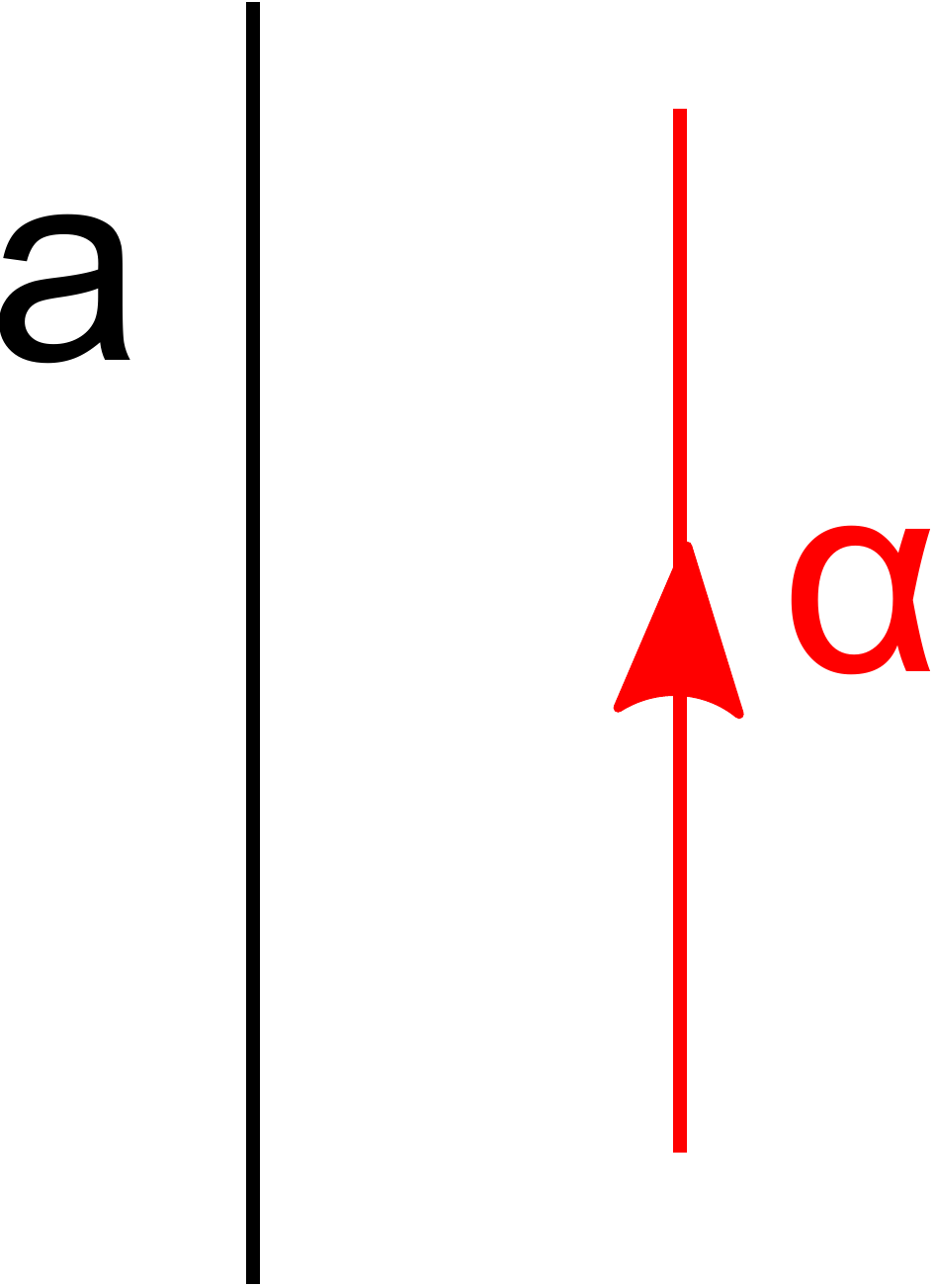}} \middle|\Phi\right\> \label{stringc}.
	\end{align}
	\label{pathind1}
\end{subequations}
Algebraically, these graphical relations are expressed as:  
\begin{subequations}
	\label{weqs}
\begin{gather}
	\sum_{a'}{\Omega}^{a,rsa'}_{\alpha}(F^{rab}_{c'a'c})^*F^{asb}_{c'a'b'}
	=	\sum_{t}{\Omega}^{c,rtc'}_{\alpha}\bar{\Omega}^{b,tsb'}_{\alpha}F^{abt}_{c'cb'} \label{weq1}\\
	\bar{\Omega}^{a,rsa'}_{\alpha}=(\Omega^{a,sra'}_{\alpha})^* \label{weq2}\\
	\sum_{s}\bar{\Omega}^{a,rsa'}_{\alpha}{\Omega}^{a,sta'}_{\alpha}=\delta_{rt}  \label{weq3}
\end{gather}
\end{subequations}
where we have used the local rules (\ref{consistency}), as well as unitarity of $F$'s. In terms of the diagrams above, Eq.~(\ref{stringa}) gives (\ref{weq1}). 
Likewise, Eq.~(\ref{stringc}) gives (\ref{weq3}). As for Eq.~(\ref{stringb}), this condition gives an equation which is the complex conjugate of (\ref{weq1}) with $\Omega_{\alpha}^{a,tsb}$ and $\bar{\Omega}_{\alpha}^{a,stb}$ interchanged. Therefore we can ensure (\ref{stringb}) if (\ref{weq2}) holds together with (\ref{weq1}).
Note that Eqs.~(\ref{weqs}) are matrix equations, with products between matrices taken over the indices as in Eq.~(\ref{Eq:MatrixMult}).

 Every solution $(\Omega_\alpha, \bar{\Omega}_\alpha, n_{\alpha})$ to (\ref{weqs}) defines a string operator $W_\alpha$. Thus, our task is find all possible solutions to (\ref{weqs}). We note that for any pair of solutions $\Omega_{\alpha}$ and $\Omega_{\beta}$ to Eqs.~(\ref{weqs}), we can always construct another solution $(\Omega, \bar{\Omega}, n)$ by taking the direct sum: $\Omega = \Omega_{\alpha} \oplus \Omega_{\beta}$ and $\bar{\Omega} = \bar{\Omega}_{\alpha} \oplus \bar{\Omega}_{\beta}$ and finally $n = n_\alpha + n_\beta$. Thus in practice one need only find solutions that are \emph{irreducible}, in the sense that they cannot be decomposed in this way. Though we do not undertake to prove it here, we conjecture that the irreducible solutions to (\ref{weqs}), and the associated string operators $W_\alpha(P)$, are sufficient to construct \emph{every} quasiparticle excitation in our models.

Before we discuss the nature of these quasiparticles, it is useful to construct closed string operators following a similar logic.  To this end, we first define the downward $\alpha$-string operator via:
\begin{align}
	\begin{split}
	\left<\raisebox{-0.22in}{\includegraphics[height=0.5in]{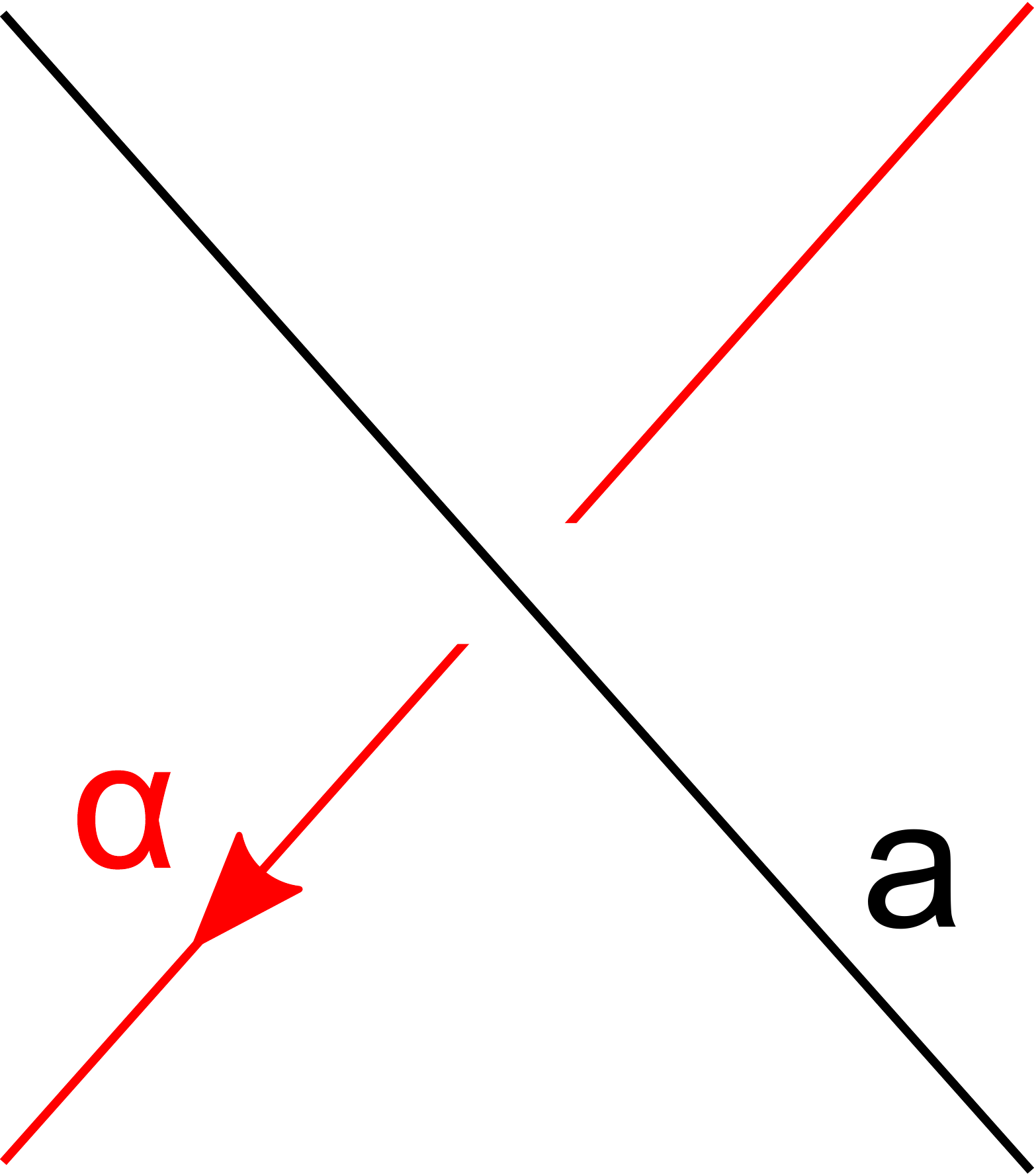}} \right|=&\sum_{b,s,r} ({\Omega}^{a,rsb}_{\bar{\alpha}})_{\sigma_r\sigma_s} \sqrt{\frac{d_b}{d_a \sqrt{d_r d_s}}} \left<\raisebox{-0.22in}{\includegraphics[height=0.5in]{alpha2b.pdf}} \right|\\
	\left<\raisebox{-0.22in}{\includegraphics[height=0.5in]{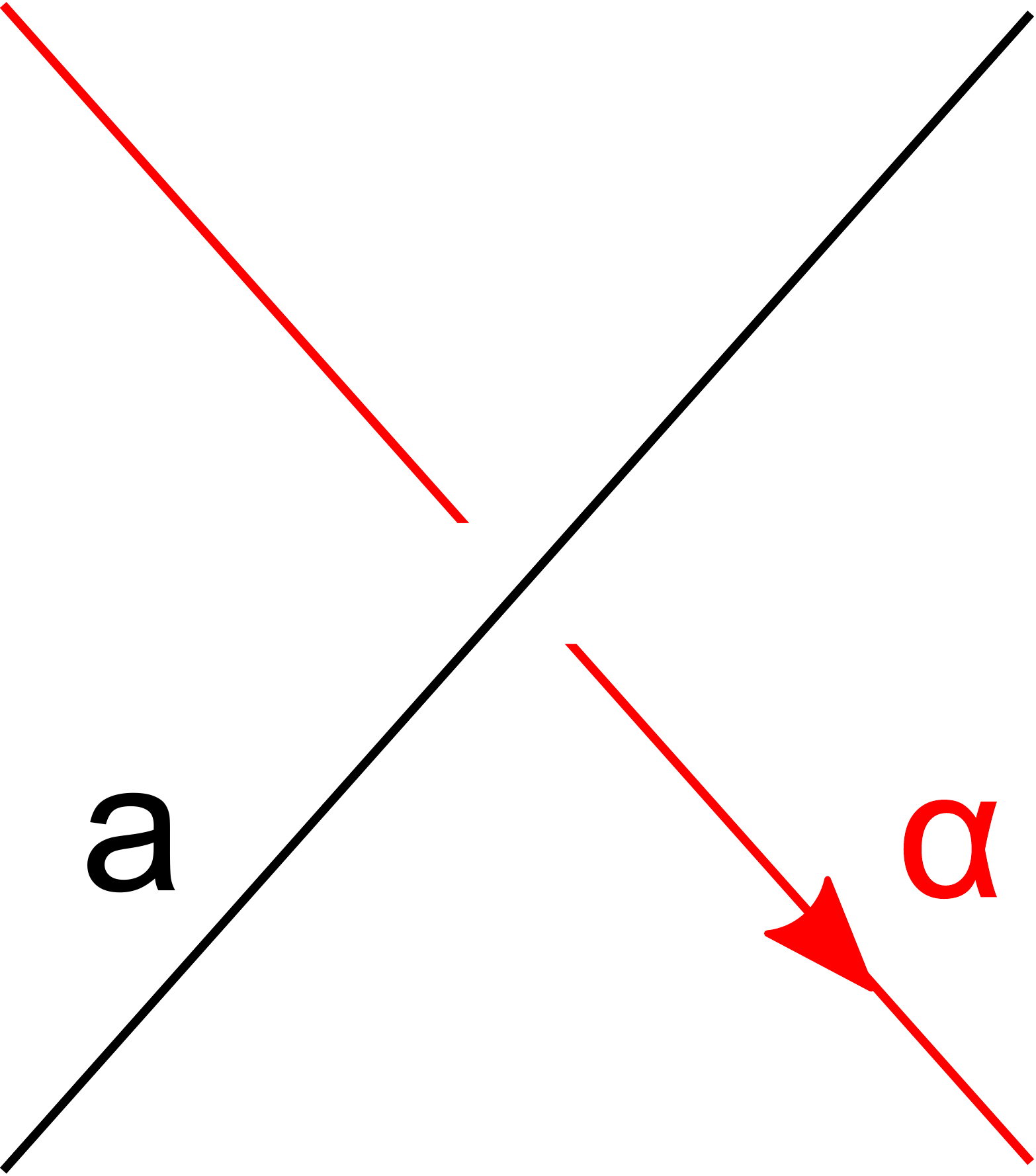}} \right|=&\sum_{b,s,r} ({\bar{\Omega}}^{a,rsb}_{\bar{\alpha}})_{\sigma_r\sigma_s} \sqrt{\frac{d_b}{d_a \sqrt{d_r d_s}}} \left<\raisebox{-0.22in}{\includegraphics[height=0.5in]{alpha1b.pdf}} \right|
\end{split}
	\label{stringrules1}
\end{align}
In other words, we define a downward-$\alpha$ string operator to be equivalent to an upward-$\bar{\alpha}$ string operator, where $\bar{\alpha}$ is the anti-particle associated with $\alpha$. 
The anti-particle $\bar{\alpha}$ is defined by the property that it can annihilate with $\alpha$, leaving only the string-net vacuum.  
In practice, this means that an upward $\alpha$ string running from $i$ to $f$ can be joined to an upward $\bar{\alpha}$ string connecting the same two points, such that the resulting closed string operator leaves the string-net in its ground state.  This joining can be done 
in the ``obvious'' way, i.e. near points $i$ and $f$, we connect the string labeled $r$ from $W_{\alpha}$ to the string labeled $s$ from $W_{\bar{\alpha}}$, impose the condition $r=s$, and contract the corresponding matrix indices.  The resulting joint between upward and downward oriented strings is path independent if:
\begin{subequations}
		\begin{align}
			\left< \raisebox{-0.12in}{\includegraphics[height=0.4in]{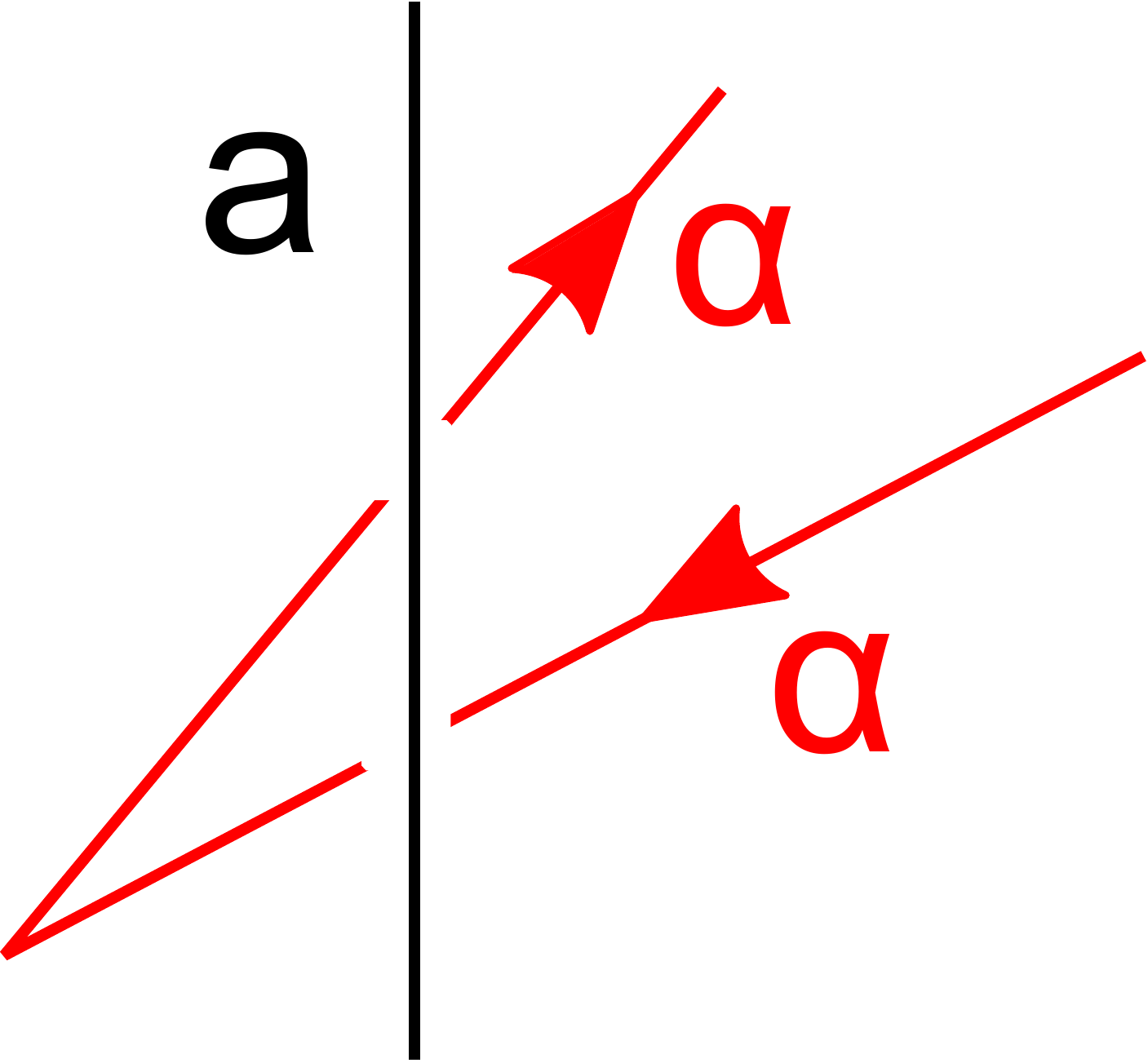}} \middle|\Phi \right \>&=\left< \raisebox{-0.12in}{\includegraphics[height=0.4in]{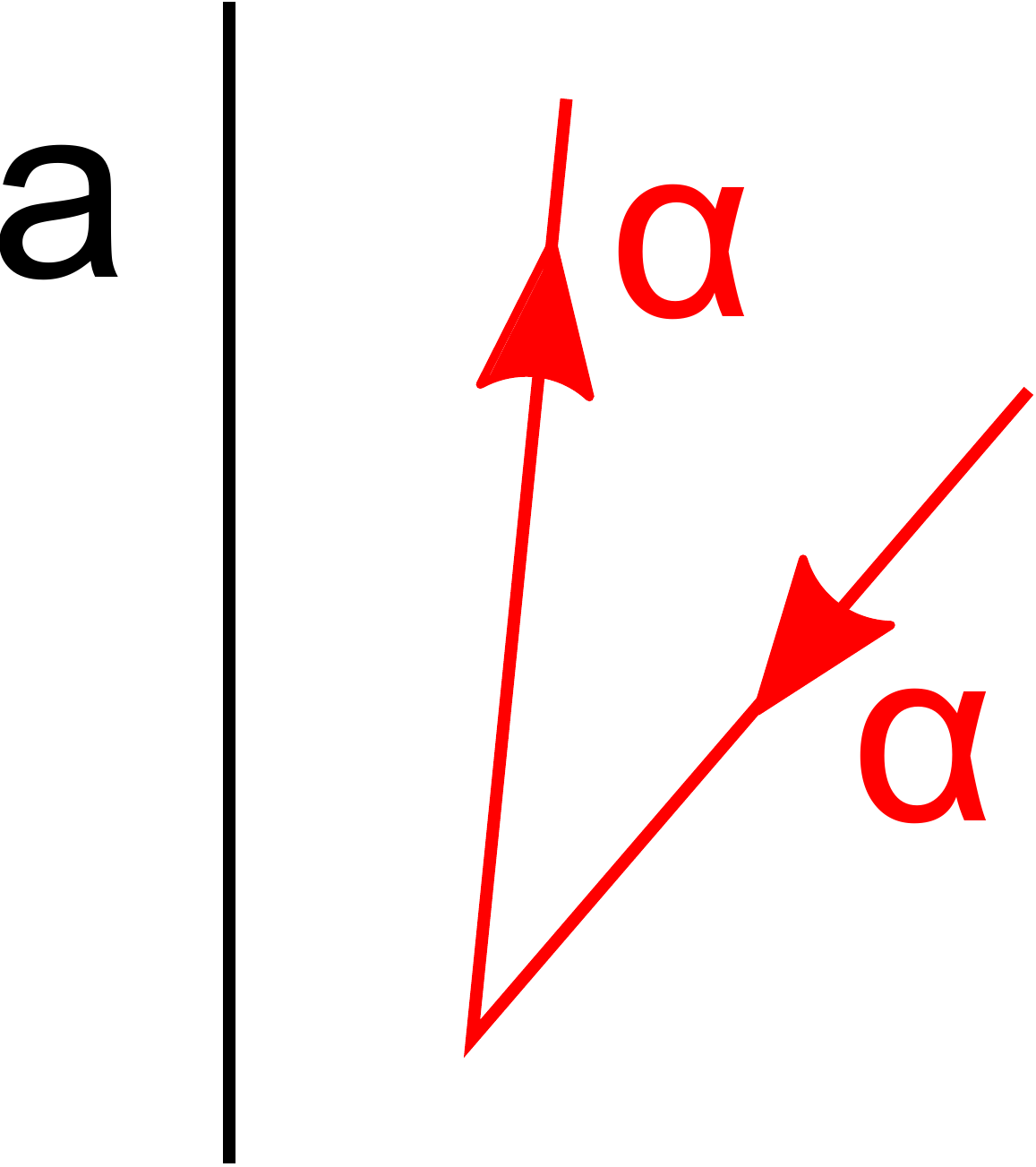}} \middle|\Phi\right\> \label{stringe} \\
			\left< \raisebox{-0.12in}{\includegraphics[height=0.4in]{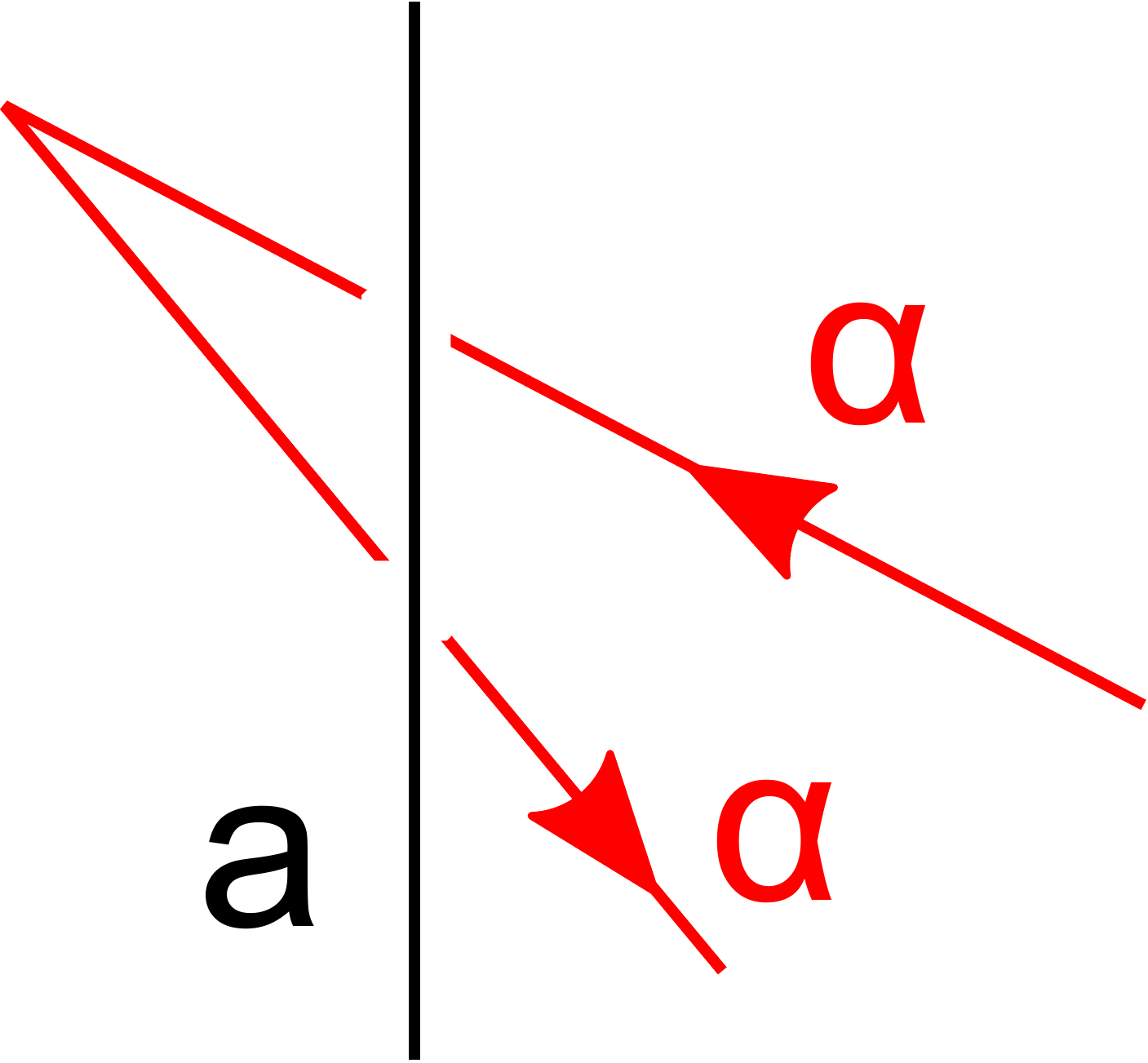}} \middle|\Phi \right \>&=\left< \raisebox{-0.12in}{\includegraphics[height=0.4in]{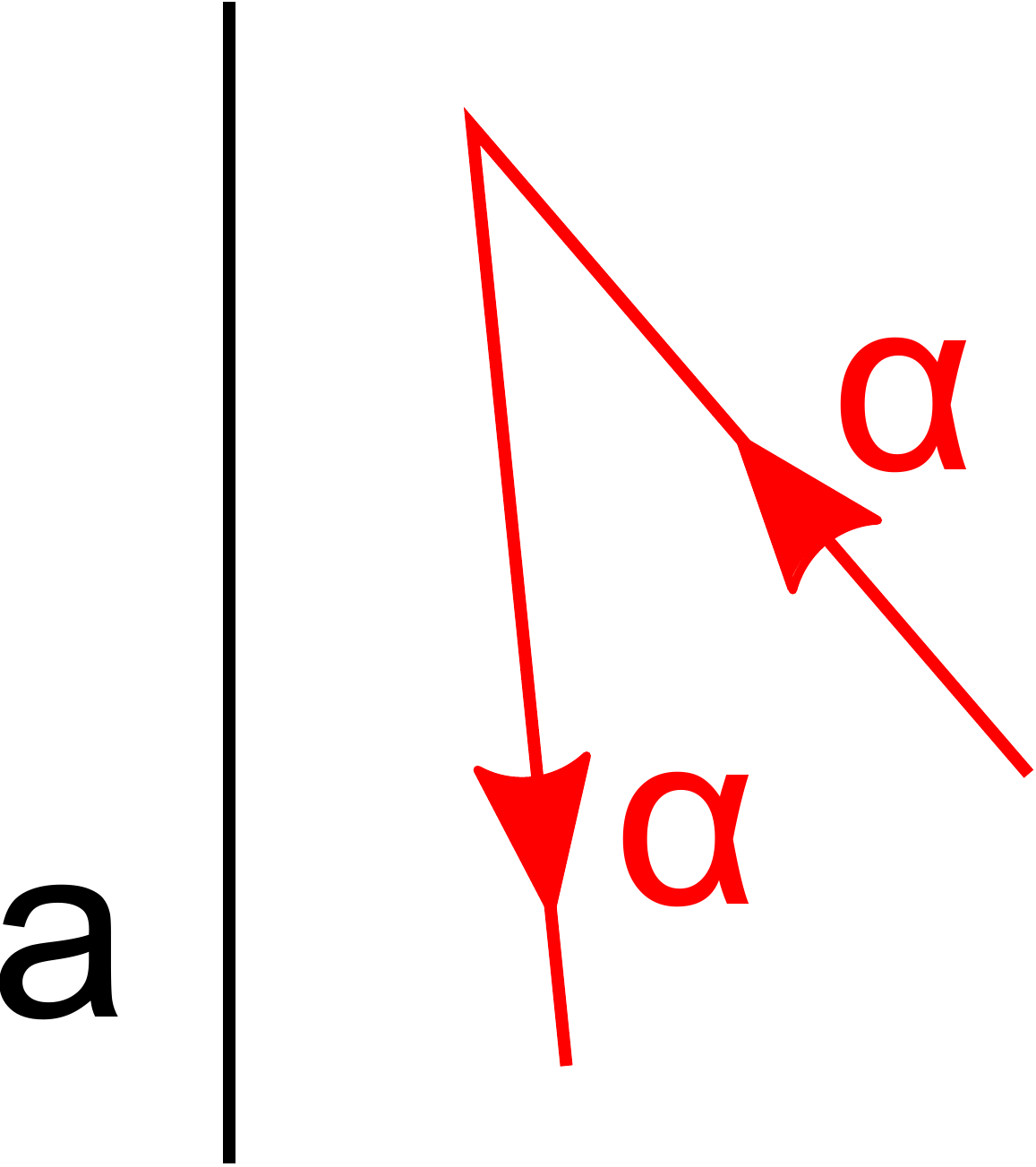}} \middle|\Phi\right\> \label{stringd}\ \ .
		\end{align}
	\label{pathind2}
\end{subequations}
Algebraically, this implies
\begin{equation}
	\sum_{a',r}\Omega^{a,rsa''}_{\alpha}  \Omega^{a,\bar{r}ta'}_{\bar{\alpha}}  F^{r\bar{r}a}_{a0a'} 
	 (F^{rat}_{aa''a'})^* \sqrt{\frac{d_r}{d_t}}= (F^{a\bar{t}t}_{aa''0})^* \delta_{s \bar{t}} \label{weq4}.
\end{equation}

If there exists an antiparticle $\bar{\alpha}$ for which $\Omega_{\bar{\alpha}}$ satisfies (\ref{weq4}), the closed $W_\alpha$ string operators obtained by joining upward $W_\alpha$ and $W_{\bar{\alpha}}$ strings are path independent in the sense of Eq.~(\ref{pathind})  at all points, and thus does not create any excitation when applied to the ground state.   Though it is not obvious from the discussion here, on general grounds\cite{KitaevHoneycomb} such a solution should always exist, provided that all quasiparticles in the theory can be created by string operators of the form described here.

\begin{figure}[ptb]
\begin{center}
\includegraphics[width=0.7\columnwidth]{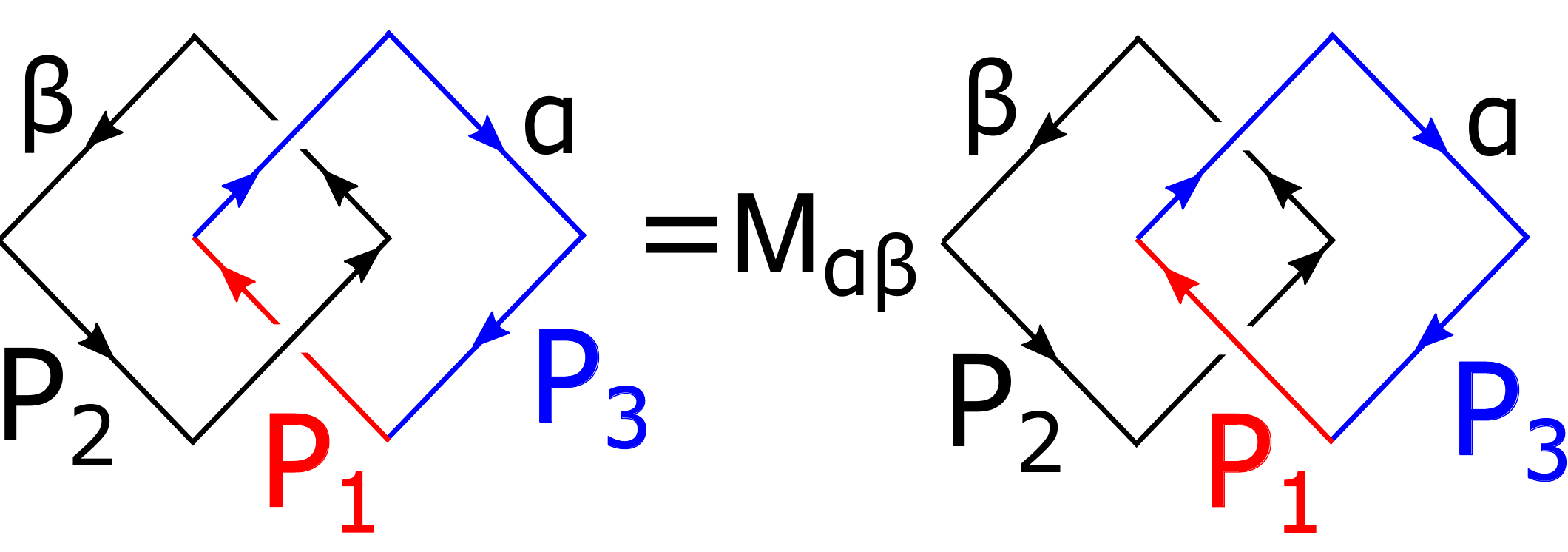}
\end{center}
\caption{
	The $S$ matrix is computed by by comparing the action of $W_\alpha(P_3) W_\beta(P_2) W_\alpha(P_1)$ and $W_\alpha(P_3) W_\alpha(P_1) W_\beta(P_2)$ where $P_1\cup P_3$ forms a closed loop.  Specifically, $S_{\alpha \beta} = \frac{d_\alpha d_\beta}{ D} M_{\alpha\beta}$.  
} 
\label{fig:statistics}
\end{figure}

\subsection{Braiding statistics of quasiparticles}

After finding the quasiparticles, we are now ready to compute their braiding statistics. Specifically, we will compute the $S$ matrix $S_{\alpha\beta}$ and the  topological spins $\theta_\alpha$ and express them in terms of string operators. 

 Before we compute the $S$ matrix, it is convenient to first compute the monodromy matrix $M_{\alpha\beta}$ which is related to $S_{\alpha \beta}$ via some normalization factors:
\begin{equation}
S_{\alpha \beta}= M_{\alpha\beta}  \frac{d_\alpha d_\beta}{ D} \ .
\label{smrel}
\end{equation} 
Here $d_\alpha$ is the quantum dimension of the quasiparticle $\alpha$ and $D = \sqrt{\sum_\alpha d_\alpha^2}$.  (For the definition of ``quantum dimension'' of quasiparticles see Ref.~\onlinecite{KitaevHoneycomb}. For an explicit formula for $d_\alpha$ in the context of string-net models, see Eq.~\ref{Eq:dalpha} below).

The monodromy matrix $M_{\alpha\beta}$ is defined in terms of a three step process in which (1) two particle-antiparticle pairs ($\alpha,\bar{\alpha},\beta,\bar{\beta}$) are created from the vacuum; (2) the particle $\alpha$ is braided around the particle $\beta$; and (3) each pair ($\alpha,\bar{\alpha},\beta,\bar{\beta}$) is re-annihilated to the vacuum (left panel of Fig.~\ref{fig:statistics}).  To define $M_{\alpha \beta}$, consider the probability amplitude for the above braiding process, divided by the probability amplitude of another process in which each pair of particles individually follows the same trajectory in space and time, but the pair ($\alpha,\bar{\alpha}$) is re-annihilated before the pair $(\beta,\bar{\beta}$) is created (right panel of Fig.~\ref{fig:statistics}). The monodromy matrix $M_{\alpha \beta}$ is defined to be this ratio of probability amplitudes. 

Equivalently, in the language of string operators, $M_{\alpha \beta}$ is given by the ratio
\begin{equation}
	M_{\alpha\beta}=\frac{\<\Phi|W_{\bar{\beta}}(P_3) W_\alpha(P_2) W_\beta(P_1)|\Phi\>}
	{\<\Phi|W_{\bar{\beta}}(P_3) W_\beta(P_1) W_\alpha(P_2) |\Phi\>}
	\label{Eq:Stime}
\end{equation}
where $P_1$ and $P_3$ are paths connecting two points $i$ and $f$, and $P_2$ is a third path that encircles the point $f$ (Fig.~\ref{fig:statistics}). Here, the numerator of Eq.~(\ref{Eq:Stime}) describes a process in which we first create a pair of quasiparticles $\beta,\bar{\beta}$ from the vacuum at positions $f$ and $i$ respectively, then act with a closed $\alpha$-string operator encircling $\beta$, and finally annihilate the $\beta,\bar{\beta}$ pair. The denominator describes a process in which we first act with the closed $\alpha$-string operator, and then create and re-annihilate the $\beta,\bar{\beta}$ pair.

 To proceed further, we join the string operators in (\ref{Eq:Stime}) into closed loops, which gives the following graphical expression for $M_{\alpha \beta}$:
\begin{equation}
	M_{{\alpha}{\beta}}=\frac{\< \Phi \left| 
	\raisebox{-0.22in}{\includegraphics[height=0.5in]{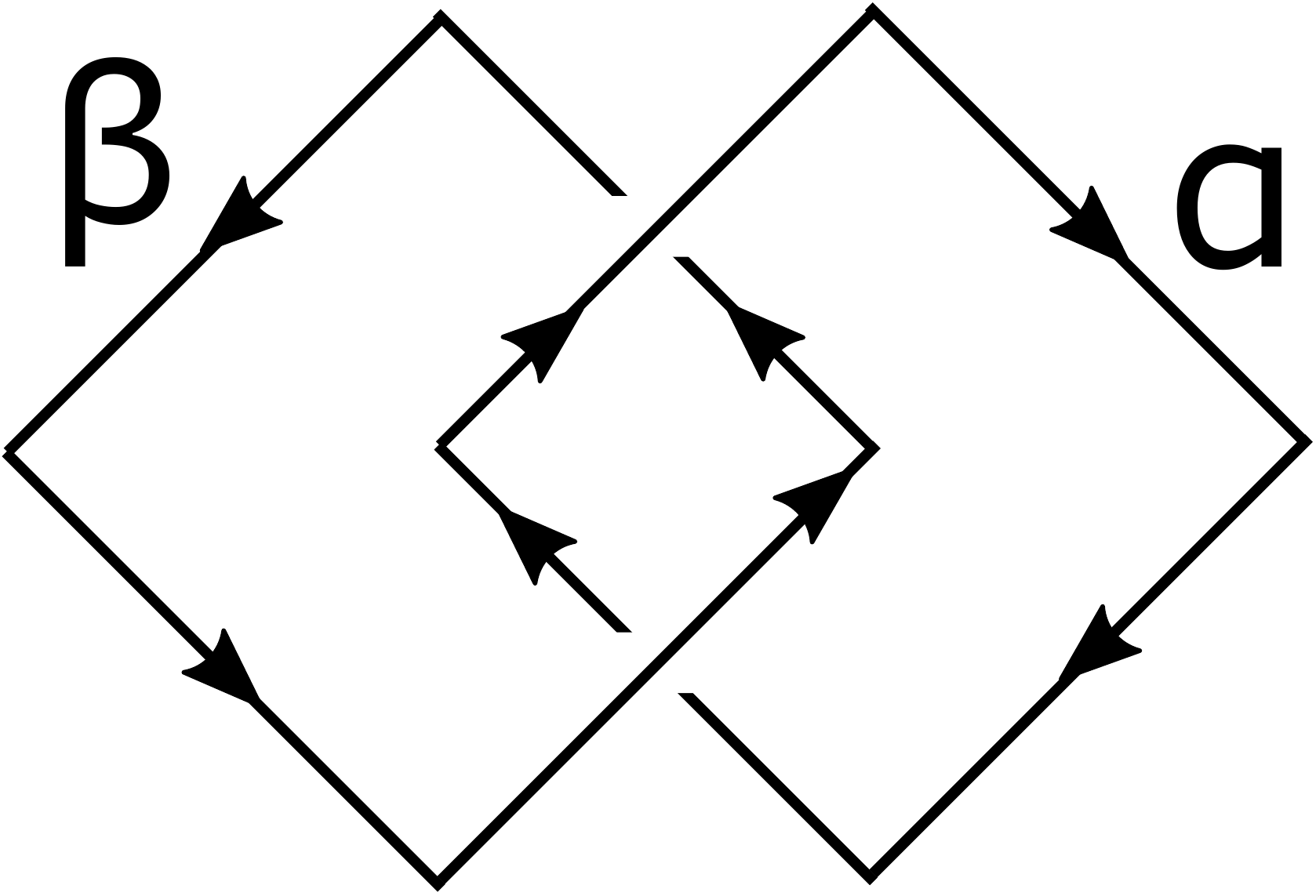}} 
\right| \Phi \>}
{\< \Phi \left| 
	\raisebox{-0.22in}{\includegraphics[height=0.5in]{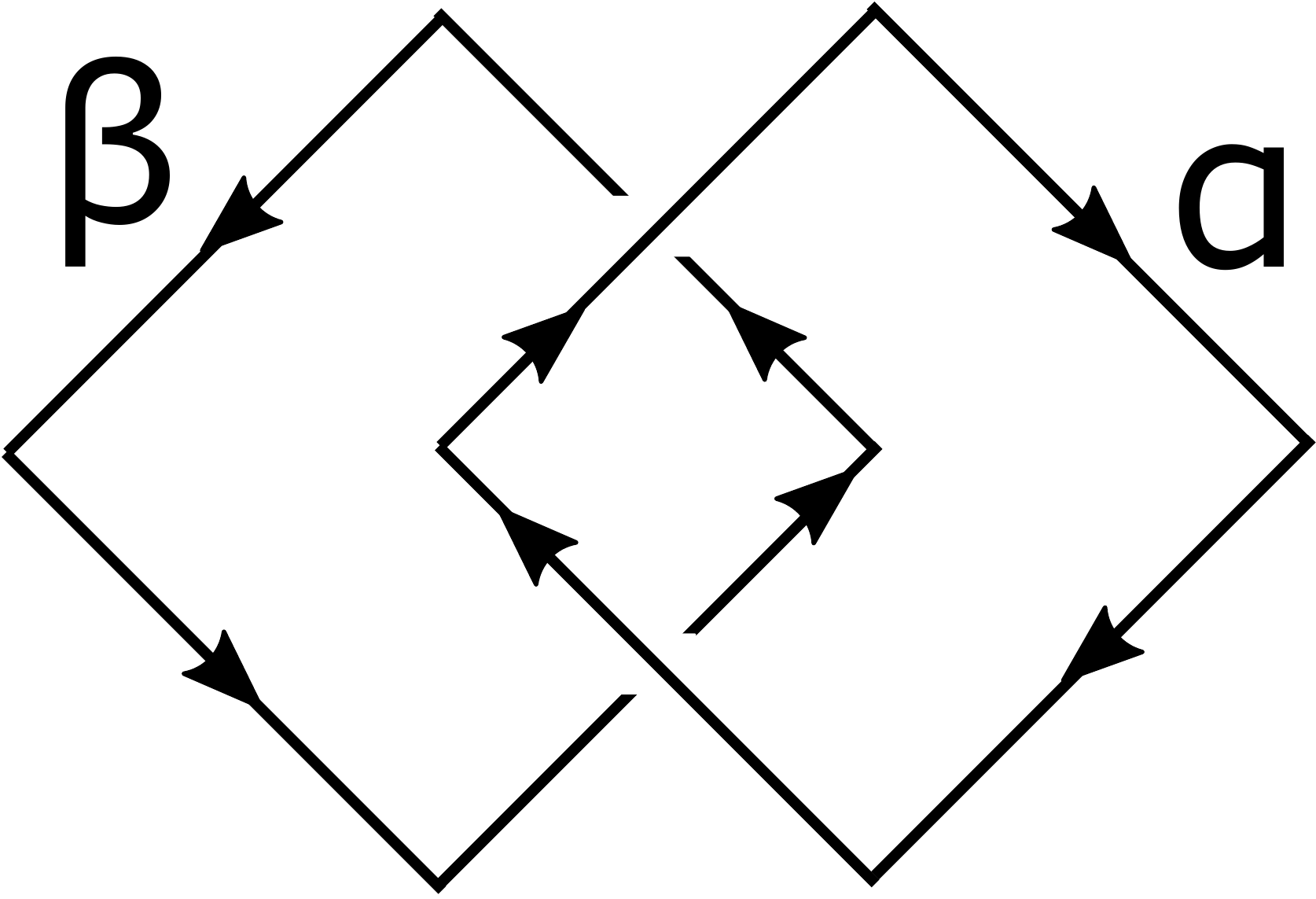}} 
\right| \Phi \>} 
	\label{smat}
\end{equation}
Here we have two closed string operators acting along two linked paths in the numerator and the same two closed string operators acting along corresponding unlinked paths in the denominator. We have used the convention that strings that act earlier (later) appear  under (over) other strings at crossings.

We now proceed to evaluate the numerator and denominator of (\ref{smat}). To evaluate the denominator, it is useful to first consider the action of a closed string operator $\alpha$ on the vacuum (empty) state:
\begin{equation}
\left< \text{vacuum} \right|
	\raisebox{-0.22in}{\includegraphics[height=0.5in]{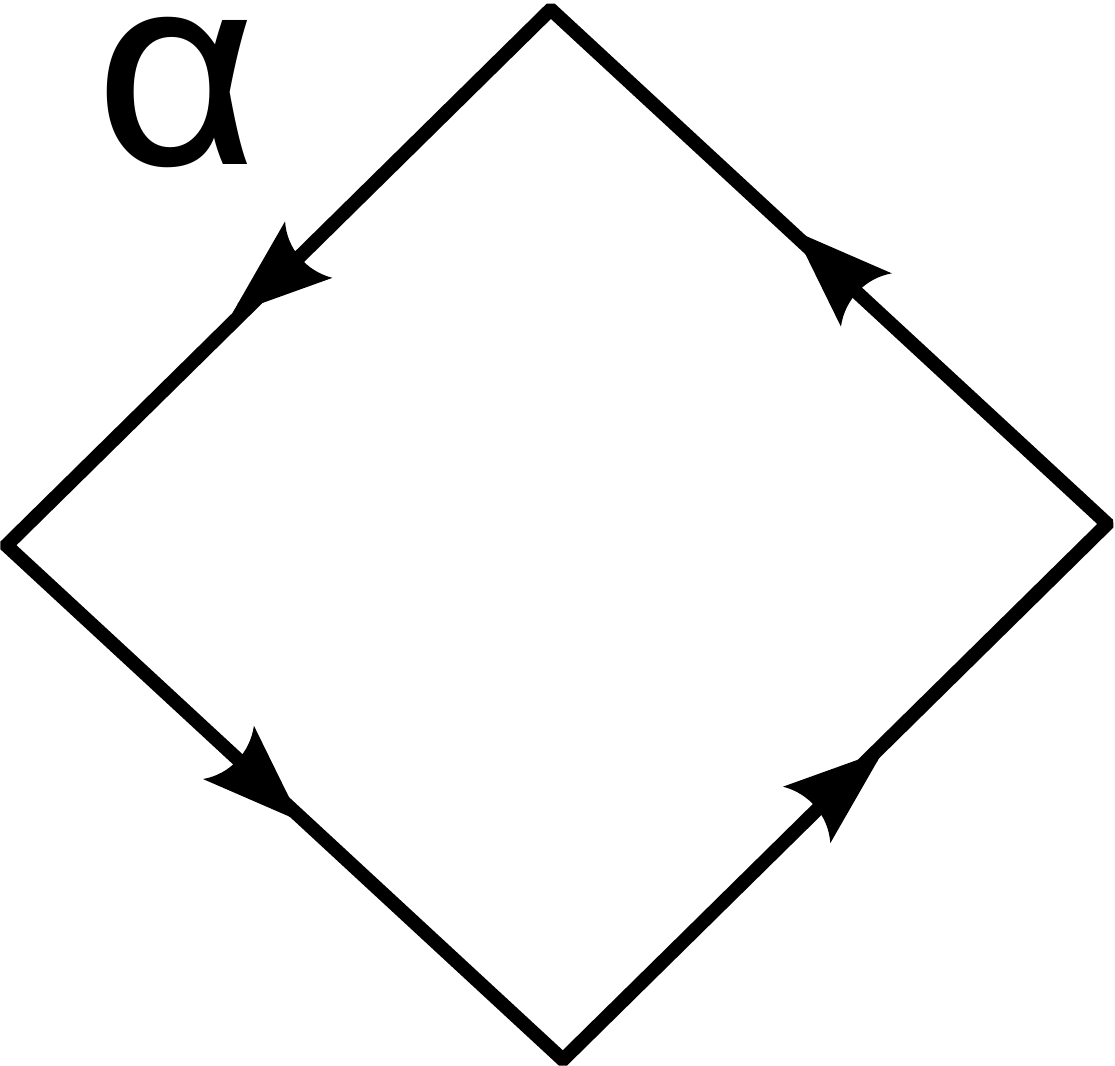}} 
	=\sum_{s} n_{\alpha,s}   \left<\raisebox{-0.22in}{\includegraphics[height=0.5in]{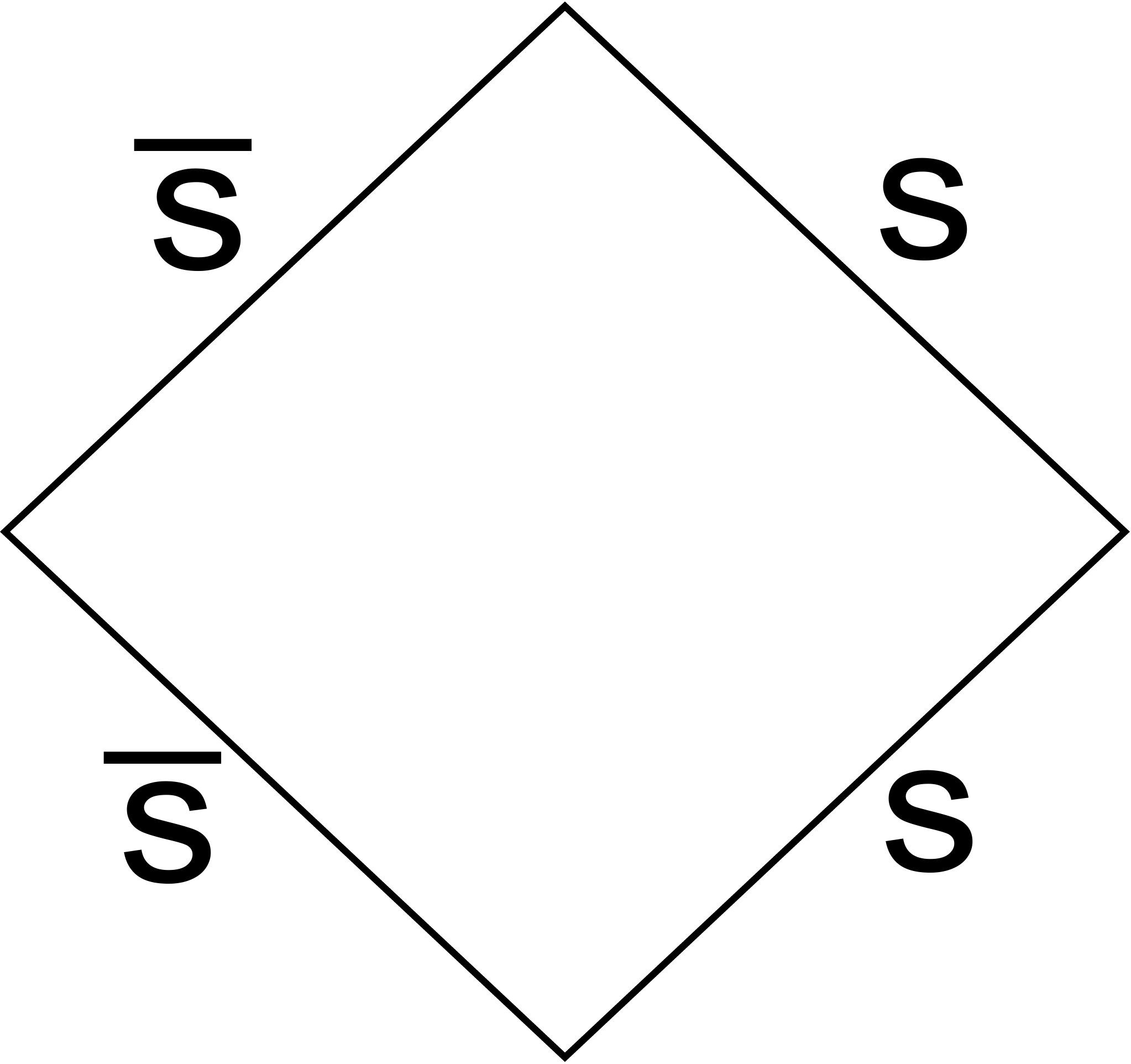}} \right|
	\label{Eq:ClosedAct}
\end{equation}
Multiplying both sides of (\ref{Eq:ClosedAct}) by $|\Phi\>$ and using the fact that $W_\alpha(P) |\Phi\> \propto |\Phi\>$ for any closed string operator $W_\alpha(P)$, we deduce that
\begin{equation}
	\raisebox{-0.22in}{\includegraphics[height=0.5in]{t2.pdf}} \left| \Phi \right>
	=\sum_{s} n_{\alpha,s} d_s \left| \Phi \right>
\label{alphaphi}
\end{equation}
Hence, the denominator of $M_{\alpha \beta}$ is: 
\begin{equation}
\< \Phi \left| 
	\raisebox{-0.22in}{\includegraphics[height=0.5in]{s2.pdf}} 
\right| \Phi \> =  \sum_{st} n_{\alpha, s} n_{\beta, t} d_s d_t
\end{equation}

To evaluate the numerator, we use the same strategy: we first consider the action of the linked string operators on the vacuum state and then deduce their action on  $|\Phi\>$ using the fact that $|\Phi\>$ is an eigenstate of these operators. In this way, we obtain
\begin{equation}
\< \Phi \left| 
	\raisebox{-0.22in}{\includegraphics[height=0.4in]{s1.pdf}} 
\right| \Phi \>=	 \sum_{stb}\text{Tr}(\bar{\Omega}_{\alpha}^{t,ssb})
	\text{Tr}(\bar{\Omega}_{{\beta}}^{s,ttb})d_b 
	\label{smatnum}
\end{equation}
 Combining the numerator and denominator of $M_{\alpha \beta}$, and substituting into (\ref{smrel}), we obtain the following general expression for the $S$-matrix, consistent with previous results\cite{LanWen13}:
\begin{equation}
	 S_{\alpha\beta}=\frac{1}{D}\sum_{stb}\text{Tr}(\bar{\Omega}_{\alpha}^{t,ssb})
	\text{Tr}(\bar{\Omega}_{{\beta}}^{s,ttb})d_b  \ . 
	\label{smat1}
\end{equation}
Here, we have used a formula that expresses the quantum dimension of $\alpha$ in terms of string operator data, namely:  
\begin{equation} \label{Eq:dalpha}
d_\alpha = \sum_s n_{\alpha, s} d_s.
\end{equation}
 We will not prove this formula here. 

Next, we compute the  topological spin of our quasiparticles, defined as the phase acquired by the wave function when a quasiparticle is rotated by $2 \pi$.  Here, we will not attempt to make a concrete connection to the associated space-time process, but rather observe that, as has been noted previously \cite{LanWen13}, 
in all known examples the topological spin can be evaluated as the ratio of amplitudes for the two processes: 
\begin{equation}
	e^{i\theta_\alpha}=
	\frac{ \< \Phi \left| 
	\raisebox{-0.22in}{\includegraphics[height=0.5in]{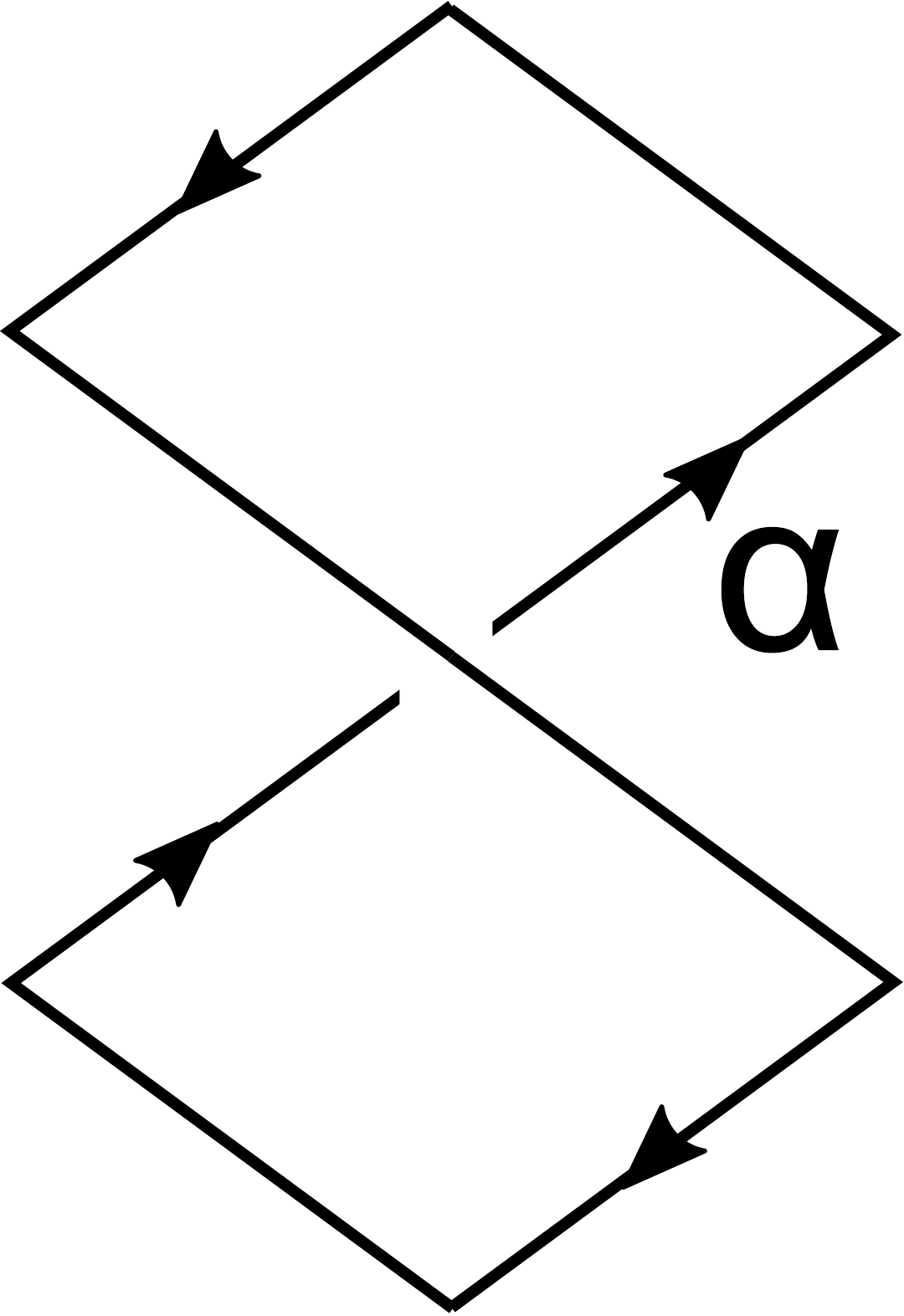}} 
	\right| \Phi \>}
	{\< \Phi \left|
	\raisebox{-0.22in}{\includegraphics[height=0.5in]{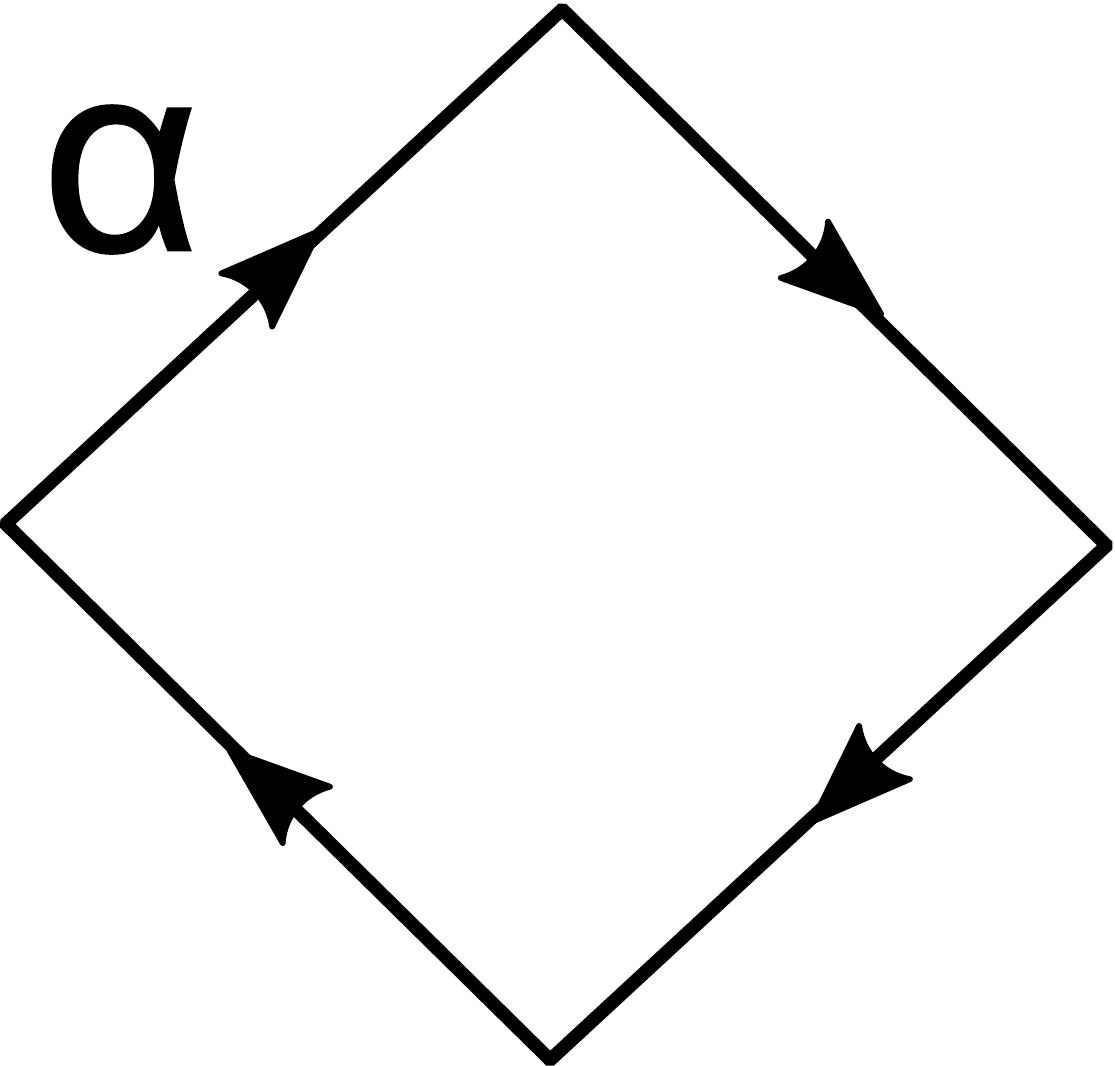}} 
	\right| \Phi \>}.
	\label{}
\end{equation} 
The two amplitudes can be expressed in terms of $(\Omega_\alpha,\bar{\Omega}_\alpha, n_\alpha)$:
\begin{equation}
	\begin{split}
	\< \Phi \left| 
	\raisebox{-0.22in}{\includegraphics[height=0.5in]{t1a.pdf}} 
	\right| \Phi \>&=\sum_{s}\text{Tr}({\Omega}^{\bar{s},ss0}_{\alpha}) d_s\\
		\< \Phi \left| 
	\raisebox{-0.22in}{\includegraphics[height=0.5in]{t2.pdf}} 
	\right| \Phi \>&=\sum_s n_{\alpha,s} d_s.
\end{split}
	\label{}
\end{equation}
Thus, the  topological spin of $\alpha$ is given by
\begin{equation}
	e^{i\theta_\alpha}=\frac{\sum_{s}\text{Tr}({\Omega}^{\bar{s},ss0}_{\alpha})d_s}{\sum_s n_{\alpha,s} d_s}.
	\label{twist}
\end{equation}

\section{Isotropic string-net models \label{sec:iso}}

One notable feature of our models is that the minimal consistency conditions (\ref{consistency}) required for the ground state wave function $\Phi$ to be well-defined are not isotropic.  
Consequently, in general two string-net configurations which can be continuously deformed into one another need not have the same ground state amplitude.
For example, while the wave function is invariant under the bendings shown in Eq.~(\ref{hbend}), it may not be invariant under vertical bendings in Eq.~(\ref{nomove}).  In addition, if $Y^{a \bar{a}}_0 \neq Y^{\bar{a}a}_0$, the corresponding string-net ground state cannot be isotropic on the sphere.   Specifically, isotropy on the sphere requires that we can pull an $(a, \bar{a})$ loop from the front of the sphere to the back of the sphere, where, when viewed from outside the sphere, it is a $(\bar{a},a)$ loop.  The two coefficients are equal if-- and only if--  $Y^{a \bar{a}} = Y^{\bar{a} a}$. 

In this section, we examine what additional conditions must be satisfied in order for our string-net ground state to be isotropic on the plane and on the sphere.  
To find these additional constraints, we first discuss how the ground state amplitude changes under planar deformations of a string-net configuration. 
Interestingly, we find that there are gauge invariant quantities that can prevent a model from being invariant under such deformations. 
We then determine the constraints that the data $\{F, Y\}$ must satisfy in order to make these amplitudes invariant under such planar deformations.  
Finally, we consider additional requirements that must be met for full isotropy on the sphere, and find that this {\it further restricts the data} $\{F, Y\}$. 
 At the end we comment on an additional tetrahedral reflection symmetry that was also required in the construction of Ref.~\onlinecite{LevinWenstrnet}.

\subsection{Bending of strings and vertices}

We first examine how deforming the string-net configuration in the plane affects the associated ground-state amplitude.  By a deformation, we mean a process in which edges and vertices can be bent, moved and twisted arbitrarily within the plane, provided that they do not intersect other segments of the string-net graph.
Any such deformation can be decomposed into a sequence of bendings of strings and vertices; thus it is thus sufficient to consider the following elementary bendings of vertices
\begin{equation}
	\begin{split}
		\Phi\left(\raisebox{-0.15in}{\includegraphics[height=0.4in]{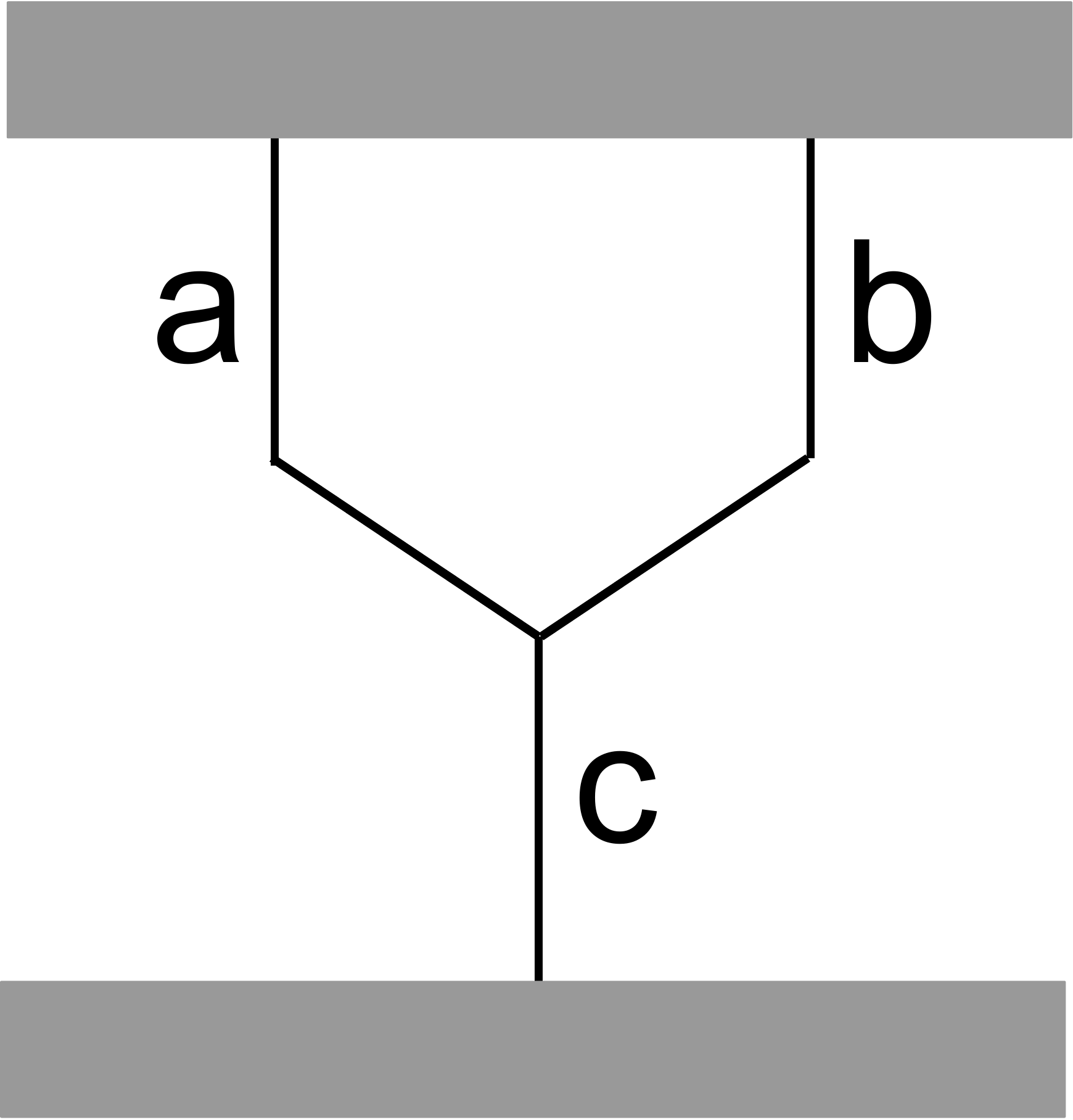}} \right)&=
		\frac{1}{[\tilde{F}^{0c}_{ab}]_{\bar{a}c}}\Phi\left(\raisebox{-0.15in}{\includegraphics[height=0.4in]{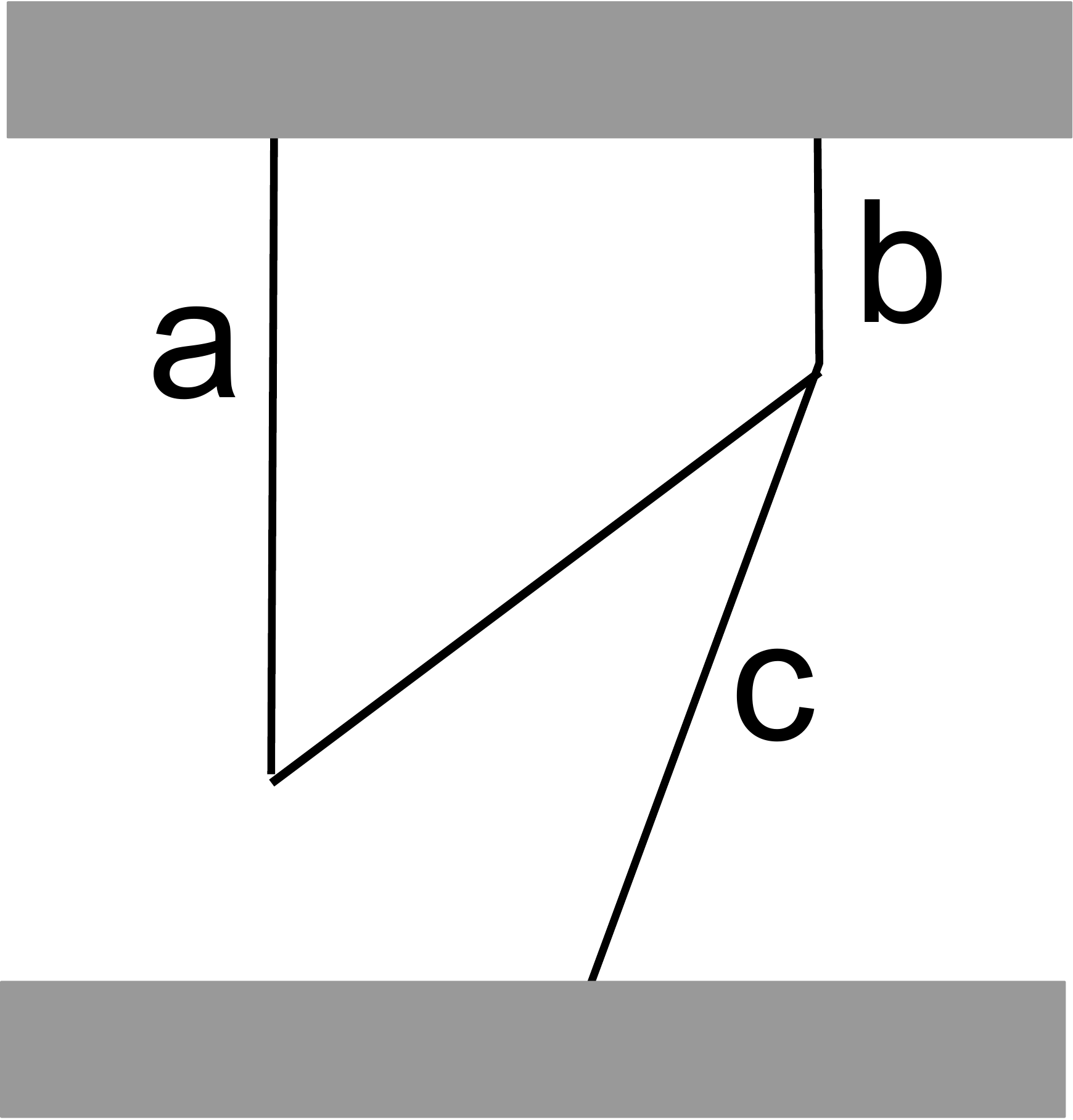}} \right)=
		\frac{1}{[F^{ab}_{c0}]_{\bar{b}c}}\Phi\left(\raisebox{-0.15in}{\includegraphics[height=0.4in]{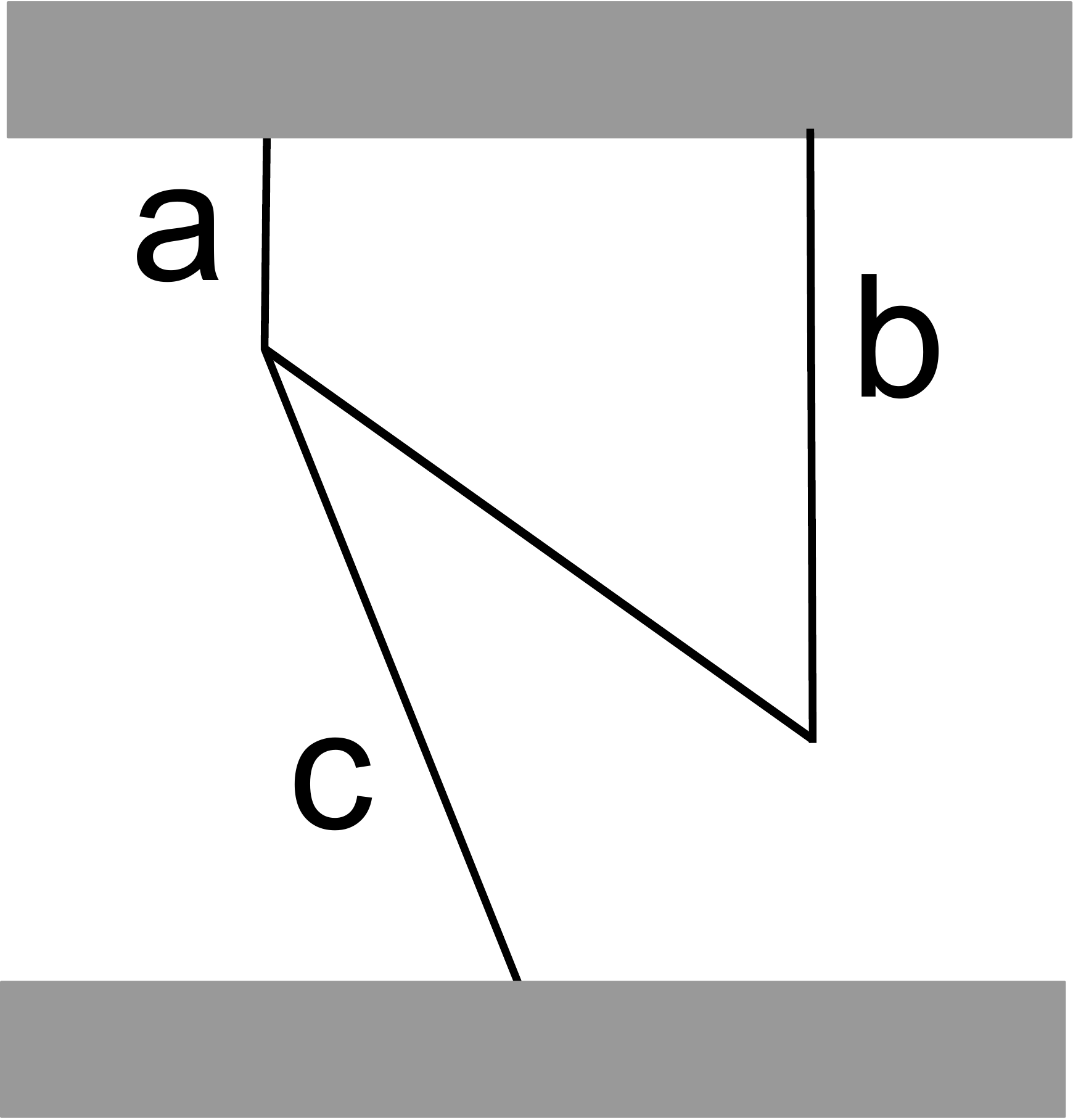}} \right)\\
		\Phi\left(\raisebox{-0.15in}{\includegraphics[height=0.4in]{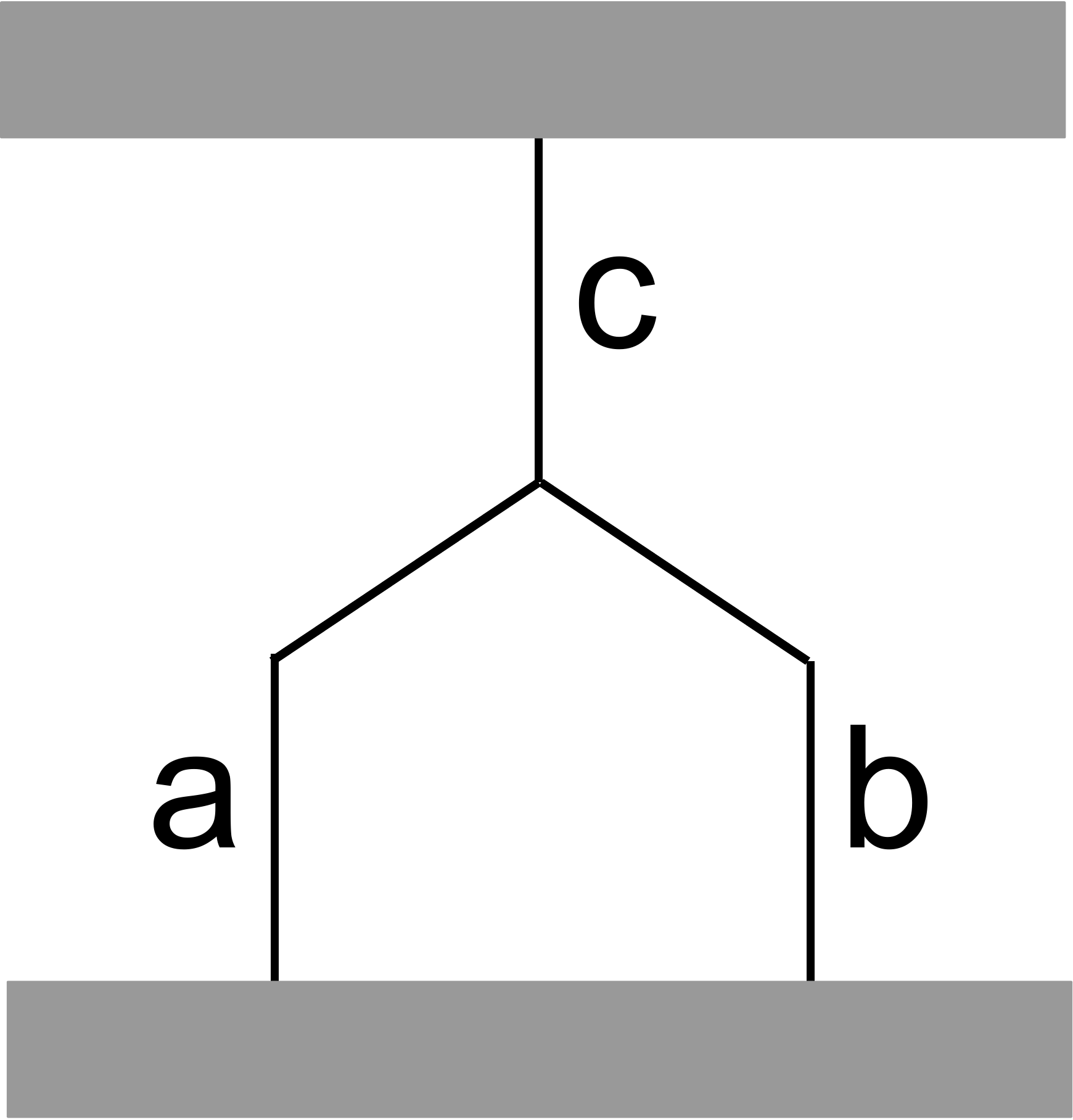}} \right)&=
		\frac{1}{[F^{0c}_{ab}]_{\bar{a}c}}\Phi\left(\raisebox{-0.15in}{\includegraphics[height=0.4in]{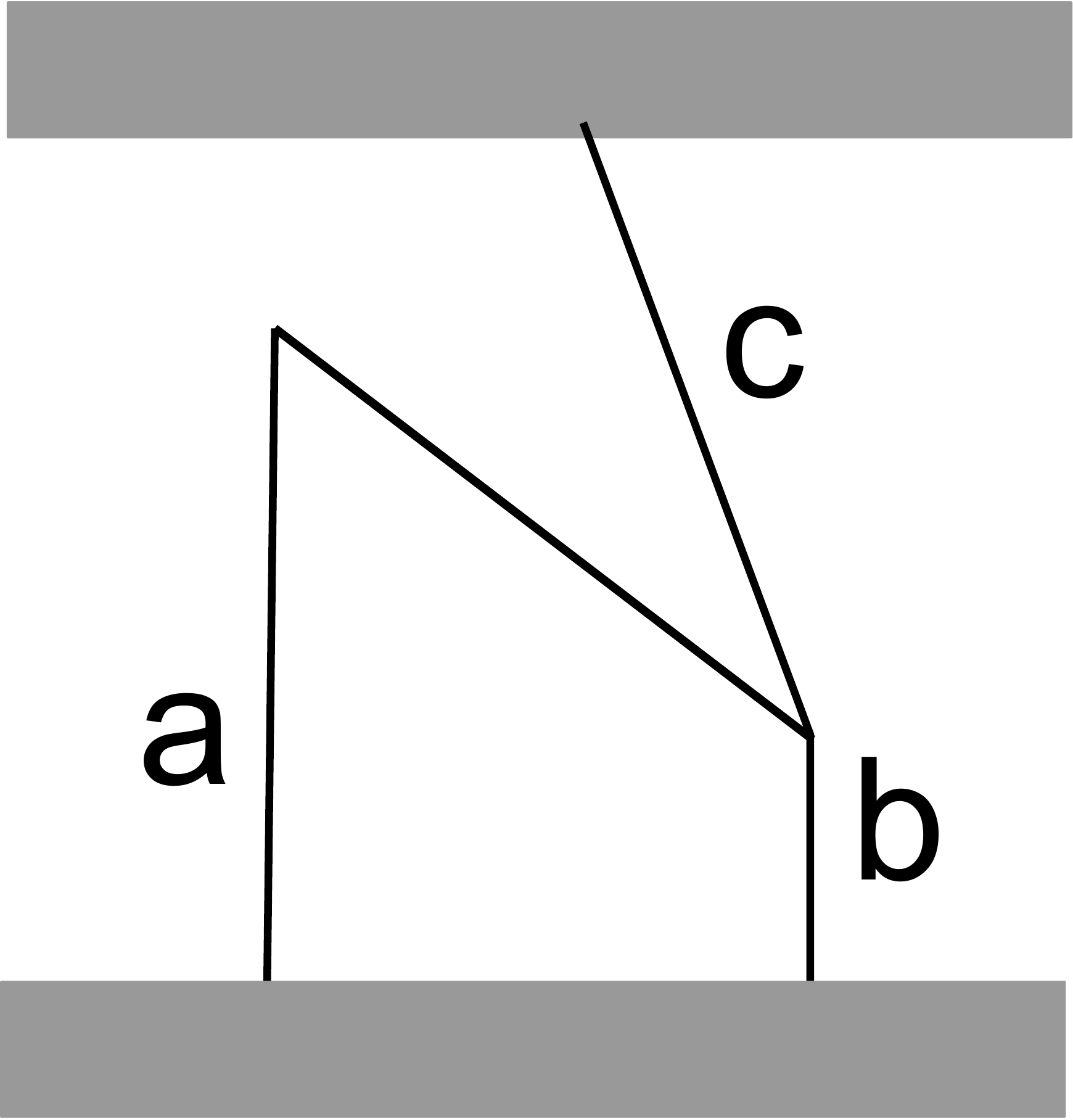}} \right)=
		\frac{1}{[\tilde{F}^{ab}_{c0}]_{\bar{b}c}}\Phi\left(\raisebox{-0.15in}{\includegraphics[height=0.4in]{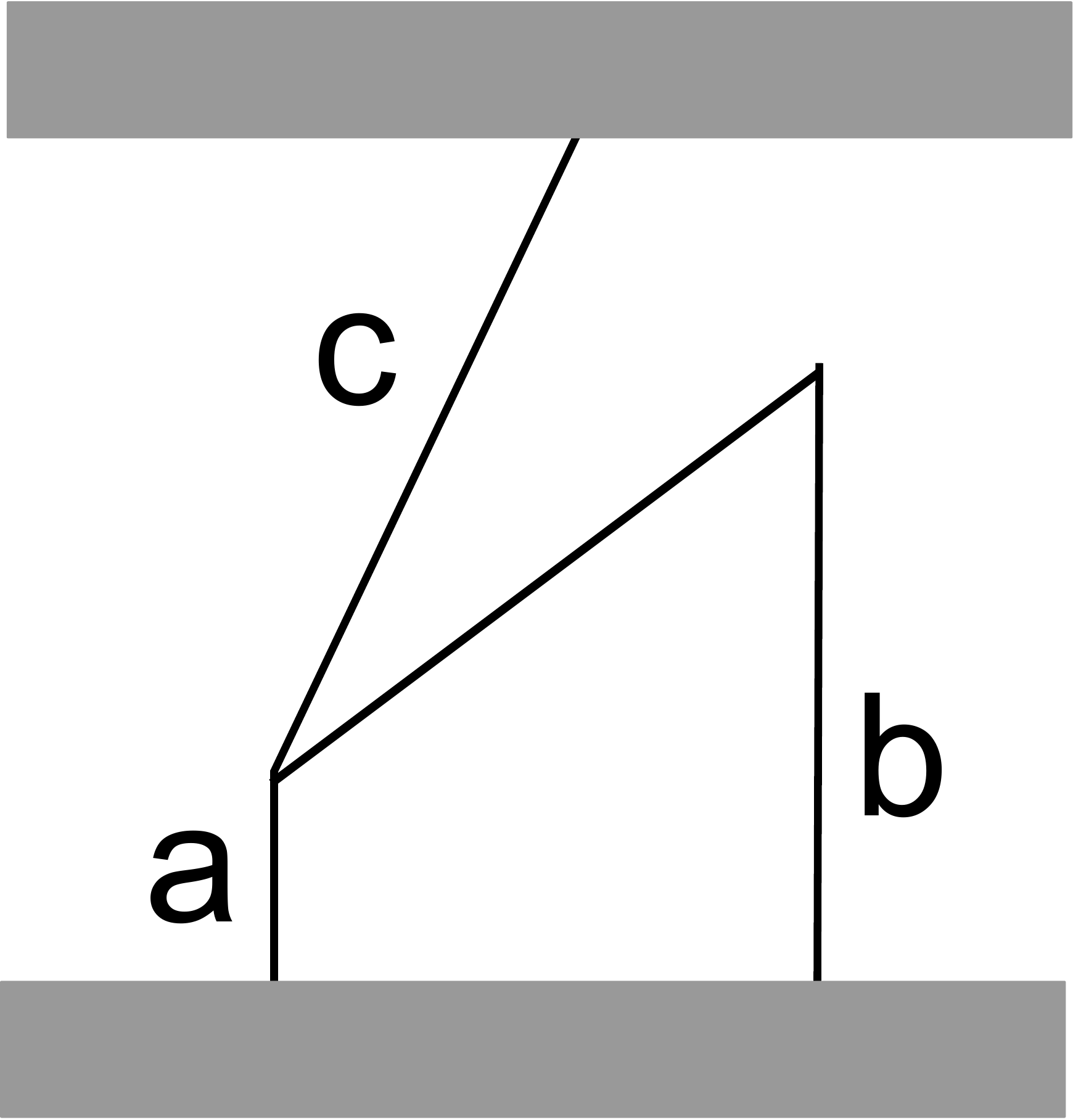}} \right).
	\end{split}
	\label{vbending}
\end{equation}
Eqs.~(\ref{vbending}) are simply special cases of (\ref{localrules1}), where one of the four external legs is the null string.
It follows from (\ref{hermicity}) that the coefficients $\{[\tilde{F}^{0c}_{ab}]_{\bar{a}c}, [F^{ab}_{c0}]_{\bar{b}c}, [F^{0c}_{ab}]_{\bar{a}c},[F^{ab}_{c0}]_{\bar{b}c}\}$ are $U(1)$ phase factors.
When these phase factors are equal to one, then two configurations which can be deformed into one another have the same ground-state amplitude, and the corresponding model is isotropic. Otherwise, the model is not isotropic.

Two comments are in order.
First, bending a string is a special case of bending a vertex (\ref{vbending}):
\begin{equation}
	\begin{split}
		\Phi\left(\raisebox{-0.22in}{\includegraphics[height=0.5in]{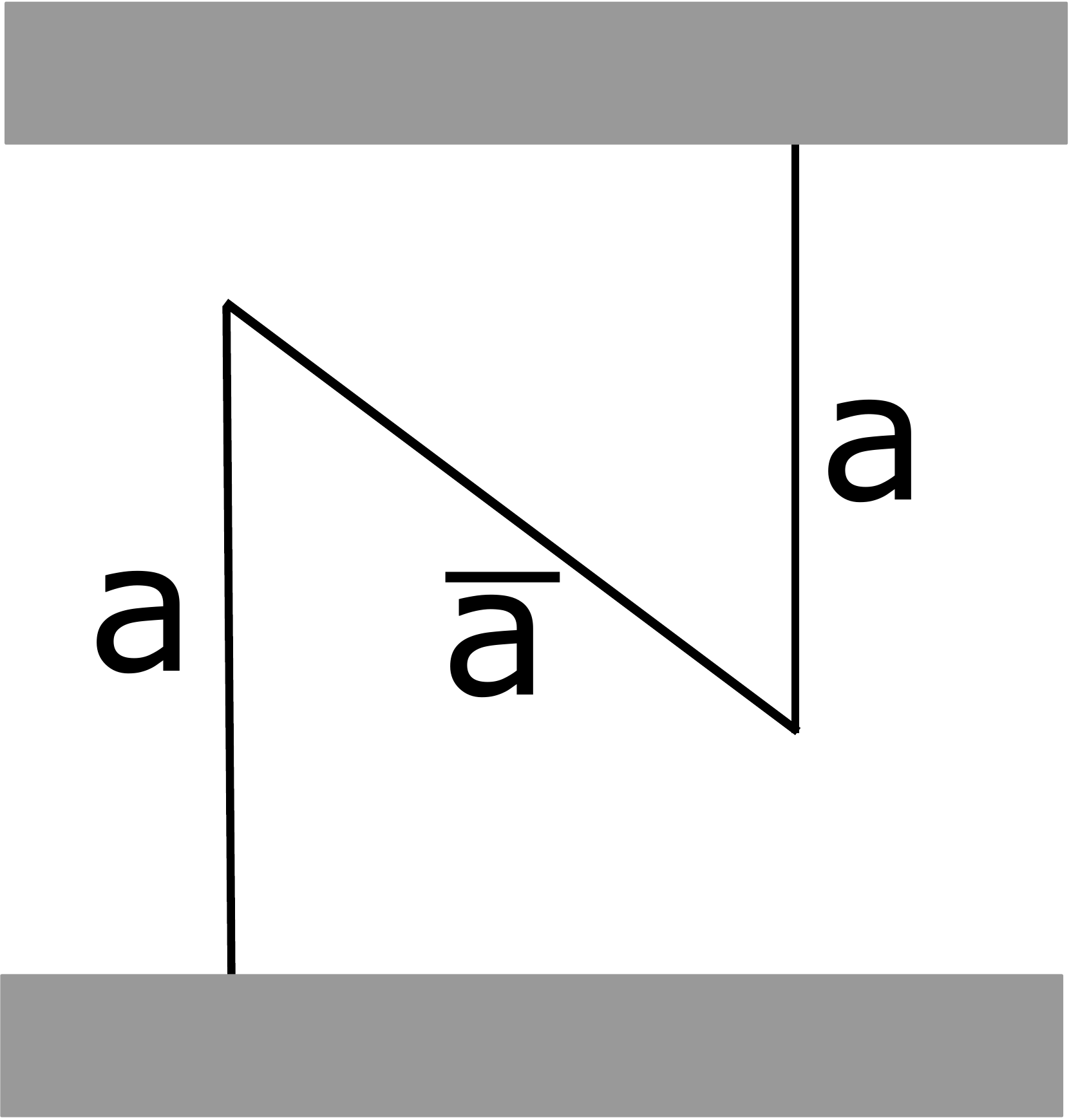}} \right)&
		=[F^{0a}_{a0}]_{\bar{a}a}
		\Phi\left(\raisebox{-0.22in}{\includegraphics[height=0.5in]{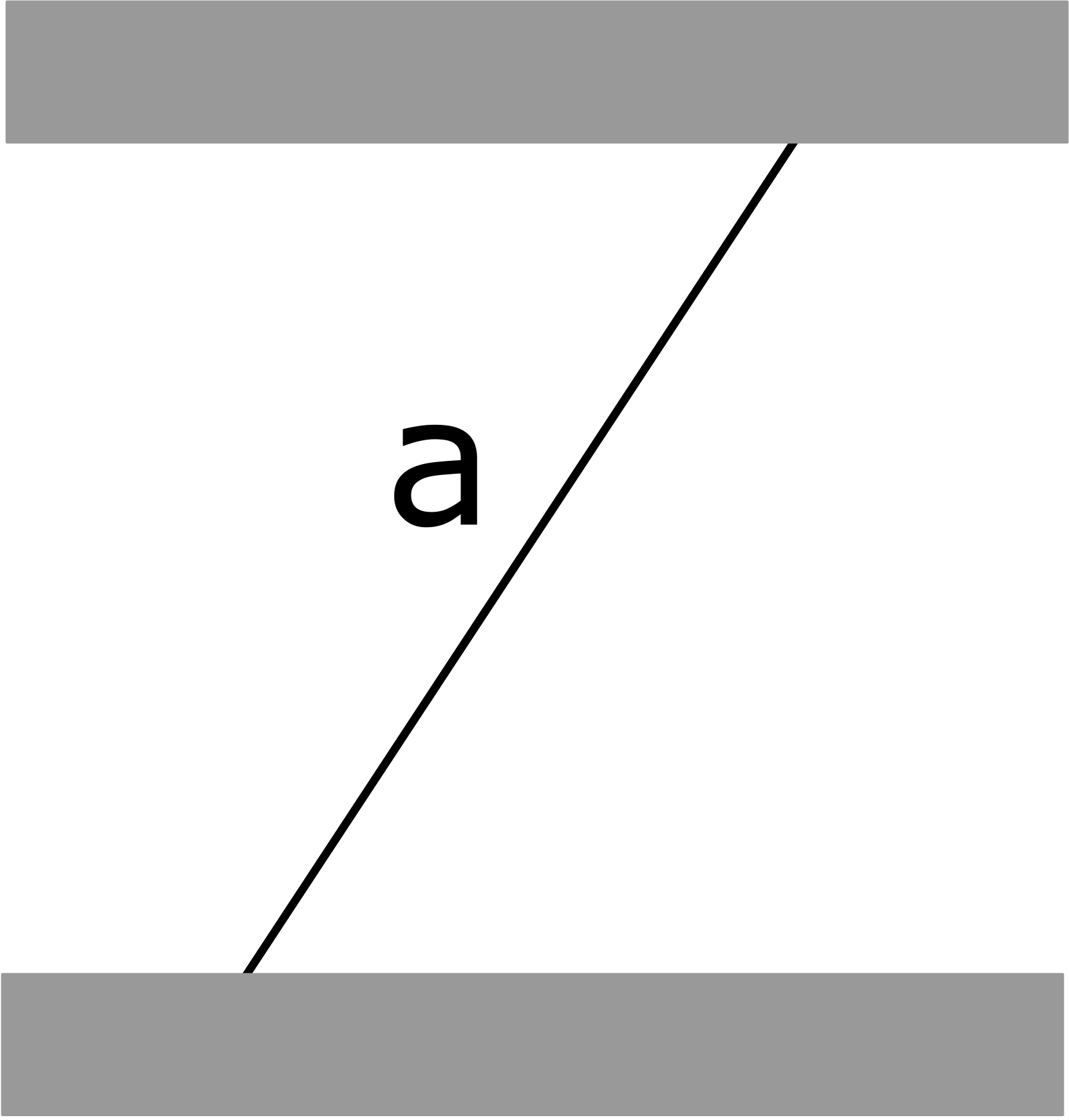}} \right)\\
		\Phi\left(\raisebox{-0.22in}{\includegraphics[height=0.5in]{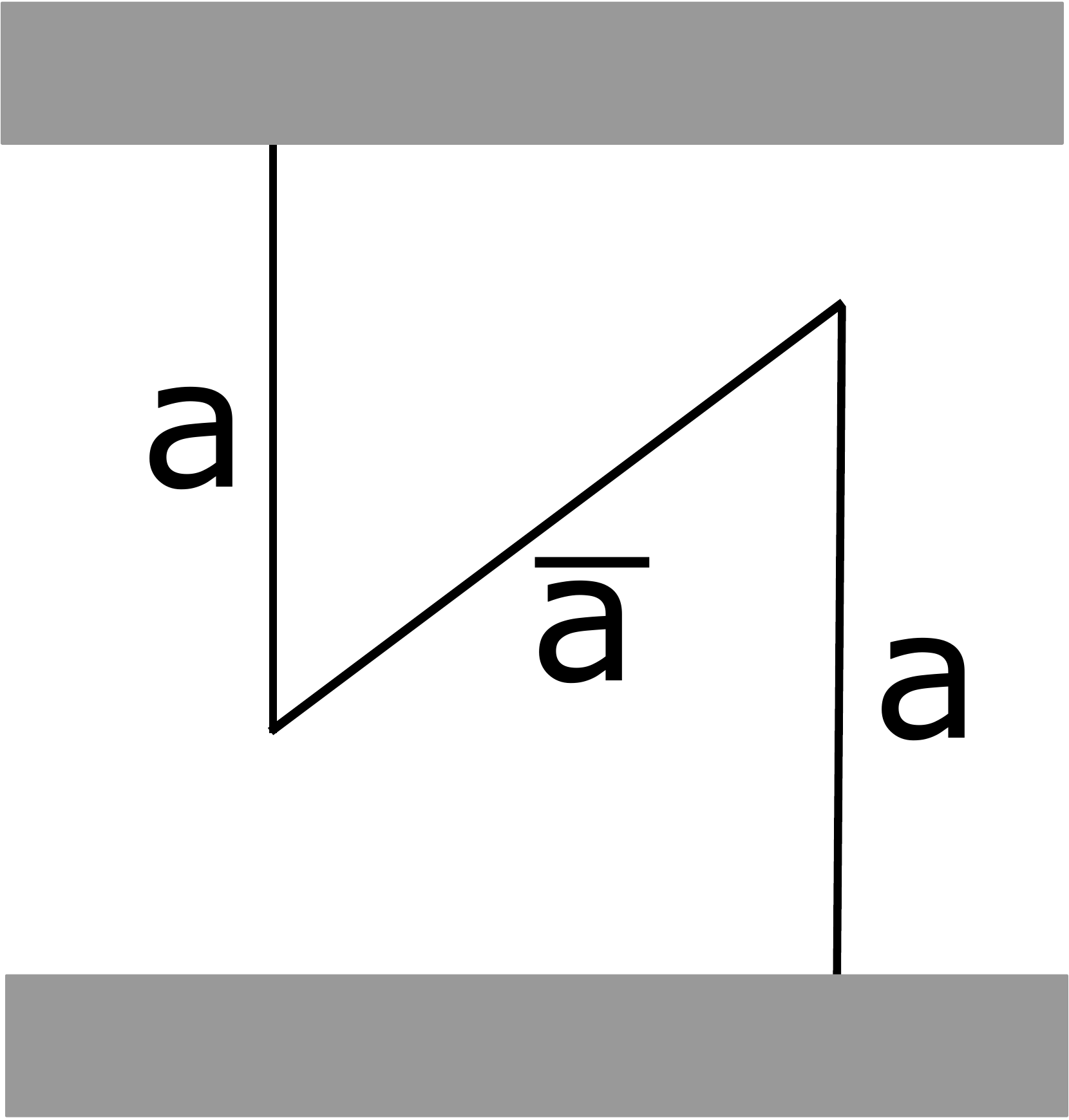}} \right)&
		=[\tilde{F}^{0a}_{a0}]_{\bar{a}a}
		\Phi\left(\raisebox{-0.22in}{\includegraphics[height=0.5in]{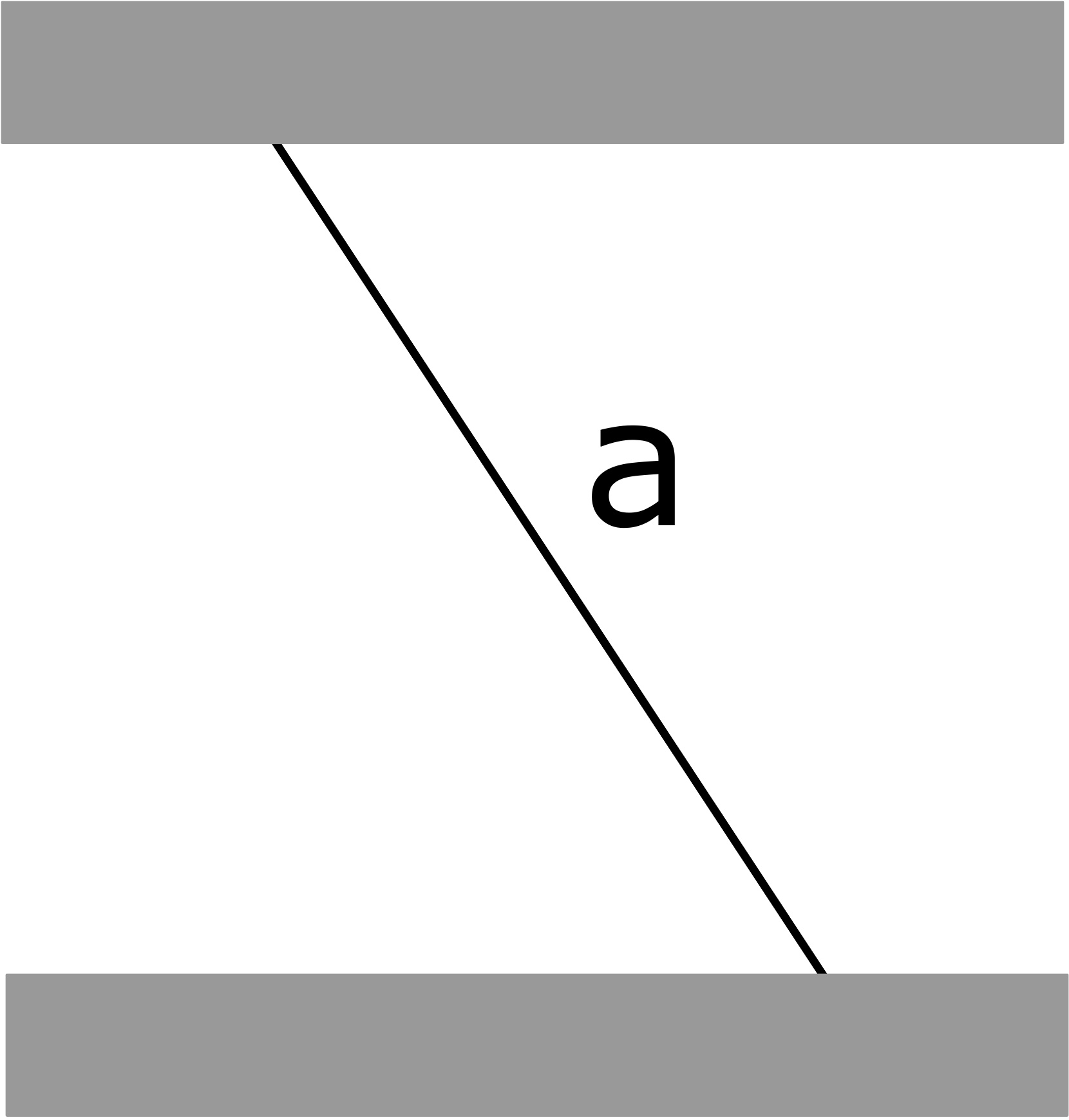}} \right).
	\label{vbend}
	\end{split}
\end{equation}
The phase factor\footnote{There are two ways to resolve an ``M'' like diagram made up of alternating ``$a$'' and ``$\bar{a}$'' strings: one can either use a $[\tilde{F}]$ rule on an $a$ line,  or an $[F]$ rule on an $\overline{a}$ line. This gives the second equality in Eq.~(\ref{gamma}).  }
\begin{equation}
	\gamma_a \equiv  [F^{0a}_{a0}]_{\bar{a}a}=[\tilde{F}^{0\bar{a}}_{\bar{a}0}]_{a\bar{a}}=F^{\bar{a}a\bar{a}}_{\bar{a}00} Y^{a\bar{a}}_0.
	\label{gamma}
\end{equation}
associated with bendings of strings is called the Frobenius-Schur indicator.
It follows from (\ref{hermicity}) that 
\begin{equation}
	|\gamma_a|=1,\quad (\gamma_a)^*= \gamma_{\bar{a}}.
	\label{Eq:gammastars}
\end{equation}
Furthermore, one can always choose the gauge function $f$ such that
\[
	\gamma_a=
	\begin{cases}
		\pm1, \quad &\text{if }a=\bar{a} \\
		1,\qquad &\text{ otherwise}.
	\end{cases}
	\label{}
\]
Then, we can use the gauge transformation $g$ to transform $\gamma$ so that $\gamma_a=1$ if $a=\bar{a}$.

Second, bending a vertex twice by (\ref{vbending}) is equivalent to rotating the vertex:
\begin{equation}
	\begin{split}
		\Phi\left(\raisebox{-0.15in}{\includegraphics[height=0.4in]{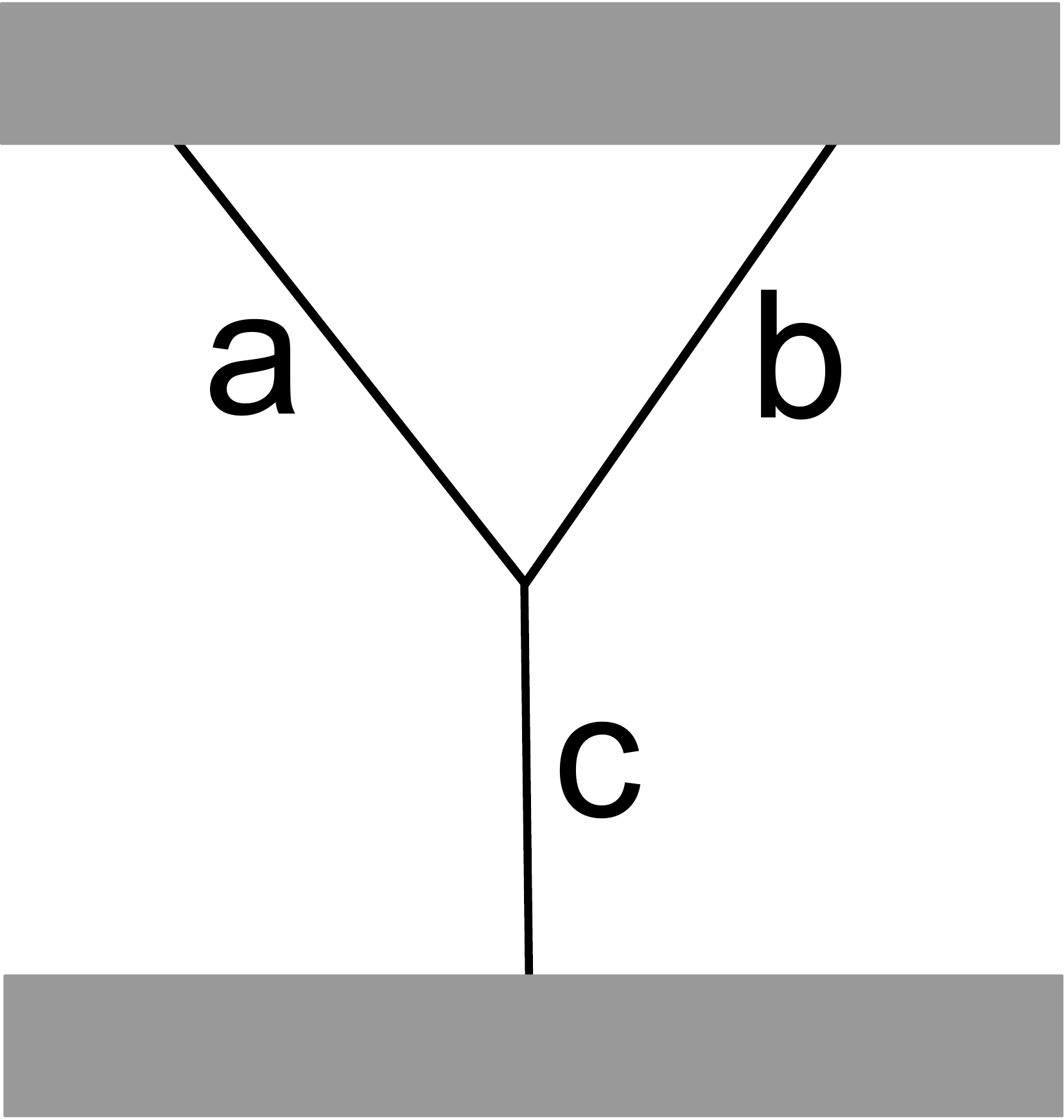}} \right)=
		\frac{1}{\alpha_{\bar{c}ab}\gamma_c}\Phi\left(\raisebox{-0.15in}{\includegraphics[height=0.4in]{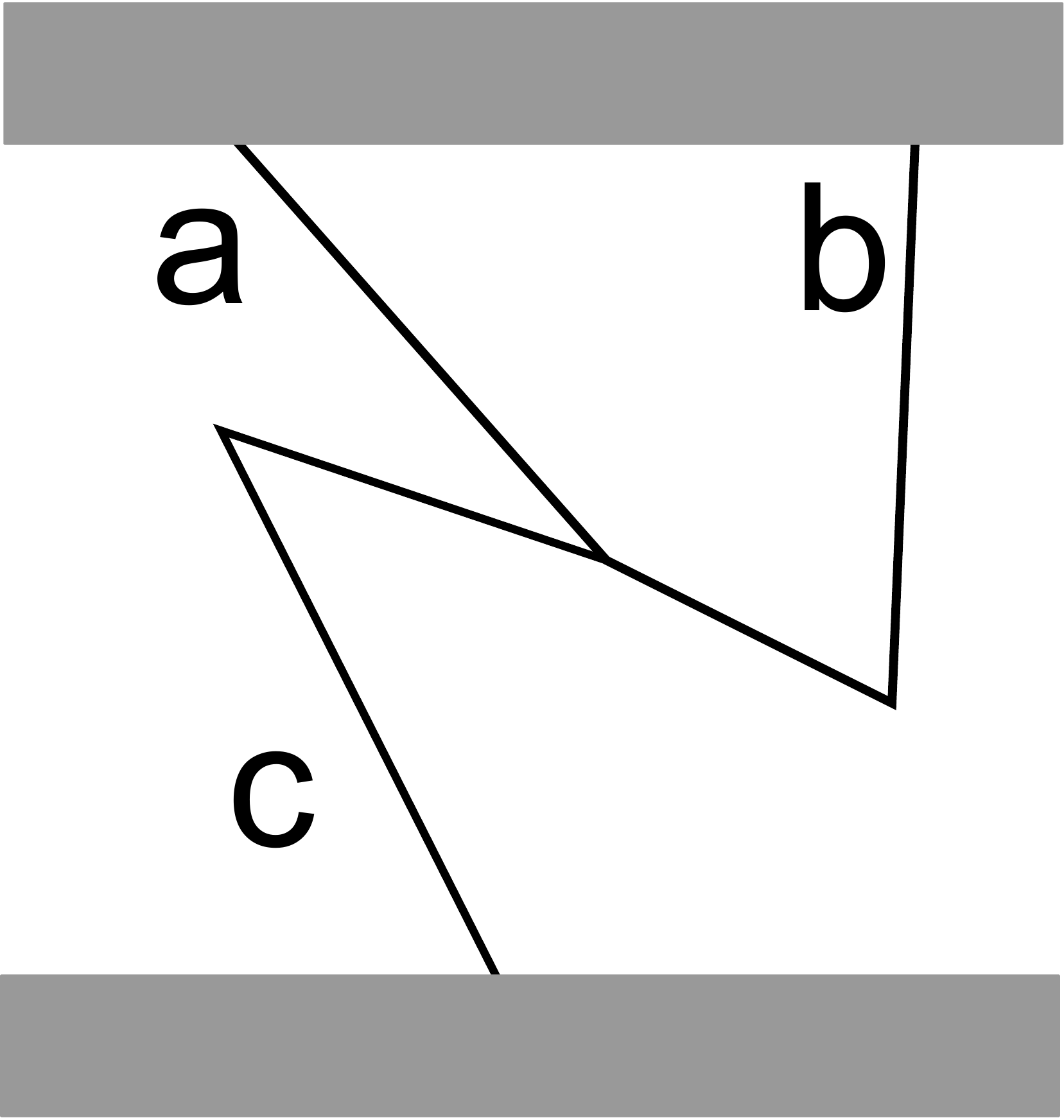}} \right)=	
		\alpha_{ab\bar{c}}\gamma_c \Phi\left(\raisebox{-0.15in}{\includegraphics[height=0.4in]{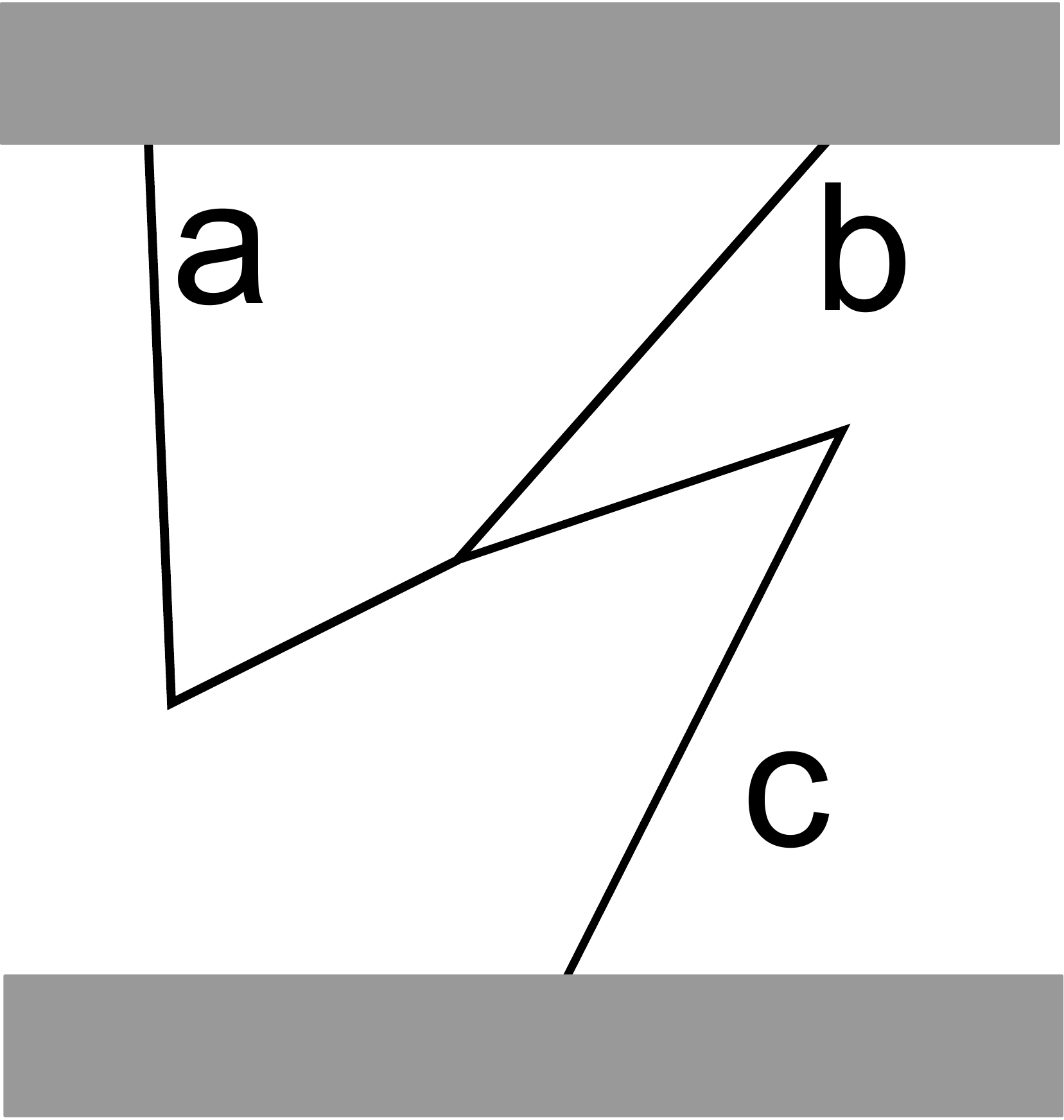}} \right) \\	
		\Phi\left(\raisebox{-0.15in}{\includegraphics[height=0.4in]{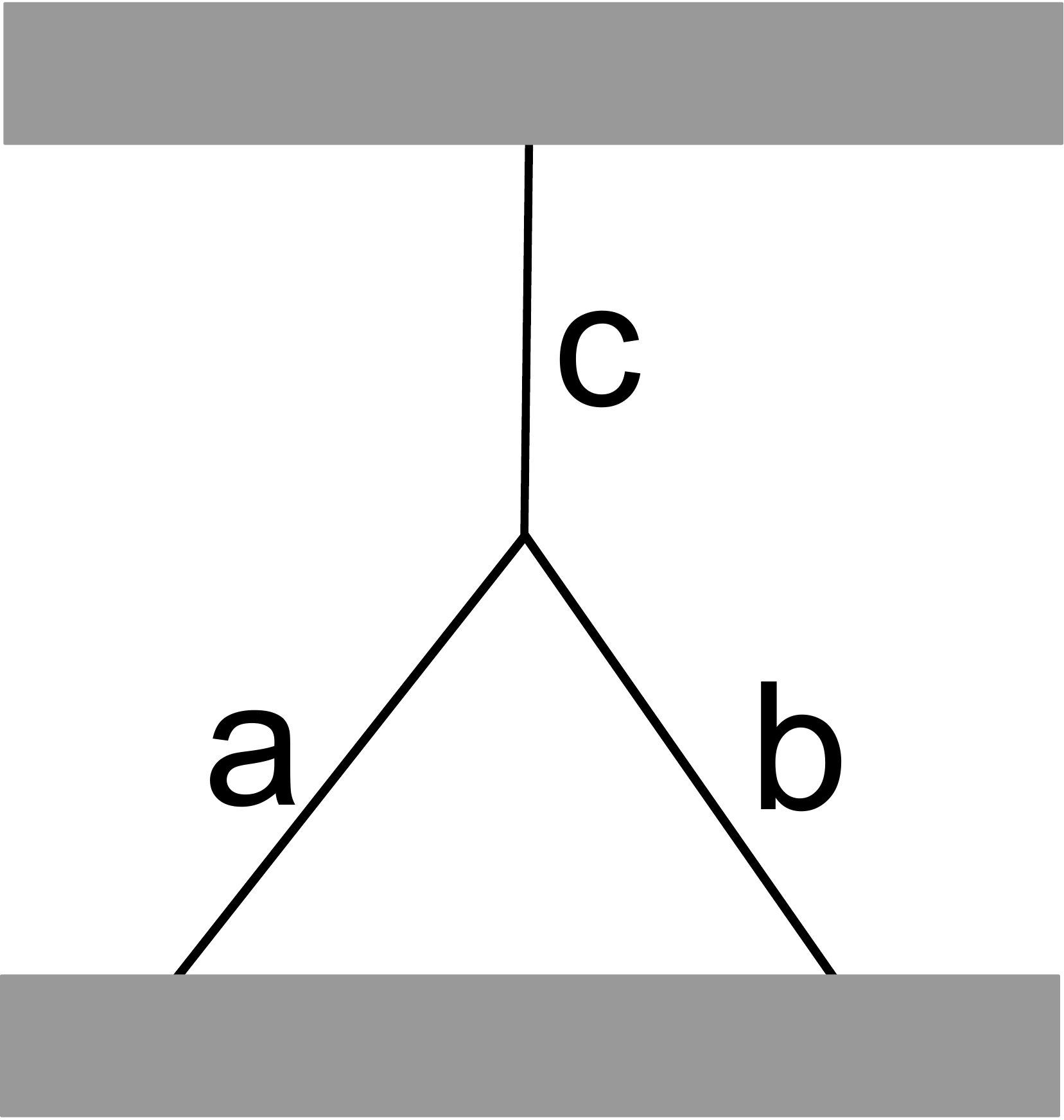}} \right)=
		 \tilde{ \alpha}_{\bar{c}ab}\gamma_c \Phi\left(\raisebox{-0.15in}{\includegraphics[height=0.4in]{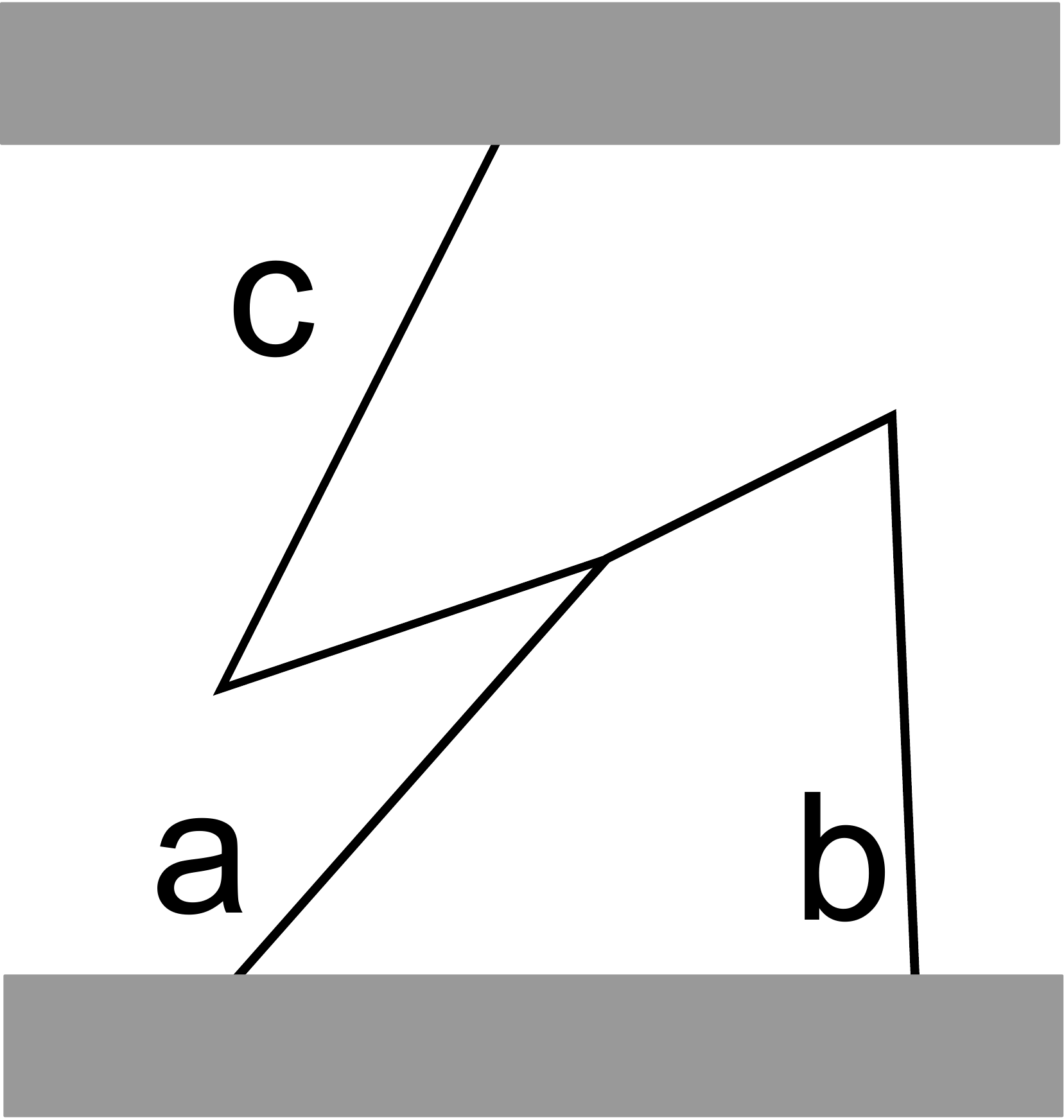}} \right)=
		\frac{1}{\tilde{ \alpha}_{ab\bar{c}}\gamma_c}\Phi\left(\raisebox{-0.15in}{\includegraphics[height=0.4in]{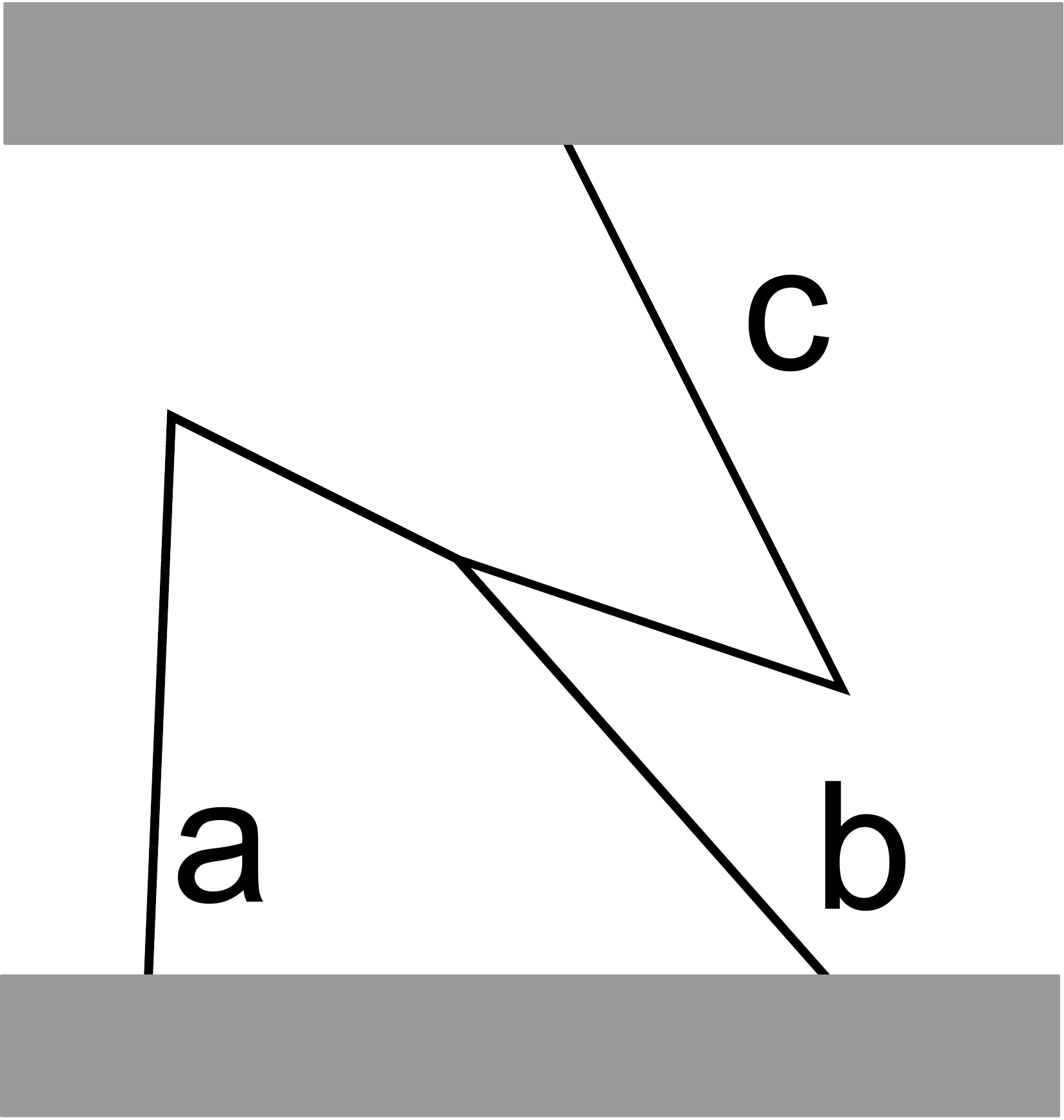}} \right)
	\end{split}
	\label{}
\end{equation}
where 
\begin{equation}
	\alpha_{abc} \equiv F^{abc}_{0\bar{c}\bar{a}} \ , \ \  \frac{1}{\tilde{\alpha}_{abc}} \equiv  \tilde{F}^{abc}_{0\bar{c}\bar{a}}.
	\label{}
\end{equation}

Notice that when $(a,a:\bar{a})$ is a valid branching, the quantity $\alpha_{aaa}\cdot\alpha_{\bar{a}\bar{a}\bar{a}}$ is gauge invariant. 
In this case, if the solution to (\ref{consistency},\ref{hermicity}) has $\alpha_{aaa}\cdot\alpha_{\bar{a}\bar{a}\bar{a}} \neq 1$, then the corresponding model is not isotropic in any gauge.

\subsection{Constraints for planar isotropy}

From Eq.~(\ref{vbending}), we see that the model is invariant under elementary bendings if: 
\begin{equation}
	\begin{split}
		[F^{0c}_{ab}]_{\bar{a}c}
		=[F^{ab}_{c0}]_{\bar{b}c}
		=[\tilde{F}^{0c}_{ab}]_{\bar{a}c}
		=[\tilde{F}^{ab}_{c0}]_{\bar{b}c}=1.	
	\end{split}
	\label{iso}
\end{equation}
or equivalently, in terms of $\{F,Y\}$,
\begin{equation}
	\begin{split}
		F^{a\bar{a}b}_{b0c}Y^{\bar{a}b}_c =1, \quad F^{ab\bar{b}}_{ac0}\frac{Y^{b\bar{b}}_0}{Y^{c\bar{b}}_a}=1 \\
		Y^{ab}_c =\left(\frac{Y^{\bar{a}a}_0}{Y^{\bar{a}c}_b} \right)^* =\left(\frac{Y^{b\bar{b}}_0}{Y^{c\bar{b}}_a}\right)^*.
		\label{}
	\end{split}
	\label{iso2}
\end{equation}
If we can find a solution to (\ref{consistency},\ref{hermicity}) and (\ref{iso2}), then the corresponding string-net will be isotropic in the plane.  
Though there are gauge-invariant obstructions to obtaining a model with planar isotopy, it is important to note that unlike the consistency conditions (\ref{consistency}), the conditions (\ref{iso2}) for isotropy are {\it not} gauge invariant.

In addition to invariance under the bending moves shown in Eqs.~(\ref{vbending}) and (\ref{vbend}), one can show that for string-nets obeying the condition (\ref{iso}),
\begin{equation}
	 [F^{ab}_{cd}]_{ef}
	= [\tilde{F}^{cd}_{ab}]_{\bar{e}f}
	\label{internal}
\end{equation}
and thus the amplitude is invariant under changes of orientation of internal legs
\begin{equation}
		\Phi\left(\raisebox{-0.15in}{\includegraphics[height=0.4in]{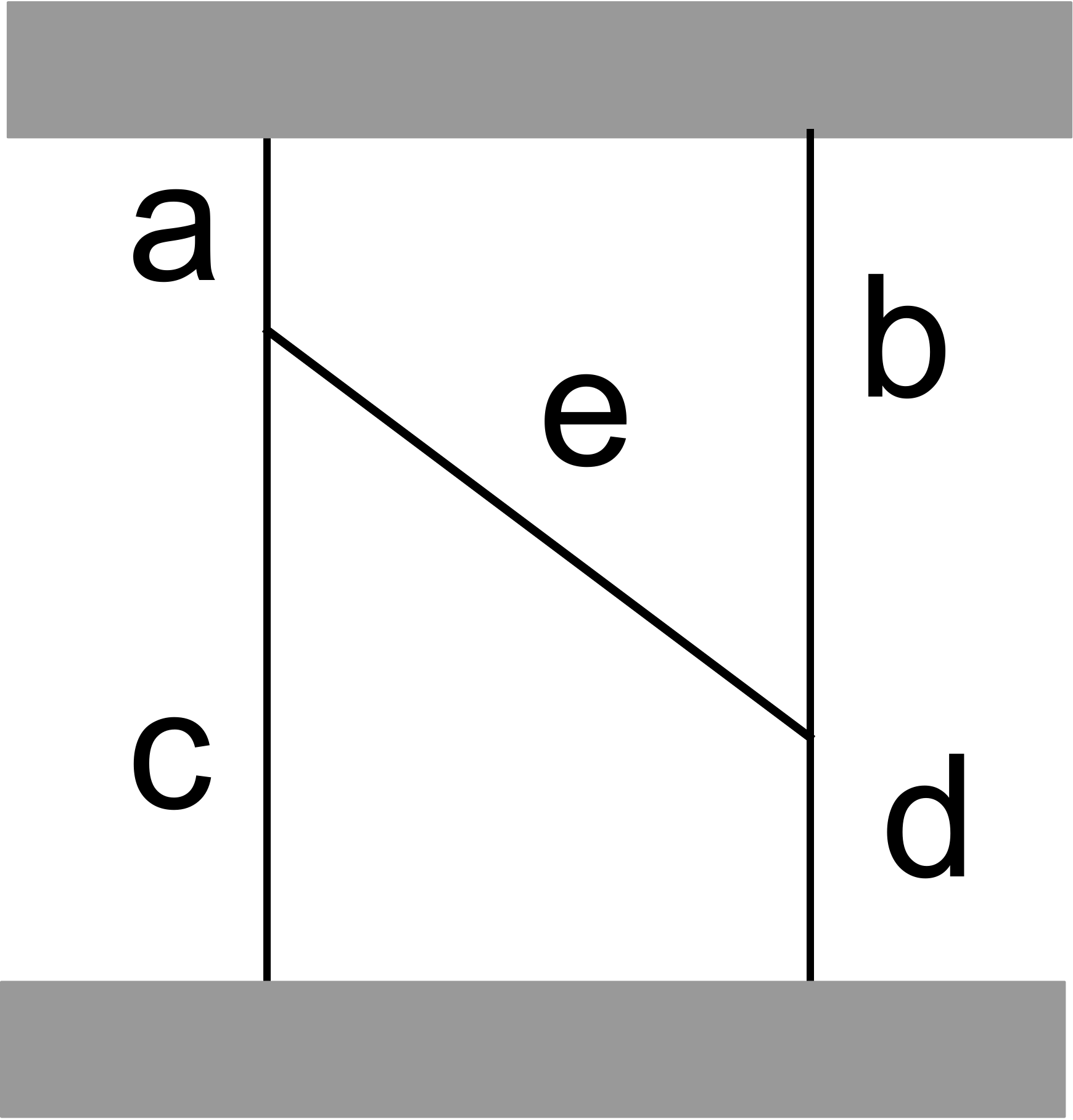}} \right)=
		\Phi\left(\raisebox{-0.15in}{\includegraphics[height=0.4in]{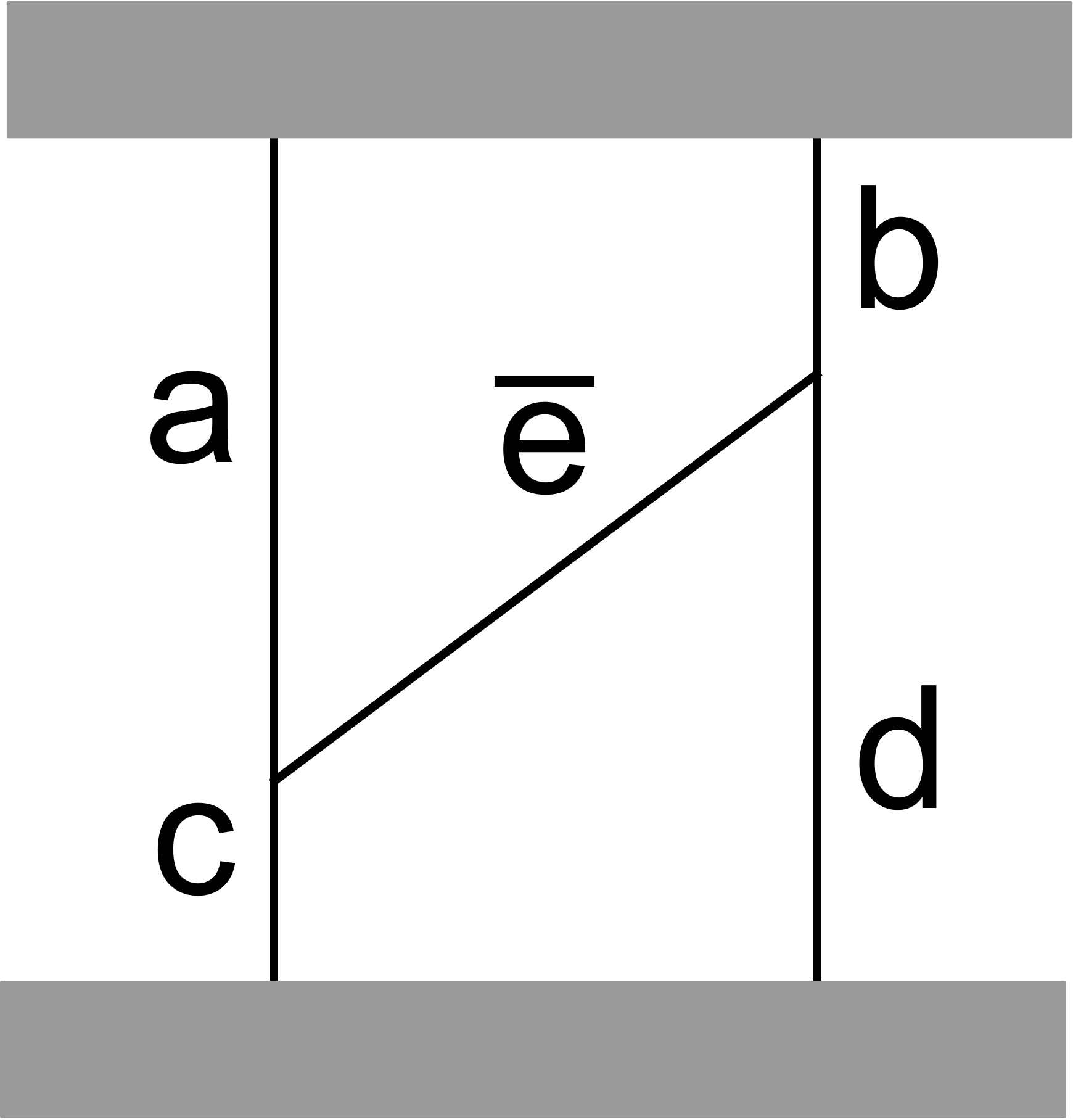}} \right).
	\label{}
\end{equation}
Eq.~(\ref{internal}) follows from (\ref{iso}) and the equality
\begin{equation}
	[\tilde{F}^{0d}_{eb}]_{\bar{e}d} [F^{ab}_{cd}]_{ef}
	=[F^{a\bar{e}}_{c0}]_{ec} [\tilde{F}^{cd}_{ab}]_{\bar{e}f}
	\label{}
\end{equation}
which can be derived from (\ref{consistency}).  \footnote{Specifically, we can use a graphical consistency condition involving $[F]$ and $[\tilde{F}]$, that relates two different paths between the same two diagrams: one with coefficient $[\tilde{F}^{0d}_{eb}]_{\bar{e}d} [F^{ab}_{cd}]_{ef}$, and one with coefficient $[F^{a\bar{e}}_{c0}]_{ec} [\tilde{F}^{cd}_{ab}]_{\bar{e}f}$.  }

Finally, with full bending invariance, we can define the amplitude of a tetrahedron, via:
\begin{equation}
		 \Phi\left(\raisebox{-0.15in}{\includegraphics[height=0.4in]{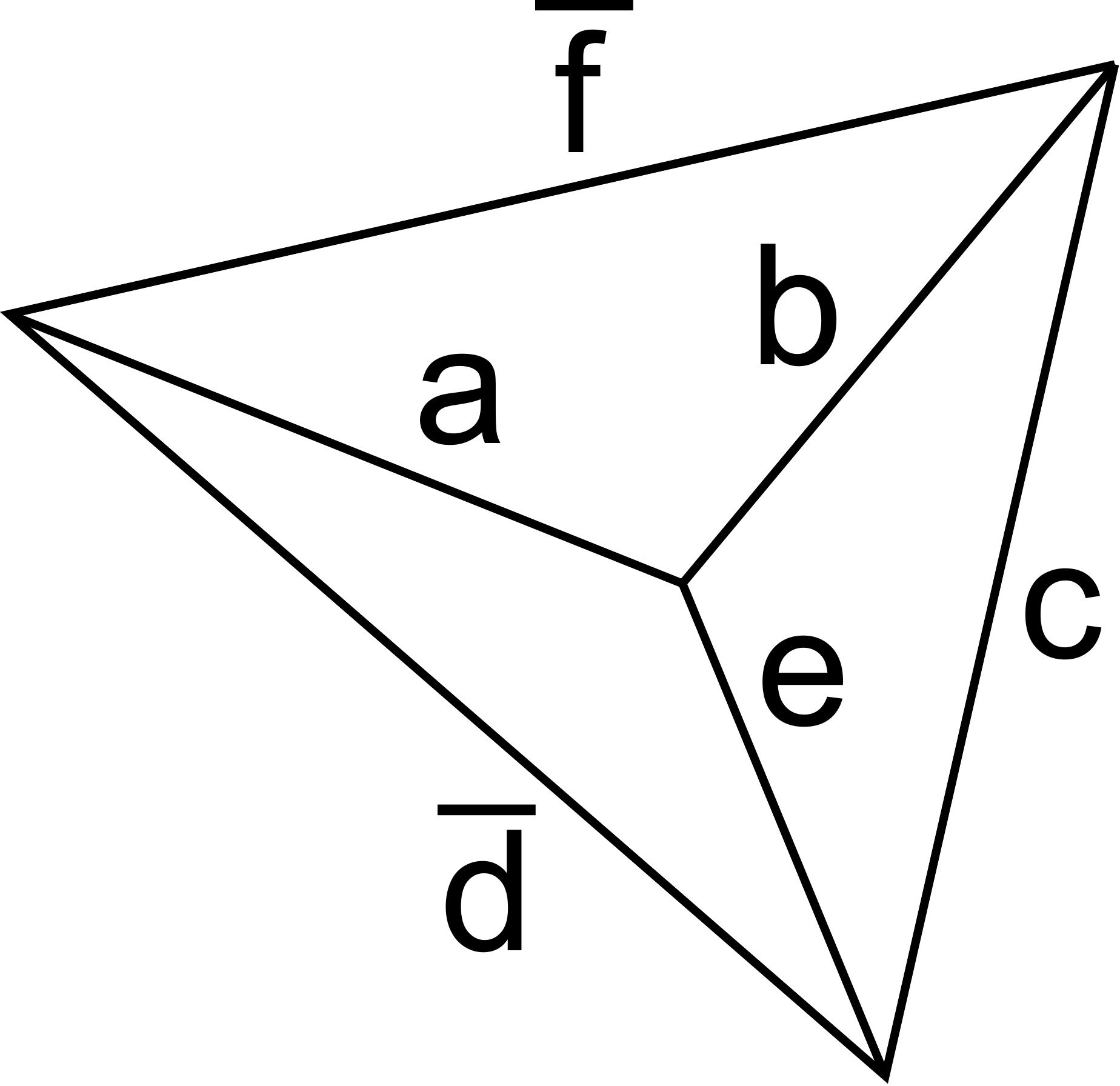}} \right) \equiv \Phi\left(\raisebox{-0.15in}{\includegraphics[height=0.4in]{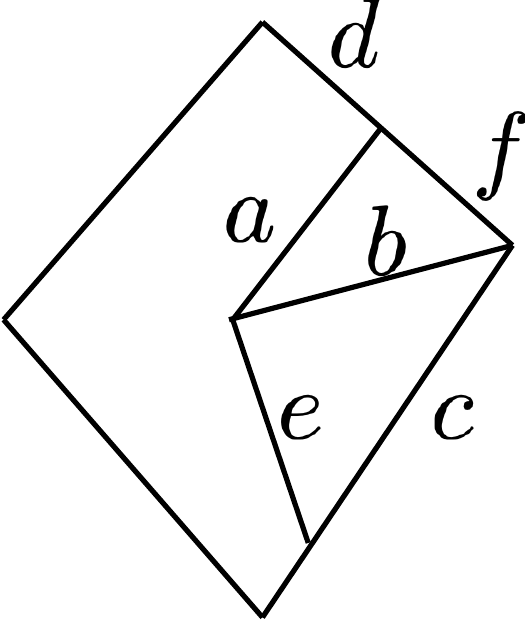}} \right) = F^{abc}_{d e f} Y^{bc}_f Y^{af}_d Y^{\bar{d} d}_0 .
	\label{tetdef}
\end{equation}
Note that the diagram on the left is not an allowed string-net diagram in our formalism, and should be interpreted as a ``shorthand" for the diagram on the right.  This shorthand makes sense in models with bending invariance, where other choices of the diagram on the right, which are related  to the one shown here by some number of bending moves, will yield the same coefficient.

One can check that in string-nets obeying (\ref{iso}),
the amplitude (\ref{tetdef}) is invariant under 3-fold rotations of the tetrahedron
\begin{equation}
	\begin{split} 
			\Phi\left(\raisebox{-0.15in}{\includegraphics[height=0.4in]{iso7a.pdf}} \right)=
		\Phi\left(\raisebox{-0.15in}{\includegraphics[height=0.4in]{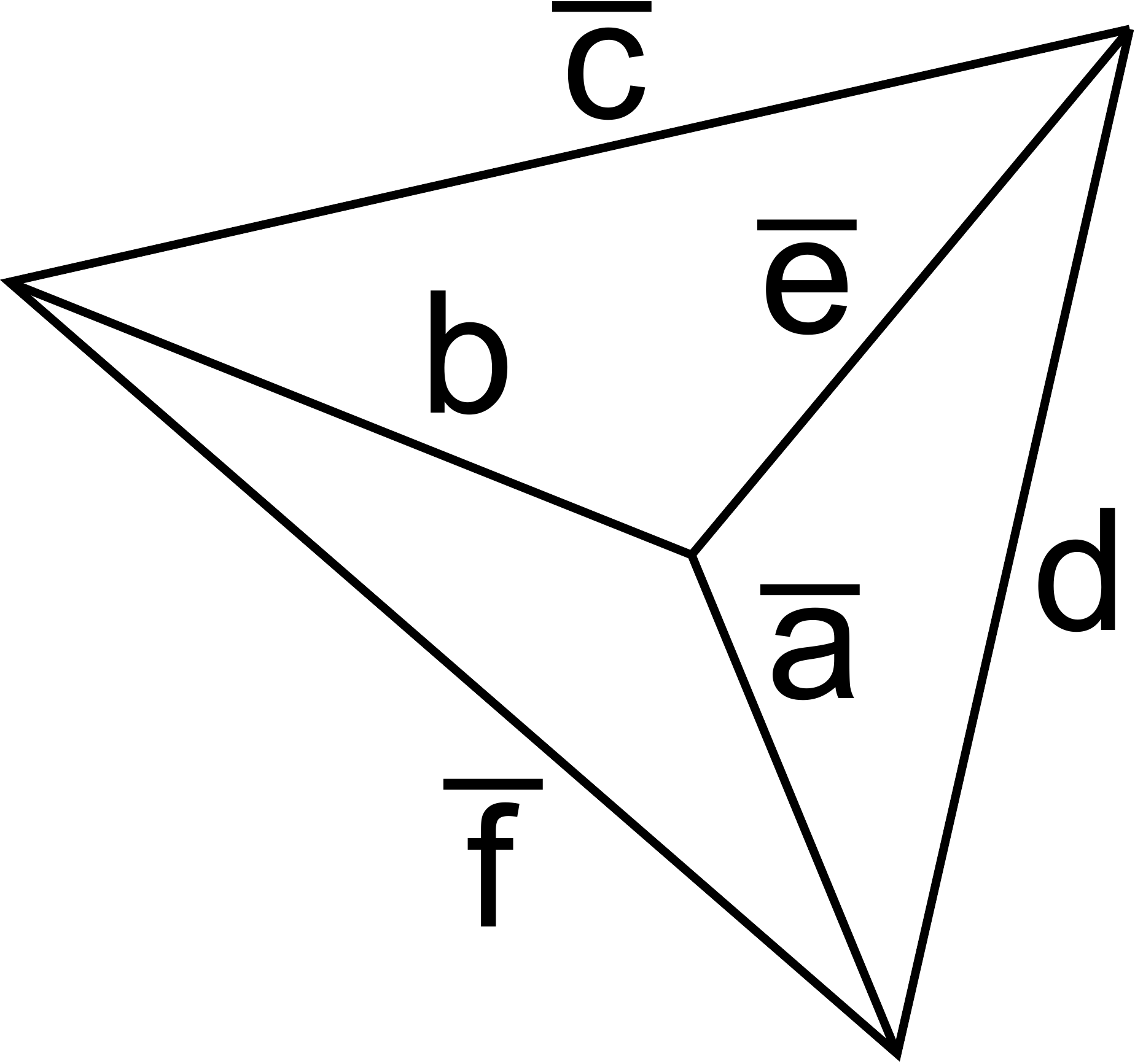}} \right).
	\end{split}
	\label{tet1}
\end{equation}
To see this, observe that we can transform the left tetrahedron into the right-hand one (as defined by Eq.~(\ref{tetdef})) through a series of  moves that bend or rotate vertices.\footnote{Specifically, we first rotate  the $(a,b;e)$ vertex to obtain a $(b, \bar{e}; \bar{a})$ vertex.  Next, bend the $e$ edge at the (upward) $(e,c; d)$ vertex downwards to obtain a (downward) $(\bar{e},d;c)$ vertex.  Then rotate the (downward) $(a,f;d)$ vertex twice, and bend the $d$ edge upwards, to give an (upward) $(\bar{a},d;f)$ vertex.  After straightening out any vertical bends in the edges, we obtain exactly the diagram corresponding to the tetrahedron on the right. }

\subsection{Isotropy on sphere}

If the ground state string-net amplitudes are to be isotropic on the sphere, we must also require invariance of our amplitudes under 
2-fold rotations of the tetrahedron:
	\begin{equation}
	\Phi\left(\raisebox{-0.15in}{\includegraphics[height=0.4in]{iso7a.pdf}} \right)=
		\Phi\left(\raisebox{-0.15in}{\includegraphics[height=0.4in]{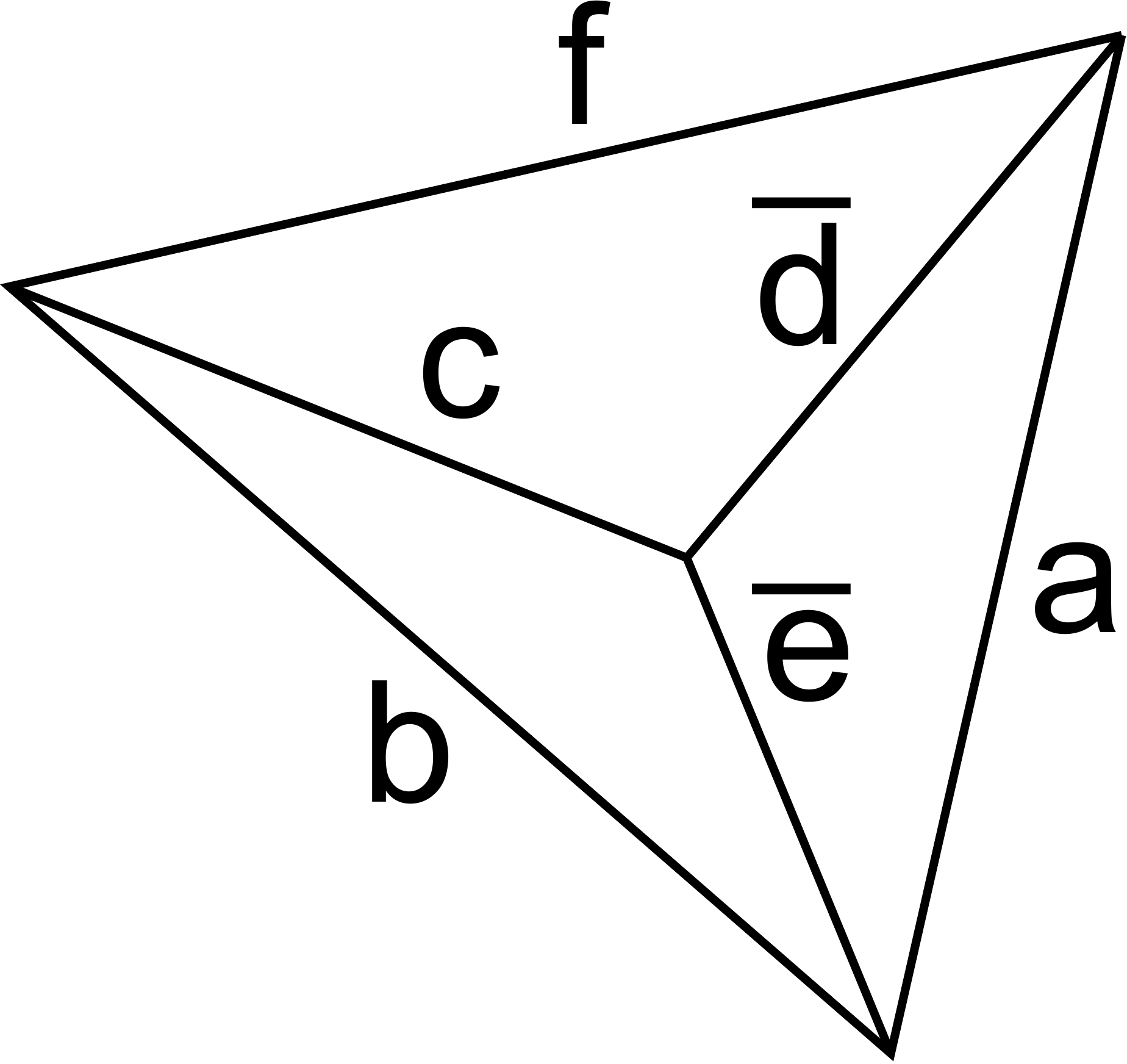}} \right)
	\end{equation}
which can be expressed as
\begin{equation}
	F^{abc}_{def}Y^{bc}_f Y^{af}_d Y^{\bar{d}d}_0
	=F^{c\bar{d}a}_{\bar{b}\bar{e}\bar{f}}
	Y^{\bar{d}a}_{\bar{f}} Y^{c\bar{f}}_{\bar{b}} Y^{b\bar{b}}_0.
	\label{2rot}
\end{equation}
Eq.~(\ref{2rot}) holds provided that
\begin{equation}
		Y^{a\bar{a}}_0 =Y^{\bar{a}a}_0.
		\label{yy}
\end{equation}
To show this, we use Eq.~(\ref{iso2}), 
as well as the relation 
	\begin{equation}
	F^{abc}_{def}
	=F^{c\bar{d}a}_{\bar{b}\bar{e}\bar{f}}\cdot \frac{\alpha_{b\bar{e}a}\alpha_{ab\bar{e}} \alpha_{ec\bar{d}}}
		{\alpha_{af\bar{d}}\alpha_{f\bar{d}a}\alpha_{bc\bar{f}}}
		\label{2rota}
	\end{equation} 
which can be derived from Eq.~(\ref{consistency}). 
Thus, to have a string-net ground state that is isotropic on the sphere, in addition to (\ref{iso2}) we must also require (\ref{yy}).\footnote{Though we do not undertake to show that these conditions are also sufficient for a fully isotropic wave-function on the sphere, we expect that this is the case.}

Interestingly, the condition (\ref{yy}) can always be met by making an appropriate choice of $g$-gauge transformation.  However, this gauge choice may not be compatible with the conditions (\ref{iso}) for planar isotropy, even if there exists a gauge in which those conditions can be met.  We discuss an example in which we must choose between planar isotropy and the condition (\ref{yy}) in Sec. \ref{sec:exp}.  
	
\subsection{Tetrahedral reflection symmetry}

The original string-net construction\cite{LevinWenstrnet} required, in addition to the conditions discussed above, that ground state amplitudes also be invariant under the tetrahedral reflection: 
	\begin{equation}
	\Phi\left(\raisebox{-0.15in}{\includegraphics[height=0.4in]{iso7a.pdf}} \right)=
		\Phi\left(\raisebox{-0.15in}{\includegraphics[height=0.4in]{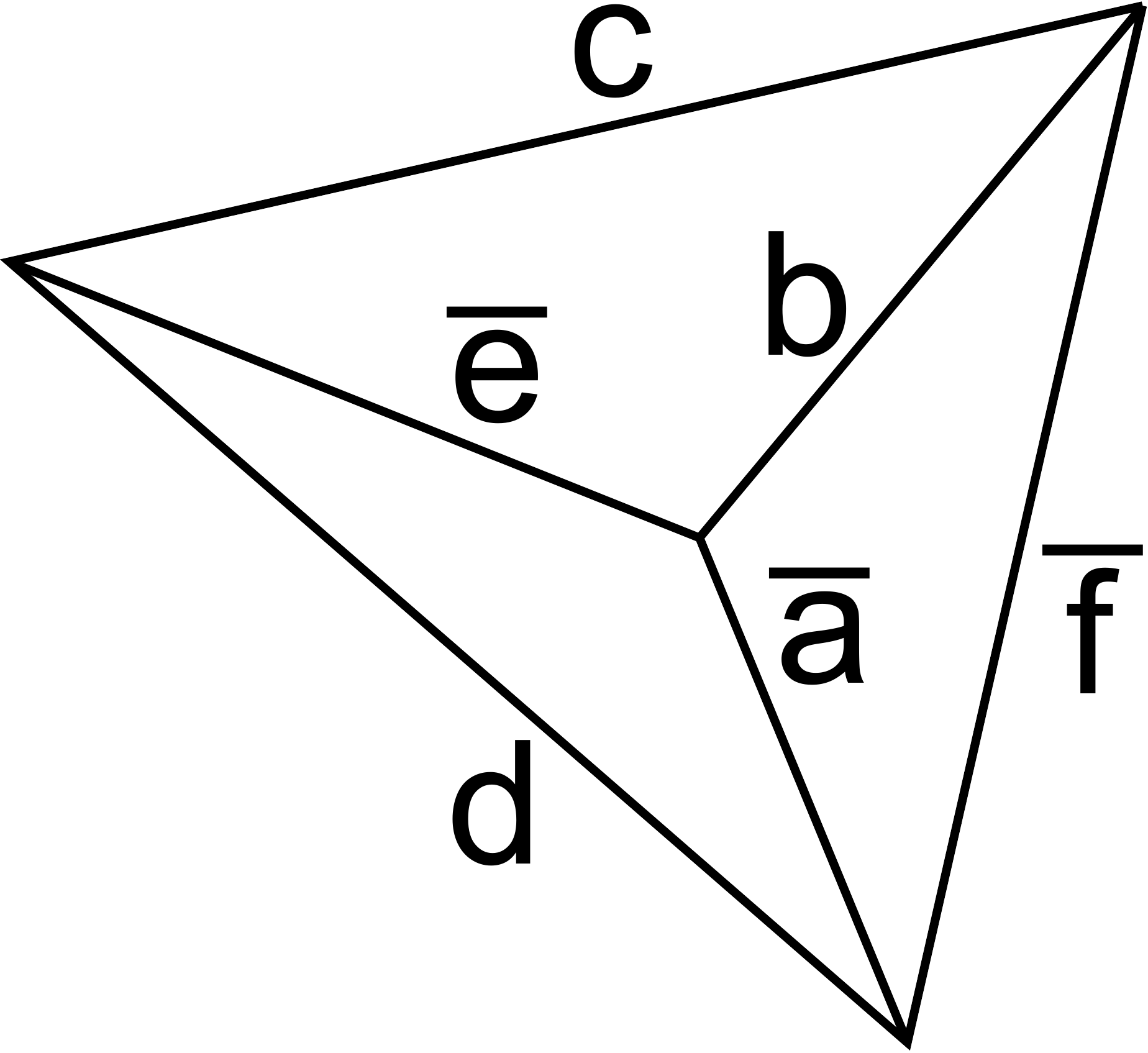}} \right) \ \ .
	\end{equation}
Algebraically, this means that
\begin{equation}
	F^{abc}_{def}Y^{bc}_f Y^{af}_d Y^{\bar{d}d}_0 
	=F^{\bar{e}b\bar{f}}_{\bar{d}\bar{a}\bar{c}} Y^{b\bar{f}}_{\bar{c}} Y^{\bar{e}\bar{c}}_{\bar{d}} Y^{d\bar{d}}_0.
	\label{reflection}
\end{equation}
By using (\ref{consistency}), we can derive 
\begin{equation}
	F^{abc}_{def}Y^{bc}_f = \gamma_{\bar{b}}[\tilde{F}^{e\bar{b}}_{a0}]_{ba} [\tilde{F}^{0c}_{\bar{b}f}]_{bc} (F^{e\bar{b}f}_{dac})^* Y^{e\bar{b}}_a.
	\label{reflection1}
\end{equation}
Using this, together with (\ref{iso}),  Eq.~(\ref{reflection}) can be simplified to
\begin{equation}
	(F^{e\bar{b}f}_{dac})^* Y^{e\bar{b}}_a Y^{af}_d Y^{\bar{d}d}_0
	=F^{\bar{e}b\bar{f}}_{\bar{d}\bar{a}\bar{c}} Y^{b\bar{f}}_{\bar{c}}
	Y^{\bar{e}\bar{c}}_{\bar{d}} Y^{d\bar{d}}_0 \ \ .
	\label{reflection2}
\end{equation}
In the gauge where
\begin{equation}
	Y^{ab}_c = \sqrt{\frac{d_a d_b}{d_c}} \frac{w_a w_b}{w_c}
	\label{Eq:gagew}
\end{equation}
with the $U(1)$ phases obeying $w_a=w_{\bar{a}}$,
Eq.~(\ref{reflection2}) further simplifies to the following condition on our $F$'s:
\begin{equation}
	(F^{abc}_{def})^* = F^{\bar{a}\bar{b}\bar{c}}_{\bar{d}\bar{e}\bar{f}}.
	\label{reflection3}
\end{equation}

If the model is isotropic in the plane, as well as invariant under 2-fold rotations and reflections of the tetrahedron, 
in the gauge (\ref{Eq:gagew}) we find that
\begin{equation}
	F^{abc}_{def} = F^{\bar{e}b\bar{f}}_{\bar{d}\bar{a}\bar{c}}\sqrt{\frac{d_e d_f}{d_a d_c}} \frac{w_e w_f}{w_a w_c} = F^{ba\bar{d}}_{\bar{c}e\bar{f}} = F^{\bar{d}cb}_{\bar{a}\bar{e}f}.
	\label{oldrules}
\end{equation}
The first equality follows from the tetrahedral reflection symmetry (\ref{reflection}), the second equality follows from the first equality and the 3-fold rotational symmetry (\ref{tet1}) while the third equality follows from the second equality and 2-fold rotational symmetry (\ref{2rot}).
We will see in the next section that these correspond exactly to the conditions imposed by Ref.~\onlinecite{LevinWenstrnet} on the original string-net models.

\section{Relationship with original string-net construction\label{oldmodels}}
In this section, we discuss the relationship between our construction and the original string-net construction of Ref.~\onlinecite{LevinWenstrnet}. Our main result is that the string-net models discussed in Ref.~\onlinecite{LevinWenstrnet} correspond to a \emph{subset} of the models constructed in this paper, and we discuss the properties of this subset.

The first step is to find the dictionary between the input data $\{F,Y\}$ that defines a string-net model in this paper and the input data $\{\bar{F},\bar{d}\}$ that was used in the original string-net construction of Ref.~\onlinecite{LevinWenstrnet}. To derive this dictionary, we compare the ``old'' local rules in Ref.~\onlinecite{LevinWenstrnet} to the ``new'' local rules in this paper, namely (\ref{localrules},\ref{localrules1}). From this comparison, it is easy to see that if a string-net state obeys the old local rules for some $\{\bar{F},\bar{d}\}$, then it obeys the new local rules with $\{F,Y\}$ given by
\begin{subequations}
	\begin{gather}
	F^{abc}_{def}=\bar{F}^{\bar{b}\bar{a}e}_{d\bar{c}f} \\
	\tilde{F}^{abc}_{def} = \bar{F}^{ab\bar{e}}_{c\bar{d}f} \\
	(Y^{ab}_c)^{-1}=\bar{F}^{\bar{a}a0}_{b\bar{b}c} \\
	Y^{ab}_c=\bar{F}^{b\bar{c}a}_{c\bar{b}0}\bar{d}_{\bar{b}} \label{3}\\
	[F^{ab}_{cd}]_{ef} = \bar{F}^{\bar{a}ce}_{d\bar{b}f} \\
	[\tilde{F}^{ab}_{cd}]_{ef} = \bar{F}^{\bar{c}a\bar{e}}_{b\bar{d}f}\\
	d_a = |\bar{d}_a| .
	\label{dabar}
	\end{gather}
	\label{match}
\end{subequations}
The above equations provide the desired dictionary between the ``old'' data $\{\bar{F},\bar{d}\}$ and the ``new'' data $\{F,Y\}$.

Next, we recall that Ref.~\onlinecite{LevinWenstrnet} imposed several self-consistency conditions on the old data $\{\bar{F},\bar{d}\}$. The first condition is that
\begin{equation}
	\bar{F}^{abe}_{cdf}=1 \text{ if }a\text{ or }b\text{ or }c\text{ or }d=0. \label{oldf1}
\end{equation}
Substituting this condition into the dictionary in Eq.~(\ref{match}), it follows that the new data satisfy (\ref{iso}). Thus the original string-net models are all \emph{isotropic}. 

In addition to (\ref{oldf1}), Ref.~\onlinecite{LevinWenstrnet} imposed the conditions
\begin{subequations}
	\begin{gather}
		\bar{F}^{abc}_{\bar{b}\bar{a}0}=\frac{v_c}{v_a v_b} 
		\text{ with }v_a=v_{\bar{a}}= \sqrt{\bar{d}_a}\\
		\bar{F}^{\bar{b}\bar{a}e}_{d\bar{c}f}=\bar{F}^{\bar{c}d\bar{e}}_{\bar{a}\bar{b}f}
		=\bar{F}^{\bar{a}\bar{b}e}_{\bar{c}d\bar{f}} =\bar{F}^{\bar{b}e\bar{a}}_{\bar{d}f\bar{c}}\frac{v_e v_f}{v_a v_c} \\ 
		\sum_n \bar{F}^{mlq}_{k\bar{p}n}\bar{F}^{jip}_{mn\bar{s}}\bar{F}^{j\bar{s}n}_{lk\bar{r}}=\bar{F}^{jip}_{\bar{q}k\bar{r}}\bar{F}^{ri\bar{q}}_{ml\bar{s}}.
	\end{gather}
	\label{oldconsistency}
\end{subequations}
Substituting (\ref{oldconsistency}) into (\ref{match}), we find that the new data satisfies the usual consistency conditions (\ref{consistency}) as well as the following additional constraints:
\begin{subequations}
	\begin{gather}
		Y^{ab}_c = \frac{v_a v_b}{v_c} \label{ydic}\\
		F^{a\bar{b}b}_{ac0}=
		\frac{v_c}{v_a v_b}
		\label{new1}
	\\
	F^{abc}_{def}=F^{\bar{e}b\bar{f}}_{\bar{d}\bar{a}\bar{c}}
	\frac{v_e v_f}{v_a v_c}=F^{\bar{d}cb}_{\bar{a}\bar{e}f} 
	= F^{ba\bar{d}}_{\bar{c}e\bar{f}} .
	\label{new2}
	\end{gather}
	\label{new}
\end{subequations}
These equations have simple physical interpretations. Eq.~(\ref{ydic}) is simply a special case of the gauge choice (\ref{Eq:gagew}), with $w_a \sqrt{d_a}= v_a$.  Eq.~(\ref{new1}) follows from this gauge choice, together with the conditions (\ref{iso}) for planar isotropy. Eq.~(\ref{new2}) is exactly the condition (\ref{oldrules}) that the string-net model is invariant under all reflections and rotations of the tetrahedron, which the original construction explicitly assumes.  Thus the extra conditions we must impose on the new data amount to requiring that, in an appropriate gauge, the string-net is isotropic on the sphere and invariant under tetrahedral reflections.

 Finally, Ref.~\onlinecite{LevinWenstrnet} imposed the following condition in order to guarantee that the string-net Hamiltonian was Hermitian:
\begin{equation}
	\bar{F}^{\bar{a}\bar{b}\bar{c}}_{\bar{d}\bar{e}\bar{f}}=(\bar{F}^{abc}_{def})^*.
	\label{unitary0}
\end{equation}
Substituting Eq.~(\ref{unitary0}) into (\ref{match}) and using (\ref{new}), one can show that the new data satisfies the condition (\ref{unitary})
\begin{equation}
	(F^{abc}_d)^{-1}_{fe}=(F^{abc}_{def})^*.
	\label{new3}
\end{equation}
The reverse is also true. One the one hand, we have
$(F^{abc}_d)^{-1}_{fe}=\tilde{F}^{abc}_{def}=\bar{F}^{ab\bar{e}}_{c\bar{d}f}$. On the other hand, we have $(F^{abc}_{def})^*=(F^{ba\bar{d}}_{\bar{c}e\bar{f}})^* =(\bar{F}^{\bar{a}\bar{b}e}_{\bar{c}d\bar{f}})^*$. Thus (\ref{unitary0}) follows from (\ref{new3}).
Similarly, the other conditions (\ref{y0}---\ref{y1}) also follow from (\ref{match}, \ref{new}).

Putting everything together, we conclude that the original string-net models of Ref.~\onlinecite{LevinWenstrnet} correspond to a \emph{subset} of the models discussed in this paper, namely the subset of models that obey the constraints (\ref{iso2}) and (\ref{new}), in addition to the usual conditions (\ref{consistency},\ref{hermicity}).

\section{Examples \label{sec:exp}}
In this section, we work out some illustrative examples.  We begin with the abelian $\mathbb{Z}_2,\mathbb{Z}_3$ and $\mathbb{Z}_4$ string-nets.  These are instructive in understanding how our construction captures models realized by the original string-net construction\cite{LevinWenstrnet}.  They also contain some models which cannot be realized by the original string-net framework because they cannot be made isotropic on the plane ($\mathbb{Z}_3$), or on the sphere ($\mathbb{Z}_4$).   Note that all of our abelian examples give topological orders that can also be realized by the twisted quantum double models of Ref.~\onlinecite{HuWanWu12}, as well as the string-net construction of Ref.~\onlinecite{LinLevinstrnet}.  For this reason we do not list the quasiparticle types or string operators in these cases.  

We then discuss two non-abelian examples: the  Fibonacci and $TY_3$ string-net models. 
The Fibonacci model is an example that can be obtained from the original string-net construction, and is included here to illustrate how our construction reduces to that of Ref.~\onlinecite{LevinWenstrnet} in this case. Finally, the $TY_3$ model is an example of a non-abelian string-net that cannot be realized without our generalized construction.

\subsection{$\mathbb{Z}_2$ string-net models}
\label{z2example}
The $\mathbb{Z}_2$ string-net models describe two string types $\{0,1\}$ where $0$ is the vacuum string and $1=\bar{1}$ is self dual with the branching rules $\{(0,0:0),(0,1:1),(1,0:1),(1,1,:1)\}$. These branching rules require that the strings form closed loops so the Hilbert space is the set of all possible closed loops.

Next, to construct the Hamiltonian and wave functions, we have to solve the consistency conditions (\ref{consistency},\ref{hermicity}) for $\{F,Y\}$. There are two distinct solutions, parameterized by an integer $p=0,1$:
\begin{equation}
	F^{111}=(-1)^p, \quad Y^{11}=1.
	\label{z2sol}
\end{equation}
where here and for our other string-net models with abelian branching rules, we use the simplified notation  
\begin{equation}
F^{abc}\equiv F^{abc}_{(a+b+c)(a+b)(b+c)}, \quad Y^{ab} \equiv Y^{ab}_{a+b} \ .
\end{equation} 
With the solutions (\ref{z2sol}) in hand, we can construct the wave functions and Hamiltonian using (\ref{localrules}) and (\ref{hsn0}). 
For the $p=0$ solution, the wave function is
\begin{equation}
	\Phi(X) =1
	\label{wfz2a}
\end{equation}
for any closed string-net configuration $X$. The corresponding Hamiltonian realizes the the toric code topological phase\cite{LevinWenstrnet,KitaevToric}. On the other hand, for the $p=1$ solution, 
we need to keep track of the vertical kinks because $\gamma_1=F^{111}Y^{11}=-1$.
The wave function is
\begin{equation}
	\Phi(X) = (-1)^{\text{loop}(X)}(-1)^{\text{vkink}(X)/2}
	\label{wfz2b}
\end{equation}
with $\text{loop}(X)$ meaning the total number of closed loops in the configuration $X$ and $\text{vkink}(X)$ meaning the total number of vertical kinks (upward and downward vertices with $c=0$ in Eq.~(\ref{vertex3})) in $X$. The corresponding Hamiltonian realizes the same phase as the doubled semion model of Ref.~\onlinecite{LevinWenstrnet}.

While the solutions (\ref{z2sol}) are sufficient for constructing exactly soluble models, it is desirable to have solutions which lead to simpler models. Specifically, we can make $\gamma_1=1$ using  the gauge transformation $g_{11}^0=(-1)^p$. After this gauge transformation we have
\begin{equation}
	F^{111}=(-1)^p,\quad Y^{11}=(-1)^p.
	\label{z2sola}
\end{equation}
The solutions (\ref{z2sola}) satisfy (\ref{iso2}) and thus the models are isotropic.
In this gauge, the wave function for the $p=0$ case is the same as (\ref{wfz2a}) while the wave function for the $p=1$ case becomes
\begin{equation}
	\Phi(X) = (-1)^{\text{loop}(X)}.
	\label{}
\end{equation}

\subsection{$\mathbb{Z}_3$ string-net model}
\label{z3example}
The $\mathbb{Z}_3$ models have three types of strings $\{0,1,2\}$ with $\bar{0}=0,\bar{1}=2,\bar{2}=1$. The branching rules are $\{(a,b:[a+b]_3)\}$ with $a,b\in\{0,1,2\}$ and $[a+b]_3=a+b \text{ mod }3$ which takes values in $\{0,1,2\}$.

To construct the Hamiltonians and wave functions for the $\mathbb{Z}_3$ models, we solve the consistency conditions for $\{F,Y\}$. There are three distinct solutions\cite{PropitiusThesis} labeled by $p=0,1,2$
\begin{equation}
	F^{abc}=e^{i\frac{2\pi p a}{9}(b+c-[b+c]_3)},\quad Y^{ab}=1.
	\label{z3sola}
\end{equation}
As in the previous example, it is instructive to ask whether we can use appropriate gauge transformations to put this data into a form where Eq.~(\ref{iso2}) is satisfied.  
However, when $p=1,2$ no such gauge transformation exists. To see this, recall that the quantity $\alpha_{111}\alpha_{222}$ is gauge invariant under $f,g$ transformation. Since $\alpha_{111}\alpha_{222}\neq 1$ in $p=1,2$ solutions, we have no hope to make (\ref{z3sola}) satisfy (\ref{iso2}) by any gauge transformation. Thus the $p=1,2$ models will not be isotropic on the plane, in any gauge.

\subsection{$\mathbb{Z}_4$ string-net model}
\label{z4example}
The string types for the $\mathbb{Z}_4$ model are $\{0,1,2,3\}$ with $\bar{0}=0,\bar{1}=3,\bar{2}=2,\bar{3}=1$. The branching rules are $\{(a,b:[a+b]_4)\}$ with $a,b\in\{0,1,2,3\}$ and $[a+b]_4=a+b \text{ mod }4$ which takes values in $\{0,1,2,3\}$.

To construct the Hamiltonians and wave functions for the $\mathbb{Z}_4$ models, we solve the consistency conditions for $\{F,Y\}$. There are four distinct solutions labeled by $p=0,1,2,3:$
\begin{equation}
	F^{abc}=e^{i\frac{2\pi p a}{16}(b+c-[b+c]_4)},\quad Y^{ab}=1.
	\label{z4a}
\end{equation}
While it is sufficient to construct the Hamiltonians and wave functions by using (\ref{z4a}), the $p=1,2,3$ models are not isotropic because the corresponding solutions (\ref{z4a}) do not satisfy (\ref{iso2}). Thus it is desirable to find proper gauge transformations $f,g$ to have simpler models if possible. To this end, we first apply the $f$-gauge transformation with $f^{32}_1=(-i)^p,f^{33}_2=(-1)^p$ followed by a $g$-gauge transformation $g_{ab}^{[a+b]_4}=F^{ab\bar{b}}$. The result is
\begin{equation}
	\begin{split}
		F^{113}&=F^{331}=F^{232}=F^{212}=F^{131}=i^p, \\
		F^{133}&=F^{311}=F^{123}=F^{321}=F^{313}=(-i)^p, \\
		F^{122}&=F^{231}=F^{223}=F^{312}=F^{222}=F^{333}=(-1)^p,\\
		Y^{ab}&=F^{ab\bar{b}}.
	\end{split}
	\label{z4b}
\end{equation}
The solutions (\ref{z4b}) satisfy (\ref{iso2}) and the corresponding models are isotropic on plane.
However, $p=1,3$ solutions do not satisfy Eq.~(\ref{yy}) and thus the $p=1,3$ models are examples which are isotropic on the plane but not on the sphere.  These models also do not satisfy the tetrahedral reflection symmetry condition (\ref{reflection3}).  The corresponding quasi-particle spectra break time-reversal symmetry\cite{LinLevinstrnet}, and thus these models cannot be realized by the original construction.

\subsection{Fibonacci string-net model} \label{FibonacciExample}

We now turn to our non-abelian examples.   We first discuss the Fibonacci string-net, which was also discussed by Ref.~\onlinecite{LevinWenstrnet}. We include it here partly to provide a simple example of the non-abelian construction, and partly to correct a minor error in the data for the string operators in Ref.~\onlinecite{LevinWenstrnet}.
 
The string types  in the Fibonacci string-net are $\{0,1\}$ where $0$ is the vacuum string and $1=\bar{1}$ is self dual. The allowed branching rules are $\{(0,0:0),(0,1:1),(1,0:1),(1,1:0),(1,1,:1)\}$.
The solution to (\ref{consistency},\ref{hermicity}) is given by
\begin{equation}
	\begin{split}
		[F^{111}_1]_{ef} &= \left[ \begin{array}{cc}
				\frac{1}{d} & \frac{1}{\sqrt{d}} \\
				\frac{1}{\sqrt{d}}  & -\frac{1}{d}
		\end{array} \right]_{ef},\\
		\quad Y^{11}_0=d,\quad Y^{11}_1&=\sqrt{d}, \text{ other }F,Y=1,\\ 
		d&=\frac{1+\sqrt{5}}{2}
	\end{split}
	\label{Ffib}
\end{equation}
where $e,f=0,1$. By using the data (\ref{Ffib}), we can construct the ground state wave function and the Hamiltonian.
Notice that (\ref{Ffib}) satisfies (\ref{iso2},\ref{yy},\ref{reflection2}) so the corresponding model is fully isotropic on the sphere, and also obeys tetrahedral reflection symmetry.  This is expected, as the Fibonacci string-net can be realized by the original construction \cite{LevinWenstrnet}.

To find the quasiparticle excitations, we need to solve (\ref{weqs}). There are four irreducible solutions to (\ref{weqs}) which correspond to four distinct quasiparticles: 
\begin{equation}
	\begin{split}
		\alpha=1:& (n_{\alpha,0},n_{\alpha,1})=(1,0) \\
		 & \Omega_{\alpha}^{1,001}=1\\
		\alpha=2:& (n_{\alpha,0},n_{\alpha,1})=(0,1) \\
		&\Omega_{\alpha}^{1,110}=e^{-i 4\pi/5}
		,\quad \Omega_{\alpha}^{1,111}=e^{i 3\pi/5}\\
		\alpha=3:& (n_{\alpha,0},n_{\alpha,1})=(0,1) \\
		& \Omega_{\alpha}^{1,110}=e^{i 4\pi/5}
		,\quad \Omega_{\alpha}^{1,111}=e^{-i 3\pi/5}\\
		\alpha=4:& (n_{\alpha,0},n_{\alpha,1})=(1,1) \\
		&\Omega_{\alpha}^{1,110}=1,\quad \Omega_{\alpha}^{1,001}=-d^{-2},\quad \Omega_{\alpha}^{1,111}=d^{-2}\\
		&\Omega_{\alpha}^{1,101}=(\Omega_{\alpha}^{1,011})^*=\sqrt{3d-4}e^{-i3\pi/10}.
	\end{split}
	\label{stringFib}
\end{equation}
Here, we omit the value of $\Omega_\alpha^{0,sss}$, since this matrix element is always fixed at $1$.
Note that in Ref.~\onlinecite{LevinWenstrnet}, it is claimed that $\bar{\Omega}=\Omega^*$ which is correct only in the gauge where $\Omega$ are chosen to be real numbers. In Eq.~(\ref{stringFib}), $\bar{\Omega}^{1,101}_{4}=\Omega^{1,101}_{4}\neq (\Omega^{1,101}_{4})^*$. However, if we choose $\Omega^{1,101}_{4}=(\Omega^{1,011}_{4})=\sqrt{3d-4}$, then in that gauge $\bar{\Omega}^{1,101}_{4}= (\Omega^{1,101}_{4})^*$.

From (\ref{weq4}), we find all quasiparticles are self-dual $\alpha=\bar{\alpha}$. Also, we can see that the quantum dimensions of the quasiparticles are $d_1 = 1$ and $d_2 = d_3 = d$ and $d_4 = d^2$. 
The topological spins and the S matrix can be computed from (\ref{twist},\ref{smat}). We find
\begin{equation}
	\begin{split}
	e^{i\theta_1}=1,\quad e^{i\theta_2}&=e^{-i 4\pi/5},\quad  e^{i\theta_3}=e^{i 4\pi/5},\quad  e^{i\theta_4}=1 \\
	S&= \frac{1}{1+d^2}\left[ \begin{array}{cccc}
				1 & d & d & d^2 \\
				d & -1 & d^2 & -d \\
				d & d^2 & -1 & -d \\
				d^2 & -d & -d & 1
				\end{array} \right].
	\end{split}
	\label{}
\end{equation}
The same result was found in Ref.~\onlinecite{LevinWenstrnet}.

\subsection{$TY_3$ string-net model}
Our final example is the string-net model associated with the Tambara-Yamagami category for $\mathbb{Z}_3$ ($TY_3$)\cite{TYcatST,TYcat}.  This category can be obtained\cite{WilliamsonWang} by taking the $\mathbb{Z}_3$ model with $F^{abc}\in \{0,1 \}$ described above, with labels $\{0,1,2\}$, 
together with a label $\sigma$ with non-abelian branching rules and $\bar{\sigma}=\sigma$. The full branching rules are
\begin{equation}
	\begin{split}
	\{(0,0:0),(0,1:1),(0,2:2),(0,\sigma:\sigma),(1,1:2),\\
	(1,2:0),(2,2:1),(1,\sigma:\sigma),(2,\sigma:\sigma),(\sigma,\sigma:0), \\
	(\sigma,\sigma:1),(\sigma,\sigma:2),(a,b:c)=(b,a:c). \}
	\end{split}
	\label{}
\end{equation}

The solution to (\ref{consistency},\ref{hermicity}) is given by
\begin{equation}
	\begin{split}
		&F^{a\sigma b}_{\sigma\sigma\sigma} =F^{\sigma a \sigma}_{b \sigma \sigma}=e^{\frac{2\pi i a b}{3}},\quad F^{\sigma \sigma \sigma}_{\sigma ab}=\frac{p}{\sqrt{3}}e^{-\frac{2\pi i ab}{3}},\\
		&Y^{\sigma \sigma}_0 = Y^{\sigma \sigma}_1=Y^{\sigma \sigma}_2 = d_\sigma,\quad Y^{\sigma \sigma}_\sigma=\sqrt{d_\sigma}, \text{ other }F,Y=1 \\
		&d_0=d_1=d_2=1,\quad d_\sigma=\sqrt{3} 
	\end{split}
	\label{ty3sol}
\end{equation}
where $a,b$ take values in $\{0,1,2\}$ and $p=1,-1$ parametrizes two different solutions.
As written, the $p=-1$ solution does not satisfy (\ref{iso2}), and hence is not isotropic in the plane.  However, this can be resolved using an $f$-gauge transformation with $f^{\sigma \sigma}_0=p$. In addition, neither solution obeys the tetrahedral reflection symmetry condition (\ref{reflection3}). Thus these models cannot be realized by the original construction. 

We now find the quasiparticles. For each of the two models parametrized by $p=\pm1$, we find $15$ irreducible solutions to (\ref{weqs}), corresponding to $15$ quasiparticles. For the $p=1$ model, we find
\begin{equation}
	\begin{split}
		\alpha=&1,2: (n_{\alpha,0},n_{\alpha,1},n_{\alpha,2},n_{\alpha,\sigma})=(1,0,0,0)\\ 
		&\Omega^{1,001}_{\alpha}=\Omega^{2,002}_{\alpha}=1,\quad \Omega^{\sigma,00\sigma}_{\alpha}=\pm 1\\
		\alpha=&3,4: (n_{\alpha,0},n_{\alpha,1},n_{\alpha,2},n_{\alpha,\sigma})=(0,1,0,0)\\
		&\Omega^{1,112}_{\alpha}=e^{i2\pi/3},\ \Omega^{2,110}_{\alpha}=e^{-i2\pi/3},
		\ \Omega^{\sigma,11\sigma}_{\alpha}=\pm e^{i 2\pi/3}\\	
		\alpha=&5,6: (n_{\alpha,0},n_{\alpha,1},n_{\alpha,2},n_{\alpha,\sigma})=(0,0,1,0)\\
		&\Omega^{1,220}_{\alpha}=e^{-i2\pi/3}, \ \Omega^{2,221}_{\alpha}=e^{i2\pi/3},
		\ \Omega^{\sigma,22\sigma}_{\alpha}=\pm e^{i 2\pi/3}\\	
             \alpha=&7: (n_{\alpha,0},n_{\alpha,1},n_{\alpha,2},n_{\alpha,\sigma})=(1,1,0,0)\\
		&\Omega^{1,001}_{\alpha}=e^{i2\pi/3},\quad \Omega^{2,002}_{\alpha}=e^{-i2\pi/3},\\
		&\Omega^{\sigma,00\sigma}_{\alpha}=\Omega^{\sigma,11\sigma}_{\alpha}=0,\quad  
		\Omega^{1,112}_{\alpha}=\Omega^{2,110}_{\alpha}=1  \\
		&\Omega^{\sigma,01\sigma}_{\alpha}=e^{i\phi_1},\quad \Omega^{\sigma,10\sigma}_{\alpha}=e^{-i\phi_1} \\
		\alpha=&8: (n_{\alpha,0},n_{\alpha,1},n_{\alpha,2},n_{\alpha,\sigma})=(1,0,1,0)\\
		&\Omega^{1,001}_{\alpha}=e^{-i2\pi/3},\quad \Omega^{2,002}_{\alpha}=e^{i2\pi/3},\\
		&\Omega^{\sigma,00\sigma}_{\alpha}=\Omega^{\sigma,22\sigma}_{\alpha}=0,\quad
		\Omega^{1,220}_{\alpha}=\Omega^{2,221}_{\alpha}=1 \\
		&\Omega^{\sigma,02\sigma}_{\alpha}=e^{i\phi_2},\quad \Omega^{\sigma,20\sigma}_{\alpha}=e^{-i\phi_2} \\
		\alpha=&9: (n_{\alpha,0},n_{\alpha,1},n_{\alpha,2},n_{\alpha,\sigma})=(0,1,1,0)\\
		&\Omega^{1,112}_{\alpha}=\Omega^{2,221}_{\alpha}=e^{-i2\pi/3},\
		\Omega^{1,220}_{\alpha}=\Omega^{2,110}_{\alpha}=e^{i2\pi/3},\\
		&\Omega^{\sigma,11\sigma}_{\alpha}=\Omega^{\sigma,22\sigma}_{\alpha}=0 \\
		&\Omega^{\sigma,12\sigma}_{\alpha}=e^{i2\pi/3}e^{i\phi_3},\quad \Omega^{\sigma,21\sigma}_{\alpha}=e^{-i\phi_3} 
\end{split}
\end{equation}
and
\begin{equation}
\begin{split}		
		\alpha=&10,11: (n_{\alpha,0},n_{\alpha,1},n_{\alpha,2},n_{\alpha,\sigma})=(0,0,0,1)\\
		&\Omega^{1,\sigma\sigma\sigma}_{\alpha}=\Omega^{2,\sigma\sigma\sigma}_{\alpha}=e^{-i2\pi/3},\quad 
		\Omega^{\sigma,\sigma\sigma 0}_{\alpha}=\pm e^{i 3\pi/4},\\
		&\Omega^{\sigma,\sigma\sigma 1}_{\alpha}= \Omega^{\sigma,\sigma\sigma 2}_{\alpha}=\mp e^{i 5\pi/12}\\
		\alpha=&12,13: (n_{\alpha,0},n_{\alpha,1},n_{\alpha,2},n_{\alpha,\sigma})=(0,0,0,1)\\
		&\Omega^{1,\sigma\sigma\sigma}_{\alpha}=1,\quad \Omega^{2,\sigma\sigma\sigma}_{\alpha}=e^{i2\pi/3}, \\
		&\Omega^{\sigma,\sigma\sigma 0}_{\alpha}= \Omega^{\sigma,\sigma\sigma 2}_{\alpha}=\mp e^{i \pi/12},\quad \Omega^{\sigma,\sigma\sigma 1}_{\alpha}=\pm e^{i 5\pi/12}\\
		\alpha=&14,15: (n_{\alpha,0},n_{\alpha,1},n_{\alpha,2},n_{\alpha,\sigma})=(0,0,0,1)\\
		&\Omega^{1,\sigma\sigma\sigma}_{\alpha}=e^{i2\pi/3},\quad \Omega^{2,\sigma\sigma\sigma}_{\alpha}=1, \\
		&\Omega^{\sigma,\sigma\sigma 0}_{\alpha}=\Omega^{\sigma,\sigma\sigma 1}_{\alpha}=\mp e^{i \pi/12},\quad \Omega^{\sigma,\sigma\sigma 2}_{\alpha}=\pm e^{i 5\pi/12}
	\end{split}
	\label{}
\end{equation}
where $\phi_1,\phi_2,\phi_3$ are three $U(1)$ gauge phases.
Evidently there are 6 abelian quasiparticles with $d_\alpha=1$, for $\alpha=1,\dots,6$, and 9 nonabelian quasiparticles with $d_\alpha=2$ for $\alpha=7,8,9$ and $d_\alpha=\sqrt{3}$ for $\alpha=10,\dots,15$.
From (\ref{weq4}), we can identify the particle-antiparticle pairs:
\begin{equation}
	\begin{split}
	1=\bar{1},\quad 2=\bar{2},\quad 3=\bar{5},\quad 4=\bar{6}, \quad 7=\bar{8}, \quad 9=\bar{9},\\
	10=\bar{10},\quad 11=\bar{11},\quad 12=\bar{14},\quad 13=\bar{15}.
	\end{split}
	\label{}
\end{equation}

The topological spins of each of these quasiparticles can be computed from (\ref{twist}):
\begin{equation}
	\begin{split}
		&\{e^{i\theta_1},\dots,e^{i\theta_{15}}\}=\\
		&\{1,1,e^{-i\frac{2\pi}{3}},e^{-i\frac{2\pi}{3}},e^{-i\frac{2\pi}{3}},e^{-i\frac{2\pi}{3}},1,1,e^{i\frac{2\pi}{3}}, \\
		& e^{i\frac{3\pi}{4}},e^{-i\frac{\pi}{4}},e^{-i\frac{11\pi}{12}},e^{i\frac{\pi}{12}},e^{-i\frac{11\pi}{12}},e^{i\frac{\pi}{12}}\}.
\end{split}
	\label{}
\end{equation}

As for the $p=-1$ model, we do not include explicit expressions for the $\Omega_{\alpha}$ here, for brevity. Instead we skip directly to the topological spins of the quasiparticles: 
\begin{equation}
	\begin{split}
		&\{e^{i\theta_1},\dots,e^{i\theta_{15}}\}=\\
		&\{1,1,e^{-i\frac{2\pi}{3}},e^{-i\frac{2\pi}{3}},e^{-i\frac{2\pi}{3}},e^{-i\frac{2\pi}{3}}
		,1,1,e^{i\frac{2\pi}{3}}, \\
		& e^{i\frac{\pi}{4}},e^{-i\frac{3\pi}{4}},e^{-i\frac{5\pi}{12}},e^{i\frac{7\pi}{12}},e^{-i\frac{5\pi}{12}},e^{i\frac{7\pi}{12}}\}.
\end{split}
\end{equation}

As can be seen from the topological spins of the quasiparticles, both models break time reversal symmetry, as one might expect given that the string-net data does not have reflection symmetry. Thus the $TY_3$ string-net is an example of a non-abelian model that cannot be realized with the original construction of Ref.~\onlinecite{LevinWenstrnet}, which implicitly assumed time reversal symmetry.

\section{Conclusion}

In this paper, we have given a detailed description of how to construct generalized string-net models. Importantly, our construction works for any unitary fusion category; unlike the original models proposed by Levin and Wen\cite{LevinWenstrnet}, we do not impose additional requirements on this category that ensure the invariance of the string-net ground state under planar or spherical isotropy or tetrahedral reflections. (Note that the construction in the main text works only for the case of no fusion multiplicities; the construction in Appendix \ref{App:FusionMults} must be used for the case with fusion multiplicities.)

 In addition to providing a detailed discussion of string-net ground states and Hamiltonians, we have also described an approach for constructing string operators and for computing quasiparticle statistics -- in particular, the $S$ and $T$ matrices. Finally, we have analyzed the conditions under which the generalized string-net models are isotropic on the plane or on the sphere, and we have discussed the relationship between generalized string-net models and the original models of Ref.~\onlinecite{LevinWenstrnet}.


\section*{acknowledgments}
We thank Tian Lan, Chenjie Wang, and Yidun Wan for helpful discussions. F. J. B. and C.-H. L. acknowledge support from NSF-DMR 1352271 and the Sloan Foundation (FG-2015-65927). F. J. B. is grateful for the financial support of NSF DMR 1928166, the Carnegie Corporation of New York, and the Institute for Advanced Study.  

\appendix

\section{Derivation of self-consistency conditions \label{app:sfc}}
\begin{figure}[ptb]
\begin{center}
\includegraphics[width=0.7\columnwidth]{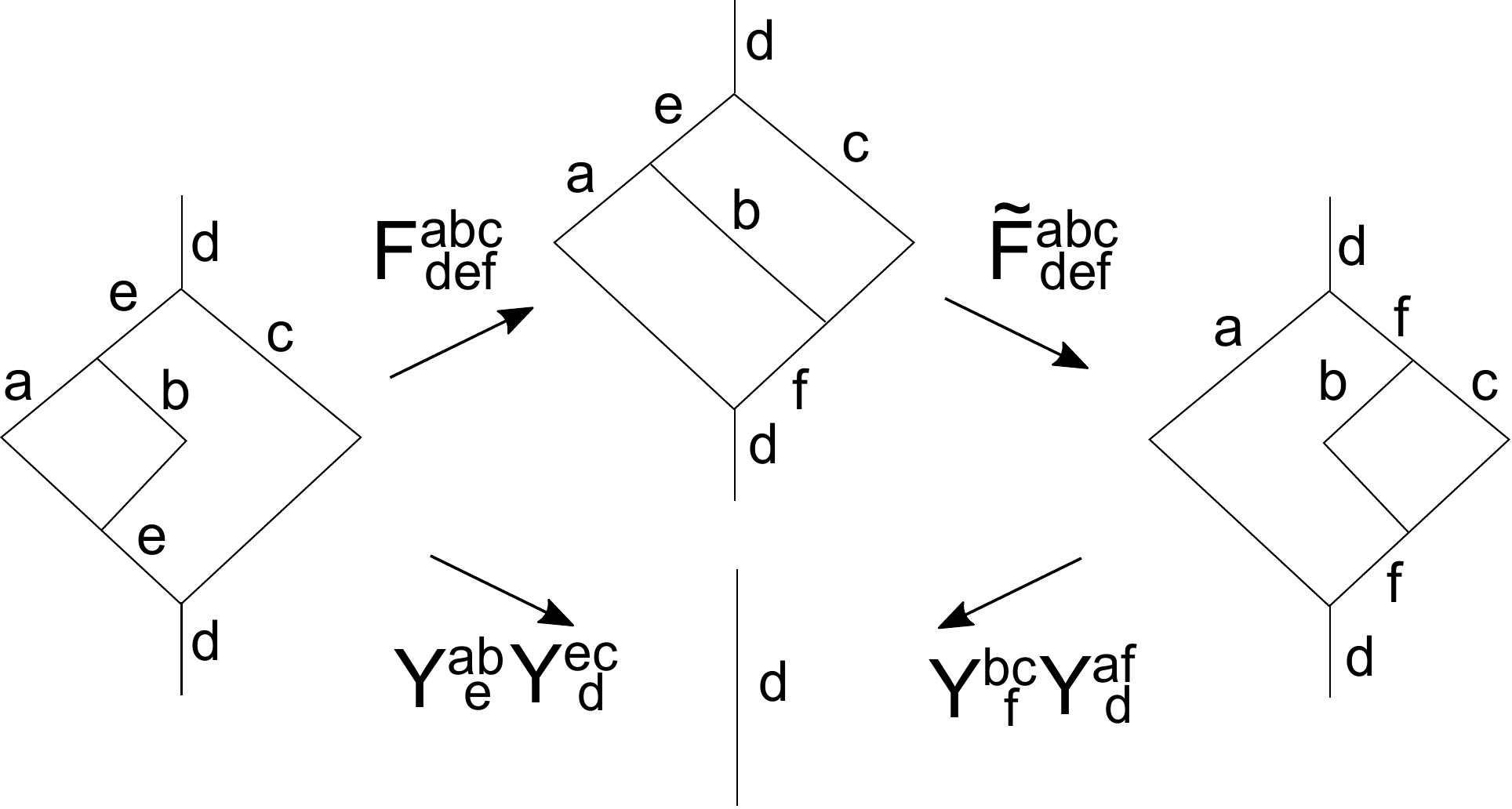}
\end{center}
\caption{Self-consistency requires the conditions (\ref{eq2a}).
} 
\label{fig:pi1}
\end{figure}

In this  appendix, we show that the parameters $\{F^{abc}_{def}, \tilde{F}^{abc}_{def}, [F^{ab}_{cd}]_{ef},[\tilde{F}^{ab}_{cd}]_{ef} \}$ must satisfy the following equations if the local rules (\ref{localrules},\ref{localrules1}) are self-consistent:
\begin{subequations}
	\begin{align}
		\tilde{F}^{abc}_{def} & =(F^{abc}_d)^{-1}_{fe}\frac{Y^{ab}_e Y^{ec}_d}{Y^{bc}_f Y^{af}_d}  \label{eq2a}\\
		[F^{ab}_{cd}]_{ef} &=  \tilde{F}^{ceb}_{fad} \frac{Y^{eb}_d}{Y^{ab}_f} \label{eq3a}\\
		[\tilde{F}^{ab}_{cd}]_{ef} &= F^{ceb}_{fad} \frac{Y^{eb}_d}{Y^{ab}_f} \label{eq4a} 
	\end{align}
	\label{consistencyapp}
\end{subequations}
Note that the above conditions are a \emph{subset} of the identities in Eq.~\ref{consistency1} and the self-consistency conditions listed in Eq.~\ref{consistency}: the remaining self-consistency conditions are derived in the main text. 

The  first condition (\ref{eq2a}) can be derived by considering the sequences in Fig. \ref{fig:pi1}:
\begin{equation}
	Y^{ab}_e Y^{ec}_d =\sum_f F^{abc}_{def} \tilde{F}^{abc}_{def} Y^{bc}_f Y^{af}_d.  
	\label{eq2a1}
\end{equation}
Then Eq.~(\ref{eq2a}) follows from (\ref{eq2a1}).

To derive (\ref{eq3a}), we consider the sequence
\begin{equation}
	\begin{split}
		\Phi\left(\raisebox{-0.22in}{\includegraphics[height=0.5in]{rule5a.pdf}} \right)&=\sum_f \frac{1}{Y^{ab}_f} \Phi\left(\raisebox{-0.22in}{\includegraphics[height=0.5in]{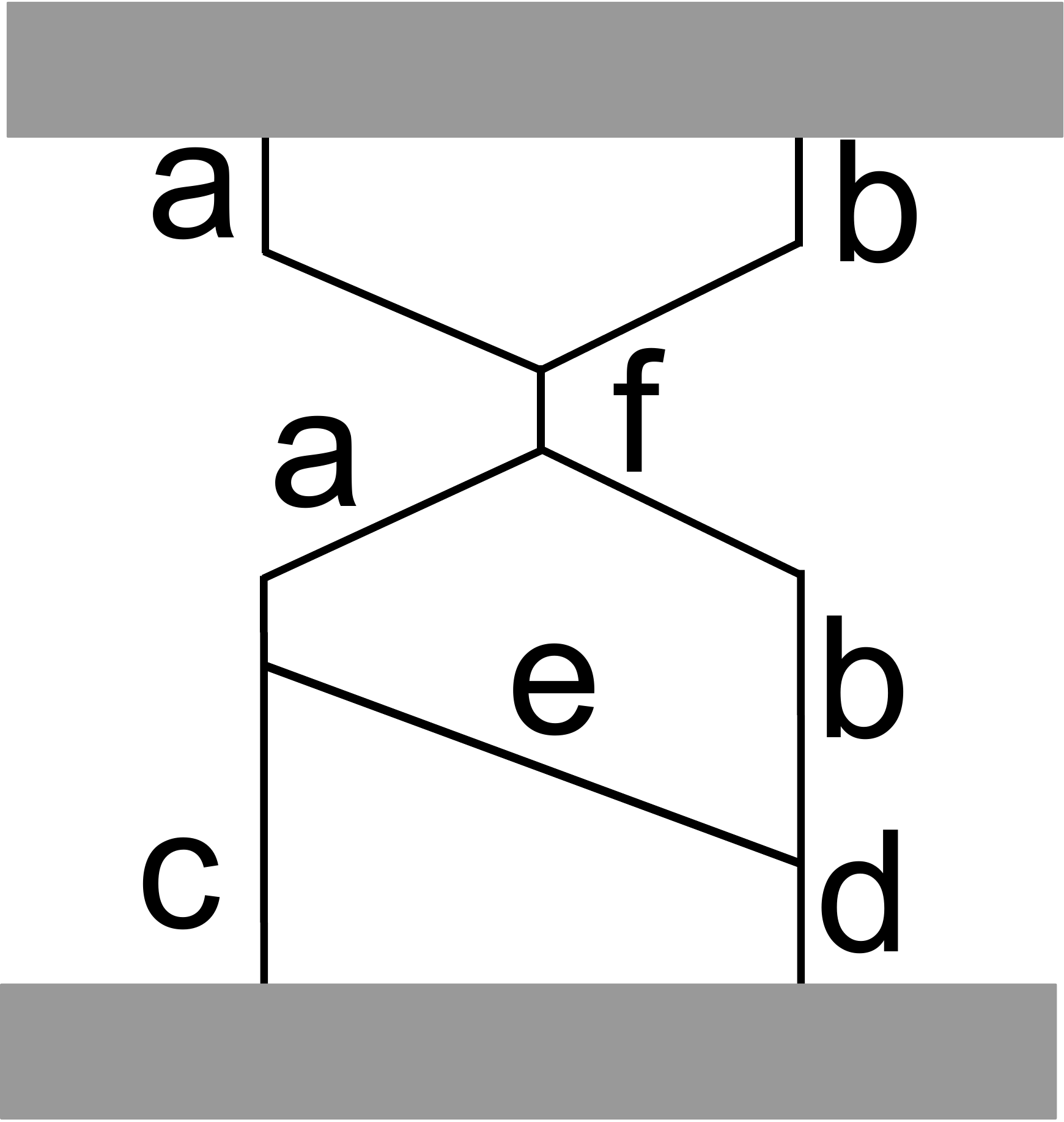}} \right)\\
		&=\sum_f \frac{1}{Y^{ab}_f} \tilde{F}^{ceb}_{fad} \Phi\left(\raisebox{-0.22in}{\includegraphics[height=0.5in]{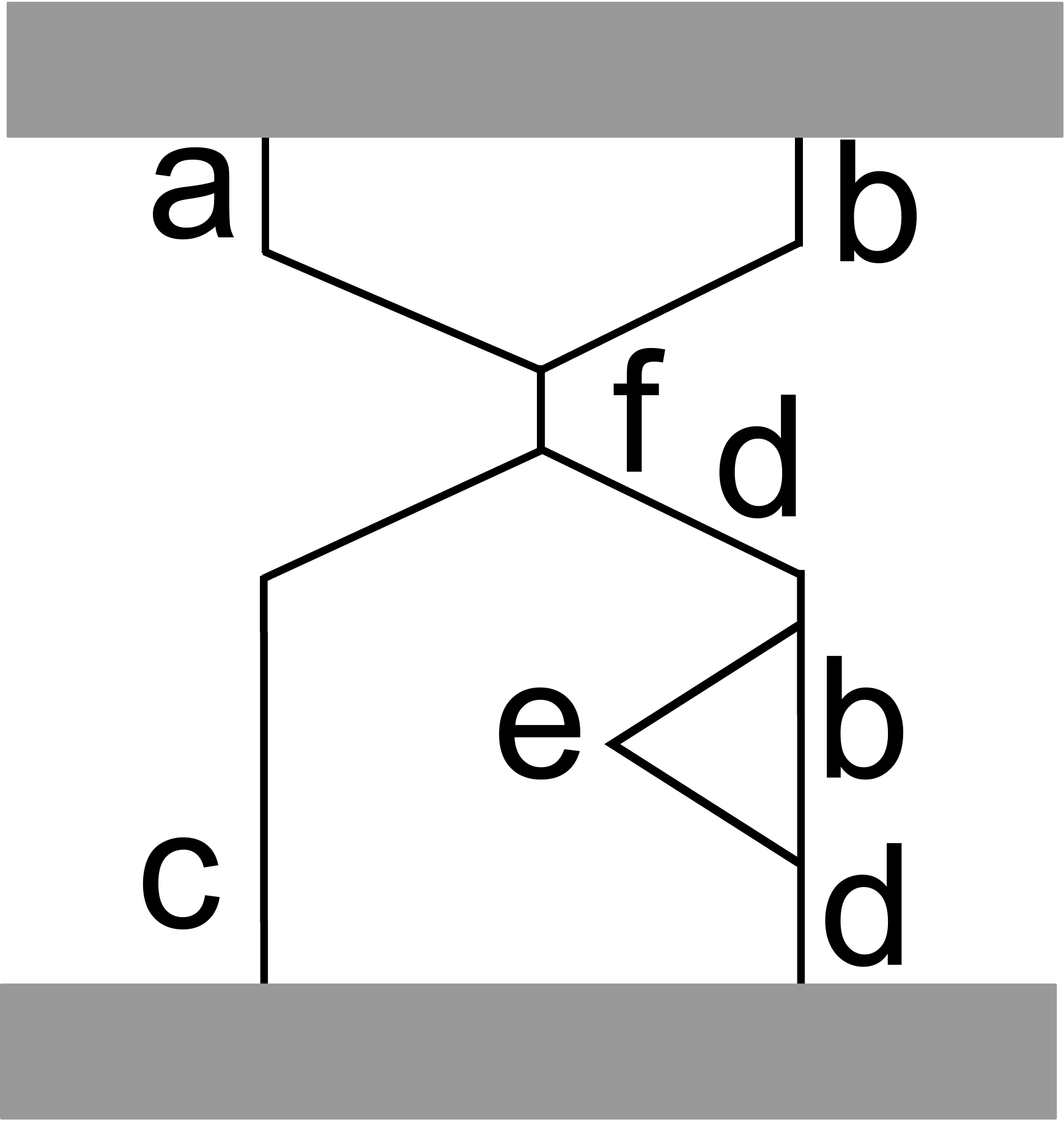}} \right) \\
		&=\sum_f \frac{1}{Y^{ab}_f} \tilde{F}^{ceb}_{fad} Y^{eb}_d\Phi\left(\raisebox{-0.22in}{\includegraphics[height=0.5in]{rule5b.pdf}} \right) \\
	&\equiv \sum_f [{F}^{ab}_{cd}]_{ef} \Phi\left(\raisebox{-0.22in}{\includegraphics[height=0.5in]{rule5b.pdf}} \right).
	\end{split}
	\label{}
\end{equation}
We use the local rules (\ref{1b},\ref{1d},\ref{1c}) in the first three equalities sequentially. Thus we have (\ref{eq3a}).

Similarly, to derive (\ref{eq4a}), we follow the same logic by considering the sequence
\begin{equation}
	\begin{split}
		\Phi\left(\raisebox{-0.22in}{\includegraphics[height=0.5in]{rule6a.pdf}} \right)&=\sum_f \frac{1}{Y^{ab}_f} \Phi\left(\raisebox{-0.22in}{\includegraphics[height=0.5in]{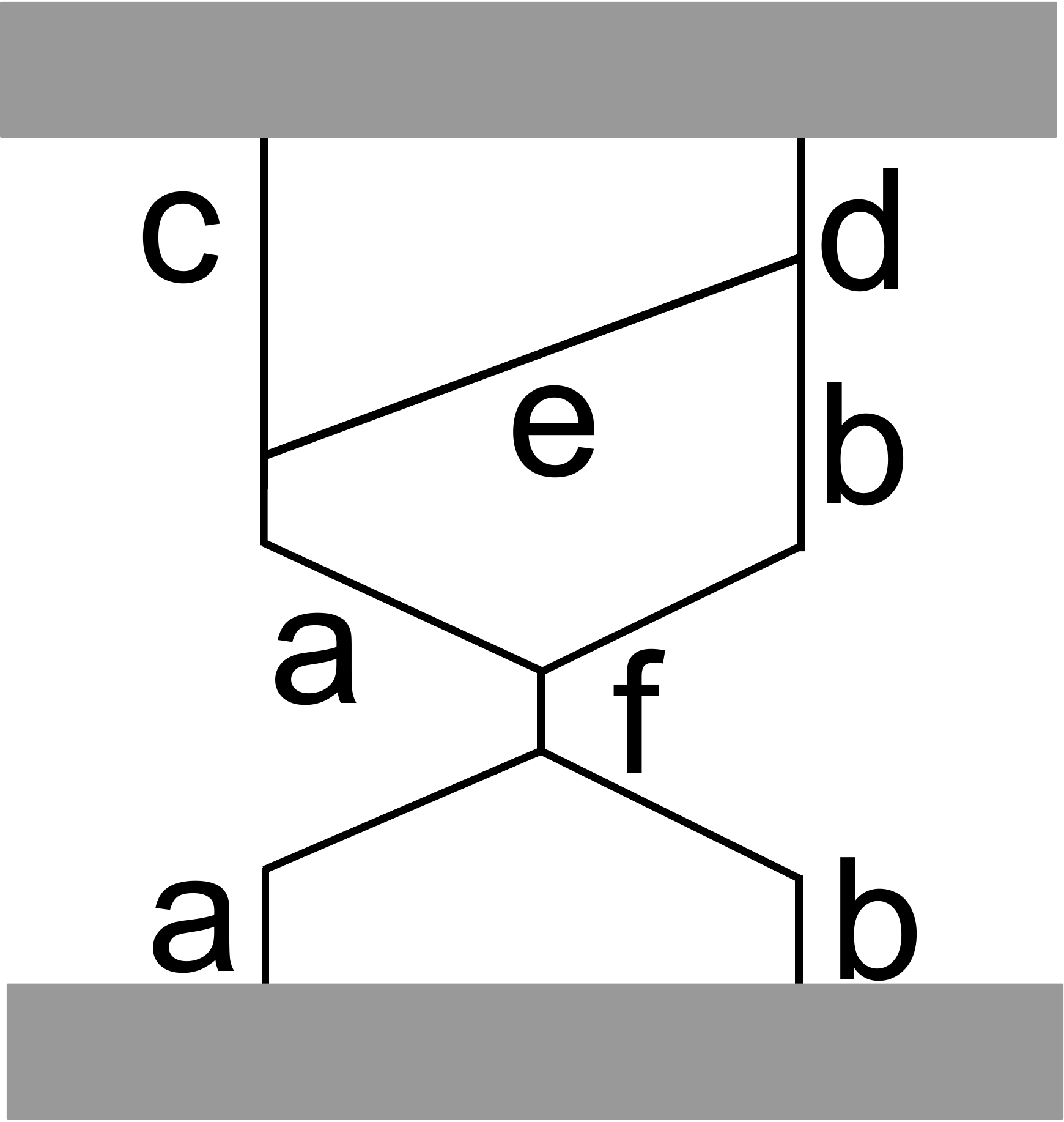}} \right)\\
		&=\sum_f \frac{1}{Y^{ab}_f} F^{ceb}_{fad} \Phi\left(\raisebox{-0.22in}{\includegraphics[height=0.5in]{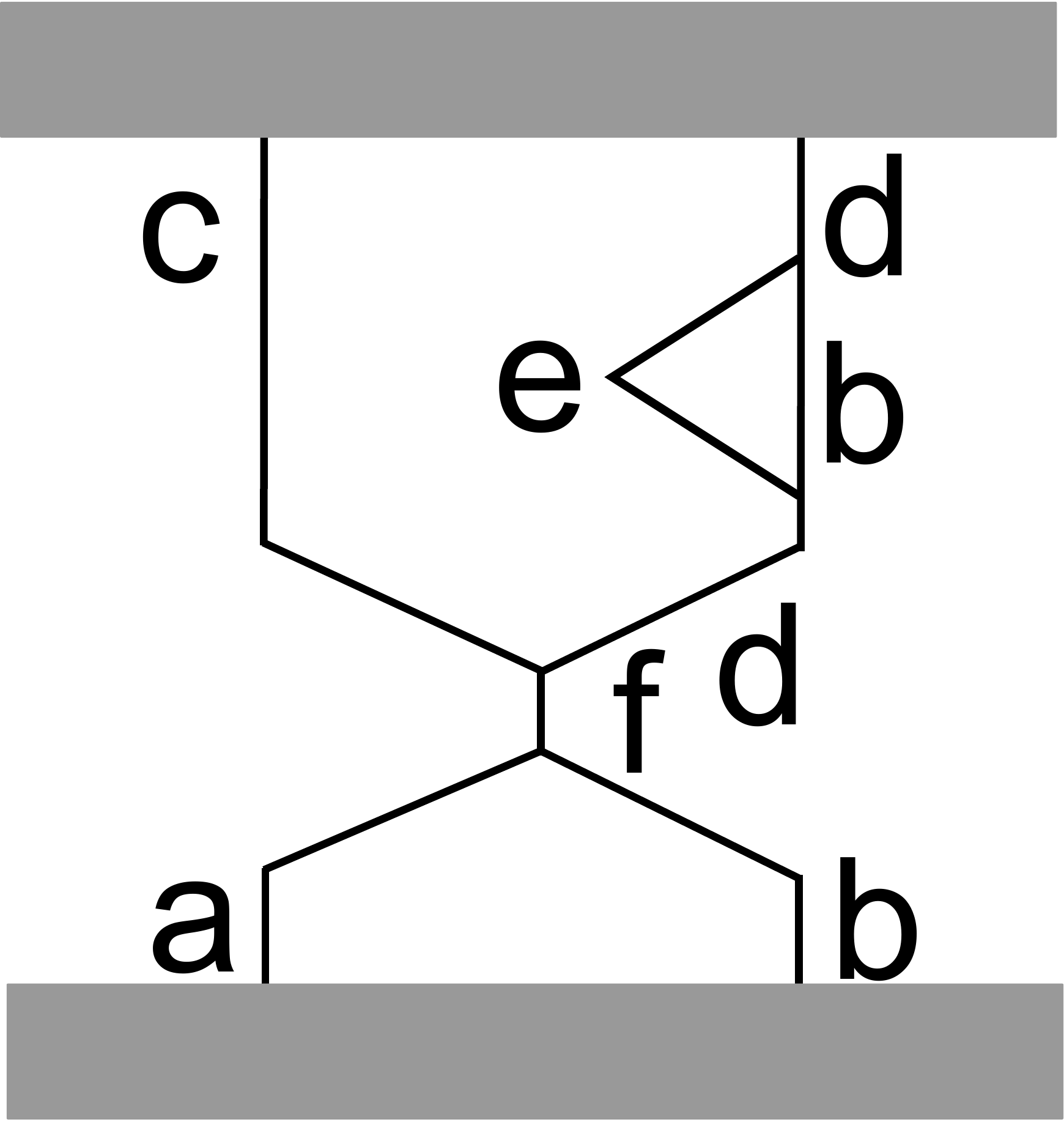}} \right) \\
		&=\sum_f \frac{1}{Y^{ab}_f} F^{ceb}_{fad}  Y^{eb}_d \Phi\left(\raisebox{-0.22in}{\includegraphics[height=0.5in]{rule6b.pdf}} \right) \\
	&\equiv \sum_f [\tilde{F}^{ab}_{cd}]_{ef} \Phi\left(\raisebox{-0.22in}{\includegraphics[height=0.5in]{rule6b.pdf}} \right).
	\end{split}
	\label{}
\end{equation}
We use the local rules (\ref{1b},\ref{1a},\ref{1c}) in the first three equalities sequentially. Thus we have (\ref{eq4a}).

\begin{figure}[ptb]
\begin{center}
\includegraphics[width=1\columnwidth]{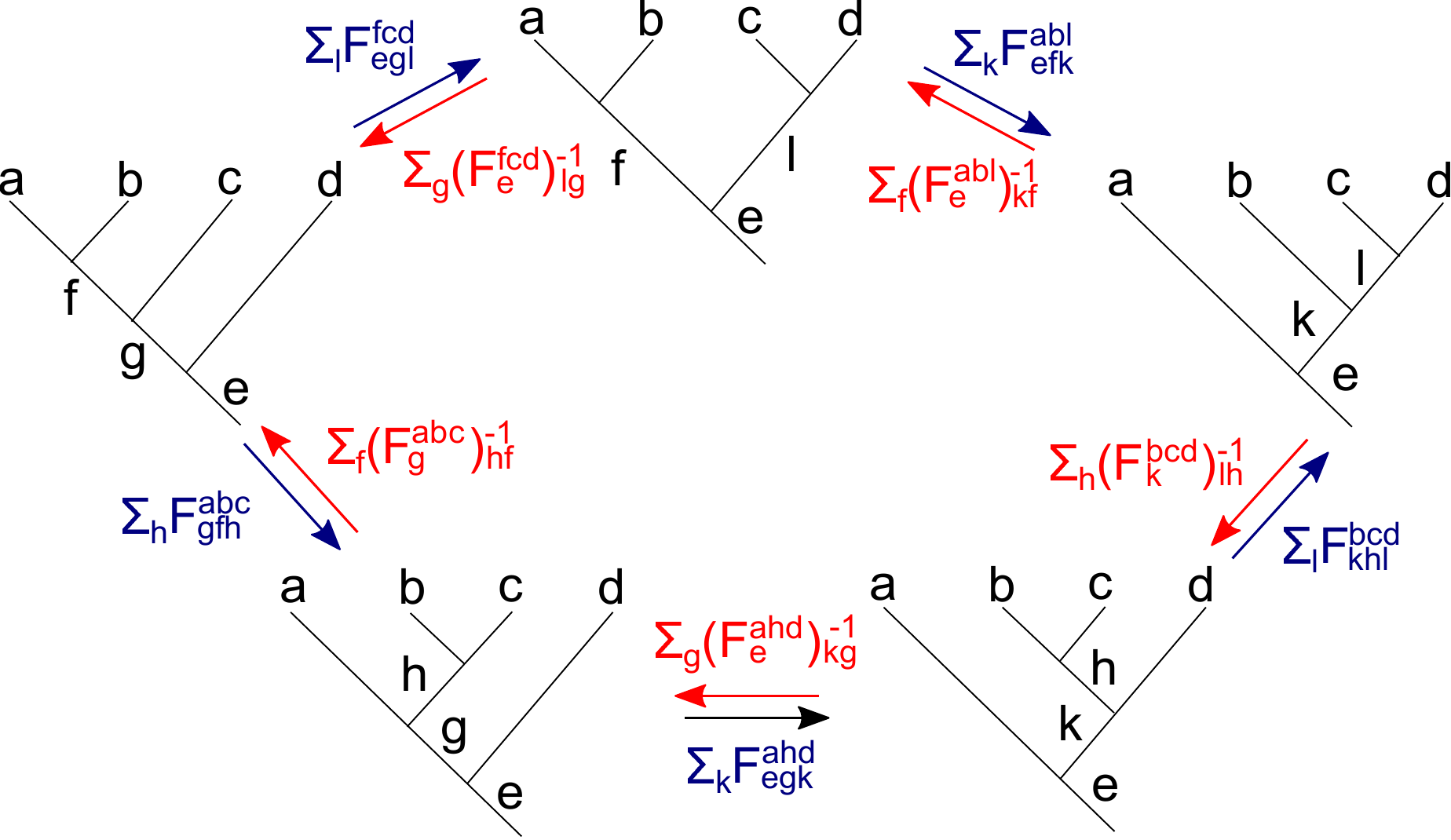}
\end{center}
\caption{Variants of the pentagon identity.
} 
\label{fig:pi2}
\end{figure}

 So far we have not discussed the most important self-consistency condition of all: the pentagon identity (\ref{3a}). The reason for this omission is that this identity is derived in the main text. Here we would like to point out that there are actually many variants of the pentagon identity which follow from similar consistency requirements. Each of these variants can be derived graphically by relating the amplitude of two of the five configurations in Fig. \ref{fig:pi2}  by sequences of $F$ and $(F)^{-1}$ operations. Likewise, these variants can be derived algebraically by multiplying both sides of (\ref{3a}) by appropriate $(F)^{-1}$ operations.
For example, by considering two sequences relating the amplitude of the top configuration and the bottom left configuration in Fig. \ref{fig:pi2}, we have
\begin{equation}
	\sum_k F^{ahd}_{egk} F^{bcd}_{khl} (F^{abl}_e)^{-1}_{kf}
	=F^{fcd}_{egl}(F^{abc}_g)^{-1}_{hf}.
	\label{piv}
\end{equation}
Eq.~(\ref{piv}) can be derived by multiplying both sides of (\ref{3a}) by $(F^{abl}_e)^{-1}_{kf}, (F^{abc}_g)^{-1}_{hf}$.

We can derive useful identities from  these variants of the pentagon identity. For example, by setting $e=0$ in (\ref{3a}) and using the fact that $F^{abc}_{0ef}=w^{abc} \delta_{e,\bar{c}} \delta_{f,\bar{a}} \delta^{ab}_{\bar{c}}$ is a complex number depending on three string types $a,b,c$, we obtain
\begin{equation}
	(F^{abc}_g)^{-1}_{hf} = 
	F^{bc\bar{g}}_{\bar{a}h\bar{f}} 
	\frac{w^{ah\bar{g}}}{w^{ab\bar{f}} w^{fc\bar{g}}}
	=F^{\bar{g}ab}_{\bar{c}\bar{h}f} \frac{w^{\bar{g}fc}}{w^{\bar{g}ah} w^{\bar{f}bc}}.
	\label{}
\end{equation}
By setting $h=0$ in (\ref{piv}), we obtain
\begin{equation}
	F^{b\bar{b}d}_{d0l} (F^{abl}_e)^{-1}_{df} = F^{f\bar{b}d}_{eal} (F^{ab\bar{b}}_{a})^{-1}_{0f}.
	\label{piv1}
\end{equation}

\section{Gauge choices  for $Y^{ab}_c$ \label{app:yabc}}

In this appendix, we discuss three different gauge choices for $Y^{ab}_c$. The first gauge choice is to take $Y^{ab}_c$ of the special form
\begin{equation}
	Y^{ab}_c =\frac{y_a y_b}{y_c}
	\label{yabc}
\end{equation}
with
\begin{equation}
	y_a=\sqrt{d_a}e^{i\phi_a}
	\label{ya}
\end{equation}
where $d_a=d_{\bar{a}}$ is the quantum dimension of the string-$a$ and $\phi_a$ is a $U(1)$ phase.
In this parametrization, the condition (\ref{y1}) requires
\begin{subequations}
	\begin{gather}
		\phi_a+\phi_{\bar{a}} =0 \text{ (mod $2\pi$)}. 
	\end{gather}
\end{subequations}

In the gauge (\ref{yabc}),
$\tilde{F}^{abc}_{def}=(F^{abc}_d)^{-1}_{fe}$ and 
the amplitude of a loop-$a$ is real: $Y^{a\bar{a}}_0=d_a.$
A special case of (\ref{yabc}), namely
\begin{equation}
	Y^{ab}_c=\sqrt{\frac{d_a d_b}{d_c}},
	\label{}
\end{equation}
is used in Refs.~\onlinecite{BondersonThesis, LevinWenstrnet, KitaevKong, HahnWolf, LakeWu}.

 Another gauge choice that is worth mentioning is
\begin{equation}
	Y^{ab}_c =1
	\label{gauge2}
\end{equation}
which satisfies (\ref{ynorm},\ref{y1}) trivially.
This choice  is appealing since it allows us to drop all the $Y$ factors.  However, this choice is \emph{not} allowed in our construction as it does not satisfy 
(\ref{y0}), and the corresponding Hamiltonian (\ref{hsn0}) is not Hermitian.

The third gauge choice which we would like to mention is restricted to Abelian string-net models with the Abelian branching rules $\{(a,b;a+b)\}$. To explain this gauge choice, it is convenient to suppress indices that can be deduced from the branching rules and define
\begin{equation}
	F^{abc}\equiv F^{abc}_{(a+b+c)(a+b)(b+c)} \quad Y^{ab}\equiv Y^{ab}_{a+b}.
	\label{}
\end{equation}
In this notation, the gauge choice corresponds to taking $Y^{ab}_c$ to be
\begin{equation}
	Y^{ab} =F^{ab\bar{b}}
	\label{yab}
\end{equation}
which can not be factorized to the form (\ref{yabc}). One can check that (\ref{yab}) satisfies (\ref{hermicity}).  This gauge has the advantage that the the Frobenius-Schur indicator $\gamma_a = 1$ (\ref{gamma}), but the disadvantage that $Y^{a\bar{a}}$ can be complex, and $\tilde{F}^{abc}$ is no longer the inverse of $F^{abc}$.

\section{Showing that $B_{p_1}^{t_1},B_{p_2}^{t_2}$ commute \label{app:commute}}
In this  appendix, we show that the operators $B^{t_1}_{p_1}$ and $B^{t_2}_{p_2}$ commute with one another for $p_1\neq p_2$. 
We only need to consider the case when $p_1$ and $p_2$ are adjacent since two operators will commute if $p_1$ and $p_2$ are further apart.

\begin{figure}[ptb]
\begin{center}
\includegraphics[width=0.6\columnwidth]{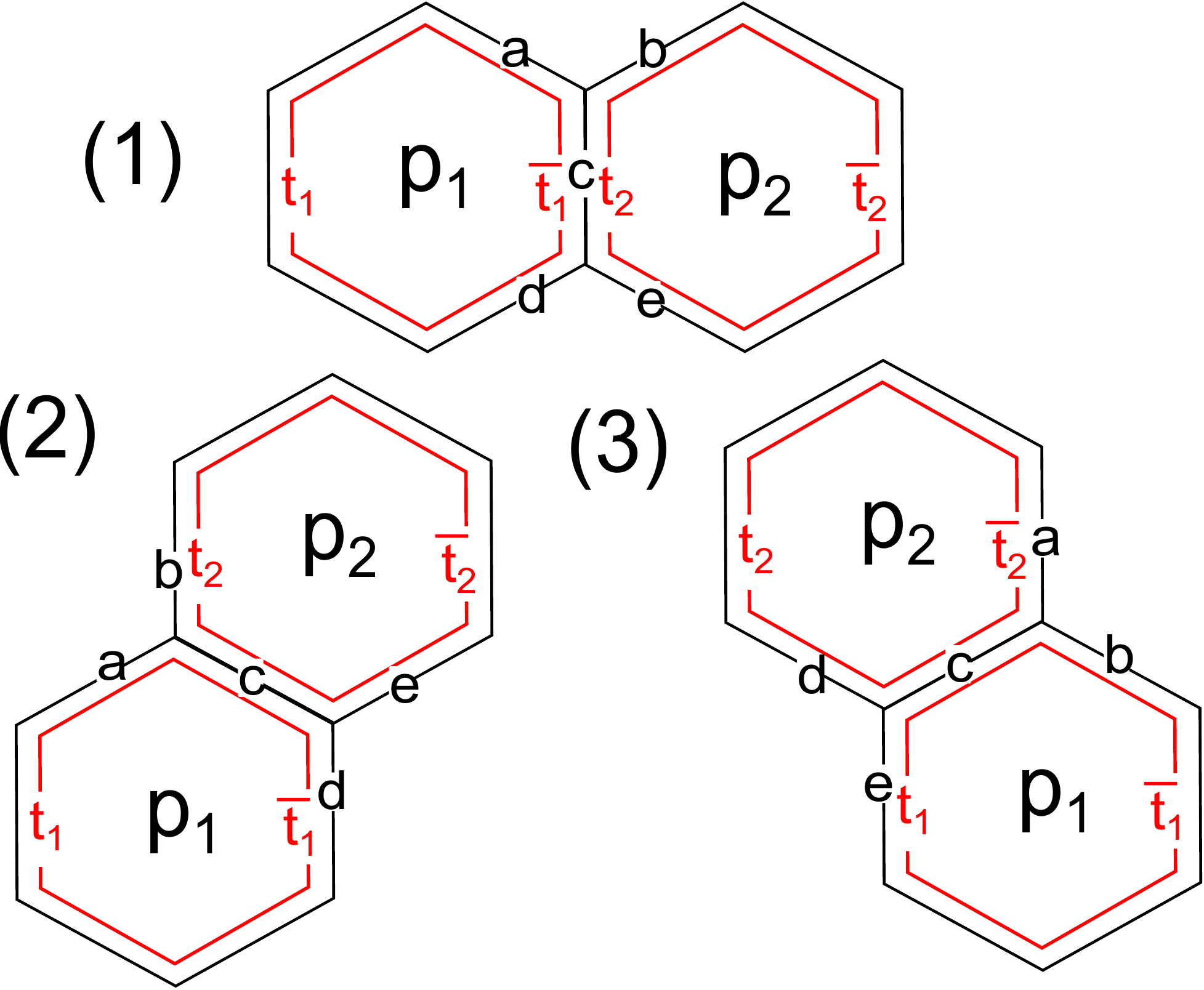}
\end{center}
\caption{The action of $B_{p_1}^{t_1}B_{p_2}^{t_2}$ and $B_{p_2}^{t_2}B_{p_1}^{t_1}$ on the shared boundary. 
} 
\label{fig:2bp}
\end{figure}

Let $B_{p_1}^{t_1}, B_{p_2}^{t_2}$ act on two adjacent plaquettes $p_1,p_2$. 
The two adjacent plaquettes can be in three possible relative positions shown in Fig. \ref{fig:2bp}. We want to show the $B_{p_1}^{t_1} B_{p_2}^{t_2} = B_{p_2}^{t_2} B_{p_1} ^{t_1}$ in these three cases. To show this, we compare the matrix elements of $B_{p_1}^{t_1} B_{p_2}^{t_2}$ and $B_{p_2}^{t_2} B_{p_1} ^{t_1}$ and show they are the same. We find that it is sufficient to compare the factors associated with the shared boundary which are different. We discuss these three cases in order.

 For case (1), we  need to show
\begin{equation}
	\begin{split}
		&\sum_{c_1 } F^{a_1bt_2}_{c_3c_1b_2} (F^{\bar{t}_1 ab}_{c_1})^{-1}_{ca_1}  
		{F}^{\bar{t}_1 de}_{c_1d_1c} (F^{d_1et_2}_{c_3})^{-1}_{e_2c_1} \\
		&\qquad \qquad =\sum_{c_2 } F^{abt_2}_{c_2cb_2} (F^{\bar{t}_1ab_2}_{c_3})^{-1}_{c_2a_1}
		{F}^{\bar{t}_1de_2}_{c_3 d_1 c_2} ({F}^{det_2}_{c_2})^{-1}_{e_2c}  
	\end{split}
	\label{case1-1}
\end{equation}
To show (\ref{case1-1}), it is sufficient to show 
\begin{equation}
	\begin{split}
		 F^{a_1 bt_2}_{c_3 c_1 b_2} (F^{\bar{t}_1ab}_{c_1})^{-1}_{ca_1}
		&=\sum_{c_2} F^{abt_2}_{c_2 c b_2} (F^{\bar{t}_1 ab_2}_{c_3})^{-1}_{c_2 a_1}  F^{\bar{t}_1 c t_2}_{c_3 c_1 c_2} \\
		{F}^{\bar{t}_1 de}_{c_1d_1c} ({F}^{d_1 et_2}_{c_3})^{-1}_{e_2c_1} &=\sum_{c_2'}   {F}^{\bar{t}_1 de_2}_{c_3d_1c'_2}  ({F}^{det_2}_{c_2'})^{-1}_{e_2c} ({F}^{\bar{t}_1 c t_2}_{c_3 })^{-1}_{c'_2c_1}
	\end{split}
	\label{case1-2}
\end{equation}
To see this, one can insert (\ref{case1-2})  into the left hand side of (\ref{case1-1}) and simplify the expression to obtain the right hand side of (\ref{case1-1}).
We can show (\ref{case1-2}) by identifying Eq.~(\ref{case1-2}) as  one of the variants of the pentagon identity (see Appendix \ref{app:sfc}).
This completes the proof of (\ref{case1-1}).

 For case (2), we  need to show 
\begin{equation}
	\begin{split}
		&\sum_{c_1} (F^{a_1 \bar{t}_1 c}_{b})^{-1}_{c_1 a} (F^{a_1 c_1 t_2}_{b_2})^{-1}_{c_3 b} 
		 F^{c_3 \bar{t}_2 e}_{d_1c_1 e_2} (F^{\bar{t}_1 ce}_{d_1})^{-1}_{dc_1} (F^{c_1t_2 \bar{t}_2}_{c_1})^{-1}_{0c_3}  \\
		&=\sum_{c_2}
		 (F^{act}_{b_2})^{-1}_{c_2b} (F^{a_1 \bar{t}_1 c}_{b_2})^{-1}_{c_3 a}
		F^{c_2 \bar{t}_2 e}_{dce_2}
		(F^{ct_2 \bar{t}_2}_c)^{-1}_{0c_2} 
		(F^{\bar{t}_1 c_2 e_2}_{d_1})^{-1}_{dc_3}
	\end{split}
	\label{case2-1}
\end{equation}
To show (\ref{case2-1}), it is sufficient to show
\begin{subequations}
	\begin{align}
		&(F^{a_1\bar{t}_1c}_b)^{-1}_{c_1a} (F^{a_1c_1t_2}_{b_2})^{-1}_{c_3b} =
		\sum_{c_2}  (F^{act_2}_{b_2})^{-1}_{c_2b} (F^{a_1 \bar{t}_1 c_2}_{b_2})^{-1}_{c_3a} 
		F^{\bar{t}_1 ct_2}_{c_3c_1c_2}
		\label{case2-2b}
		\\
		&  F^{c_3 \bar{t}_2 e}_{d_1 c_1 e_2}
		(F^{c_1 t_2 \bar{t}_2}_{c_1})^{-1}_{0c_3}
		(F^{\bar{t}_1 ce}_{d_1})^{-1}_{dc_1}  = \nonumber \\
		& \qquad \qquad \qquad \sum_{c_2'} 
		F^{c'_2 \bar{t}_2 e}_{dce_2} 
		(F^{ct_2 \bar{t}_2}_c)^{-1}_{0c'_2} 
		(F^{\bar{t}_1 c'_2 e_2}_{d_1})^{-1}_{dc_3}
		(F^{\bar{t}_1 c t_2}_{c_3})^{-1}_{c'_2 c_1} 
		\label{case2-2c}
	\end{align}
	\label{case2-2}
\end{subequations}
To see this, we insert (\ref{case2-2})  into the left hand side of (\ref{case2-1}) and simplify the expression to obtain the right hand side of (\ref{case2-1}). 
What remains is to show (\ref{case2-2}).
Eq.~(\ref{case2-2b}) is  a variant of the pentagon identity. To show (\ref{case2-2c})  we need to do more work.

First, to show (\ref{case2-2c}), it is sufficient to show
\begin{subequations}
	\begin{align}
	& (F^{c_1t_2 e_2}_{d_1})^{-1}_{ec_3} (F^{\bar{t}_1 ce}_{d_1})^{-1}_{dc_1}  =  \nonumber \\
		&\qquad \qquad \qquad\sum_{c_2'} 
		(F^{ct_2e_2}_d)^{-1}_{ec'_2}
		(F^{\bar{t}_1ct_2}_{c_3})^{-1}_{c'_2 c_1} 
		(F^{\bar{t}_1 c'_2 e_2}_{d_1})^{-1}_{dc_3}  \label{case2-2c3}\\
		& F^{c_3 \bar{t}_2 e}_{d_1 c_1 c_2} (F^{c_1 t_2 \bar{t}_2}_{c_1})^{-1}_{0c_3}  = F^{t_2 \bar{t}_2 e}_{e0e_2} (F^{c_1 t_2 e_2}_{d_1})^{-1}_{ec_3} \label{case2-2c1}\\
		&  F^{c_2 \bar{t}_2 e}_{dce_2} (F^{ct_2 \bar{t}_2}_c)^{-1}_{0c_2} = F^{t_2 \bar{t}_2 e}_{e0e_2} (F^{ct_2 e_2}_{d})^{-1}_{ec_2} \label{case2-2c2}
		\end{align}
	\label{case2-3}
\end{subequations}
To see this, we multiply both sides of (\ref{case2-2c3}) by $F^{t_2\bar{t}_2 e}_{e0e_2}$ and use (\ref{case2-2c1},\ref{case2-2c2}) to simplify the expression to obtain (\ref{case2-2c}). 
What remains is to show (\ref{case2-3}).
The first equation (\ref{case2-2c3}) is a variant of the pentagon identity and the last two equations follow from (\ref{piv1}).
This completes the proof for case (2).

 For case (3), we arrive at a similar equation as (\ref{case2-1}) with $(F)$ replaced by $(F)^{-1}$. Thus, an dentical proof as in case (2) goes through by changing $(F)$ by $(F)^{-1}$ in Eq.~(\ref{case2-2},\ref{case2-3}). This completes the proof for case (3).

\section{Properties of the Hamiltonian (\ref{hsn0}) \label{app:property}}
In this  appendix, we establish the following properties of the Hamiltonian (\ref{hsn0}):
\begin{enumerate}
	\item $(B_p^s)^\dagger= B_p^{\bar{s}}$
	\item $B_p$ is a  projection operator, i.e. $B_p^2=B_p$
	\end{enumerate}

To show the first property, we first use the pentagon identity to derive 
\begin{subequations}
	\begin{align}
		\frac{[F^{a'b}_{sc}]_{ac'}}{[\tilde{F}^{ab}_{\bar{s}c'}]_{a'c}}
		&=\frac{[\tilde{F}^{0a}_{\bar{s}a'}]_{sa}}{[\tilde{F}^{0c}_{\bar{s}c'}]_{sc}}
		\frac{Y^{ab}_c}{Y^{a'b}_{c'}} \label{e1a}\\
		\frac{[F^{cs}_{ab'}]_{bc'}}{[\tilde{F}^{c'\bar{s}}_{ab}]_{b'c}}
		&=\frac{[{F}^{cs}_{c'0}]_{\bar{s}c'}}{[{F}^{bs}_{b'0}]_{\bar{s}b'}}
		\frac{Y^{ab}_c}{Y^{ab'}_{c'}} \label{e1b}\\
		\frac{[F^{ab}_{a'b'}]_{\bar{s}c}}{[\tilde{F}^{a'b'}_{ab}]_{sc}}
		&=\frac{[{F}^{0b}_{sb'}]_{\bar{s}b}}{[\tilde{F}^{a'\bar{s}}_{a0}]_{sa}} \label{e1c}
	\end{align}
	\label{feqs}
\end{subequations}
Eq.~(\ref{feqs}) follows from variants of the pentagon identity.
Specifically, Eq.~(\ref{e1a}--\ref{e1c}) can be obtained respectively from
\begin{equation}
	\begin{split}
	\sum_l F^{fcd}_{egl} F^{abl}_{efk} (F^{bcd}_k)^{-1}_{lh}
	=F^{abc}_{gfh} F^{ahd}_{egk}\\
	\sum_f (F^{abl}_e)^{-1}_{kf} (F^{fcd}_e)^{-1}_{lg} F^{abc}_{gfh}
	=(F^{bcd}_k)^{-1}_{lh} (F^{ahd}_e)^{-1}_{kg} \\
	\sum_g (F^{fcd}_e)^{-1}_{lg} F^{abc}_{gfh} F^{ahd}_{egk}
	=F^{abl}_{efk} (F^{bcd}_{k})^{-1}_{lh}
	\end{split}
	\label{}
\end{equation}
by setting $f=0, l=0, h=0$ in  the first, second and third equation above.

By using (\ref{unitary}) and (\ref{feqs}), it is straightforward to show that
\begin{equation}
	\begin{split}
	&\frac{B^{s,i_1i_2\dotsi_6}_{p,i'_1i'_2\dots i'6}(e_1e_2\dots e_6)}{(B^{\bar{s},i'_1i'_2\dots i'_6}_{i_1i_2\dots i_6}(e_1 e_2 \dots e_6))^*} = \\
	&\frac{|F^{\bar{i}_1s\bar{s}}_{\bar{i}_1\bar{i}_1'0}|^2 
	|F^{\bar{i}_2'\bar{s}s}_{\bar{i}_2'\bar{i}_2 0}|^2
	|F^{\bar{i}_3s\bar{s}}_{\bar{i}_3\bar{i}_3'0}|^2}
	{|F^{i_4'\bar{s}s}_{i'_4i_40}|^2
	|F^{i_5s\bar{s}}_{i_5i'_50}|^2
	|F^{i'_6\bar{s}s}_{i'_6i_60}|^2} 
	\frac{Y^{s \bar{s}}_0}{(Y^{\bar{s}s}_0)^*}
	\frac{|Y^{i_6 i_1}_{e_1}Y^{i_3 e_3}_{i_2} Y^{e_5 i_4}_{i_5}|^2}
	{|Y^{i_6' i_1'}_{e_1}Y^{i_3' e_3}_{i_2'} Y^{e_5 i_4'}_{i_5'}|^2}
\end{split}
	\label{bratio}
\end{equation}
Thus, to show the first property is equivalent to show
\begin{equation}
	(\ref{bratio})=1.
	\label{bratio1}
\end{equation}
 This identity follows immediately by substituting (\ref{y0}-\ref{y1}) into the right hand side of (\ref{bratio}) and simplifying the resulting expression. 

In fact, we can also show that (\ref{y0}-\ref{y1}) are \emph{necessary} conditions for (\ref{bratio1}) to hold. To see this, we consider some simple cases. First, we consider the case when $e_1=e_2=\dots=0$ and $i_1=i_2=i_3=\bar{i}_4=\bar{i}_5=\bar{i}_6=i$. In this case, (\ref{bratio1}) reduces to
\begin{equation}
	\frac{|F^{\bar{i}s\bar{s}}_{\bar{i}\bar{i}'0}|^2 Y^{s\bar{s}}_0
	|Y^{\bar{i}i}_0|^2}{|F^{\bar{i}'\bar{s}{s}}_{\bar{i}'\bar{i}0}|^2 (Y^{\bar{s}{s}}_0)^*
	|Y^{\bar{i}'i'}_0|^2}=1.
	\label{case1}
\end{equation}
When $i=s$,  and $\bar{i}' = 0$, (\ref{case1}) becomes
\begin{equation}
	|F^{\bar{s}s\bar{s}}_{\bar{s}00}|^2=\frac{1}{Y^{s\bar{s}}_0 Y^{\bar{s}s}_0}.
	\label{case1a}
\end{equation}

Second, when $e_3=e_4=\dots=e_6=0$ and $i_1=i_3=\bar{i}_4=\bar{i_5}=\bar{i_6}=i,i_2=j$, (\ref{bratio1}) reduces to
\begin{equation}
	\frac{|F^{\bar{i}s\bar{s}}_{\bar{i}\bar{i}'0}|^2}{|F^{\bar{j}'\bar{s}s}_{\bar{j}'\bar{j}0}|^2} = \frac{(Y^{\bar{s}s}_0)^* |Y^{\bar{j}'i'}_{e_1}|^2}{Y^{s\bar{s}}_0 |Y^{\bar{j}i}_{e_1}|^2}.
	\label{case2}
\end{equation}
By comparing (\ref{case2}) with $i=\bar{j}'=s,i'=j=0$ and (\ref{case1a}), we find
\begin{equation}
	Y^{a\bar{a}}_0 = (Y^{\bar{a}a}_0)^*.
	\label{case2b}
\end{equation}
By using (\ref{case1a}) and (\ref{case2b}), we find that (\ref{case2}) with $j=0,\bar{j}'=s$ reduces to
\begin{equation}
	|F^{ab\bar{b}}_{ac0}|=\frac{|Y^{b\bar{c}}_{\bar{a}}|}{|Y^{b\bar{b}}_0|}.
	\label{case2a}
\end{equation}
Plugging (\ref{case2a}) and (\ref{case2b}) to (\ref{case2}), we have
\begin{equation}
	|Y^{ab}_c||Y^{cd}_f|=|Y^{ae}_f||Y^{bd}_e|.
	\label{case2c}
\end{equation}

Similarly, by considering (\ref{bratio1}) when $e_1=e_2=e_5=e_6=0$ and $e_1=e_2=e_3=e_6=0$, we obtain
\begin{equation}
	\begin{split}
	|F^{\bar{i}_3 s \bar{s}}_{\bar{i}_3 \bar{i}_3' 0}|=
	\frac{|Y^{i'_3 \bar{i}_3}_{\bar{s}}|} {|Y^{i_3\bar{i}_3}_0|},
	\quad 
	|F^{{i}_4' \bar{s} {s}}_{{i}_4' {i}_4 0}|=
	\frac{|Y^{\bar{i}'_4 {i}_4}_{\bar{s}}|} {|Y^{\bar{i}_4' {i}_4'}_0|}.
\end{split}
	\label{case2d1}
\end{equation}
From (\ref{case2a},\ref{case2d1}), we find that
\begin{equation}
	|Y^{ab}_c|= 
	|Y^{\bar{b}\bar{a}}_{\bar{c}}| =
	|Y^{b\bar{c}}_{\bar{a}}| \frac{|Y^{a\bar{a}}_0|}{|Y^{c\bar{c}}_0|}.
	\label{case2d}
\end{equation}
Then, from (\ref{case2c}) with $e=0,d=\bar{b},f=a$ and using (\ref{case2d}), we find that
\begin{equation}
	|Y^{ab}_c| = \sqrt{\frac{d_a d_b}{d_c}} \delta^{ab}_c.
	\label{yabcapp}
\end{equation}
Combining (\ref{yabcapp}), (\ref{case2b}) and (\ref{case2a}), we derive conditions (\ref{y0}--\ref{y1}). This completes our discussion of the first property of the Hamiltonian, i.e. $(B_p^s)^\dagger = B_p^{\bar{s}}$.

 We now move on to the second property, i.e. $B_p^2 = B_p$. To prove this result, we use the identity 
\begin{align}
	{B}_p^{t_1} {B}_p^{t_2}&=\sum_{u}  M^{t_1 t_2}_u B_p^u 
\end{align}
with
\begin{align}
M^{t_1 t_2}_u &= 
	F^{t_2 \bar{t}_2 \bar{t}_1}_{\bar{t}_1 0 \bar{u}} (F^{t_2 \bar{t}_2 \bar{t}_1}_{\bar{t}_1})^{-1}_{\bar{u}0}	\frac{ Y^{t_1 \bar{t}_1}_0 Y^{t_2\bar{t}_2}_0}{ Y^{u\bar{u}}_{0}},
	\label{bp1bp2}
\end{align}
which we will derive below. From (\ref{bp1bp2}), we can derive two other useful identities:
\begin{align}
\sum_u M^{t_1 t_2}_u a_{\bar{u}}  &= a_{\bar{t}_1} a_{\bar{t}_2} \sum_s d_s^2 \label{MYid} \\
M^{t_1 t_2}_u &= M^{\bar{u} t_1}_{\bar{t}_2} \label{Mcyc}
\end{align}
Here (\ref{MYid}) follows immediately from the expression for $a_s$ (\ref{as}). As for (\ref{Mcyc}), this follows from
two other identities:
\begin{align}
M^{\bar{u} u}_0 M^{t_1 t_2}_u &= M^{\bar{u} t_1}_{\bar{t}_2} M^{\bar{t_2} t_2}_0 \label{assoc} \\
M^{\bar{s} s}_0  &= 1 \label{Maid} 
\end{align}
Here  (\ref{assoc}) follows from comparing the coefficient of $B_p^0$ that appears in the two (identical) products 
$B_p^{\bar{u}} (B_p^{t_1} B_p^{t_2})$ and $ (B_p^{\bar{u}} B_p^{t_1}) B_p^{t_2}$. Eq.~(\ref{Maid}) follows from (\ref{bp1bp2}) combined with (\ref{unitary}-\ref{y1}).

We are now ready to derive $B_p^2 = B_p$. Proving this relation is equivalent to showing
\begin{align}
\sum_{t_1 t_2} M^{t_1 t_2}_u a_{t_1} a_{t_2} = a_u
\label{Bp2}
\end{align}
We will prove this in three steps. First we use (\ref{Mcyc}) and (\ref{MYid}) in succession to derive
\begin{align}
\sum_{t_1 t_2} M^{t_1 t_2}_u a_{t_1} a_{t_2} 
&= \sum_{t_1 t_2} M^{\bar{u} t_1}_{\bar{t}_2} a_{t_1} a_{t_2} \nonumber \\
&= \sum_{t_1} a_{t_1} a_{\bar{t}_1} a_{u}  \cdot \sum_s d_s^2 
\label{step1}
\end{align}
Next, we note that
\begin{align}
\sum_{t_1} a_{t_1} a_{\bar{t}_1} = \left(\sum_s d_s^2 \right)^{-1}
\label{step2}
\end{align}
Combining (\ref{step1}) and (\ref{step2}), we derive (\ref{Bp2}).

All that remains is to show (\ref{bp1bp2}).
To this end, we consider
\begin{equation}
	\begin{split}
		&\left\< \raisebox{-0.22in}{\includegraphics[height=0.5in]{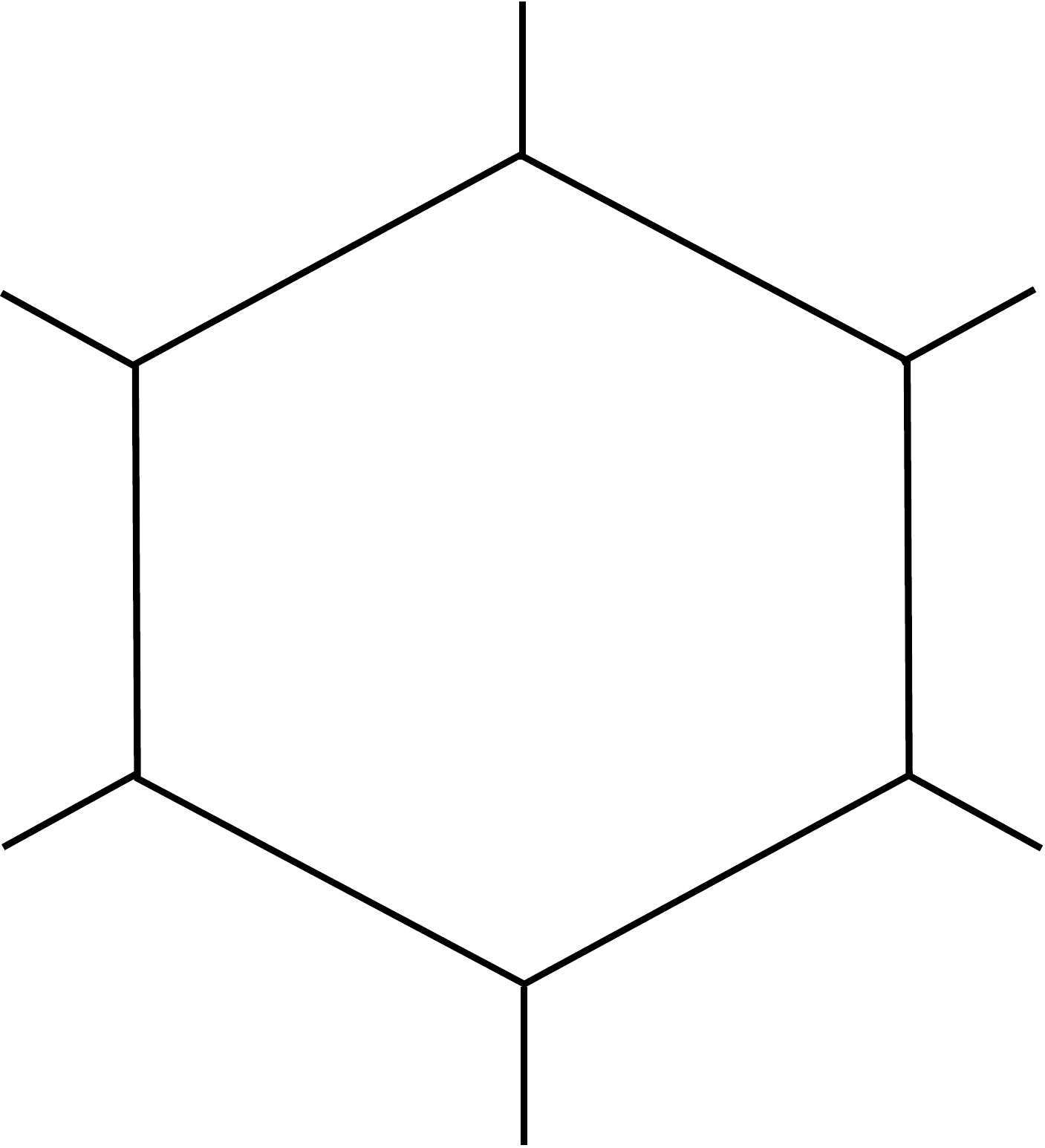}}\right|  {B}_p^{t_1} {B}_p^{t_2}
		=\left\<\raisebox{-0.22in}{\includegraphics[height=0.5in]{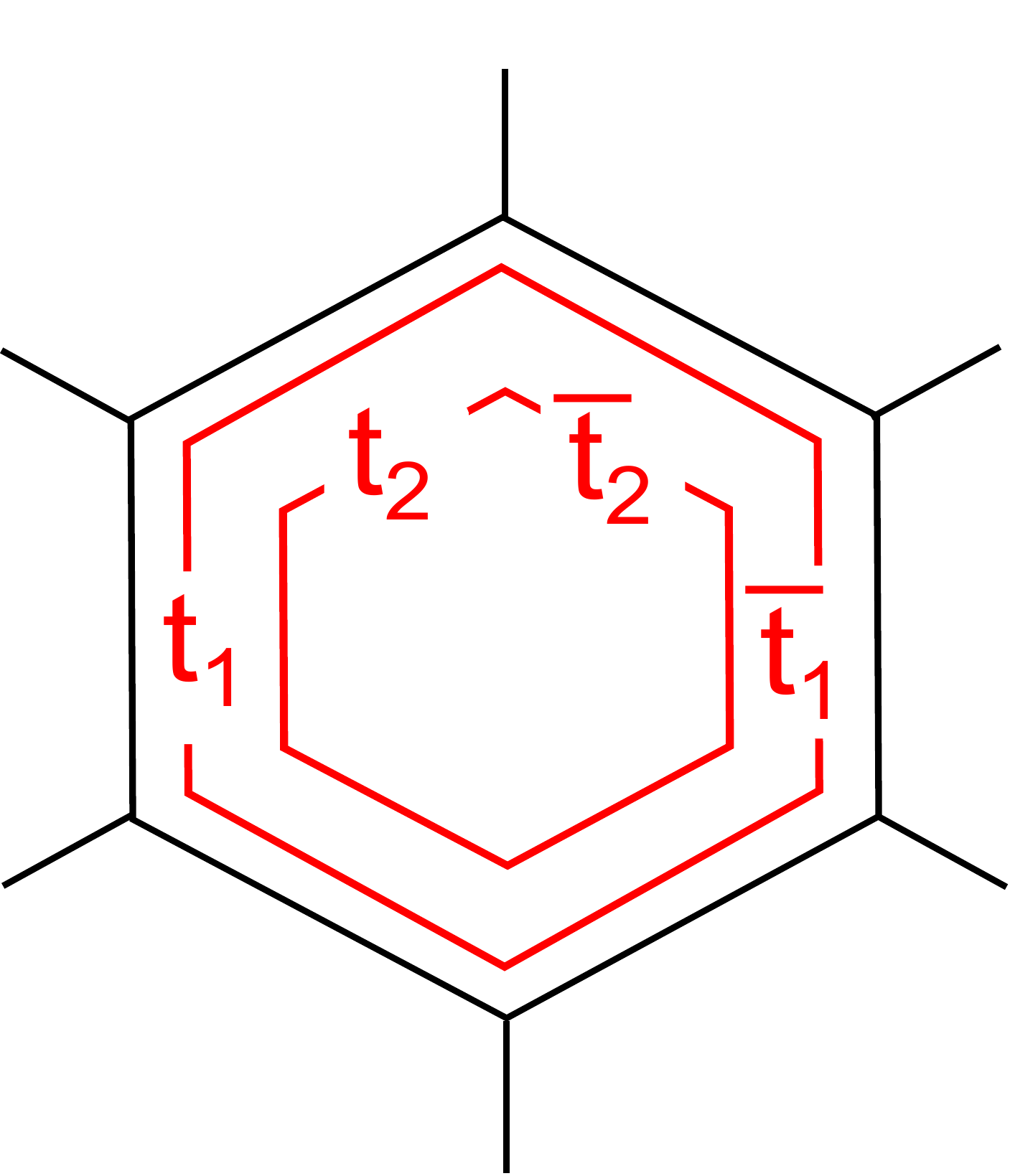}}\right| \\
		&=\sum_{u}\left\<\raisebox{-0.22in}{\includegraphics[height=0.5in]{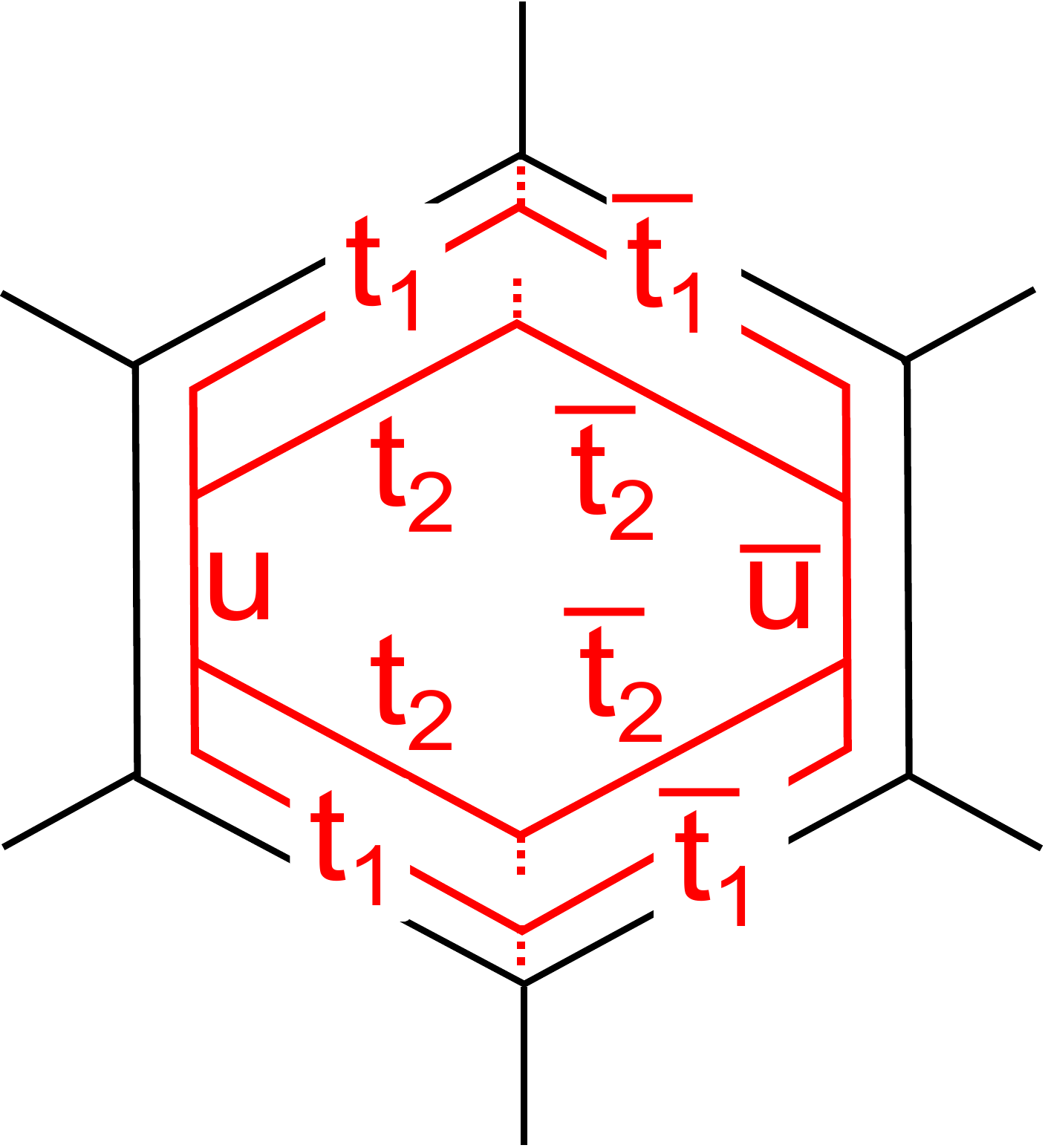}}\right| \frac{1}{Y^{t_1t_2}_u Y^{\bar{t}_2\bar{t}_1}_{\bar{u}}}
		\\
		&=\sum_{u} \left\<\raisebox{-0.22in}{\includegraphics[height=0.5in]{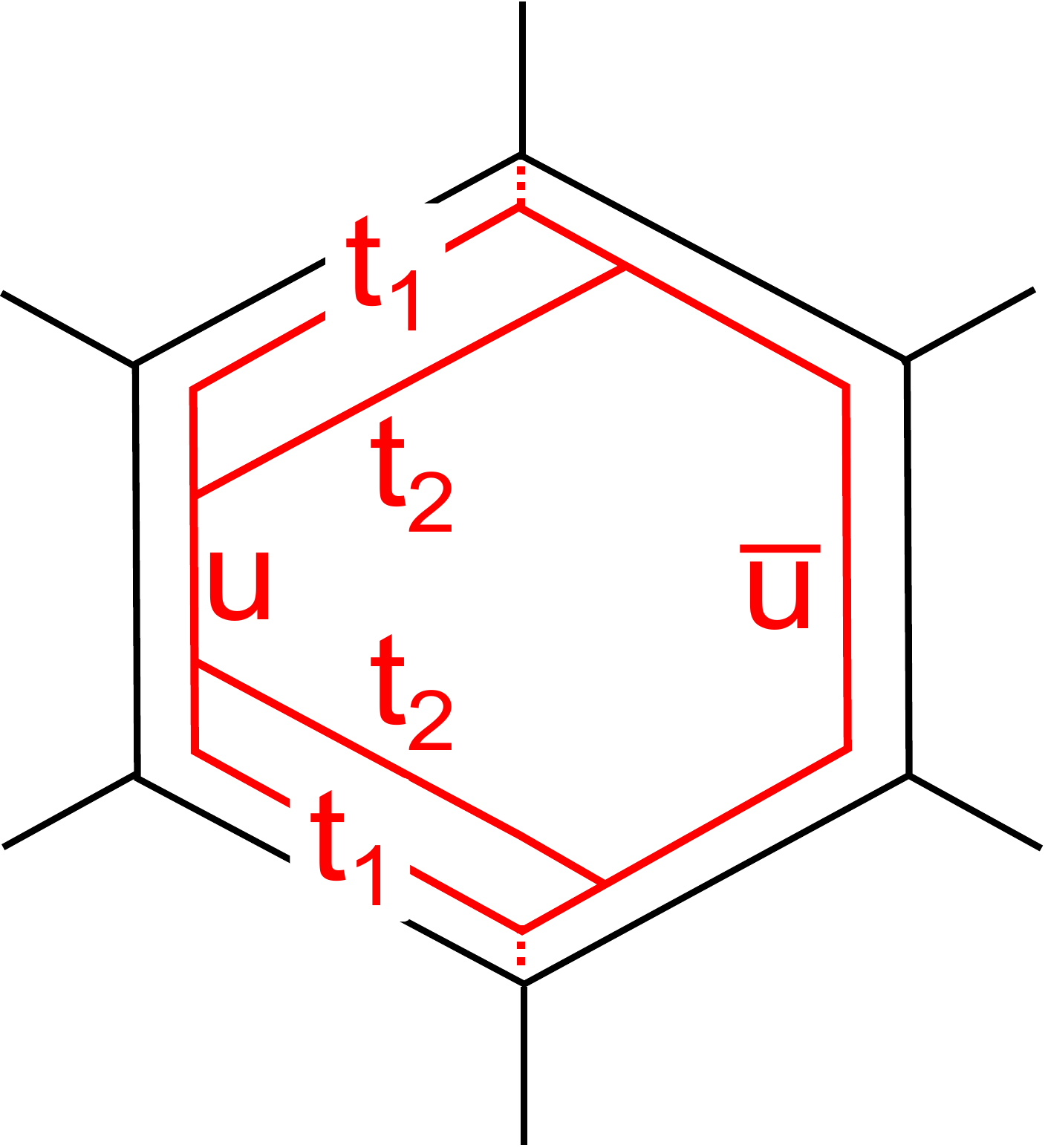}}\right|
		\frac{[F^{0\bar{t}_1}_{t_2\bar{u}}]_{\bar{t}_2\bar{t}_1}
		[\tilde{F}^{0\bar{t}_1}_{t_2\bar{u}}]_{\bar{t}_2\bar{t}_1} }{Y^{t_1t_2}_u Y^{\bar{t}_2\bar{t}_1}_{\bar{u}}} \\
		&=\sum_{u} \left\<\raisebox{-0.22in}{\includegraphics[height=0.5in]{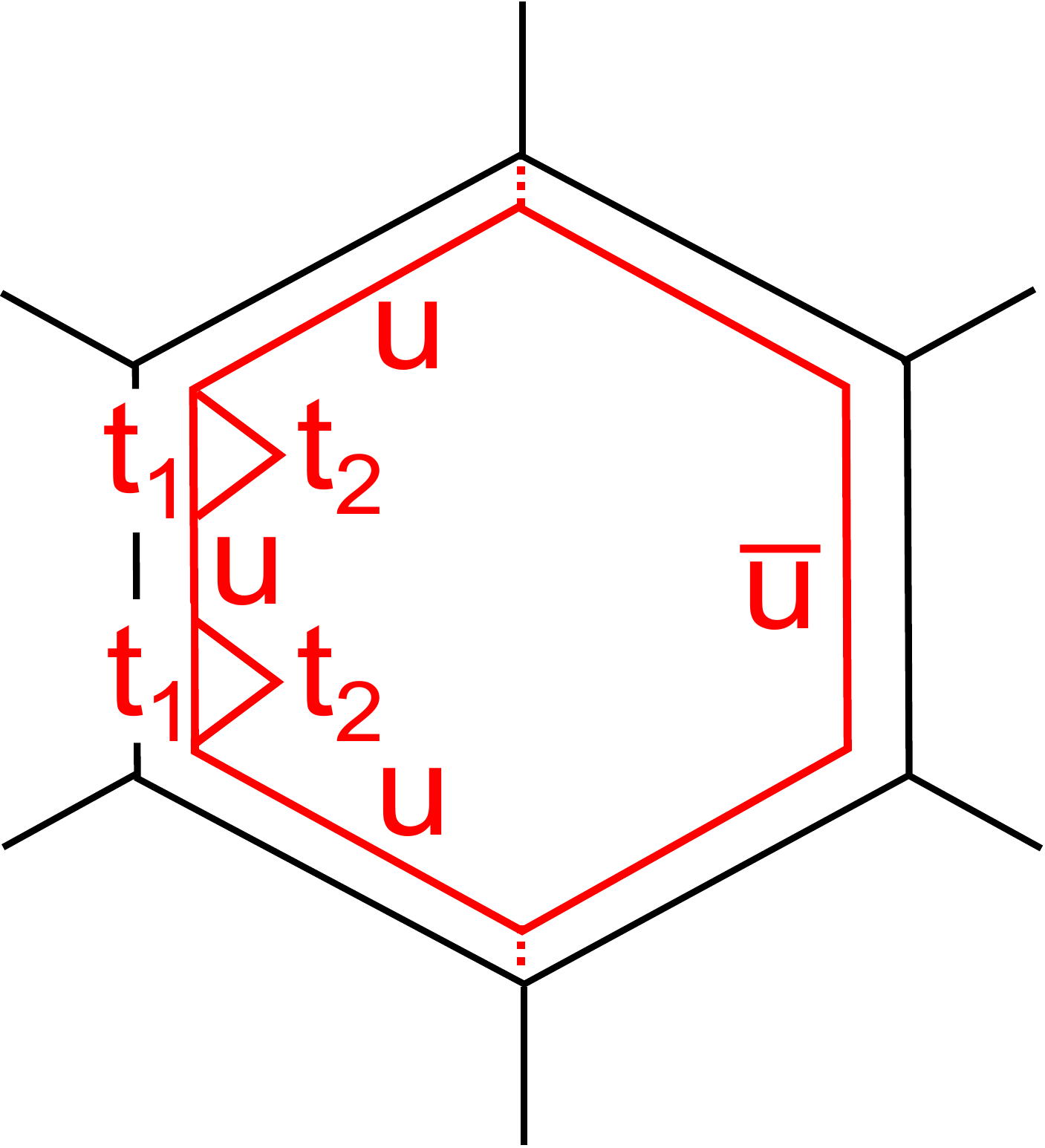}}\right|
		\frac{(F^{t_1t_2\bar{u}}_{0})^{-1}_{\bar{t}_1u}
		(\tilde{F}^{t_1t_2\bar{u}}_{0})^{-1}_{\bar{t}_1u}
		[F^{0\bar{t}_1}_{t_2\bar{u}}]_{\bar{t}_2\bar{t}_1}
		[\tilde{F}^{0\bar{t}_1}_{t_2\bar{u}}]_{\bar{t}_2\bar{t}_1} }{Y^{t_1t_2}_u Y^{\bar{t}_2\bar{t}_1}_{\bar{u}}}\\
		&=\sum_{u} \left\<\raisebox{-0.22in}{\includegraphics[height=0.5in]{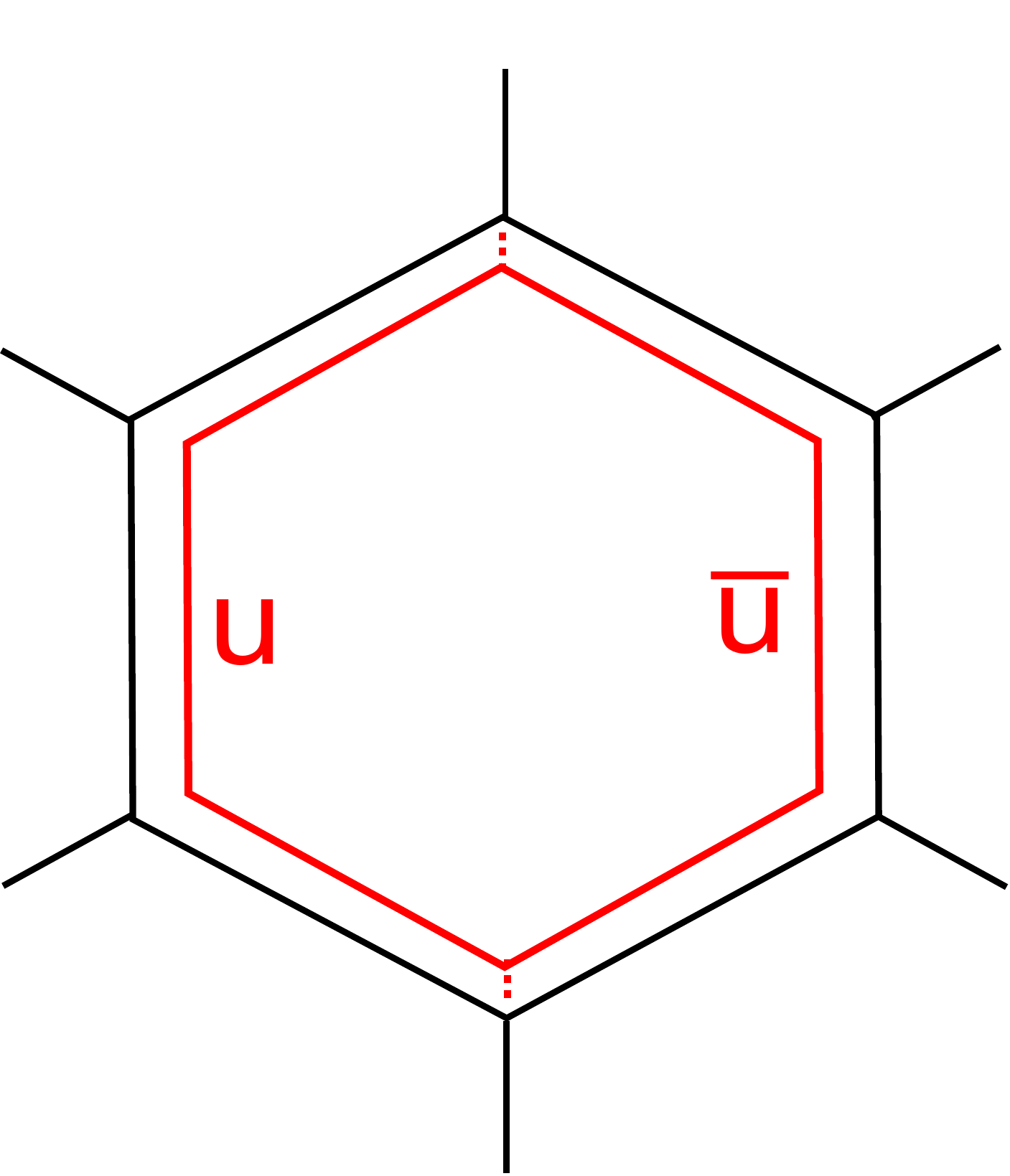}}\right|
		F^{t_2 \bar{t}_2 \bar{t}_1}_{\bar{t}_1 0 \bar{u}} (F^{t_2 \bar{t}_2 \bar{t}_1}_{\bar{t}_1})^{-1}_{\bar{u}0}
		\frac{ Y^{t_1 \bar{t}_1}_0 Y^{t_2\bar{t}_2}_0}{ Y^{u\bar{u}}_{0}}\\
		&=\sum_u \left\< \raisebox{-0.22in}{\includegraphics[height=0.5in]{bp3a0.pdf}}\right|  	F^{t_2 \bar{t}_2 \bar{t}_1}_{\bar{t}_1 0 \bar{u}} (F^{t_2 \bar{t}_2 \bar{t}_1}_{\bar{t}_1})^{-1}_{\bar{u}0}
		\frac{ Y^{t_1 \bar{t}_1}_0 Y^{t_2\bar{t}_2}_0}{ Y^{u\bar{u}}_{0}} {B}_p^{u} 
	\end{split}
	\label{bp1bp20}
\end{equation}
 Here, we have used (\ref{consistency1},\ref{consistency}) to simplify (\ref{bp1bp20}). 
Thus, we obtain (\ref{bp1bp2}).

\section{ Showing the ground state obeys the local rules} \label{app:localrules}
 In this appendix, we show that any state $|\Phi\>$ such that $Q_I |\Phi\> = B_p |\Phi\> = |\Phi\>$ obeys a lattice version of the local rules (\ref{localrules}). We also discuss some implications of this result.

The lattice local rules are as follows:
\begin{subequations}
	\begin{align}
		\Phi\left(\raisebox{-0.22in}{\includegraphics[height=0.5in]{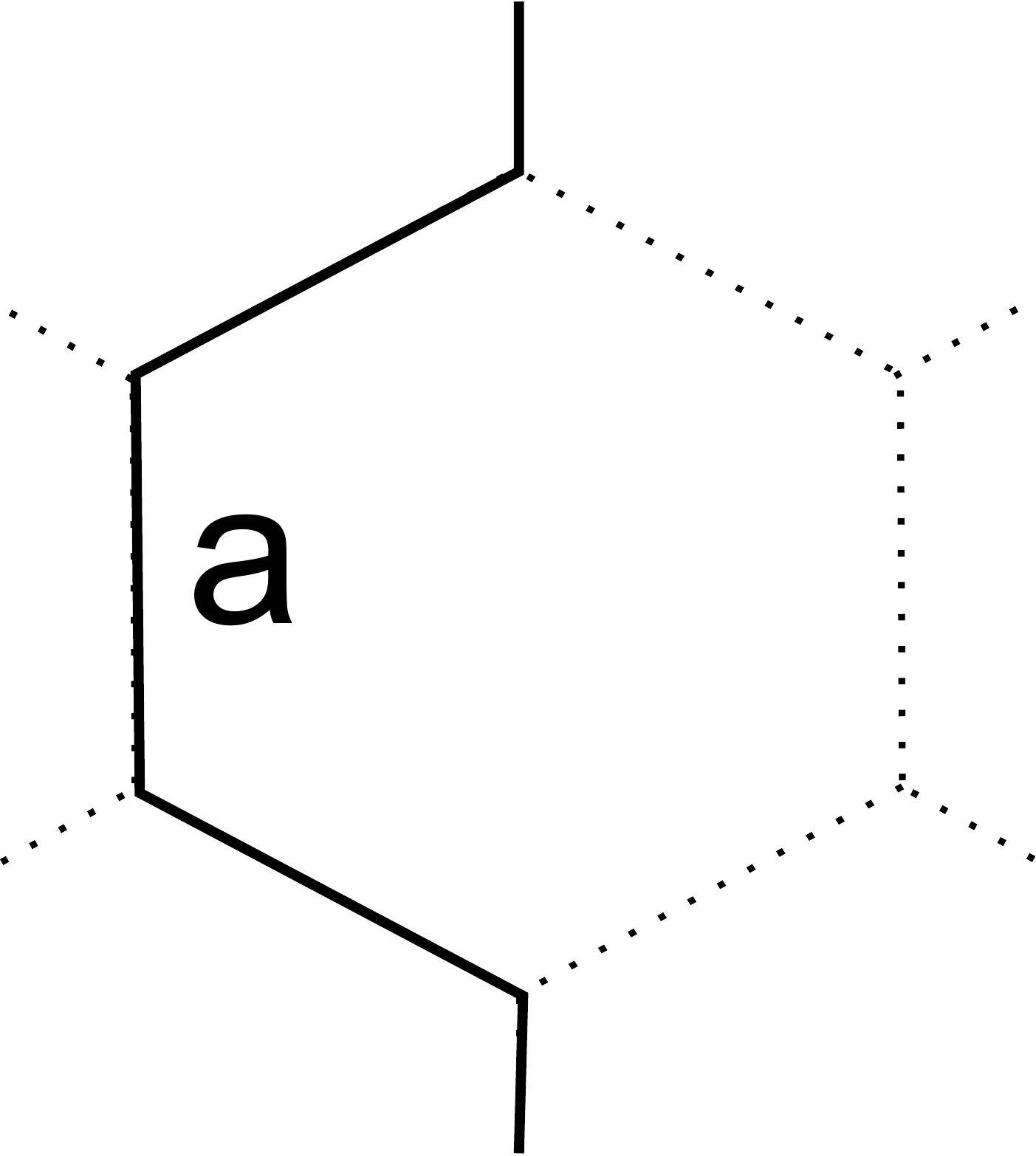}} \right)&=\Phi \left(\raisebox{-0.22in}{\includegraphics[height=0.5in]{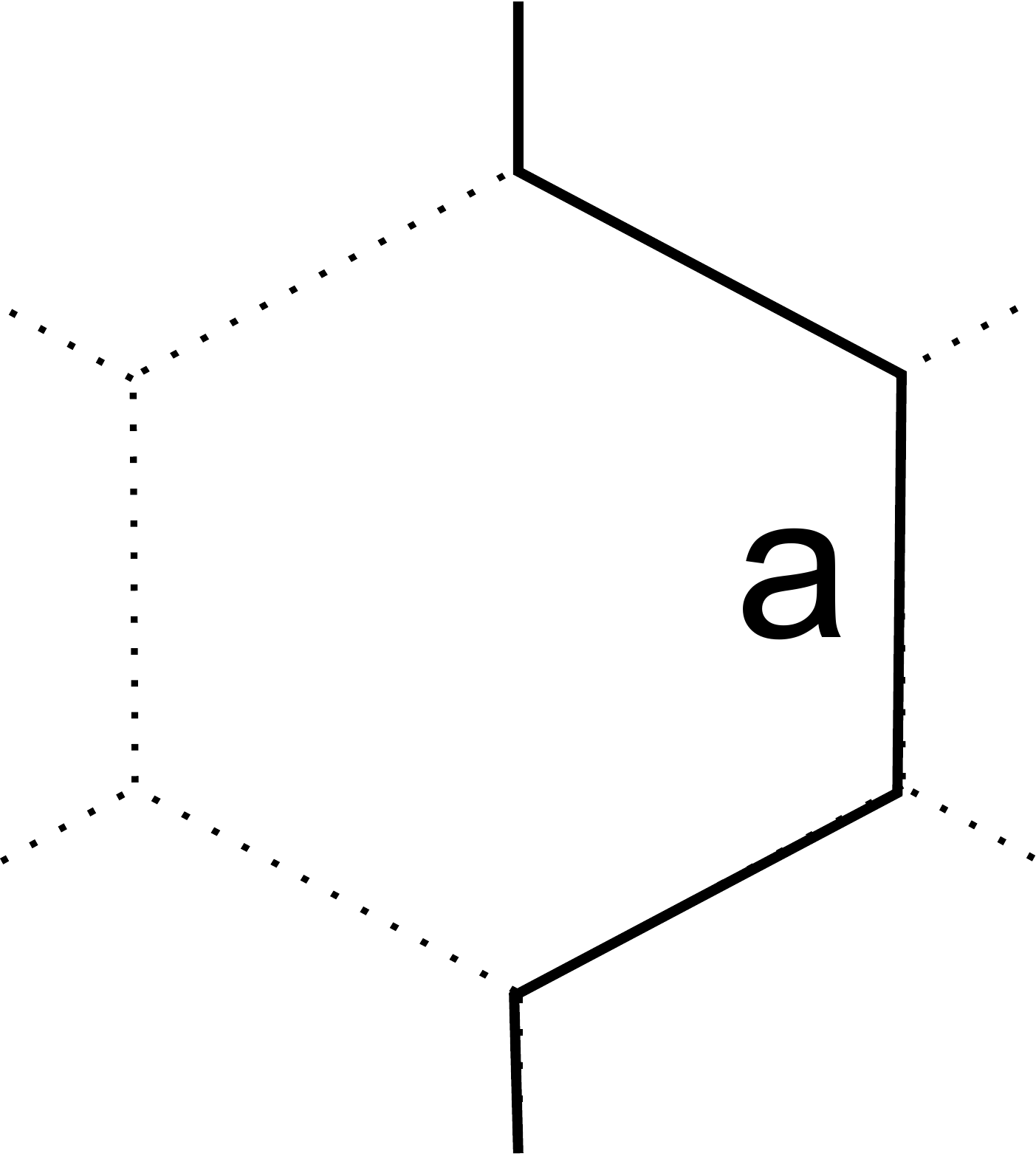}} \right) \label{lattrules10}\\
		\Phi \left(\raisebox{-0.22in}{\includegraphics[height=0.5in]{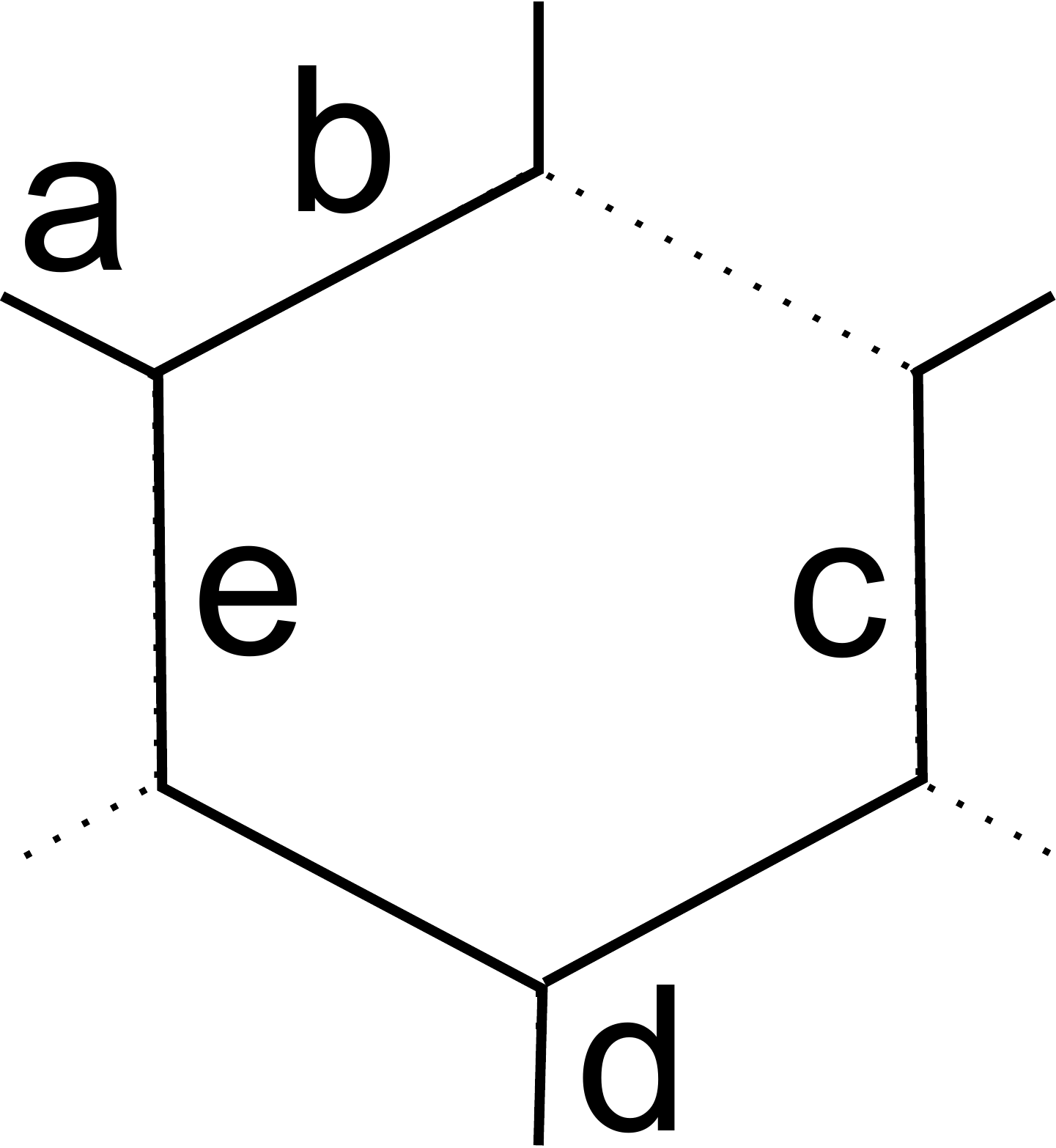}} \right)&=\sum_f F^{abc}_{def} \Phi \left(\raisebox{-0.22in}{\includegraphics[height=0.5in]{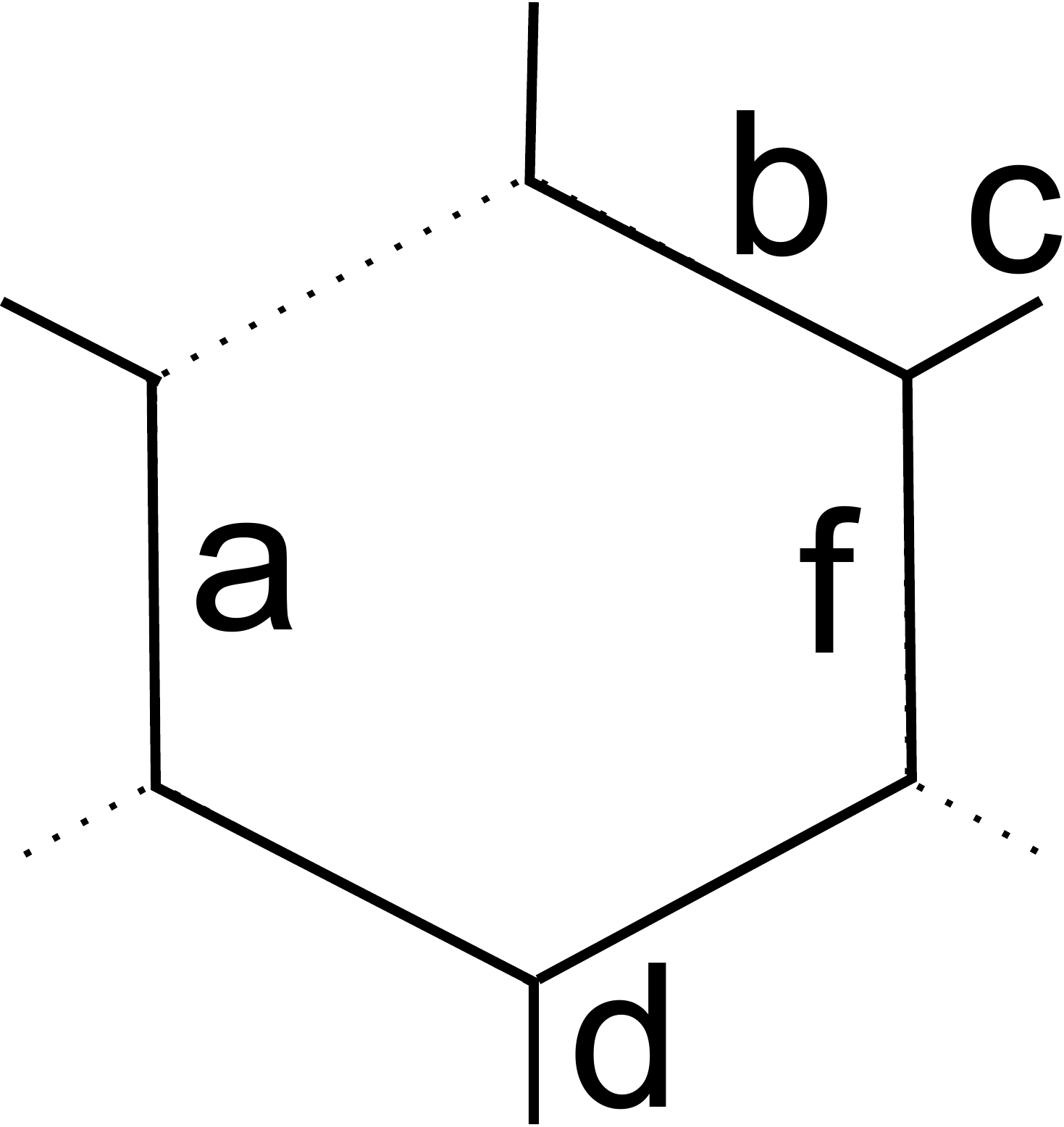}} \right) \\
		\Phi \left(\raisebox{-0.22in}{\includegraphics[height=0.5in]{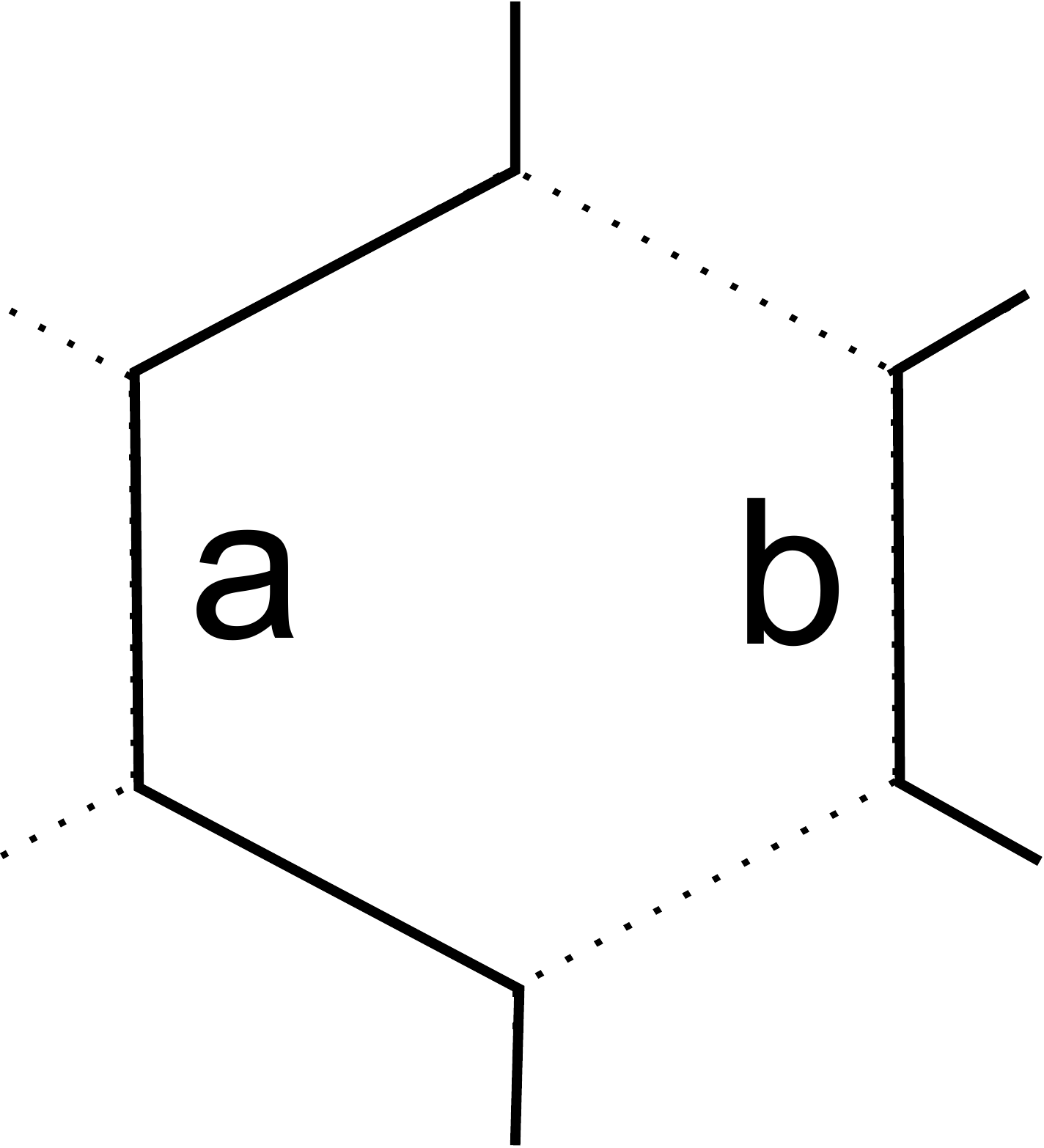}} \right)&=\sum_c \frac{1}{Y^{ab}_c} \Phi \left(\raisebox{-0.22in}{\includegraphics[height=0.5in]{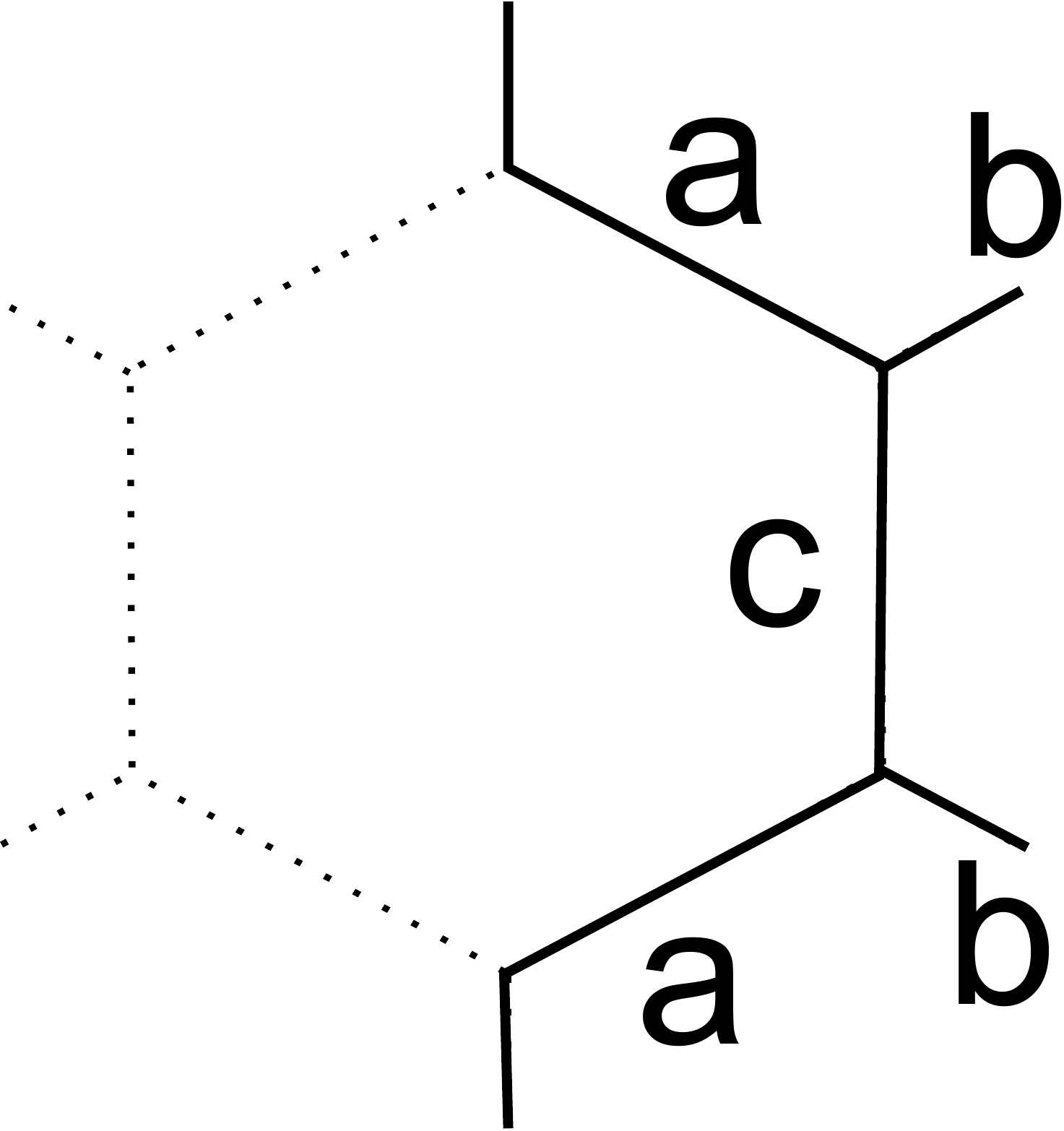}} \right) \\
		\Phi \left(\raisebox{-0.22in}{\includegraphics[height=0.5in]{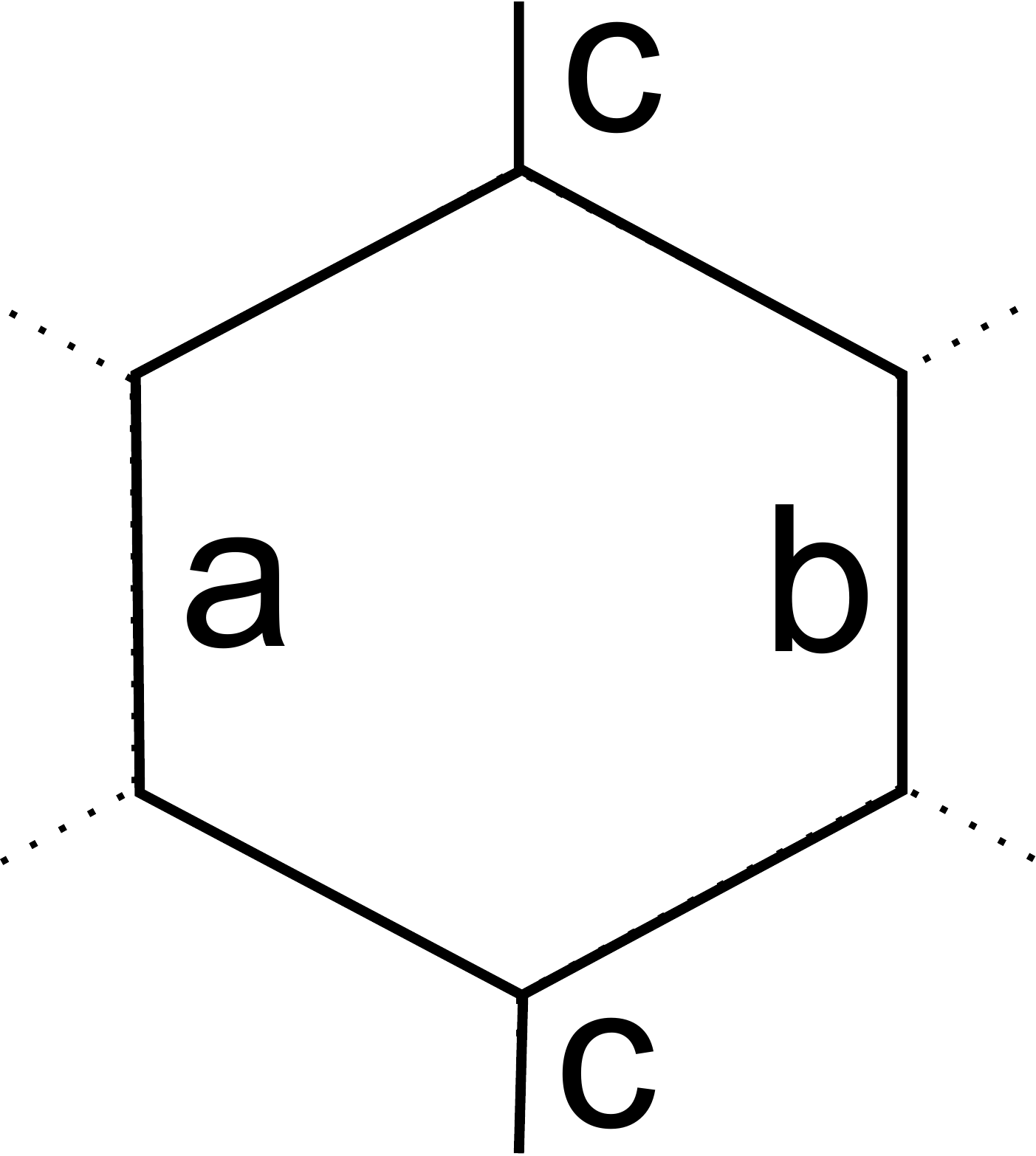}} \right)&= Y^{ab}_c \Phi \left(\raisebox{-0.22in}{\includegraphics[height=0.5in]{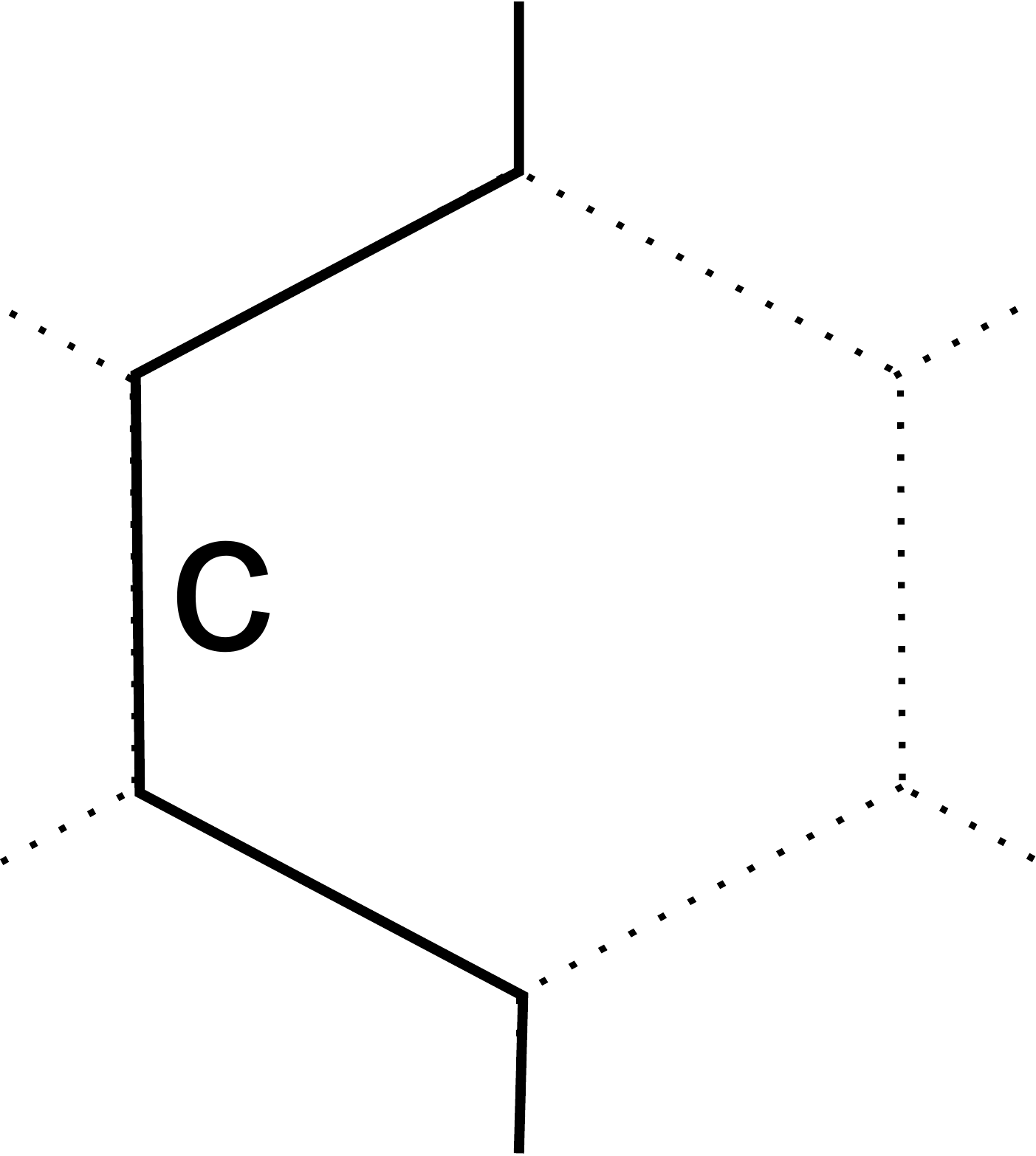}} \right) 
	\end{align}
	\label{lattrules1}
\end{subequations}
 Our strategy for deriving these rules is to use the fact that $B_p|\Phi\> = |\Phi\>$ together with the following relations:
\begin{subequations}
	\begin{align}
		\left\< \raisebox{-0.22in}{\includegraphics[height=0.5in]{lrule0a.pdf}} \right|B_p &=\left\< \raisebox{-0.22in}{\includegraphics[height=0.5in]{lrule0b.pdf}} \right|B_p \label{lrule0}\\
	\left\< \raisebox{-0.22in}{\includegraphics[height=0.5in]{lrule1a.pdf}} \right|B_p&=\sum_f F^{abc}_{def} \left\<\raisebox{-0.22in}{\includegraphics[height=0.5in]{lrule1b.pdf}} \right|B_p \label{lrule1}\\
	\left\<\raisebox{-0.22in}{\includegraphics[height=0.5in]{lrule2a.pdf}} \right|B_p&=\sum_c \frac{1}{Y^{ab}_c} \left\<\raisebox{-0.22in}{\includegraphics[height=0.5in]{lrule2b.pdf}} \right|B_p \label{lrule2}\\
	\left\<\raisebox{-0.22in}{\includegraphics[height=0.5in]{lrule3a.pdf}} \right|B_p&=Y^{ab}_c \left\<\raisebox{-0.22in}{\includegraphics[height=0.5in]{lrule3b.pdf}} \right|B_p \label{lrule3}
	\end{align}
	\label{lattrules2}
\end{subequations}
Multiplying these equations  by $|\Phi\>$, we can see that the wave function defined by $\Phi(X)=\<X|\Phi\>$ satisfies the local rules (\ref{lattrules1}).  

The relations (\ref{lattrules2}) can be shown using the expression for the matrix elements of $B_p^s$ in (\ref{bps0}) together with the pentagon identity and (\ref{hermicity}).
For example, to show (\ref{lrule2}), we expand out the left hand side as 
\begin{equation}
	\begin{split}
	&\left\< \raisebox{-0.22in}{\includegraphics[height=0.5in]{lrule1a.pdf}} \right|B_p \\
	&=\sum_{sb'c'e'} F^{abs}_{e'eb'}F^{e'\bar{s}c}_{dec'}F^{bs\bar{s}}_{bb'0} (F^{es\bar{s}}_{e})^{-1}_{0e'} \frac{d_s^2}{D Y^{b'\bar{s}}_b} \left\< \raisebox{-0.22in}{\includegraphics[height=0.5in]{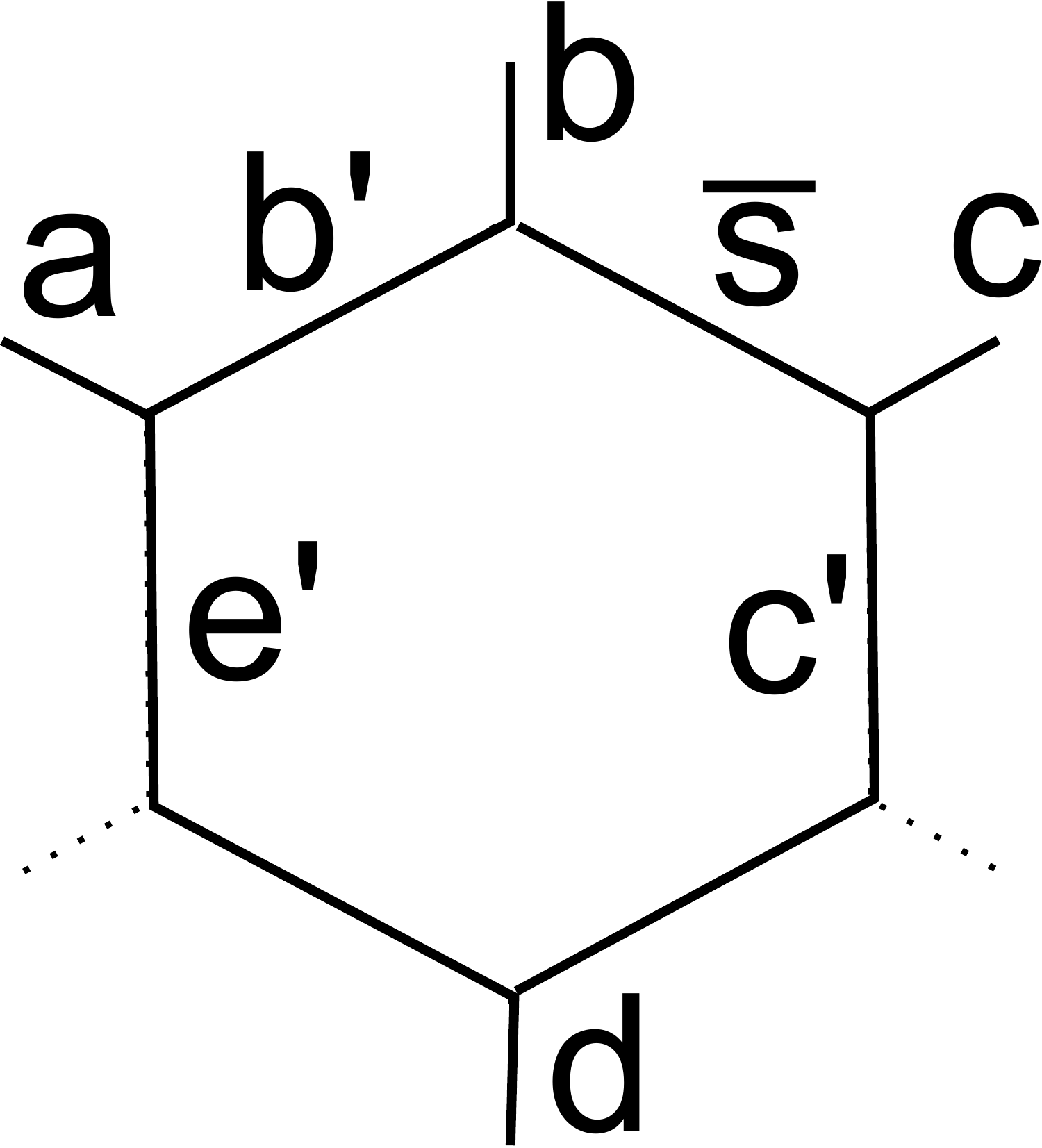}} \right|
\end{split}
	\label{}
\end{equation}
and the right hand side as
\begin{equation}
	\begin{split}
	&\sum_f F^{abc}_{def} \left\<\raisebox{-0.22in}{\includegraphics[height=0.5in]{lrule1b.pdf}} \right|B_p  = \sum_{fsb'f'a'}F^{abc}_{def} \\
	&\cdot F^{a'\bar{s}f}_{daf'}(F^{\bar{s}bc}_{f'})^{-1}_{fb'}(F^{as\bar{s}}_{a})^{-1}_{0a'} (F^{s\bar{s}b}_{b})^{-1}_{b'0} \frac{d_s^2}{DY^{sb'}_b}	
	\left\< \raisebox{-0.22in}{\includegraphics[height=0.5in]{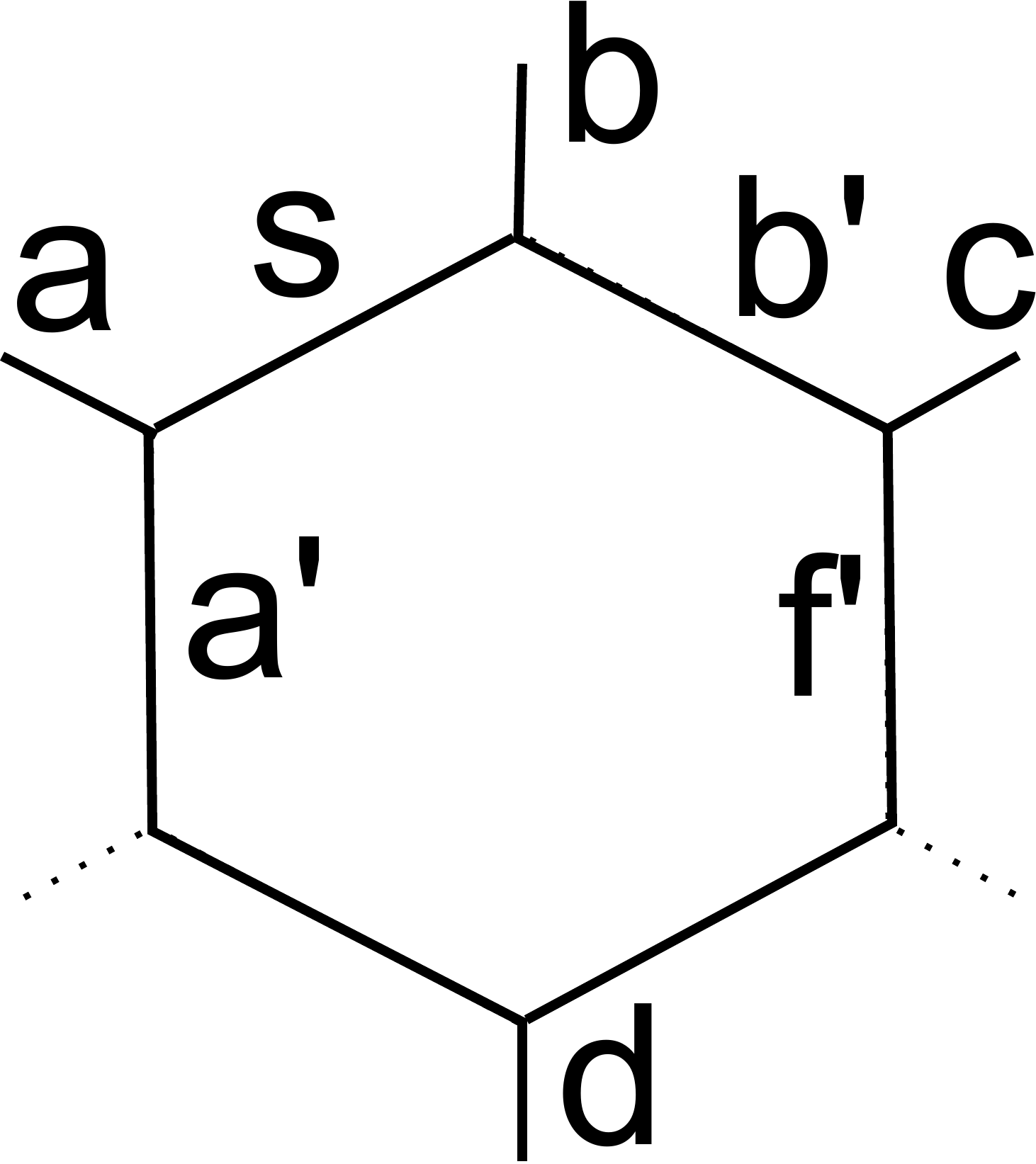}} \right|
	\end{split}
	\label{}
\end{equation}
where $D = \sum_s d_s^2$. Changing the dummy variables $b'\rightarrow \bar{s}, f'\rightarrow c',a'\rightarrow e',s\rightarrow b'$ in the second expression, and  matching coefficients, we see that showing (\ref{lrule1}) is equivalent to showing
\begin{equation}
	\begin{split}
		&d_s^2 F^{abs}_{e'eb'} (F^{es\bar{s}}_{e})^{-1}_{0e'}
		F^{e'\bar{s}c}_{dec'} F^{bs\bar{s}}_{bb'0} \\
		&=d_{b'}^2 (F^{ab'\bar{b}'}_{a})^{-1}_{0e'}
		(F^{b'\bar{b}'b}_{b})^{-1}_{\bar{s}0}
		\sum_f F^{abc}_{def}F^{e'\bar{b}'f}_{dac'} (F^{\bar{b}'bc}_{c'})^{-1}_{f\bar{s}}
\end{split}
	\label{e5}
\end{equation}
Next we use the following three variants of the pentagon identity:
\begin{equation}
	\begin{split}
		F^{abs}_{e'eb'} (F^{es\bar{s}}_{e})^{-1}_{0e'}&= (F^{bs\bar{s}}_{b})^{-1}_{0b'} (F^{ab'\bar{s}}_{e})^{-1}_{be'} \\
		\sum_f F^{abc}_{def}F^{e'\bar{b}'f}_{dac'} (F^{\bar{b}'bc}_{c'})^{-1}_{f\bar{s}} &= F^{e'\bar{b}'b}_{ea\bar{s}} F^{e'\bar{s}c}_{dec'}. \\
F^{b'\bar{b}'b}_{b0\bar{s}} (F^{ab'\bar{s}}_e)^{-1}_{be'} &=
	F^{e'\bar{b}'b}_{ea\bar{s}} (F^{ab'\bar{b}'}_a)^{-1}_{0e'}
	\end{split}
	\label{}
\end{equation}
With these identities, (\ref{e5}) reduces to proving
\begin{align}
d_s^2  (F^{bs\bar{s}}_{b})^{-1}_{0b'} F^{bs\bar{s}}_{bb'0} =d_{b'}^2 F^{b'\bar{b}'b}_{b0\bar{s}} (F^{b'\bar{b}'b}_{b})^{-1}_{\bar{s}0}
\label{g2cred}
\end{align}

To prove the above identity, we first prove the following auxiliary identities:
\begin{align}
(F^{b' \bar{b}' b'}_{b'})^{-1}_{00} F^{b s \bar{s}}_{bb' 0} F^{\bar{b}' b s}_{0 \bar{s} b'} &=(F^{s \bar{s} s}_{s} )^{-1}_{00} (F^{b' \bar{b}' b}_b)^{-1}_{\bar{s} 0} 
\label{id1} \\
 F^{b' \bar{b}' b'}_{b' 0 0} (F^{b s \bar{s}}_b)^{-1}_{0b'} &=F^{s \bar{s} s}_{s 00} F^{b' \bar{b}' b}_{b 0 \bar{s}} F^{\bar{b}' b s}_{0 \bar{s} b'}
\label{id2}
\end{align}
Once we prove these two auxiliary identities, we will be done since multiplying them together gives the desired identity, (\ref{g2cred}).

To prove (\ref{id1}), we substitute $l = 0$, $d = b$, $f=e$ into Eq.~(\ref{piv1}). This gives
\begin{align}
F^{b \bar{b} b}_{b00} = F^{e \bar{b} b}_{ea0} (F^{a b \bar{b}}_a)^{-1}_{0e}
\label{id10}
\end{align}
We then make the following change of variables: $b \rightarrow \bar{s}$, $e \rightarrow b$, $a \rightarrow b'$. The result is
\begin{align}
F^{\bar{s} s \bar{s}}_{\bar{s} 0 0} = F^{b s \bar{s}}_{b b' 0} (F^{b' \bar{s} s}_{b'})^{-1}_{0b}
\label{id11}
\end{align}
Similarly, we substitute $f=0$, $b = \bar{a}$, $l = e$ into Eq.~(\ref{piv1}). This gives
\begin{align}
F^{\bar{a} a d}_{d 0 e} (F^{a \bar{a} e}_e)^{-1}_{d0} = (F^{a \bar{a} a}_a)^{-1}_{00}
\label{id120}
\end{align}
Making the change the variables $a \rightarrow b'$, $e \rightarrow b$, $d \rightarrow \bar{s}$, we obtain
\begin{align}
F^{\bar{b}' b' \bar{s}}_{\bar{s} 0 b} (F^{b' \bar{b}' b}_b)^{-1}_{\bar{s} 0} = (F^{b' \bar{b}' b'}_{b'})^{-1}_{00}
\label{id12}
\end{align}
Multiplying (\ref{id11}) and (\ref{id12}) gives
\begin{align}
F^{\bar{s} s \bar{s}}_{\bar{s} 0 0} F^{\bar{b}' b' \bar{s}}_{\bar{s} 0 b} (F^{b' \bar{b}' b}_b)^{-1}_{\bar{s} 0} =
F^{b s \bar{s}}_{b b' 0} (F^{b' \bar{s} s}_{b'})^{-1}_{0b} (F^{b' \bar{b}' b'}_{b'})^{-1}_{00}
\label{id13}
\end{align}
To proceed further, we consider the version of the pentagon identity in Fig.~\ref{fig:pi2}, which relates the diagram at the top to the diagram in the bottom right:
\begin{align}
(F^{bcd}_k)^{-1}_{lh} F^{abl}_{efk} = \sum_g (F^{fcd}_e)^{-1}_{lg} F^{abc}_{gfh} F^{ahd}_{egk}
\end{align}
We set $f = l = e = 0$, $b = \bar{a}$, $d = \bar{c}$, $g=c$, $k = \bar{a}$. With these substitutions the pentagon identity reduces to
\begin{align}
(F^{\bar{a} c \bar{c}}_{\bar{a}})^{-1}_{0h} = F^{a \bar{a} c}_{c 0 h} F^{a h \bar{c}}_{0 c \bar{a}}
\end{align}
Next we make the following change of variables: $a \rightarrow \bar{b}'$, $c \rightarrow \bar{s}$, $h \rightarrow b$. This gives the identity
\begin{align}
(F^{b' \bar{s} s}_{b'})^{-1}_{0b} = F^{\bar{b}' b' \bar{s}}_{\bar{s} 0 b} F^{\bar{b}' b s}_{0 \bar{s} b'}
\label{id14}
\end{align}
Substituting (\ref{id14}) into (\ref{id13}), we obtain
\begin{align}
F^{\bar{s} s \bar{s}}_{\bar{s} 0 0} (F^{b' \bar{b}' b}_b)^{-1}_{\bar{s} 0} =
F^{b s \bar{s}}_{b b' 0}  (F^{b' \bar{b}' b'}_{b'})^{-1}_{00} F^{\bar{b}' b s}_{0 \bar{s} b'}
\end{align}
This is \emph{almost} the desired identity (\ref{id1}): all that is left is to show that 
\begin{align}
F^{\bar{s} s \bar{s}}_{\bar{s} 0 0} = (F^{s \bar{s} s}_{s} )^{-1}_{00}
\label{sssid}
\end{align}
Conveniently, this follows immediately from (\ref{id11}), by setting $a = 0$, $a' = s$.

We now move on to prove the second identity (\ref{id2}). The proof is very similar to that of (\ref{id1}). The first step is to take (\ref{id10}), and make the change of variables $a \rightarrow s$, $e \rightarrow b'$. This gives
\begin{align}
F^{s \bar{s} s}_{s 00} = F^{b' \bar{s} s}_{b' b 0} (F^{b s \bar{s}}_b)^{-1}_{0b'}
\label{id21}
\end{align}
Next we take (\ref{id120}) and make the change of variables $a \rightarrow \bar{b}'$, $d \rightarrow b$, $e \rightarrow \bar{s}$. This gives:
\begin{align}
F^{b' \bar{b}' b}_{b 0 \bar{s}} (F^{\bar{b}' b' \bar{s}}_{\bar{s}})^{-1}_{b0} = (F^{\bar{b}' b' \bar{b}'}_{\bar{b}'})^{-1}_{00}
\label{id22}
\end{align}
Multiplying (\ref{id21}) and (\ref{id22}) gives
\begin{align}
F^{s \bar{s} s}_{s 00} F^{b' \bar{b}' b}_{b 0 \bar{s}} (F^{\bar{b}' b' \bar{s}}_{\bar{s}})^{-1}_{b0} =
F^{b' \bar{s} s}_{b' b 0} (F^{b s \bar{s}}_b)^{-1}_{0b'} (F^{\bar{b}' b' \bar{b}'}_{\bar{b}'})^{-1}_{00}
\label{id23}
\end{align}
Next we take Eq.~(\ref{piv}) and set  $f = l = e = 0$, $b = \bar{a}$, $d = \bar{c}$, $g=c$, $k = \bar{a}$. The result is
\begin{align}
F^{a h \bar{c}}_{0 c  \bar{a}} F^{\bar{a} c \bar{c}}_{\bar{a} h 0} = (F^{a \bar{a} c}_c)^{-1}_{h0}
\end{align}
We then make the change of variables, $a \rightarrow \bar{b}'$, $c \rightarrow \bar{s}$, $h \rightarrow b$. This gives
\begin{align}
F^{\bar{b}' b s}_{0 \bar{s} b'} F^{b' \bar{s} s}_{b' b 0} = (F^{\bar{b}' b' \bar{s}}_{\bar{s}})^{-1}_{b0}
\label{id24}
\end{align}
Substituting (\ref{id24}) into (\ref{id23}), we obtain:
\begin{align}
F^{s \bar{s} s}_{s 00} F^{b' \bar{b}' b}_{b 0 \bar{s}} F^{\bar{b}' b s}_{0 \bar{s} b'} =
(F^{b s \bar{s}}_b)^{-1}_{0b'} (F^{\bar{b}' b' \bar{b}'}_{\bar{b}'})^{-1}_{00}
\end{align}
Again, this is \emph{almost} the desired identity (\ref{id2}): to get there, we simply make the substitution
$(F^{\bar{b}' b' \bar{b}'}_{\bar{b}'})^{-1}_{00} = F^{b' \bar{b}' b'}_{b'00}$ which follows from (\ref{sssid}). This completes our proof of Eq.~(\ref{lrule1}).
The other local rules, (\ref{lrule2},\ref{lrule3}) can be shown in a similar manner, while Eq.~(\ref{lrule0}) follows from (\ref{lrule1}) by setting $a=c=0$. 

We now move on to discuss some of the implications of (\ref{lattrules1}). One implication is that the lattice local rules (\ref{lattrules1}) are \emph{self-consistent} in a disk geometry. Indeed, there is always at least one state $|\Phi\>$ with $Q_I |\Phi\> = B_p |\Phi\> = |\Phi\>$ in such a geometry (see Sec.~\ref{propHsect}), which means there is always at least one solution to the lattice local rules. 

Going a step further, this result suggests that the \emph{continuum} local rules (\ref{localrules}) are self-consistent, since any inconsistency in the continuum rules would presumably also show up on the lattice for a fine enough discretization.\footnote{To make this argument solid, we would need to find a set of lattice moves that are sufficiently general that they can be used to connect any two string-net configurations that can be deformed into each other in the continuum.  We would then have to show that $\Phi$ is invariant under these moves, as in Eq.~(\ref{lattrules10}).}
In fact, we believe that this line of reasoning can be used to \emph{prove} that the conditions (\ref{consistency}) are sufficient to ensure that the continuum local rules (\ref{localrules}) are consistent in a disk geometry: the idea of the argument is to establish three claims: (i) the conditions (\ref{consistency}) are sufficient for constructing commuting projectors $Q_I$, $B_p$; (ii) there is always at least one state $|\Phi\>$ with $Q_I |\Phi\> = B_p |\Phi\> = |\Phi\>$ in a disk geometry; (iii) any state with $Q_I |\Phi\> = B_p |\Phi\> = |\Phi\>$ obeys the lattice local rules. In this paper we have sketched proofs of all three of these claims, but in some of the steps we have used the Hermiticity conditions (\ref{hermicity}) in addition to (\ref{consistency}). That said, we believe that the proofs can be modified so that they do not use the Hermiticity conditions. Assuming this is correct, the above argument can be used to prove that the self-consistency conditions (\ref{consistency}) are sufficient.

\section{General string-net models}  \label{App:FusionMults}
 In this appendix, we discuss how to extend our construction to the most general class of string-net models, in which the string types have fusion multiplicities.

The main new element in these general models is that the Hilbert space associated with the vertex $(a,b:c)$ is not one-dimensional, as we assumed in the main body, but rather has dimension $N^{ab}_c$, where $N^{ab}_c$ is a non-negative integer. To describe this Hilbert space, we add an index $\sigma$ at each vertex of the string-net. At a vertex $(a,b; c)$, $\sigma$ ranges over the set $\sigma = 1, ... N^{ab}_c$. 

The non-negative integers $N^{ab}_c$ can be thought of as a generalization of the branching rules $\delta^{ab}_c$, and like $\delta^{ab}_c$, we require that $N^{ab}_c$ obeys the associativity condition:
\begin{align}
\sum_e N^{ab}_e N^{ec}_d = \sum_f N^{bc}_f N^{af}_d
\label{fusionassoc2}
\end{align}
We also require that the null-string obeys the same kind of branching rules as in the main text: $N^{a0}_a = N^{0a}_a = N^{a \bar{a}}_0 = 1$.

The local rules for general string-net models are similar to the local rules (\ref{localrules}) in the main body except for extra indices at each vertex:
\begin{subequations}
\begin{align}
	\Phi\left(\raisebox{-0.22in}{\includegraphics[height=0.5in]{rule0a.pdf}} \right)&=\Phi \left(\raisebox{-0.22in}{\includegraphics[height=0.5in]{rule0b.pdf}} \right) \label{1g1}\\
	\Phi\left(\raisebox{-0.22in}{\includegraphics[height=0.5in]{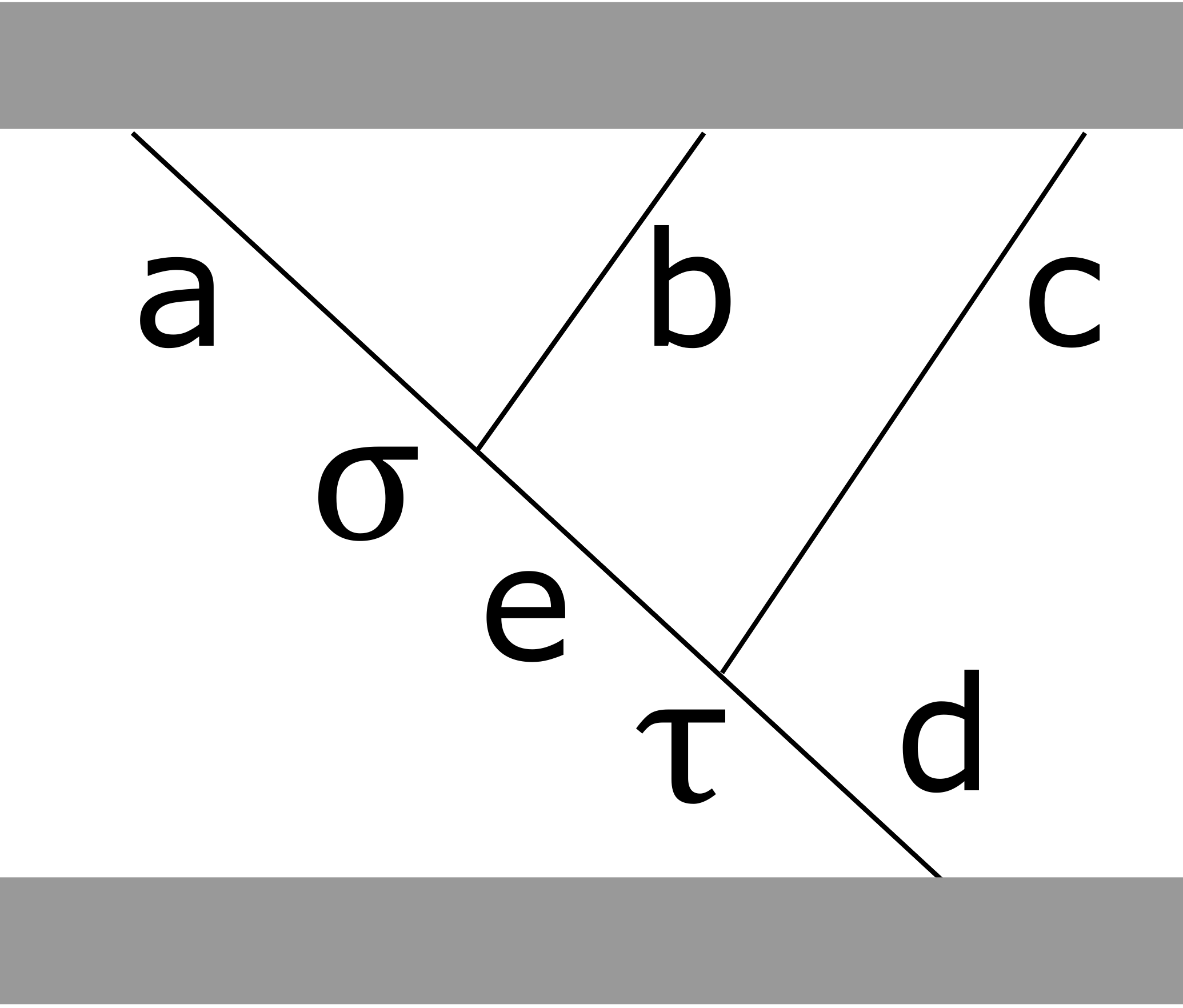}} \right)&=\sum_{f\mu\nu} F^{abc,\sigma\tau}_{def,\mu\nu}\Phi \left(\raisebox{-0.22in}{\includegraphics[height=0.5in]{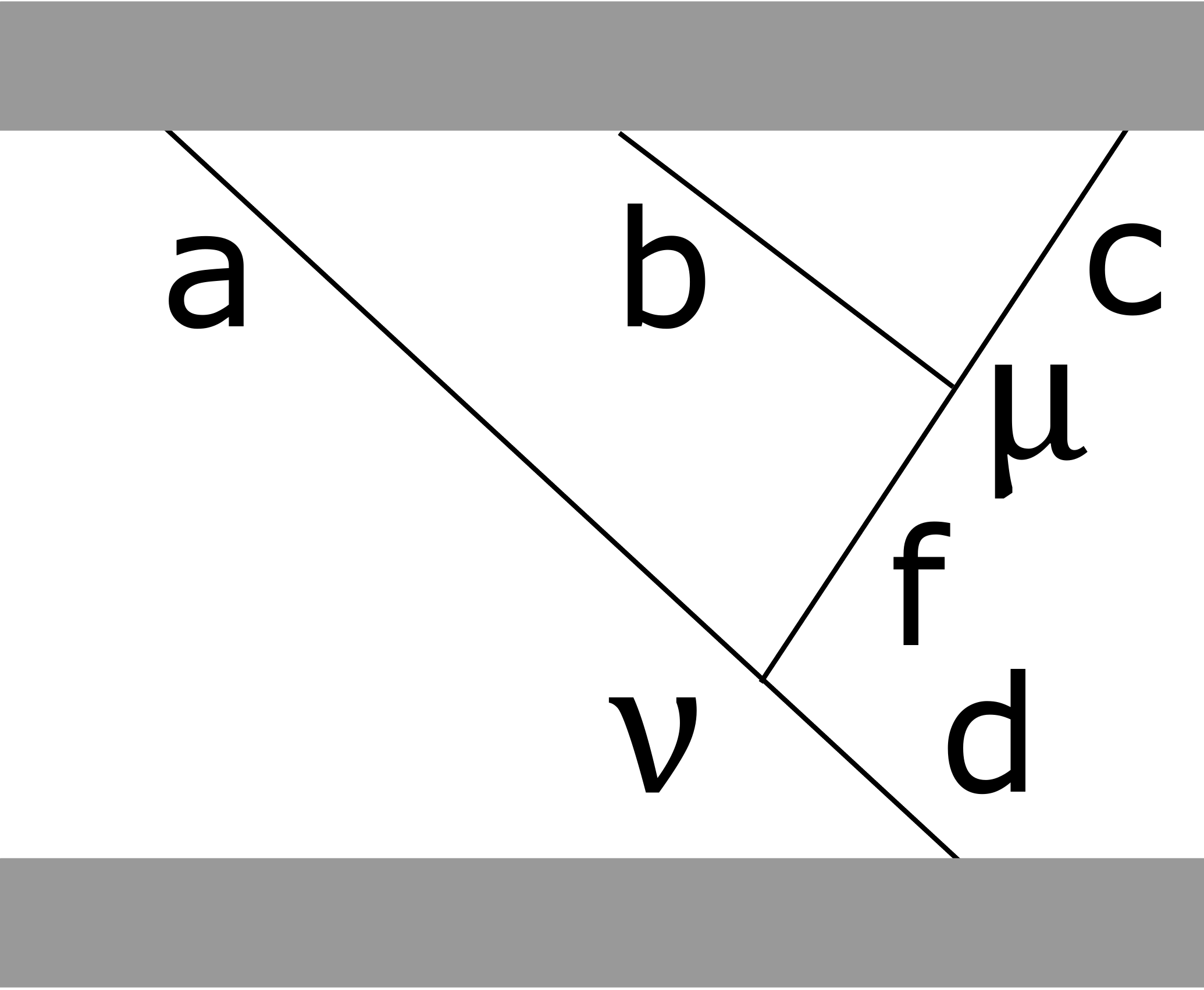}} \right) \label{1a1}\\
	\Phi\left(\raisebox{-0.22in}{\includegraphics[height=0.5in]{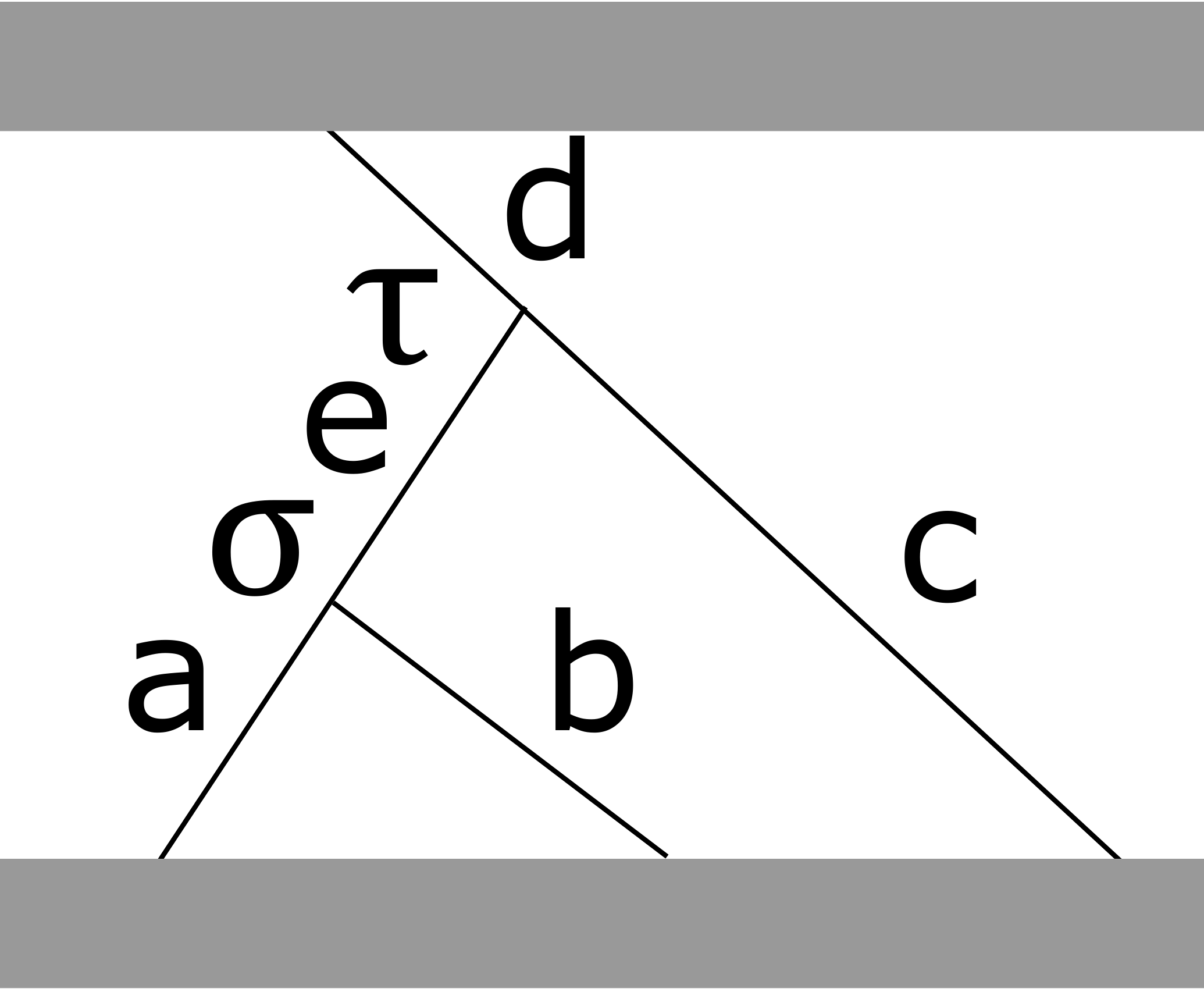}} \right)&=\sum_{f\mu\nu} \tilde{F}^{abc,\sigma\tau}_{def,\mu\nu}\Phi \left(\raisebox{-0.22in}{\includegraphics[height=0.5in]{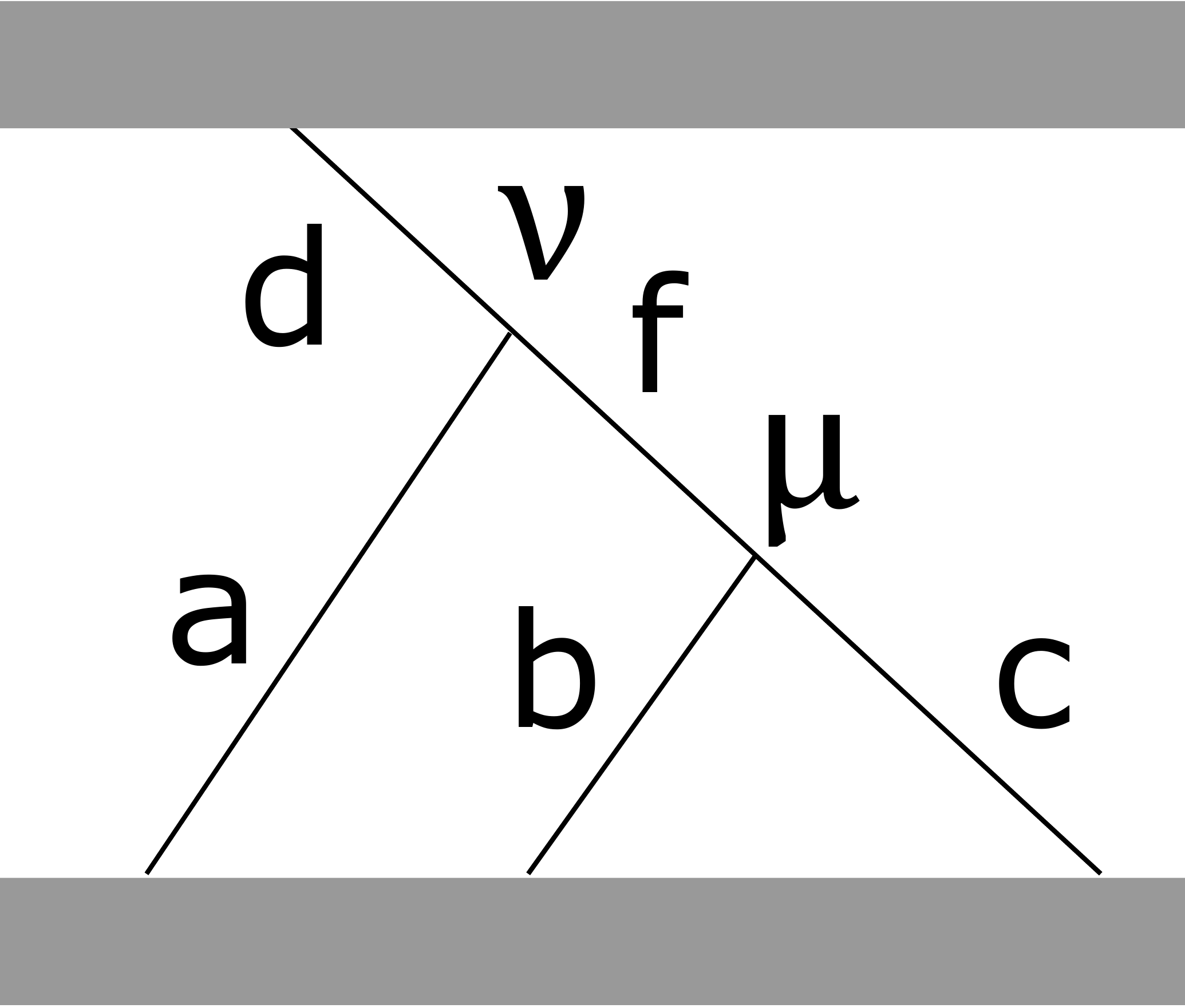}} \right) \label{1d1} \\	
	\Phi\left(\raisebox{-0.22in}{\includegraphics[height=0.5in]{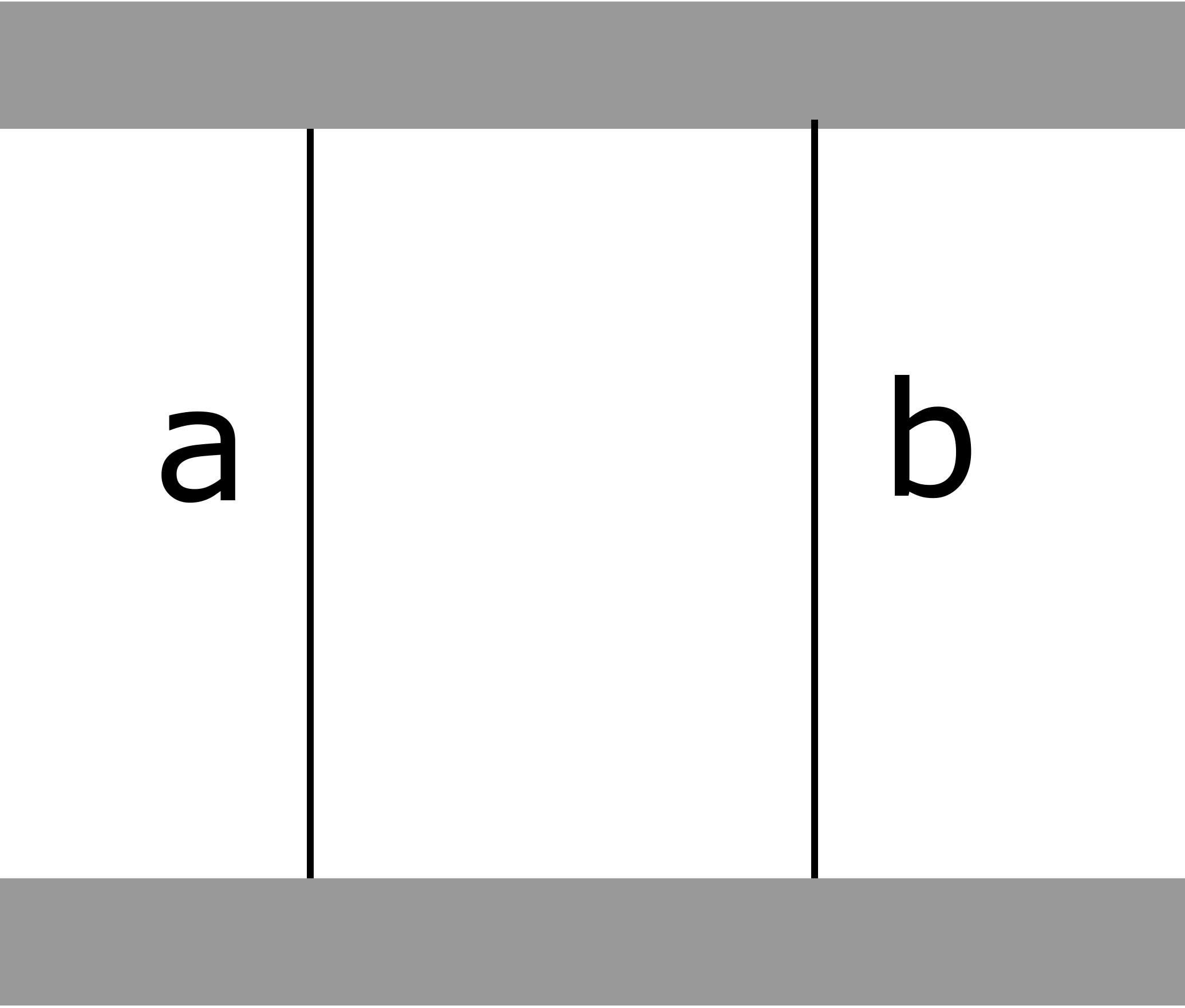}} \right)&=\sum_{c\sigma}  \frac{1}{Y^{ab}_c}
	\Phi \left(\raisebox{-0.22in}{\includegraphics[height=0.5in]{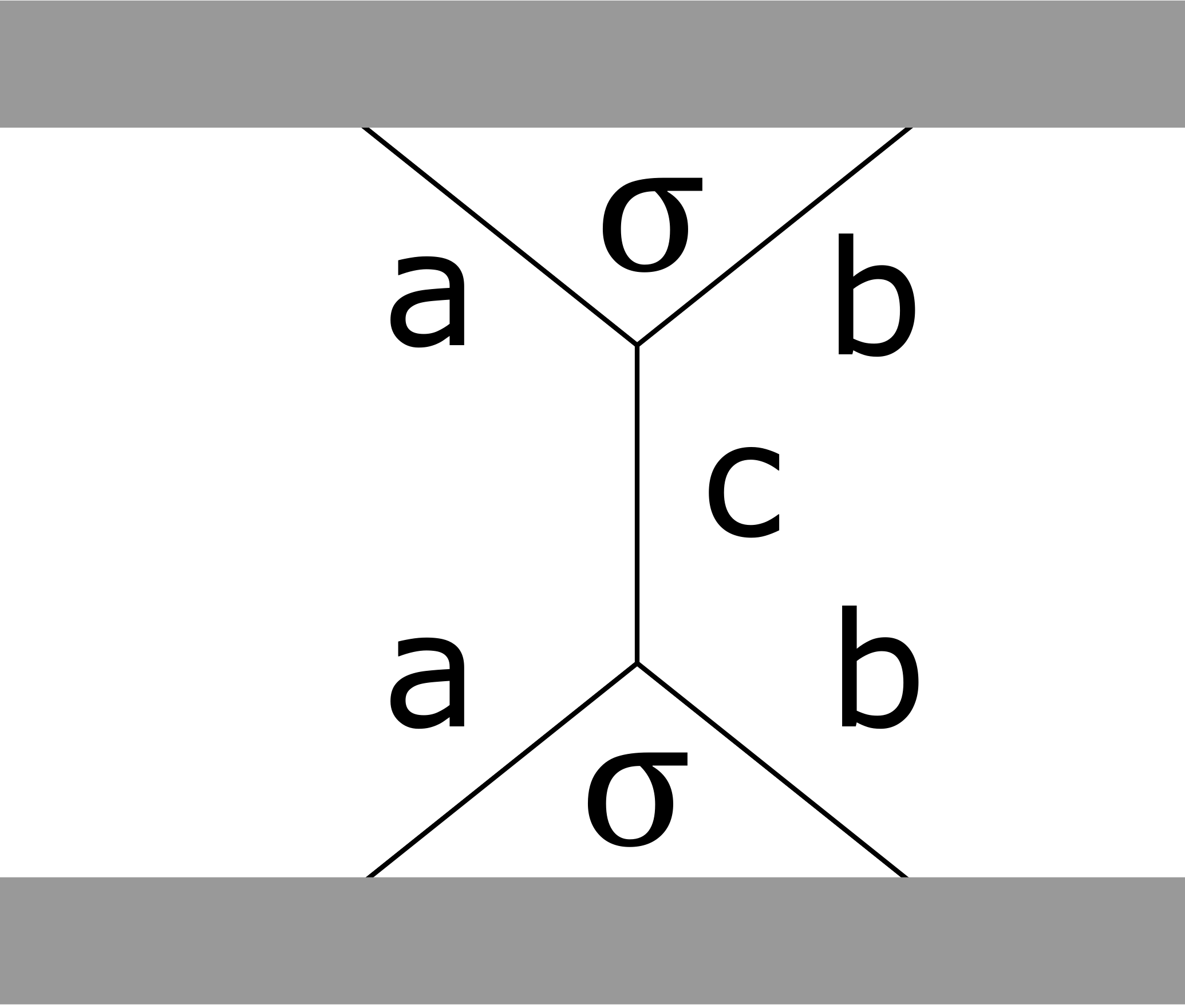}} \right) \label{1b1} \\
	\Phi\left(\raisebox{-0.22in}{\includegraphics[height=0.5in]{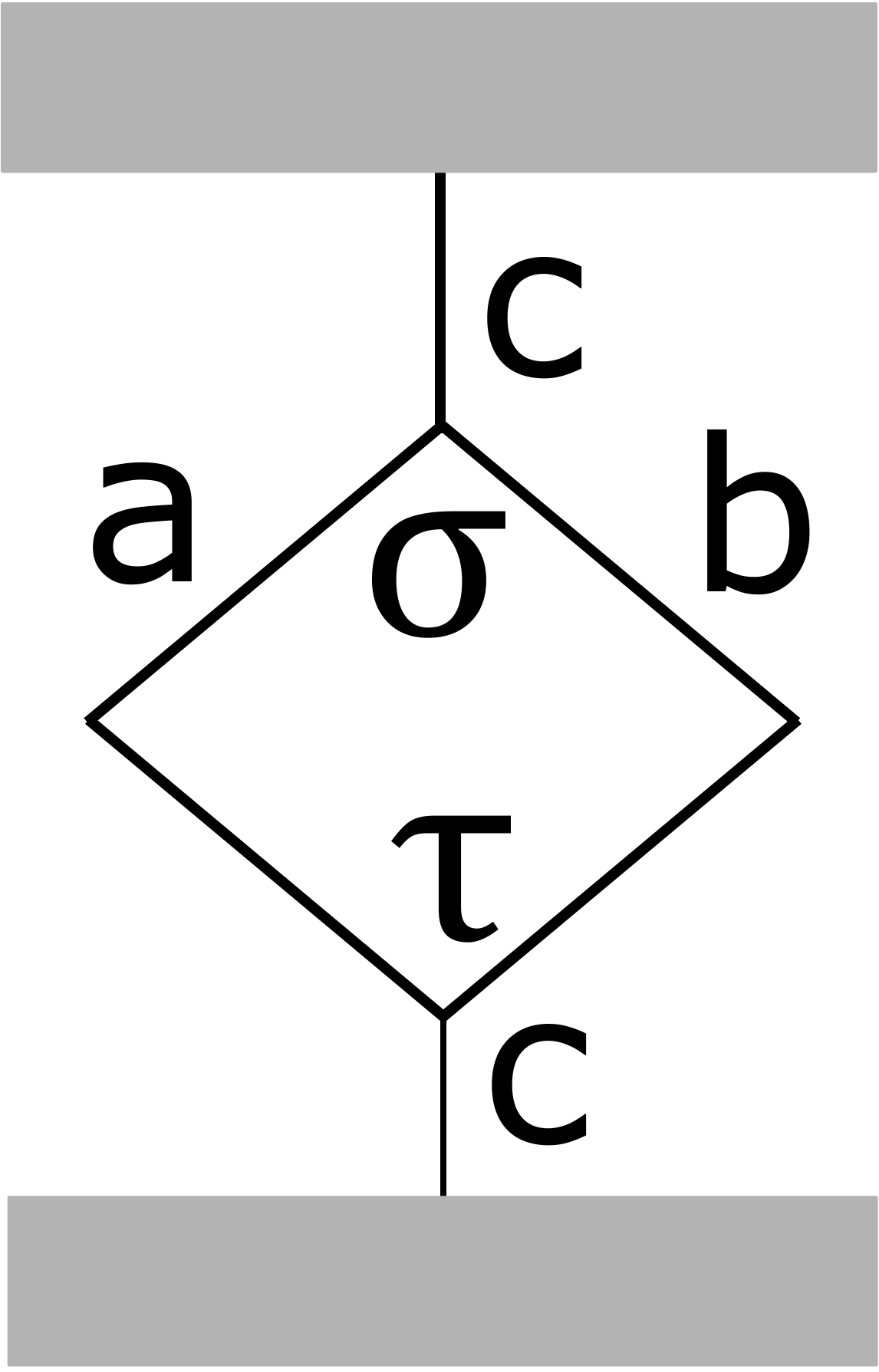}} \right)&=\delta_{c,d} \delta_{\sigma,\tau}Y^{ab}_c
	\Phi\left(\raisebox{-0.22in}{\includegraphics[height=0.5in]{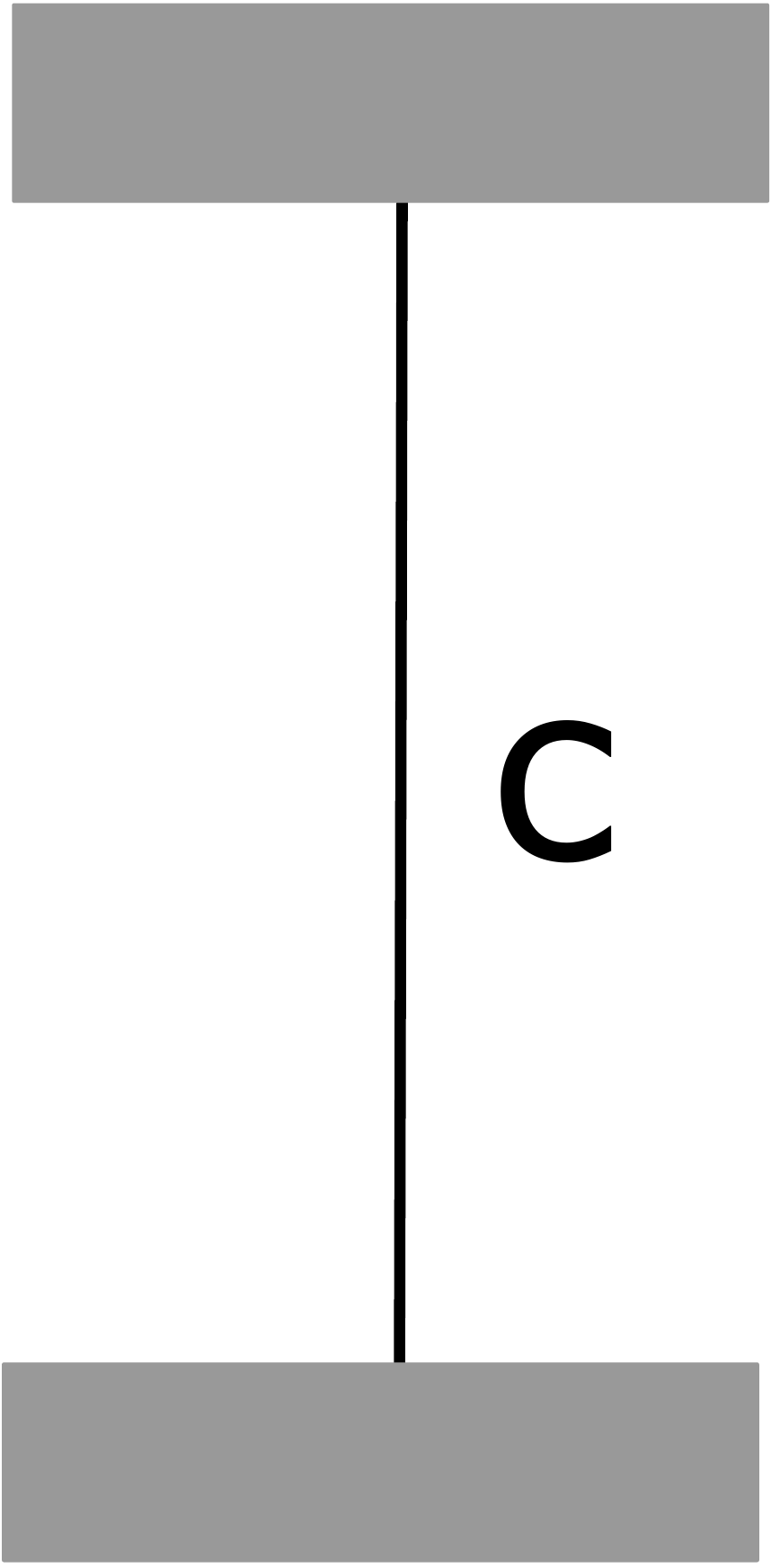}} \right). \label{1c1}
\end{align}
\label{localrules2}
\end{subequations}
For fixed string types $(a,b,c,d,e,f)$, the F-symbol becomes a complex tensor $F^{abc,\sigma\tau}_{def,\mu\nu}$ of dimension $N^{ab}_e\times N^{ec}_d \times N^{bc}_f \times N^{af}_d$.

The self-consistency conditions (\ref{consistency}), the Hermiticity conditions (\ref{hermicity}), and the Hamiltonian (\ref{hsn0}) can also be generalized straightforwardly; we will not write down the explicit formulas here as they are not particularly illuminating.

\bibliography{stringnet2}

\end{document}